\documentclass[a4paper,12pt]{article}

\usepackage{psfrag}
\usepackage{mathrsfs} 

\usepackage{todonotes}
\usepackage{graphicx}
\usepackage{subfigure}
\usepackage[margin=10pt,font=small,labelfont=bf,
labelsep=period]{caption}
\usepackage{amsmath}
\usepackage{amssymb}
\usepackage{dsfont} 
\usepackage[bottom]{footmisc} 
\usepackage{indentfirst} 
\usepackage{upref} 
\usepackage{cite} 
\usepackage{bbm}
\usepackage{rotating}

\numberwithin{equation}{section}

\newcommand{\ddx}{\!\textrm{d}^4 x\ }
\newcommand{\Tr}{\text{Tr}\:}

\newcommand{\eu}{\mbox{e}}

\newcommand{\Gr}{{\rm G}}
\newcommand{\Br}{{\rm B}}
\newcommand{\Dr}{{\rm D}}
\newcommand{\Lr}{{\rm L}}
\newcommand{\dr}{{\rm d}}

\newcommand{\openone}{\mathbbm 1}
\newcommand{\Ss}{\scriptstyle}
\newcommand{\Sss}{\scriptscriptstyle}

\def\Xint#1{\mathchoice
   {\XXint\displaystyle\textstyle{#1}}%
   {\XXint\textstyle\scriptstyle{#1}}%
   {\XXint\scriptstyle\scriptscriptstyle{#1}}%
   {\XXint\scriptscriptstyle\scriptscriptstyle{#1}}%
   \!\int}
\def\XXint#1#2#3{{\setbox0=\hbox{$#1{#2#3}{\int}$}
     \vcenter{\hbox{$#2#3$}}\kern-.5\wd0}}
\def\ddashint{\Xint=}

\renewcommand{\baselinestretch}{1.5}
\begin{document}
\begin{titlepage}
\renewcommand{\baselinestretch}{1.1}
\title{\begin{flushright}
\normalsize{MITP/13-007}
\bigskip
\vspace{1cm}
\end{flushright}
Einstein--Cartan gravity, Asymptotic Safety,\\ and the running Immirzi parameter}
\date{}
\author{J.-E. Daum and M. Reuter\\
{\small Institute of Physics, University of Mainz}\\[-0.2cm]
{\small Staudingerweg 7, D-55099 Mainz, Germany}}
\maketitle\thispagestyle{empty}
\begin{abstract} 
In this paper we analyze the functional renormalization group flow of quantum gravity on the Einstein-Cartan theory space. The latter consists of all action functionals depending on the spin connection and the vielbein field (co-frame) which are invariant under both spacetime diffeomorphisms and local frame rotations. In the first part of the paper we develop a general methodology and corresponding calculational tools which can be used to analyze the flow equation for the pertinent effective average action for any truncation of this theory space. In the second part we apply it to a specific three-dimensional truncated theory space which is parametrized by Newton's constant, the cosmological constant, and the Immirzi parameter. A comprehensive analysis of their scale dependences is performed, and the possibility of defining an asymptotically safe theory on this hitherto unexplored theory space is investigated. In principle Asymptotic Safety of metric gravity (at least at the level of the effective average action) is neither necessary nor sufficient for Asymptotic Safety on the Einstein-Cartan theory space which might accommodate different ``universality classes'' of microscopic quantum gravity theories. Nevertheless, we do find evidence for the existence of at least one non-Gaussian renormalization group fixed point which seems suitable for the Asymptotic Safety construction in a setting where the spin connection and the vielbein are the fundamental field variables.
\end{abstract}
\end{titlepage}
\newpage 
\section{Introduction}
Coarse graining flows are a powerful tool for the exploration of
complex interacting systems in both quantum field theory and
statistical physics. In the case of Einstein gravity their
implementation has led to the construction of the gravitational
average action \cite{mr}. It can be employed for the quantization of
fundamental theories; in this case the coarse graining flow
provides the information of how to take the continuum limit of 
the
pertinent functional integral. It may also be seen as a
tool for evolving effective theories from one scale to another
without invoking a continuum limit. In either case, the
perturbative nonrenormalizability of quantized General Relativity
poses no conceptual or computational problems of principle.

So far the functional renormalization group flows of the
gravitational average action have mostly been used within the
Asymptotic Safety program \cite{livrev}. Its key idea is S.~Weinberg's 
proposal
that quantum gravity might be nonperturbatively renormalizable if
its renormalization group (RG) flow possesses a nontrivial
fixed point at which the infinite
cutoff limit can be taken \cite{wein}. Then, the degree of predictivity is the higher 
the smaller is the dimensionality of the fixed point's
ultraviolet critical manifold (the set of points mapped onto the
fixed point under the inverse flow). In fact, every complete RG
trajectory running entirely within this manifold defines one specific
quantum theory. In the ultraviolet (UV), it hits the fixed point
and, as a consequence, has a comparatively simple and easy to control, well behaved short distance
behavior. In the framework of the average action \cite{avact} the
fixed point is closely related to the bare action, 
while its infrared (IR) limit equals the ordinary effective action
$\Gamma$.

During the past 15 years the nonperturbative RG flow of the
gravitational average action has been investigated within
approximations (``truncations'') of increasing complexity \cite{mr, percadou, oliver1, frank1, oliver2, oliver3, oliver4, souma, frank2, prop, perper1, codello, litimgrav,r6,MS1,Codello:2008vh, creh1,creh2,creh3,HD1,JEUM,Benedetti:2010nr,frank+friends,Saueressig:2011vn, Eichhorn:2009ah,Groh:2010ta,Eichhorn:2010tb,Codello:2010mj,elisa1,elisa2,MRS,e-only,oliverfrac,frankfrac,maxpert}. By now
there is a significant body of evidence suggesting that Quantum
Einstein Gravity (QEG) does indeed have a non-Gaussian fixed 
point
(NGFP), suitable for the Asymptotic Safety construction, and with
a low dimensional ultraviolet critical manifold \cite{NJP,reviews}.

While contrary to other approaches to quantum gravity such as Loop Quantum
Gravity (LQG) or string theory for instance, asymptotically safe gravity
does not leave the realm of quantum field theory, it is, in at least one
respect, fundamentally different from basically all quantum
systems we are familiar with: In fact, in the conventional way of thinking
about quantum theory the process of ``quantization'' plays a 
central
role. Usually we start out from a given classical dynamical system, in
particular a Hamiltonian encoding its dynamics, then apply to it a set
of heuristic ``quantization rules'' in order to find a 
quantum system reproducing it as its classical limit, and finally work out the
properties of the quantum theory away from the classical limit.
Clearly the predictive power of this procedure is limited by the
fact that right from the start we need to have a ``prejudice'' or
``educated guess'', perhaps inspired by experiment, about the
Hamiltonian operator. Moreover, it is well known that in general
the way from the classical to the quantum system is far from
unique.

In a sense, the Asymptotic Safety program, at least when
formulated in the framework of the gravitational average action, can be
seen as the attempt of inverting this procedure: The quantum
theory one is after is not found by quantizing an a priori given
bare action or Hamiltonian, but rather by a {\it selection 
process}
directly at the quantum level, the key requirement being that of
nonperturbative renormalizability at a suitable NGFP. Somewhat
idealistically, the necessary steps can be described as follows.

Starting from the functional RG equation (FRGE) which governs the
scale dependence of the running action we are given a vector 
field on the space of all action functionals whose components have the
interpretation of beta functions for the infinitely many couplings
that parametrize a generic action functional. The first step
consists in finding the zeros of this vector field, that is, the
fixed points of the RG flow. A priori there could be one, or 
many, or none.

In the favorable case there exists at least one NGFP which we 
then can declare to be the UV limit of all admissible RG trajectories. The
totality of all those trajectories sweeps out the fixed point's 
UV critical manifold, henceforth denoted $\mathscr{S}_{\rm UV}$. If dim
$\mathscr{S}_{\rm UV}\equiv s < \infty$ we then pick one specific trajectory in
$\mathscr{S}_{\rm UV}$ by fixing the $s$ parameters that amount to local
coordinates on $\mathscr{S}_{\rm UV}$. Every such trajectory of running 
actions,
usually denoted  $\Gamma_{k}$, where $k$ is the coarse graining
(mass) scale, defines a quantum field theory which is
``asymptotically safe'' in the UV, i.\,e.
$\lim_{k\longrightarrow\infty}$ $\Gamma_{k}\equiv \Gamma_{\ast}$
equals the fixed point action.

Thus, up to this point, the result of the computations is a
trajectory $\{\Gamma_{k}, 0\leq k<\infty \}$ which emanates from
the NGFP action $\Gamma_{\ast}$ in the UV and connects
it to the ordinary effective action $\Gamma=\Gamma_{k=0}$ in which the IR cutoff $k$ is
removed. Knowing $\Gamma$ we know basically everything about the
{\it quantum} theory defined by this trajectory; in particular, 
all $n$-point functions obtain as multiple functional derivatives of
$\Gamma$ simply. In principle one could stop at this stage since 
all possible ``output'' we may expect from the quantum theory is given
in terms of $\Gamma$ .

However, by this FRGE-based construction the resulting quantum
field theory is not presented to us as the ``quantization'' of any
obvious, let alone unique classical system. 

Nevertheless one may ask whether there
exists a regularized functional integral representation of the trajectory
$\{\Gamma_{k}, 0\leq k<\infty\}$. In \cite{frank2} this question was 
answered in the affirmative and an explicit construction was presented. 
But we emphasize again that in the Asymptotic
Safety program based on the average action this step is actually
redundant and only a matter of convenience; it does not lead to
further predictions. In particular if one also allows for changes of
the field variables, the way from the scale dependent effective
action $\Gamma_k$ to the functional integral is highly non-unique. As an
additional ingredient one has to specify a regularized measure on
field space. Then, by the general method described in \cite{elisa1},
one can deduce a bare action, to be used under the functional
integral which reproduces $\Gamma_k$. The bare action $S\equiv S_{\Lambda}$
depends on a UV cutoff, and under appropriate conditions the
$k$-dependence of $\Gamma_k$ dictates how $S$ has to be tuned in the limit
$\Lambda\rightarrow\infty$ . (See \cite{elisa1} for further details.) 

Given
the ``classical'' action under the (Lagrangian) functional integral
one can try to rewrite it in Hamiltonian form and to read off the
corresponding Hamiltonian and symplectic structure. Since $S$, when 
written in terms of the original field variables,
is likely to contain higher derivatives and nonlocalities this
last step might involve introducing auxiliary fields in order to
display the underlying canonical system in a transparent (local) way. It
is this resulting system that, implicitly, was ``quantized'' when we 
picked one of the eligible special RG trajectories.

We can summarize this discussion by saying that {\it once a
concrete functional RG flow equation is specified} there exists 
in principle a ``canonical'' procedure one can try to follow in order
to search for an asymptotically safe field theory. Its only
ambiguities are related to the dimensionality of $\mathscr{S}_{\rm UV}$ and the
possibility that there might exist several suitable fixed points. In this
sense the fundamental dynamics, encoded in a certain action or
Hamiltonian, is a {\it prediction} in Asymptotic Safety, not an
input as in standard quantum mechanics.

So, if the Hamiltonian is an output, what is actually the {\it
input} into the above chain of steps which decides about whether
we end up with a quantum theory of gravity rather than an
asymptotically safe matter field theory, say? 

The answer to this question lies entirely in
the specification of a concrete functional RG equation, or more
precisely, the nature of the so-called ``theory space'' of action
functionals on which this FRGE is defined. A concrete theory space
${\cal T}\equiv\{A[\Psi], {\bf G}\}$ is fixed by selecting a set of fields
$\Psi$ on which the actions $A$ depend, a group {\bf G} of (gauge) symmetry
transformations under which all $A[\Psi]$ are required to be invariant,
and possibly certain regularity properties they must have. Given ${\cal T}$, 
and leaving technical issues aside, it is then possible to straightforwardly 
construct a coarse graining flow for the given theory space and to set up 
the corresponding FRGE. It is important to note that, besides the details 
of the coarse graining scheme, {\it the only nontrivial input which determines 
the structure of the FRGE and its flow is the underlying theory space.}

Combined with the above remarks the last statement implies that
the only choice we have in our search for nonperturbatively
renormalizable theories is that of theory space. Once ${\cal T}$ is
fixed, everything else, in particular the number and properties 
of fixed points of the resulting RG flow, follows in principle
straightforwardly. In a slight abuse of language\footnote{Strictly 
speaking only the actions within the basin of attraction of a given
fixed point form a universality class in the sense of critical 
phenomena.} we shall henceforth refer to the actions of a 
given theory space as forming a ``universality class''.

As to yet, all searches for asymptotically safe theories of (pure)
quantum gravity adopted the same choice of theory space: The 
field variable $\Psi$  was taken to be the spacetime metric
$g_{\mu\nu}$, and the gauge symmetry requirement imposed on the
action functionals was that of diffeomorphism invariance. This
setting is usually referred to as Quantum Einstein Gravity or
``QEG''. This name is supposed to indicate that, as in classical
general relativity, the field variable is $g_{\mu\nu}$. The
pertinent action may be different though.

In the present paper we report on a first exploration of another
``universality class'' which contains possibly inequivalent quantum
gravity theories in 4 dimensions. Rather than the metric, we take
the vielbein (or co-frame) field $e^{a}{}_{\mu}$ and the spin
connection $\omega^{ab}{}_{\mu}$ as the fundamental fields, and we 
enlarge
the group {\bf G} of gauge transformations to contain also local 
Lorentz
transformations (frame rotations) besides the diffeomorphisms.
Considering the Euclidean form of the theory this will lead to the
semidirect product structure ${\bf G}={\sf Diff}({\cal M})\ltimes {\sf O}(4)_{\rm loc}$, where
${\sf Diff}({\cal M})$ stands for the diffeomorphisms of the spacetime 
manifold, henceforth denoted ${\cal M}$. While, in 4 dimensions, the choice
$\Psi=g_{\mu\nu}$ gives rise to 10 field variables, their number
increases to 40 for the pair
$\Psi=(e^{a}{}_{\mu},\omega^{ab}{}_{\mu})$. Thanks to its enlarged
field content, field configurations in the new universality class
can carry spacetime torsion, for instance, while this was not
possible in metric gravity \cite{hehl,peldan}. 

We shall refer to all theories defined via ${\sf Diff}({\cal M}) 
\ltimes {\sf O}(4)_{\rm loc}$-invariant functionals of
$e^{a}{}_{\mu}$ and $\omega^{ab}{}_{\mu}$ as a Quantum Einstein-Cartan
Gravity or ``QECG''. As in the case of QEG, this name is just
meant to specify the field content and the gauge group, not the
dynamics.

As to yet, nothing is known about the nonperturbative RG flow on
the theory space ${\cal T}\equiv{\cal T}_{\rm EC}$ of Einstein-Cartan
theory. In particular the (non-)existence of fixed points 
suitable
for defining a fundamental theory is an open question. We
emphasize that the NGFP which is likely to exist on the theory
space of metric gravity (${\cal T}_{\rm E}$) has no direct implications
for the Einstein-Cartan setting. A priori there is no general
principle that would forbid the quantum properties of metric and
$(e,\omega)$-gravity to be quite different.

At the purely classical level, Einstein-Cartan gravity, equipped
with the Hilbert-Palatini action $S_{\rm HP}[e,\omega]$ to define the
dynamics, is a well established alternative to General Relativity
\cite{peldan}. In absence of spinning matter its equations of motion imply
Einstein's equation for the composite metric field
$g_{\mu\nu}=e^{a}{}_{\mu}e^{b}{}_{\nu}\eta_{ab}$, along with the
statement that ``on shell'' torsion is always zero. It can be made
non-zero though by coupling spinors to gravity.

Even in the vacuum sector there is an interesting difference
between the two classical theories, namely with respect to the
possibility of consistently incorporating degenerate geometries.
While the entire framework of Riemannian geometry underlying
standard General Relativity breaks down for degenerate metrics 
which
are not invertible, the field configuration $(e^{a}{}_{\mu}=0,
\omega^{ab}{}_{\mu}=0)$ is a well defined solution to the field
equations obtained by varying $S_{\rm HP}[e,\omega]$. This difference
might be important for understanding a
conjectural phase of quantum gravity in which the vielbein has a
vanishing vacuum expectation value, a ``phase of unbroken
diffeomorphism invariance'' \cite{Giddings, Witten}.

In the literature many generalizations of classical
Einstein-Cartan theory with actions more complicated than
$S_{\rm HP}[e,\omega]$ have been considered 
\cite{Robertobook,peldan}. In particular in the context of Loop Quantum Gravity
(LQG) the so-called Holst action $S_{\rm Ho}[e,\omega]$ plays an
important role \cite{soeren, barbero}. It contains an additional term that exists only 
in
4 dimensions; its prefactor is the dimensionless Immirzi 
parameter
$\gamma$. This term is typical of Einstein-Cartan theory; it
vanishes for vanishing torsion and, as a result, does not exist 
in
metric gravity. Remarkably, the vacuum field equations implied by
$S_{\rm Ho}[e,\omega]$ do not depend on  $\gamma$, even though the
part of the action it multiplies is not a surface term. Indeed, 
in
presence of fermions coupled to gravity in a non-minimal way, the
Immirzi term induces a CP violating four-fermion interaction that
might be interesting for phenomenological reasons, in the
cosmology of the early universe, for instance \cite{nieh-yan, freidel2, shapiro1}.

The Holst action is of central importance for several modern
approaches to the quantization of gravity \cite{kiefer}. This includes 
canonical
quantum gravity on the basis of Ashtekar's variables \cite{A}, Loop
Quantum Gravity \cite{T}, spin foam models \cite{Perez}, and group 
field theory \cite{oriti}. In LQG, for instance, $\gamma$ makes 
its appearance in the spectrum of area and volume operators. 
It was also believed to determine 
the entropy of black holes since the standard semiclassical result
($S=A/4G$) obtained for a single value of $\gamma$ only. This picture was questioned recently, however \cite{bianchi}. At least
the kinematical level of LQG suggests that  $\gamma$ constitutes 
a fixed parameter which labels physically distinct quantum 
theories. In this respect  $\gamma$ might be comparable to the
$\Theta$-parameter of QCD which, too, is absent from the 
classical equations of motion, but nevertheless leads to observable 
quantum effects. Contrary to the Immirzi parameter, $\Theta$ does 
however multiply a topological invariant which spoils the analogy to some 
extent.

There is an obvious tension between this picture of a universal,
constant value of $\gamma$, fixing for instance the absolute 
size
of quantized areas of volumes, on the one hand, and the framework
of RG flow equations and Asymptotic Safety on the other. Setting
up a FRGE for the theory space  ${\cal T}_{\rm EC}$, one of the
infinitely many couplings parametrizing a generic action is the
Immirzi parameter. A priori it must be treated as a ``running'',
i.\,e. scale-dependent quantity $\gamma\equiv\gamma_{k}$; there is
no obvious general principle (nonrenormalization theorem) that
would forbid such a scale dependence.\footnote{See however 
\cite{percaccisezgin} for an example of a parameter non-renormalization 
in a similar theory, topologically massive 3D gravity.} For this reason 
the renormalization behavior of the Immirzi parameter will be one 
of the main themes in the following.

\vspace{4mm}
The purpose of the present paper is twofold: First, we are going 
to construct a general framework which allows 
the nonperturbative calculation of coarse graining flows in 
Einstein-Cartan gravity; it will be based on a suitable variant 
of the gravitational average action. Hereby various new 
difficulties not present in the metric case must be overcome. In 
particular a careful analysis of the background and quantum gauge 
algebra and their respective implementations is necessary to make 
sure that the effective average action is background gauge 
invariant on all scales. Furthermore, we shall describe a general 
technique for the computation of the functional traces which make 
their appearance in the corresponding functional RG equation. 
These general developments are not related to any special 
truncation. They provide the tools necessary for any future FRGE 
computation on ${\cal T}_{EC}$ or a subspace thereof.

The second purpose of this paper is to test the general framework 
we have developed by applying it to a first explicit example of a 
truncated RG flow for the average action of $(e,\omega)$-gravity.

Concretely, we are going to project the flow on the 3-dimensional 
subspace of ${\cal T}_{EC}$ which is spanned by the field monomials that
appear in the classical Holst action already. Treating their 
coefficients as running couplings we thus obtain approximations 
for the beta functions of the running Newton constant $G_{k}$, 
cosmological constant $\lambda_{k}$, and, most interestingly, the 
Immirzi parameter $\gamma_{k}$.

To avoid any misunderstanding we emphasize that, even within this 
``Holst truncation'', the known RG flows of (truncated) metric 
gravity have no obvious implications for the flow in the 
$(e,\omega)$-case. While it is true that, when the vielbein is 
assumed to be invertible, the pure gravity theory, based upon the 
Holst action, is {\it on-shell} equivalent to metric gravity with 
the Einstein-Hilbert action, the beta functions of the average 
action are {\it off-shell} quantities, and the classical 
equivalence is not directly relevant to them. In fact, solving 
the FRGE is tantamount to performing a certain functional 
integral, and almost all $(e,\omega)$-configurations contributing to it 
are off-shell and carry non-zero torsion in particular. Thus, 
even leaving the running $\gamma_{k}$ aside, as to yet nothing is 
known about the Wilsonian RG behavior of the Holst action. In 
this sense the results of the present paper are new and 
independent of the properties found in the metric theory, QEG.

In analyzing the Holst flow we shall focus on two central issues, 
namely on whether $\gamma_{k}$ does have a nontrivial RG running 
at all, and on the question of fixed points that would allow for 
an asymptotically safe UV limit. In particular we shall be interested
in how the presence of ``off shell'' torsion affects the renormalization 
properties.

The rest of this paper is organized as follows. In Section 2 we 
prepare the stage by reviewing those aspects of classical first-order 
gravity that will be needed later on. Then, in Section 3, we 
discuss its formal quantization by means of a functional integral, 
construct the related effective average action pertaining to the 
Einstein-Cartan theory space, and set up both the exact FRGE and 
its proper-time approximation. Section 4 deals with the ghost sector 
arising from the semidirect product of diffeomorphisms and local 
${\sf O}(4)$ transformations which requires some care if one 
wants the average action to be a background gauge invariant 
functional of its arguments. Beginning with Section 5, we switch 
from the exact setting to the example of a truncated flow 
which we work out explicitly, namely the ``running'' 3-parameter 
Holst action.

Section 5 introduces the truncation ansatz for $\Gamma_k$ and 
gives a brief outline of our computational strategy for finding 
its RG flow. From there, the reader who is mainly interested in 
the results can jump directly to Section 8 where the main results 
concerning the RG flow of the Holst action are displayed and discussed.

Readers who are also interested in the more technical aspects of 
our work will find in Sections 6 and 7 a detailed account of the 
tools we developed in order to deal with flow equations on the 
Einstein-Cartan theory space. The ``tool kit'' presented in these 
two sections is useful in its own right and can be used also in 
future explorations of more general truncations. Section 9 finally 
contains a summary of our results.

Short accounts of the present work appeared in refs. \cite{eomega-plb} 
and \cite{eomega-corfu} already.

\section{The Classical Theory}
Throughout the whole paper, we will deal with gravity in the 
Euclidean formulation of the theory, basically because the 
numerical data to which we would like to ultimately compare 
our results is obtained by Euclidean methods as well (Monte 
Carlo simulations, etc.). This allows us, at least in principle, 
to search for similarities of the different approaches. For 
the same reason also metric gravity has been treated almost 
exclusively this way up to now. As we are primarily interested 
in comparing the RG behavior of Einstein-Cartan gravity with 
its counterpart derived from the metric theory, this is a second 
reason to employ the Euclidean formulation. Furthermore, recent 
investigations \cite{frank+friends} indicate that the results 
obtained for metric gravity in the Euclidean formulation might 
indeed carry over to the Lorentzian signature almost unaltered. 
For this reason we do not expect the choice of spacetime signature 
to be essential for the UV behavior of the quantum theory of 
gravity under consideration.

\subsection{Field content}
The first basic field we want to employ is the {\it vielbein} 
$\hat{e}^a_{\ \mu}(x)$ that provides a local isomorphism between 
the local tangent space $T_x{\cal M}$ of the four-dimensional 
spacetime manifold ${\cal M}$ and a local Minkowski frame 
$\mathds{M}^4$. (In the Euclidean formulation the local Minkowski 
frames are given by copies of $\mathds{R}^4$, of course; 
nevertheless throughout the whole paper we will use the 
standard Lorentzian terminology.) Explicitly, this isomorphism 
is given by
\begin{equation}
 T_x{\cal M} \ni v^\mu(x)\mapsto \hat{e}^a_{\ \mu}(x)v^\mu (x)\equiv v^a(x) \in \mathds{M}^4.
\end{equation}
Therein, Greek letters $\mu,\nu, \cdots$ denote spacetime indices, 
whereas Latin letters $a,b,\cdots$ denote (Lorentz) frame indices; 
both types of indices run from 1 to 4. Contravariant spacetime 
vectors can be transformed to covariant ones by means of the 
spacetime metric $\hat{g}_{\mu\nu}$, while for Lorentz vectors 
the same is achieved using the metric $\eta_{ab}=
\text{diag}(1,1,1,1)$. Demanding the vielbein $\hat{e}^a_{\ \mu}(x)$ 
to be norm-preserving $v^\mu v^\nu \hat{g}_{\mu\nu} \equiv 
v^2=v^a v^b \eta_{ab}$, i.\,e. an isometry, we can express 
the metric $\hat{g}_{\mu\nu}$ in terms of the vielbein 
$\hat{e}^a_{\ \mu}$ according to 
\begin{equation}\label{DecomposedMetric}
 \hat{g}_{\mu\nu} = \hat{e}^a_{\ \mu} \hat{e}^b_{\ \nu} \eta_{ab}.
\end{equation}

For a non-degenerate vielbein $\hat{e}^a_{\ \mu}$, the inverse map 
exists and is denoted by $\hat{e}^{\ \mu}_a$; it provides a local 
isomorphism between co-vectors. In the non-degenerate case, the 
following additional relations hold:
\begin{equation}
 \hat{g}^{\mu\nu}=\hat{e}^{\ \mu}_a \hat{e}^{\ \nu}_b \eta^{ab},
\qquad \hat{e}^a_{\ \mu} \hat{e}^{\ \nu}_a=\delta^{\ \nu}_{\mu}, 
\qquad \hat{e}^a_{\ \mu} \hat{e}^{\ \mu}_b = \delta^a_{\ b}.
\end{equation}

For a given metric $\hat{g}_{\mu\nu}$ the relation \eqref{DecomposedMetric} 
does not fix $\hat{e}^a_{\ \mu}$ uniquely, but only up to local 
{\sf O}(4) transformations. We treat this arbitrariness as an 
additional gauge freedom. It gives rise to an associated covariant 
derivative, $\hat{\nabla}_\mu$. The corresponding connection 
$\hat{\omega}^{ab}_{\ \ \mu}$, the {\it spin connection}, is the 
second fundamental field in the Cartan formulation of gravity. 
When acting on Lorentz vectors the covariant derivative $\hat{\nabla}$ 
is thus formally given by $\hat{\nabla}\equiv \partial+\hat{\omega}$.
The associated {\it field strength} $\hat{F}^{ab}_{\ \ \mu \nu}$ 
is obtained as the commutator of two covariant derivatives, yielding
\begin{equation}
 \hat{F}^{ab}_{\ \ \mu\nu}= 2\,\Big(\partial_{[\mu} \hat{\omega}^{ab}_{\ \ \nu]}+\hat{\omega}^a_{\ b[\mu}\hat{\omega}^{cb}_{\ \ \nu]}\Big).
\end{equation}

In order to define a covariant derivative $\hat{\cal D}_\mu$ that 
acts covariantly on spacetime tensors, we introduce a {\it spacetime 
connection} $\hat\Gamma$. In addition we introduce a third covariant 
derivative $\hat{D}_\mu$ that is covariant w.\,r.\,t. both spacetime 
and Lorentz indices. Demanding the vielbein to be covariantly 
constant, $\hat{D}_\mu \hat{e}^a_{\ \nu}=0$, which implies metricity 
($\hat{\cal D}_\mu g_{\rho \sigma}=0$) of the connection $\hat{\Gamma}$, 
leads to 
\begin{equation}
 \hat{D}_\mu \hat{e}^a_{\ \nu} = \partial_\mu \hat{e}^a_{\ \nu} + 
\hat{\omega}^a_{\ b \mu} \hat{e}^b_{\ \nu} - \hat{\Gamma}^{\lambda}_{\mu\nu} 
\hat{e}^a_{\ \lambda}=0.
\end{equation}
For a non-degenerate vielbein with inverse $\hat{e}^{\ \mu}_{a}$ 
we can solve this expression for $\hat{\Gamma}(e,\omega)$ according to:
\begin{equation}
 \hat{\Gamma}^\lambda_{\mu\nu}= \hat{e}^{\ \lambda}_{a} 
\big( \partial_\mu \hat{e}^a_{\ \nu} + \hat{\omega}^a_{\ b\mu} 
\hat{e}^b_{\ \nu}\big) = \hat{e}^{\ \lambda}_{a} \hat{\nabla}_\mu 
\hat{e}^a_{\ \nu}.
\end{equation}
To summarize, we can write the three different covariant derivatives 
for short:
\begin{equation}
 \hat{\nabla}=\partial+ \hat{\omega}, \qquad {\cal D}= \partial+ 
\hat{\Gamma}, \qquad \hat{D}=\partial+ \hat{\omega}+ \hat{\Gamma}.
\end{equation}
When acting on pure spacetime or Lorentz tensors the general covariant 
derivative $\hat{D}$ specializes to $\hat{\cal D}$ or $\hat\nabla$, 
respectively.

Like the Christoffel symbol, the spacetime connection $\hat{\Gamma}$ 
satisfies the metricity condition, but in contrast to the Levi-Civita 
connection its {\it torsion}
\begin{equation}
 \hat{T}^\lambda_{\mu\nu} = 2 \, \hat{\Gamma}^{\lambda}_{[\mu\nu]}
\end{equation}
does not vanish, in general.

\subsection{Classical actions for gravity}
Classical (Euclidean) Einstein-Cartan gravity is based on the 
{\it Hilbert-Palatini action}
\begin{equation}\label{HPAction}
\begin{split}
 S_{\text{HP}}[\hat{e},\hat{\omega}]&=-\frac{1}{16 \pi G} 
\int \ddx \hat{e} \Big[\hat{e}^{\ \mu}_a \hat{e}^{\ \nu}_b 
\hat{F}^{ab}_{\ \ \mu \nu}-2 \Lambda\Big]\\
&=-\frac{1}{64 \pi G} \int \ddx \varepsilon^{\mu\nu\rho\sigma}
\varepsilon_{a b c d} \bigg[  F^{ab}_{\ \ \mu \nu}- 
\frac{\Lambda}{3} \hat{e}^a_{\ \mu}\hat{e}^b_{\ \nu}\bigg]
\hat{e}^c_{\ \rho} \hat{e}^d_{\ \sigma}
\end{split}
\end{equation}
with $\hat{e}\equiv \det(\hat{e}^a_{\ \mu})$. The tensor density
\begin{equation}
 \varepsilon^{\mu\nu\rho\sigma}\equiv \hat{e}\: \hat{e}_a^{\ \mu}
\hat{e}_b^{\ \nu}\hat{e}_c^{\ \rho}\hat{e}_d^{\ \sigma}
\varepsilon^{a b c d},
\end{equation}
i.\,e. the oriented volume form on ${\cal M}$ is given by
\begin{equation}
 \text{d}x^\mu \wedge\text{d}x^\nu \wedge\text{d}x^\rho 
\wedge\text{d}x^\sigma =\varepsilon^{\mu\nu\rho\sigma}
\text{d}x^1 \wedge\text{d}x^2 \wedge\text{d}x^3 \wedge\text{d}x^4
\end{equation}
and $\varepsilon^{\mu\nu\rho\sigma}$ is independent of 
$\hat{e}^a_{\ \mu}$. Thus the second way of writing $S_{\text{HP}}$ 
in \eqref{HPAction} makes it obvious that this functional is also 
defined for singular vielbeins $\hat{e}^a_{\ \mu}$ with $\hat{e}=0$. 
In that respect, $S_{\text{HP}}[\hat{e},\hat{\omega}]$ differs from 
its metric counterpart, the classical Einstein-Hilbert action 
$S_{\text{EH}}[\hat{g}]$, that is only defined for non-degenerate metrics $\hat{g}_{\mu\nu}$.

Moreover, $S_{\text{HP}}[\hat{e},\hat{\omega}]$ can be supplemented 
by another monomial, the {\it Immirzi term}. This leads to the 
{\it Holst action} \cite{soeren}\footnote{The history of this action \cite{Mielke-2009} dates back well before the work of Holst, see in particular refs. \cite{M5,M6,M8,M9}.}
\begin{equation}\label{Holst-action}
S_{\text{Ho}}[\hat{e},\hat{\omega}]=-\frac{1}{16\pi G}
\int \ddx \hat{e} \bigg[ \hat{e}^{\ \mu}_a \hat{e}^{\ \nu}_b 
\bigg( 1 - \frac{1}{\gamma} \,\star\!\bigg) \hat{F}^{ab}_{\ \ \mu\nu}
-2\Lambda\bigg]
\end{equation}
with the {\it Immirzi parameter} $\gamma$ \cite{immirzi}, and $\star$ 
denoting the duality operator acting on the Lorentz indices according to 
\begin{equation}
\star F^{ab}_{\ \ \mu\nu}= \frac{1}{2} \varepsilon^{ab}_{\ \ cd} 
F^{cd}_{\ \ \mu\nu }.
\end{equation}

In order to determine how the Immirzi term affects the classical 
dynamics of the theory, we have to distinguish two cases:

\noindent{\bf (1) $\gamma=\mp 1$:} In this case we find $(1- \frac{1}{\gamma} \star)\hat{F}\rightarrow 2 \hat{F}^{(\pm)}$ with $\hat{F}^{(\pm)}$ being the (anti-) selfdual projection of $\hat{F}$, $\star\hat{F}^{(\pm)}=\pm \hat{F}^{(\pm)}$. Since $\hat{F}$ satisfies the relation $\hat{F}^{(\pm)}(\hat{\omega})=\hat{F}(\hat{\omega}^{(\pm)})$, for $\gamma=\mp1$ we find that the action $S_{\text{Ho}}$ only depends on one of the two chiral components $\hat{\omega}^{(\pm)}$ while the second is projected out. Moreover, the Holst action then corresponds to the Hilbert-Palatini action with the full $\hat{\omega}$ replaced by one of the two chiralities $\hat{\omega}^{(\pm)}$ (up to an overall factor of 2 that does not affect the stationary points of the functional).

\noindent{\bf (2) $\gamma\neq \mp1$:} In this case the operator $\big(\mathds{1}-\frac{1}{\gamma} \star\big)$ is not a projector. Taking the variation of $S_{\text{Ho}}$ with respect to $\hat{\omega}$ we find $\delta\hat\omega$ only in the linear combination $\hat{\zeta}^{ab}_{\ \ \mu} \equiv \frac{1}{2}(\delta^a_{\ [c}\delta^b_{\ d]}-\frac{1}{2\gamma} \varepsilon^{ab}_{\ \ cd}) \delta\hat{\omega}^{cd}_{\ \ \mu}$ that now can be solved for $\delta \hat{\omega}$, i.\,e. there is a linear one-to-one correspondence between the variations $\delta\hat{\omega}$ and $\hat{\zeta}$ for any fixed value of $\gamma \neq \mp 1$. Thus, the stationarity conditions of the action functional with respect to $\delta \hat \omega$ and $\delta\hat\zeta$ are equivalent and lead to equivalent equations of motion. Those are independent of $\gamma$ as it only occurs in the definition of $\hat\zeta$.

For $\gamma\neq\mp1$, varying $S_{\text{Ho}}$ with respect to $\hat{\omega}^{ab}_{\ \ \mu}$ leads to the equation of motion 
\begin{equation}\label{eqmomega}
 \varepsilon^{\mu\nu\rho\sigma}\varepsilon_{abcd} \hat{e}^c_{\ \rho}\hat{\nabla}_\nu \hat{e}^d_{\ \sigma}=0 \qquad \forall \mu,a,b
\end{equation}
Assuming a regular vielbein $\hat{e}^a_{\ \mu}$, these equations can be cast into the form
\begin{equation}
 \hat{\nabla}_{[\mu} \hat{e}^a_{\ \nu]}=\frac{1}{2}\hat{e}^a_{\ \lambda}\hat{T}^\lambda_{\mu\nu}=0,
\end{equation}
implying vanishing torsion. These 24 equations can be used to express the 24 components of the spin connection $\hat{\omega}$ in terms of the vielbein:
\begin{equation}\label{OmegaOfE}
 \hat{\omega}^{ab}_{\ \ \mu}(\hat{e})=\hat{e}^{a\lambda}\partial_{[\mu}\hat{e}^b_{\ \lambda]}+ \hat{e}^{b\nu} \partial_{[\nu}\hat{e}^a_{\ \mu]}-\hat{e}^{a\lambda}\hat{e}^{b\nu}\big(\partial_{[\lambda}\hat{e}^c_{\ \nu]}\big)\hat{e}_{c\mu}\:.
\end{equation}

Furthermore, varying $S_{\text{Ho}}$ with respect to $\hat{e}^a_{\ \mu}$ leads to
\begin{equation}\label{eqme}
 \varepsilon^{\mu\nu\rho\sigma}\varepsilon_{abcd}\bigg[\bigg(F^{ab}_{\ \ \mu\nu} -\frac{1}{\gamma}\star\! F^{ab}_{\ \ \mu\nu}\bigg)-\frac{2}{3}\Lambda\,\hat{e}^a_{\ \mu}\hat{e}^b_{\ \nu}\bigg]\hat{e}^c_{\ \rho}=0\qquad\forall\, \sigma, d
\end{equation}
Inserting $\hat{\omega}^{ab}_{\ \ \mu}(\hat{e})$ into \eqref{eqme} and again assuming a regular vielbein $\hat{e}^a_{\ \mu}$, these equations can be cast into the usual form of Einstein's equations familiar from metric gravity.

For $\gamma=\mp1$, we obtain equations of motion of the same form as in \eqref{eqmomega}, but $\hat{\omega}$ substituted by its (anti-)selfdual component $\hat{\omega}^{(\pm)}$. Under the same conditions this leads to an expression for $\hat{\omega}^{(\pm)}(\hat{e})$ being the (anti-)selfdual projection of \eqref{OmegaOfE}. Although this spin-connection does not give rise to vanishing torsion, as it does not coincide with the unique, torsionless Levi-Civita choice \eqref{OmegaOfE}, we nevertheless regain Einstein's equations in terms of the tetrad by employing $\hat{\omega}^{(\pm)}(\hat e)$ in \eqref{eqme} for $\gamma=\mp1$.

Despite this apparent equivalence to the metric formulation, it has to be emphasized that the equations of motion in their original form \eqref{eqmomega}, \eqref{eqme} are also solved by the degenerate configuration $\hat{e}^a_{\ \mu}=0$ with arbitrary spin connection $\hat{\omega}^{ab}_{\ \ \mu}$, a solution that has no correspondent counterpart in the metric formulation. Thus even at the classical level and in vacuo we find that the solution spaces of metric gravity and Einstein-Cartan gravity do not coincide. In presence of fermionic matter the correspondence to metric gravity breaks down completely, since the fermion current acts as a source of torsion. Hence, even ``on-shell'' the spacetime exhibits torsion in this case; a situation that cannot be described with the metric as the only fundamental field variable.

\subsection{Structural aspects of the Immirzi term}

With $\hat{e}^a=\hat{e}^a_{\ \mu} \text{d}x^\mu$ and $\hat{T}^a=\hat{\nabla}_{[\mu}\hat{e}^a_{\ \nu]}\text{d}x^\mu\wedge\text{d}x^\nu$ denoting the vielbein one-form and the torsion 2-form, respectively, the Immirzi term can be expressed as:
\begin{equation}\label{Im-1}
 S_{\text{Im}}[\hat{e},\hat{\omega}]=\frac{1}{16 \pi G}\frac{1}{\gamma} \bigg(\int_{\cal M} \hat{T}^a\wedge\hat{T}_a-\int_{\partial {\cal M}}\hat{e}^a\wedge\hat{T}_a\bigg).
\end{equation}
Thus we find that, up to a boundary term, it is given by the square of the torsion 2-form. Written in this way it is particularly obvious that there does not exist a corresponding counterpart in metric gravity. In addition it can be inferred that $S_{\text{Ho}}[\hat{e},\hat{\omega}]$, too, is well-defined for singular vielbeins.

By means of the curvature two-form $F^{ab}= \frac{1}{2!} F^{ab}_{\ \ \mu\nu} \text{d}x^\mu\wedge\text{d}x^\nu$, the Immirzi term can also be written as
\begin{equation}\label{Im-2}
 S_{\text{Im}}[\hat{e},\hat{\omega}]=\frac{1}{16 \pi G} \frac{1}{\gamma} \int_{\cal M} \hat{F}^{ab}\wedge\hat{e}_a\wedge\hat{e}_b
\end{equation}
which makes it obvious that $S_{\text{Im}}$ only exists in four dimensions. In the literature, the topological invariant
\begin{equation}
 \int_{\cal M} \Big(\hat{T}^a\wedge\hat{T}_a - \hat{F}^{ab}\wedge\hat{e}_a\wedge\hat{e}_b\Big)=\int_{\partial{\cal M}}\hat{e}^a\wedge\hat{T}_a
\end{equation}
is known as the Nieh-Yan invariant \cite{NY, zanelli,kreimer}.

Besides the three monomials contained in 
\begin{equation}
 S_{\text{Ho}}[\hat{e},\hat{\omega}]=- \frac{1}{16\pi G}\bigg(\frac{1}{2}\int \varepsilon_{abcd}\hat{F}^{ab}\wedge\hat{e}^c\wedge\hat{e}^d-\frac{1}{\gamma} \int \hat{F}^{ab}\wedge\hat{e}_a\wedge\hat{e}_b-\frac{\Lambda}{12} \int \varepsilon_{abcd}\hat{e}^a \wedge \hat{e}^b\wedge \hat{e}^c\wedge \hat{e}^d\bigg)
\end{equation}
there are only three more monomials that can be written down without explicitly assuming the existence of an inverse vielbein $\hat{e}^{\ \mu}_a$, all of them being topological: In addition to the Nieh-Yan invariant, there are only the Pontryagin index $\propto \int \hat{F}^{ab} \wedge \hat{F}_{ab}$, and the Euler number $\propto \int \varepsilon_{abcd}\hat{F}^{ab}\wedge \hat{F}^{cd}$.

Finally, when the Holst action is exponentiated in the path integral, in the limit $\gamma\rightarrow 0^+$ the Immirzi term gives rise to a $\delta$-function $\delta(\int \hat{T}^a\wedge\hat{T}_a)$ due to \eqref{Im-1}; this is similar to the Landau gauge ``$\alpha=0$'' that implements a sharp gauge fixing. The $\delta$-function in the path integral leads to a suppression of certain torsion modes, while others are not affected.\footnote{Note that $\hat{T}^a\wedge \hat{T}_a$ is proportional to $\varepsilon^{\mu\nu\rho\sigma}T^{a}{}_{\mu\nu}T_{a\rho\sigma}$ which, for a fixed value of $a$, is similar to $\varepsilon^{\mu\nu\rho\sigma}F_{\mu\nu}F_{\rho\sigma}\propto\vec{E}\cdot\vec{B}$ in electromagnetism.} Thus, the limit $\gamma\rightarrow 0^+$ only corresponds to a {\it partial suppression} of torsion. Hence, in this limit Einstein-Cartan theory based on the Holst action does {\it not} reduce to metric gravity based on the Einstein-Hilbert action; rather one ends up with a theory ``as close to metric gravity as possible''.\footnote{In contrast to the Euclidean formulation, in the Lorentzian formulation, due to an additional factor of $i$, the limits $\gamma\rightarrow0^+$ and $\gamma\rightarrow0^-$ coincide; moreover, all Gaussian integrals turn into Fresnel integrals, and employing the Riemann-Lebesgue lemma one draws the same conclusion for $\gamma\rightarrow0^\pm$ as stated above.} On the other hand, for $\gamma\rightarrow\pm \infty$ torsion fluctuates freely without any suppression at all; this case amounts to a theory ``most distant from metric gravity''.

\vspace{1cm}

\section{Effective Average Action and Theory Space}

\subsection{The formal functional integral and its associated FRGE}
Denoting the quantum vielbein and the quantum spin connection by $\hat{e}^a_{\ \mu}$ and $\hat{\omega}^{ab}_{\ \ \mu}$, respectively, the starting point for the construction of the effective average action is the generating functional 
\vspace{0.5cm}
\begin{equation}\label{FunctInt}
\begin{split}
{\cal Z} [s_a^{~\mu}, t_{ab}^{~~\mu}, \sigma^\mu, \bar{\sigma}_\mu, \rho^{ab}, \bar{\rho}_{ab}] = \int{\cal D}\hat{e}\:{\cal D}\hat{\omega}\:{\cal D}{\cal C}{\cal D}\bar{{\cal C}}\:{\cal D}\Sigma\:{\cal D}\bar{\Sigma}\hspace{4cm}\\
  {\rm exp}\Big\{ - S[\hat{e}, \hat{\omega}] - S_{\rm gf} [\hat{e}, \hat{\omega}] - S_{\rm gh} [\hat{e}, \hat{\omega}, {\cal C}, \bar{{\cal C}}, \Sigma, \bar{\Sigma}] - S_{\rm source}\Big\}
\end{split}
\end{equation}
\vspace{0.2cm}

\noindent Therein, ${\cal C}^\mu$ and $\bar{\cal C}_\mu$ denote the diffeomorphism ghost fields familiar from metric gravity whereas the ghost fields associated with the additional ${\sf O}(4)_{\text{loc}}$ gauge invariance are denoted $\Sigma^{ab}$ and $\bar{\Sigma}_{ab}$. The arguments of ${\cal Z}$ are the external sources coupled to each of the quantum fields via the source terms
\begin{equation}
 S_{\rm source} = - \int\ddx \:\hat{e}\:\big\{s_a^{~\mu}\hat{e}^a_{~\mu} + t_{ab}^{~~\mu}\hat{\omega}^{ab}_{~~\mu} + \bar{\sigma}_\mu {\cal C}^\mu + \sigma^\mu \bar{{\cal C}}_\mu + \bar{\rho}_{ab}\Sigma^{ab} + \rho^{ab} \bar{\Sigma}_{ab}\big\}
\end{equation}
with $\hat{e}\equiv \det  (\hat{e}^a_{\ \mu})$. The bare action $S[\hat{e},\hat{\omega}]$ is assumed invariant with respect to spacetime diffeomorphisms,
\begin{equation}\label{undecomposed-diff}
\delta_\Dr (w) \hat{e}^a_{~\mu} = {\cal L}_w \hat{e}^a_{~\mu}\:,\qquad
\delta_\Dr (w) \hat{\omega}^{ab}_{~~\mu} = {\cal L}_w \hat{\omega}^{ab}_{~\mu}
\end{equation}
and local Lorentz transformations,
\begin{equation}\label{undecomposed-lorentz}
 \delta_\Lr (\lambda) \hat{e}^a_{~\mu} = \lambda^a_{~b} \hat{e}^b_{~\mu}\:,\qquad
\delta_\Lr (\lambda) \hat{\omega}^{ab}_{~~\mu} = - \partial_\mu \lambda^{ab} + \lambda^a_{~c}\hat{\omega}^{cb}_{~~\mu} + \lambda^b_{~c}\hat{\omega}^{ac}_{~~\mu} \equiv - \hat{\nabla}_\mu \lambda^{ab}\:,
\end{equation}
of $\hat{e}^a_{\ \mu}$ and $\hat{\omega}^{ab}_{\ \ \mu}$. This gauge invariance has to be broken by the gauge fixing term $S_{\text{gf}}$ in order to ensure the existence of a well-defined propagator. Moreover, if the source $t^{\ \ \mu}_{ab}$ associated with the quantum spin connection transforms as a tensor, the source term will spoil this gauge invariance as well since $\hat{\omega}^{ab}_{\ \ \mu}$ transforms inhomogeneously under local ${\sf O}(4)$ rotations.

Next we perform a background-quantum field split which ultimately will allow for a convenient gauge fixing and a suitable coarse graining of the functional integral \eqref{FunctInt}. We decompose the quantum fields $(\hat{e}, \hat{\omega})$ according to 
\begin{equation}
 \hat{e}^a_{~\mu} = \bar{e}^a_{~\mu} + \varepsilon^a_{~\mu}\:,\qquad
\hat{\omega}^{ab}_{~~\mu} = \bar{\omega}^{ab}_{~~\mu} + \tau^{ab}_{~~\mu}
\end{equation}
into background fields $(\bar{e},\bar{\omega})$ and fluctuations $(\varepsilon,\tau)$. In addition, the source term is altered by coupling the fluctuations $(\varepsilon, \tau)$ instead of the full quantum fields to the sources $(s, t)$ and by choosing the volume element induced by the background vielbein $\bar{e}=\det( \bar{e}^a_{\ \mu})$. 

Here and in the following we always demand the background vielbein $\bar{e}^a_{\ \mu}$ to be non-degenerate while we do not make any assumption concerning the regularity of the full quantum vielbein $\hat{e}^a_{\ \mu}$.

Assuming a translationally invariant functional measure, we are thus led to the generating functional 
\begin{equation}\label{path-integral}
 \begin{split}
  {\cal Z} [s_a^{~\mu}, t_{ab}^{~~\mu}, \sigma^\mu, \bar{\sigma}_\mu, \rho^{ab}, \bar{\rho}_{ab}; \bar{e}^a_{~\mu}, \bar{\omega}^{ab}_{~~\mu}] = \int{\cal D}\varepsilon\:{\cal D}\tau\:{\cal D}{\cal C}\:{\cal D}\bar{{\cal C}}\:{\cal D}\Sigma\:{\cal D}\bar{\Sigma} \hspace{3.4 cm} \\
 {\rm exp}\Big\{\! -\!S[\bar{e} + \varepsilon, \bar{\omega} + \tau] -\!S_{\rm gf} [\varepsilon, \tau; \bar{e}, \bar{\omega}]
 -\!S_{\rm gh} [\varepsilon, \tau, {\cal C}, \bar{{\cal C}}, \Sigma, \bar{\Sigma}; \bar{e}^a_{~\mu}, \bar{\omega}^{ab}_{~~\mu}] -\! S^{\rm back}_{\rm source}\Big\}
 \end{split}
\end{equation}
that parametrically depends on the chosen background configuration ($\bar{e}^a_{\ \mu}$, $\bar{\omega}^{ab}_{\ \ \mu}$).

At the level of the background fields and the fluctuations, the gauge transformations \eqref{undecomposed-diff}, \eqref{undecomposed-lorentz} can now be realized in two different ways: Assuming that the background fields are invariant under gauge transformations leads to the {\it true gauge transformations} $\delta^{\text{G}}$:
\begin{equation}\label{true-transformations}
\begin{aligned}
&\delta^\Gr_\Dr (w) \bar{e}^a_{~\mu}& &\hspace{-.5cm}= 0\:,\\
&\delta^\Gr_\Dr (w) \varepsilon^a_{~\mu}& &\hspace{-.5cm}= {\cal L}_w (\bar{e}^a_{~\mu} \!+\! \varepsilon^a_{~\mu})\:,\\
&\delta^\Gr_\Dr (w) \bar{\omega}^{ab}_{~~\mu}& &\hspace{-.5cm}= 0\:,\\
&\delta^\Gr_\Dr (w) \tau^{ab}_{~~\mu}& &\hspace{-.5cm}= {\cal L}_w (\bar{\omega}^{ab}_{~~\mu} \!+\! \tau^{ab}_{~~\mu})\:,\,
\end{aligned}
\begin{aligned}
 &\delta^\Gr_\Lr (\lambda) \bar{e}^a_{~\mu}& &\hspace{-.5cm}= 0\:,\\
&\delta^\Gr_\Lr (\lambda) \varepsilon^a_{~\mu}& &\hspace{-.5cm}= \lambda^a_{~b} (\bar{e}^b_{~\mu} \!+\! \varepsilon^b_{~\mu})\:,\\
&\delta^\Gr_\Lr (\lambda) \bar{\omega}^{ab}_{~~\mu}& &\hspace{-.5cm}= 0\:,\\
&\delta^\Gr_\Lr (\lambda) \tau^{ab}_{~~\mu}& &\hspace{-.5cm}= - \partial_\mu \lambda^{ab}\!+\! \lambda^a_{~c} (\bar{\omega}^{cb}_{~~\mu} \!+\! \tau^{cb}_{~~\mu}) \!+\! \lambda^b_{~c} (\bar{\omega}^{ac}_{~~\mu} \!+\! \tau^{ac}_{~~\mu}).
\end{aligned}
\end{equation}
On the other hand demanding all fluctuations to transform homogeneously leads to the following {\it background gauge transformations} $\delta^{\text{B}}$:
\begin{equation}\label{background-transformations}
\begin{aligned}
&\delta^\Br_\Dr (w) \bar{e}^a_{~\mu}& &\hspace{-.5cm}= {\cal L}_w \bar{e}^a_{~\mu}\:,\\
&\delta^\Br_\Dr (w) \varepsilon^a_{~\mu}& &\hspace{-.5cm}= {\cal L}_w \varepsilon^a_{~\mu}\:,\\
&\delta^\Br_\Dr (w) \bar{\omega}^{ab}_{~~\mu}& &\hspace{-.5cm}= {\cal L}_w \bar{\omega}^{ab}_{~~\mu} \:,\\
&\delta^\Br_\Dr (w) \tau^{ab}_{~~\mu}& &\hspace{-.5cm}= {\cal L}_w  \tau^{ab}_{~~\mu}\;,
\end{aligned}\quad
\begin{aligned}
&\delta^\Br_\Lr (\lambda) \bar{e}^a_{~\mu}& &\hspace{-.5cm}=  \lambda^a_{~b} \bar{e}^b_{~\mu} \:, \\
&\delta^\Br_\Lr (\lambda) \varepsilon^a_{~\mu}& &\hspace{-.5cm}=  \lambda^a_{~b} \varepsilon^b_{~\mu}\:, \\
&\delta^\Br_\Lr (\lambda) \bar{\omega}^{ab}_{~~\mu}& &\hspace{-.5cm}=  - \partial_\mu \lambda^{ab} + \lambda^a_{~c} \bar{\omega}^{cb}_{~~\mu} + \lambda^b_{~c} \bar{\omega}^{ac}_{~~\mu} \equiv - \bar{\nabla}_\mu \lambda^{ab}\:, \\
&\delta^\Br_\Lr (\lambda) \tau^{ab}_{~~\mu}& &\hspace{-.5cm}=  \lambda^a_{~c} \tau^{cb}_{~~\mu} + \lambda^b_{~c} \tau^{ac}_{~~\mu} \:.
\end{aligned}
\end{equation}
In particular, the spin connection fluctuation $\tau^{ab}_{~~\mu}$ now transforms as a tensor w.\,r.\,t ${\sf O}(4)_{\text{loc}}$, whereas the background spin connection $\bar{\omega}^{ab}_{~~\mu}$ transforms inhomogeneously, i.\,e. like a connection. Both classes of transformations \eqref{true-transformations} and \eqref{background-transformations}, respectively, reproduce the former gauge transformations \eqref{undecomposed-diff} and \eqref{undecomposed-lorentz} at the level of the undecomposed quantum fields $(\hat{e},\hat{\omega})$.

For the ghost fields, no decomposition of this kind will be performed so that for them true gauge and background gauge transformations coincide. In each case, the ghost fields are supposed to transform tensorially:
\begin{equation}\label{ghost-transformations}
 \begin{aligned}
  &\delta^\Gr_\Dr (w) {\cal C}^\mu& &\hspace{-.4cm}= \delta^\Br_\Dr (w) {\cal C}^\mu = {\cal L}_w {\cal C}^\mu \:, \\
&\delta^\Gr_\Dr (w) \bar{{\cal C}}_\mu& &\hspace{-.4cm}= \delta^\Br_\Dr (w) \bar{{\cal C}}_\mu = {\cal L}_w \bar{{\cal C}}_\mu\:, \\
&\delta^\Gr_\Dr (w) \Sigma^{ab}& &\hspace{-.4cm}= \delta^\Br_\Dr (w) \Sigma^{ab} = {\cal L}_w \Sigma^{ab}\:, \\
&\delta^\Gr_\Dr (w) \bar{\Sigma}_{ab}& &\hspace{-.4cm}= \delta^\Br_\Dr (w) \bar{\Sigma}_{ab} = {\cal L}_w \bar{\Sigma}_{ab}\:,
 \end{aligned}
\quad
\begin{aligned}
 &\delta^\Gr_\Lr (\lambda) {\cal C}^\mu& &\hspace{-.4cm}= \delta^\Br_\Lr (\lambda) {\cal C}^\mu = 0 \:, \\
&\delta^\Gr_\Lr (\lambda) \bar{{\cal C}}_\mu& &\hspace{-.4cm}= \delta^\Br_\Lr (\lambda) \bar{{\cal C}}_\mu = 0 \:, \\
&\delta^\Gr_\Lr (\lambda) \Sigma^{ab}& &\hspace{-.4cm}= \delta^\Br_\Lr (\lambda) \Sigma^{ab} = \lambda^a_{~c} \Sigma^{cb} + \lambda^b_{~c} \Sigma^{ac}\:, \\
&\delta^\Gr_\Lr (\lambda) \bar{\Sigma}_{ab}& &\hspace{-.4cm}= \delta^\Br_\Lr (\lambda) \bar{\Sigma}_{ab} = \lambda_a^{~c} \bar{\Sigma}_{cb} + \lambda_b^{~c} \bar{\Sigma}_{ac}\:.
\end{aligned}
\end{equation}

For the construction of a background gauge invariant effective (average) action we need all parts of the action functional in \eqref{path-integral} to be $\delta^{\text{B}}$-invariant. With regard to the graviton sector $(\hat{e},\hat{\omega})$, the crucial idea is to choose a gauge condition that breaks true gauge invariance but retains background gauge invariance; the associated gauge fixing term $S_{\text{gf}}$ thus has to be $\delta^{\text{B}}$-invariant. In order to obtain a background gauge invariant ghost action $S_{\text{gh}}$, a reparametrization of the group of gauge transformations will be necessary as we shall see in the next section. Since the bare action $S$ in \eqref{path-integral} only depends on the sum of $(\bar{e},\bar{\omega})$ and $(\varepsilon,\tau)$, it is $\delta^{\text G}$- as well as $\delta^{\text{B}}$-invariant. Finally, demanding that the sources transform tensorially w.\,r.\,t. $\delta^{\text{B}}$ ensures the background gauge invariance of the complete action functional that is exponentiated under the functional integral in \eqref{path-integral}.

The next step in the construction of the effective average action consists in the addition of a mode suppression, or cutoff term $\Delta_k S$ to the exponent in \eqref{path-integral}. It exhibits a block structure w.\,r.\,t. the graviton and the ghost sector. Formally, it has the structure
\begin{equation}
 \label{cutoff-ECT}
\begin{split}
\Delta_k S[\varepsilon, \tau, {\cal C}, \bar{{\cal C}}, \Sigma, \bar{\Sigma}; \bar{e}, \bar{\omega}]=\hspace{9.8cm}\\
=\frac{1}{16 \pi G}\int\dr^4 x\:\bar{e}\left\{ \left( \begin{array}{c} \varepsilon^c_{~\nu} \\ \!\!\tau^{cd}_{~~\nu} \end{array} \!\!\right)^{\!\!\!\rm T} \left( \begin{array}{cc} R[\bar{e}, \bar{\omega}]^{{\rm grav}\:~\nu~\mu}_{~~~~\:c~a} & R[\bar{e}, \bar{\omega}]^{{\rm grav}\:~\nu~~\mu}_{~~~~\:c~ab} \\ R[\bar{e}, \bar{\omega}]^{{\rm grav}\:~~\nu~\mu}_{~~~~\:cd~a} & R[\bar{e}, \bar{\omega}]^{{\rm grav}\:~~\nu~~\mu}_{~~~~\:cd~ab} \end{array} \right) \left( \begin{array}{c} \varepsilon^a_{~\mu} \\ \!\! \tau^{ab}_{~~\mu} \end{array}\!\! \right) \right.\\
\hspace{2.65cm}+ \left. \left( \begin{array}{c} \bar{{\cal C}}_{\nu} \\ \!\!\bar{\Sigma}_{cd} \end{array} \!\!\right)^{\!\!\!\rm T} \left( \begin{array}{cc} R[\bar{e}, \bar{\omega}]^{{\rm gh}\:\nu}_{~~\:~\mu} & R[\bar{e}, \bar{\omega}]^{{\rm gh}\:\nu}_{~~\:~ab} \\ R[\bar{e}, \bar{\omega}]^{{\rm gh}\:cd}_{~~\:~~\mu} & R[\bar{e}, \bar{\omega}]^{{\rm gh}\:cd}_{~~\:~~ab} \end{array} \right) \left( \begin{array}{c} {\cal C}^{\mu} \\ \!\!\Sigma^{ab} \end{array} \!\!\right)\right\}
\end{split}
\end{equation}
with $k$ representing a momentum scale. By construction, it is $\delta^{\text{B}}$-invariant as well as quadratic in the graviton fluctuations and in the ghost fields, respectively. Before explaining the purpose it serves, we will proceed with the formal derivation.

First, we note that the generating functional acquires an additional $k$-dependence:
\begin{equation}
 \label{modified-path-integral}
\begin{split}
{\cal Z}_k [s_a^{~\mu}, t_{ab}^{~~\mu}, \sigma^\mu, \bar{\sigma}_\mu, \rho^{ab}, \bar{\rho}_{ab}; \bar{e}^a_{~\mu}, \bar{\omega}^{ab}_{~~\mu}]=\hspace{7.5 cm}\\ 
= \int{\cal D}\varepsilon{\cal D}\tau{\cal D}{\cal C}{\cal D}\bar{{\cal C}}{\cal D}\Sigma{\cal D}\bar{\Sigma}\; {\rm exp}\Big\{ - S[\bar{e} + \varepsilon, \bar{\omega} + \tau] - S_{\rm gf} [\varepsilon, \tau; \bar{e}, \bar{\omega}]\hspace{2.5cm}\\ - S_{\rm gh} [\varepsilon, \tau, {\cal C}, \bar{{\cal C}}, \Sigma, \bar{\Sigma}; \bar{e}^a_{~\mu}, \bar{\omega}^{ab}_{~~\mu}]
 - \Delta_k S[\varepsilon, \tau, {\cal C}, \bar{{\cal C}}, \Sigma, \bar{\Sigma}; \bar{e}, \bar{\omega}] - S^{\rm back}_{\rm source}\Big\}\:.
\end{split}
\end{equation}
We proceed as in the derivation of the standard effective action $\Gamma$, thereby always keeping track of the modifications induced by the cutoff term $\Delta_k S$. We define the `connected' generating functional
\begin{equation}
 W_k [s_a^{~\mu}, t_{ab}^{~~\mu}, \sigma^\mu, \bar{\sigma}_\mu, \rho^{ab}, \bar{\rho}_{ab}; \bar{e}^a_{~\mu}, \bar{\omega}^{ab}_{~~\mu}] \equiv  {\rm ln}\big({\cal Z}_k [s_a^{~\mu}, t_{ab}^{~~\mu}, \sigma^\mu, \bar{\sigma}_\mu, \rho^{ab}, \bar{\rho}_{ab}; \bar{e}^a_{~\mu}, \bar{\omega}^{ab}_{~~\mu}]\big)
\end{equation}
and construct the (now $k$- as well as background-dependent) vacuum expectation values: 
\begin{equation}
 \begin{aligned}
  e^a{}_\mu &\equiv \langle \hat{e}^a{}_\mu\rangle & &= \bar{e}^a{}_\mu+\bar{\varepsilon}^a{}_\mu & &\text{with}& \bar{\varepsilon}^a{}_\mu& \equiv \langle \varepsilon^a{}_\mu\rangle,\\
  \omega^{ab}{}_\mu &\equiv \langle \hat{\omega}^{ab}{}_\mu\rangle & &= \bar{\omega}^{ab}{}_\mu+\bar{\tau}^{ab}{}_\mu & &\text{with} & \bar{\tau}^{ab}{}_\mu& \equiv \langle \tau^{ab}{}_\mu\rangle,\\
\xi^{\mu}&\equiv \langle C^{\mu}\rangle,& & \bar{\xi}_{\mu} \equiv \langle \bar{C}_{\mu}\rangle, & \Upsilon^{ab} & \equiv \langle \Sigma^{ab}\rangle, & \bar{\Upsilon}^{\mu}&\equiv \langle \bar{\Sigma}_{ab}\rangle.
 \end{aligned}
\end{equation}
The fluctuation and ghost expectation values are obtained by functionally differentiating w.\,r.\,t. their associated sources:
\begin{equation}
\begin{split}
 \begin{aligned}
  \bar{\varepsilon}^a_{~\mu} (x) &\equiv \frac{1}{\bar{e} (x)} \frac{\delta W_k}{\delta s_a^{~\mu} (x)}\:,& \bar{\tau}^{ab}_{~~\mu} (x) &\equiv \frac{1}{\bar{e} (x)} \frac{\delta W_k}{\delta t_{ab}^{~~\mu} (x)}\:,\\
 \xi^\mu (x) &\equiv \frac{1}{\bar{e} (x)} \frac{\delta W_k}{\delta \bar{\sigma}_\mu (x)}\:,&\bar{\xi}_\mu (x) &\equiv \frac{1}{\bar{e} (x)} \frac{\delta W_k }{\delta \sigma^\mu (x)}\:,\\
\Upsilon^{ab} (x) &\equiv \frac{1}{\bar{e} (x)} \frac{\delta W_k }{\delta \bar{\rho}_{ab} (x)}\:, & \bar{\Upsilon}_{ab} (x) &\equiv \frac{1}{\bar{e} (x)} \frac{\delta W_k}{\delta \rho^{ab} (x)}\:.
\end{aligned}
\end{split}
\end{equation}
Assuming the Hessian of $W_k$ w.\,r.\,t. the sources to be regular, the Legendre transformation of $W_k$ w.\,r.\,t. to the sources $\{s_a^{~\mu}, t_{ab}^{~~\mu}, \sigma^\mu, \bar{\sigma}_\mu, \rho^{ab}, \bar{\rho}_{ab}\}$ defines a functional of the expectation values that again depends parametrically on the background fields and on $k$:
\begin{equation}
\begin{split}
 &\Gamma_k[\bar{\varepsilon}^a_{~\mu}, \bar{\tau}^{ab}_{~~\mu}, \xi^\mu, \bar{\xi}_\mu, \Upsilon^{ab}, \bar{\Upsilon}_{ab}; \bar{e}^a_{~\mu}, \bar{\omega}^{ab}_{~~\mu}]=\\
&= \int\dr^4 x\:\bar{e}\Big\{s_a^{~\mu}\bar{\varepsilon}^a_{~\mu} + t_{ab}^{~~\mu}\bar{\tau}^{ab}_{~~\mu} + \bar{\sigma}_\mu \xi^\mu + \sigma^\mu \bar{\xi}_\mu + \bar{\rho}_{ab} \Upsilon^{ab} + \rho^{ab} \bar{\Upsilon}_{ab} \Big\}\\
&\ - W_k [s_a^{~\mu}, t_{ab}^{~~\mu}, \sigma^\mu, \bar{\sigma}_\mu, \rho^{ab}, \bar{\rho}_{ab}; \bar{e}^a_{~\mu}, \bar{\omega}^{ab}_{~~\mu}] - \Delta_k S[\bar{\varepsilon}^a_{~\mu}, \bar{\tau}^{ab}_{~~\mu}, \xi^\mu, \bar{\xi}_\mu, \Upsilon^{ab}, \bar{\Upsilon}_{ab}; \bar{e}^a_{~\mu}, \bar{\omega}^{ab}_{~~\mu}]\\
&=  \Gamma_k [e^a_{~\mu} - \bar{e}^a_{~\mu}, \omega^{ab}_{~~\mu} - \bar{\omega}^{ab}_{~~\mu}, \xi^\mu, \bar{\xi}_\mu, \Upsilon^{ab}, \bar{\Upsilon}_{ab}; \bar{e}^a_{~\mu}, \bar{\omega}^{ab}_{~~\mu}]\\
&\equiv \Gamma_k [e^a_{~\mu}, \omega^{ab}_{~~\mu}, \bar{e}^a_{~\mu}, \bar{\omega}^{ab}_{~~\mu}, \xi^\mu, \bar{\xi}_\mu, \Upsilon^{ab}, \bar{\Upsilon}_{ab}].
\end{split}
\end{equation}
This object will be called the {\it effective average action} for the Einstein-Cartan theory space. It applies to a new theory space but is otherwise similar to the running action functional employed in virtually all recent continuum RG investigations of gauge theories and of metric gravity.

In these calculations, the bare action $S$ and truncations of $\Gamma_k$ contain a standard kinetic term such that the inverse (effective) propagator $S^{(2)}$ and $\Gamma^{(2)}_k$, respectively, constitutes a second-order differential operator with $S^{(2)}$ and $\Gamma^{(2)}_k$ denoting the Hessian w.\,r.\,t. the respective non-background fields. In this case, the defining property of the cutoff operator ${\cal R}_k \equiv Z_k \cdot k^2 R^{(0)}(p^2/k^2)$, where $Z_k$ is a matrix in field space, is the following: with $p^2 \in \text{spec}\{-\bar{D}^2\equiv - \bar{g}^{\mu\nu}\bar{D}_\mu \bar{D}_\nu\}$, the presence of $\Delta_k S$ should lead to the substitution $p^2\mapsto p^2 + k^2 R^{(0)}(p^2/k^2)$ in all eigenvalues of $\Gamma^{(2)}_k$ compared to the case where no cutoff term is present. The explicit structure of ${\cal R}_k$ always has to be adjusted to the chosen truncation, by choosing the matrix $Z_k$ such that the above rule is obeyed. For gauge theories and metric gravity with $\Gamma_k$ being a scale dependent generalization of the Yang-Mills action and the Einstein-Hilbert action, respectively, ${\cal R}_k$ constitutes a second-order operator as well.

As already said above, $k$ denotes a momentum scale, and $R^{(0)}$ is a scalar ``shape function''. Demanding the following general features 
\begin{equation}
 R^{(0)} (0) = 1\qquad\text{and} \qquad \lim_{y\to\infty} R^{(0)} (y) = 0
\end{equation}
$R^{(0)}$ acts as an {\it infrared cutoff} at the scale $k$, i.\,e. the infrared modes with $(-\bar{D})^2$-eigenvalues below $k^2$ are given a mass of order $k^2$ while the ultraviolet modes above $k^2$ are left untouched.

If a mode suppression operator ${\cal R}_k$ with these properties can be constructed, its tensor structure is completely fixed, and the only feature left to vary is the explicit profile of $R^{(0)}$. Moreover, for ``second-order theories'' of this kind, an exact RG equation of the general form \cite{avact, mr}
\begin{equation}\label{FRGE}
 \partial_t \Gamma_k = \frac{1}{2} \,\text{STr} \Big[ (\Gamma^{(2)}_k+{\cal R}_k)^{-1} \partial_t {\cal R}_k \Big]
\end{equation}
can be derived which (together with an initial condition at $k\rightarrow \infty$) fully determines the running action $\Gamma_k$.

\subsection{The ``proper-time'' approximation}
Our analysis of the RG flow of Einstein-Cartan gravity in this paper will not be based on the construction of an {\it explicit} cutoff operator. This is due to the following complications that arise in the treatment of Einstein-Cartan gravity compared to other gauge theories. 

\noindent{\bf (i)} As we shall see below explicitly, for a truncation of $\Gamma_k$ of the Holst type
\begin{equation}
 \Gamma_{k\:{\rm Ho}} [e, \omega, \bar{e}, \bar{\omega}] = - \frac{1}{16 \pi G_k}\int\dr^4 x\:e\Big\{e_a{}^\mu e_b{}^\nu\Big(F^{ab}_{~~\mu\nu} - \frac{1}{2\gamma_k} \varepsilon^{ab}_{~~cd} F^{cd}_{~~\mu\nu}\Big) - 2 \Lambda_k\Big\}
\end{equation}
the Hessian $\Gamma^{(2)}_k$ is a first-order differential operator, and in this respect Einstein-Cartan gravity is similar to fermionic quantum field theories. Therefore, the cutoff adaptation rule $p^2 \mapsto p^2+k^2 R^{(0)} (p^2/k^2)$ with $p^2\in \text{spec}\{-\bar{D}^2\}$ can only be implemented at the level of the {\it squared} inverse propagator $\big(\Gamma^{(2)}_k\big)^2$. As in the fermionic case, ${\cal R}_k$ is supposed to be a first-order operator that enters the construction of $\Gamma_k$ via $\Delta_k S$ in the way sketched above. While in fermionic theories the explicit construction of an adapted cutoff operator is feasible usually as they exhibit a rich algebraic structure like the existence of a $\gamma_5$-involution, for Einstein-Cartan gravity it is not known whether analogous structures exist which could help at this point. 

\noindent{\bf (ii)} Moreover, the squared operator $\big(\Gamma^{(2)}_k\big)^2$ cannot be solely expressed in terms of $\bar{D}_\mu$.\footnote{We implement the cutoff rule w.\,r.\,t. $\bar{D}_\mu$ since this derivative acts on {\it both} types of indices. As a result, the Laplacian $\bar{D}^2\equiv\bar{D}_\mu \bar{g}^{\mu\nu}\bar{D}_\nu$ is covariant under both diffeomorphisms and $\sf{O}(4)_{\rm loc}$.} 

\noindent{\bf (iii)} Finally, the above procedure demands for the computation of the complete spectra of $\Gamma^{(2)}_k$ and $\big(\Gamma^{(2)}_k\big)^2$. However, even in the free case ($\bar{\omega}^{ab}_{~~\mu}=0$) the largest block of the operator that needs to be diagonalized is a $7 \times 7$ matrix whose spectrum cannot be determined analytically. Therefore the construction of a fully adapted cutoff operator for Einstein-Cartan gravity seems out of reach; even if all but the last issue could be solved, at least in part of field space, one would have to rely on a brute force cutoff whose tensor structure could be given by the identity operator on field space, for example.

For these reasons we decided to analyze the renormalization behavior of Einstein-Cartan gravity by means of a simpler proper-time RG equation which obtains from the exact equations \eqref{FRGE} by a certain structural approximation over and above the truncation of theory space. An important justification is that it has been shown in the case of metric gravity that the proper-time equation leads to virtually the same results as the FRGE, both qualitatively and quantitatively, as far as the UV renormalization behavior is concerned \cite{prop}. It also performed extremely well in high precision computations of critical exponents \cite{bonannozappala, litimpawlowski}.

The essential step in turning the mode suppression into a propertime cutoff \cite{schwinger, martindittrich, mrpt} consists in replacing ${\cal R}_k(\bar{D}^2)$ with ${\cal R}_k(\Gamma^{(2)}_k)$ where the Hessian is evaluated at vanishing fluctuations. As a result, the RHS of the flow equation \eqref{FRGE}, for vanishing fluctuations\footnote{For the truncations considered in the following (``single field truncations'' \cite{elisa2,MRS}) this is general enough.}, depends only on a single operator:
$ \partial_t\Gamma_k=\frac{1}{2} {\rm STr}\,\Big[W_k\big(\Gamma^{(2)}_k\big)\Big]$. The function $W_k$ involves $R^{(0)}$ and can be read off from \eqref{FRGE}. Denoting its Laplace transform by $F_k(s)$ we get the following ``proper-time flow equation'': 
\begin{equation}
\partial_t \Gamma_k = \frac{1}{2}\int_0^\infty {\rm d}s \,F_k(s) \,{\rm STr}\, \Big[e^{-s \Gamma^{(2)}_k}\Big]. 
\end{equation}
For every admissible function $R^{(0)}$ the properties of $F_k(s)$ are such that it cuts off the $s$-integral both in the UV (for $s\rightarrow0$) and in the IR, i.\,e. for large $s$, exactly as in Schwinger's original application of this method \cite{schwinger}. Rather than specifying $R^{(0)}$, it is more convenient to directly pick a function $F_k(s)$ which has the correct general properties \cite{liao}. Indeed, our analysis of the RG flow of Einstein-Cartan gravity will be based on a proper-time equation of the form 
\begin{equation}
\label{proper-time-equation}
\partial_t \Gamma_k =  - \frac{1}{2}\int_0^\infty\frac{\dr s}{s}\,\big(\partial_t\:f_k^{\Lambda_{\rm UV}} (s)\big) \,{\rm STr} \big\{{\rm e}^{-s\:\Gamma_k^{(2)}}\}\:,
\end{equation}
where the function $f^{\Lambda_{\rm UV}}_k$ arises by a convenient redefinition of $F_k$. It is arbitrary except that it must satisfy
\begin{equation}
 \begin{aligned}
  f_k^{\Lambda_{\rm UV}} (s) \approx 0 \qquad \text{for} \qquad k^{-2}\ll s\quad \text{and}\quad s \ll \Lambda^{-2}_{\rm UV}\\[-0.2cm]
  f_k^{\Lambda_{\rm UV}} (s) \approx 1 \qquad \text{for} \qquad  k^{-2} \gg s \quad \text{and}\quad s \gg \Lambda^{-2}_{\rm UV}.
 \end{aligned}
\end{equation}
Thus the function $f_k^{\Lambda_{\rm UV}}$ acts both as a UV regulator (at the scale $\Lambda_{\rm UV}$) and as an IR regulator (at the scale $k$). We will focus on the latter property since we are interested in a flow equation w.\,r.\,t. the infrared scale. Concretely, we will study regularization schemes that lead to flow equations of the form \cite{liao, bonannozappala}
\begin{equation}
 \partial_t \Gamma_k = \text{STr}\bigg(\frac{k^2}{\Gamma_k^{(2)} + k^2}\bigg)^{m+1}\qquad\text{and} \qquad \partial_t \Gamma_k = \text{STr}\bigg(\frac{m\,k^2}{\Gamma_k^{(2)} + m\,k^2}\bigg)^{m+1}
\end{equation}
with an arbitrary integer $m\geq 1$. The second scheme contains the special case of a sharp proper-time cutoff since for $m\rightarrow \infty$ it leads to \cite{mrpt}
\begin{equation}
 \partial_t \Gamma_k=\text{STr}\Big(\text{exp}\big\{-\Gamma^{(2)}_k /k^2\big\}\Big).
\end{equation}
Nevertheless, the corresponding cutoff scale is actually given by $\sqrt{m} k$ instead of $k$.

\subsection{The Einstein-Cartan theory space}
To the end of this section let us briefly compare the theory spaces associated with Einstein-Cartan and metric gravity. The flow equations \eqref{FRGE} and \eqref{proper-time-equation} are defined on the theory space of all background gauge invariant functionals denoted by ${\cal T}_{\text{QECG}}$:
\begin{equation}
\begin{split}
 {\cal T}_{\rm QECG}&\equiv \big\{ A[\bar{\varepsilon}, \bar{\tau}, \xi, \bar{\xi}, \Upsilon, \bar{\Upsilon}; \bar{e}, \bar{\omega}]\big|\\
& \hspace{0.8cm} A[\bar{\varepsilon} + \delta^\Br_\Dr (w) \bar{\varepsilon} + \delta^\Br_\Lr (\lambda) \bar{\varepsilon}, \bar{\tau} + \delta^\Br_\Dr (w) \bar{\tau} + \delta^\Br_\Lr (\lambda) \bar{\tau}, \xi + \delta^\Br_\Dr (w) \xi + \delta^\Br_\Lr (\lambda) \xi,\\
& \hspace{1.1cm} \bar{\xi} + \delta^\Br_\Dr (w) \bar{\xi} + \delta^\Br_\Lr (\lambda) \bar{\xi}, \Upsilon + \delta^\Br_\Dr (w) \Upsilon + \delta^\Br_\Lr (\lambda) \Upsilon, \bar{\Upsilon} + \delta^\Br_\Dr (w) \bar{\Upsilon} + \delta^\Br_\Lr (\lambda) \bar{\Upsilon};\\
& \hspace{1.1cm} \bar{e} + \delta^\Br_\Dr (w) \bar{e} + \delta^\Br_\Lr (\lambda) \bar{e}, \bar{\omega} + \delta^\Br_\Dr (w) \bar{\omega} + \delta^\Br_\Lr (\lambda) \bar{\omega}]\\
& = A[\bar{\varepsilon}, \bar{\tau}, \xi, \bar{\xi}, \Upsilon, \bar{\Upsilon}; \bar{e}, \bar{\omega}]~\forall w^\mu, \:\lambda^{ab}\big\}
\end{split}
\end{equation}
The subscript ``QECG'' thereby refers to the fact that the existence of a non-Gaussian fixed point would allow for the definition of a {\it quantum} field theory of Einstein-Cartan gravity. At this point, its existence is merely hypothetical but our later results serve as a first step to suggest that such a theory called Quantum-Einstein-Cartan-gravity (QECG) can indeed be defined.

In contrast, its metric counterpart QEG is based on a non-Gaussian fixed point in the theory space 
\begin{equation}
 \begin{split}
  {\cal T}_{\rm QEG} &\equiv \big\{A[\bar{h}, \xi, \bar{\xi}; \bar{g}] \big|
 A[\bar{h} + \delta^\Br_\Dr (w) \bar{h}, \xi + \delta^\Br_\Dr (w) \xi, \bar{\xi} + \delta^\Br_\Dr (w) \bar{\xi}; \bar{g} + \delta^\Br_\Dr (w) \bar{g}] \\&\hspace{6.7cm} = A[\bar{h}, \xi, \bar{\xi}; \bar{g}]~\forall w^\mu\big\} \:.
 \end{split}
\end{equation}

Since neither the gauge groups nor the field contents coincide, the two theories belong to different ``universality classes''. In particular, the existence of a NGFP in the latter neither implies the existence of a NGFP in the former nor vice versa. Therefore all RG studies of metric gravity are conceptually independent from the investigation of ${\cal T}_{\text{QECG}}$ on which we shall embark.

\section{Gauge Fixing and Ghost Action}\label{GFandGhosts}
In order to arrive at a functional integral which can be computed (actually {\em defined}) by means of a functional RG flow we introduced arbitrary background fields 
$\bar{e}^a_{~\mu}$ and $\bar{\omega}^{ab}_{~~\mu}$, decomposed the variables of integration as $\hat{e}^a_{~\mu} \equiv \bar{e}^a_{~\mu} + \varepsilon^a_{~\mu}$, $\hat{\omega}^{ab}_{~~\mu} \equiv \bar{\omega}^{ab}_{~~\mu} + \tau^{ab}_{~~\mu}$, and performed a background covariant gauge fixing. This leads to a functional integral of the form \eqref{path-integral}.

As already introduced in the previous section $S_{\rm gf}$ and $S_{\rm gh}$ denote the gauge fixing and corresponding ghost action, respectively, ${\cal C}^\mu$ and $\bar{{\cal C}}_\mu$ are the diffeomorphism ghosts, and similarly $\Sigma^{ab}$ and $\bar{\Sigma}_{ab}$ are those related to the local ${\sf O}(4)$. With $G$ denoting Newton's constant, the gauge fixing action is of the form
\begin{equation}\label{gf}
S_{\rm gf} = \frac{1}{2 \alpha_{\rm D}\cdot 16 \pi G}\int {\rm d}^4 x\:\bar{e}\:\bar{g}^{\mu\nu}\:{\cal F}_\mu {\cal F}_\nu + \frac{1}{2 \alpha_{\rm L}} \int {\rm d}^4 x\:\bar{e}\:{\cal G}^{ab} {\cal G}_{ab}\:,
\end{equation}
where ${\cal F}_\mu$ and ${\cal G}^{ab}$ break the ${\sf Diff} ({\cal M})$ and ${\sf O}(4)_{\rm loc}$ gauge invariance, respectively. In order to ultimately arrive at a ${\sf Diff}({\cal M}) \ltimes {\sf O}(4)_{\rm loc}$ invariant effective average action we employ special gauge fixing conditions ${\cal F}_\mu$ and ${\cal G}^{ab}$ of the ``background type'' so that $S_{\rm gf} [\varepsilon, \tau; \bar{e}, \bar{\omega}]$ is invariant under the combined background gauge transformations $\delta^{\rm B}_{{\rm D,\,L}}$ acting on both $(\varepsilon, \,\tau)$ and $(\bar{e}, \,\bar{\omega})$ while, of course, it is not invariant under the ``true'' (or ``quantum'') gauge transformations, denoted by $\delta^{\rm G}_{\:{\rm D}}$ and $\delta^{\rm G}_{\:\,{\rm L}}$, respectively. 

The action of the true and background gauge transformations on the background fields $(\bar{e},\bar{\omega})$ and the fluctuations $(\varepsilon,\tau)$ is given in \eqref{true-transformations} and \eqref{background-transformations}.

Since no background split is introduced for the ghost fields, their true and background gauge transformations happen to coincide. We require a tensorial transformation law corresponding to their index structure as given in \eqref{ghost-transformations}.

\noindent{\bf (A)} The effect infinitesimal gauge transformations have on functionals $A[\hat{e}^a_{~\mu}, \hat{\omega}^{ab}_{~~\mu}, \cdots]$ of the quantum fields can be expressed in terms of the Ward operators ${\cal W}^\Br_{~\Dr}$, ${\cal W}^\Br_{~\Lr}$ for the background gauge transformations, and ${\cal W}^\Gr_{~\Dr}$, ${\cal W}^\Gr_{~\Lr}$ for the ``gauge'' or ``true'' transformations, that are applied to the functional $A$, being of the form
\begin{equation}\begin{aligned}
{\cal W}^{\rm B,G}_{\rm D}(w)&\equiv \sum_{\phi\in \Phi} \int {\rm d}^4 x\, \delta^{\rm B,G}_{\rm D}(w)\phi(x)\frac{\delta}{\delta \phi(x)},\\ {\cal W}^{\rm B,G}_{\rm L}(\lambda)&\equiv \sum_{\phi\in \Phi} \int {\rm d}^4 x\, \delta^{\rm B,G}_{\rm D}(\lambda)\phi(x)\frac{\delta}{\delta \phi(x)}\:.
\end{aligned}
\end{equation}
Here, $\Phi \equiv\{\hat{e}^a_{~\mu}, \hat{\omega}^{ab}_{~~\mu}, {\cal C}^\mu, \bar{{\cal C}}_\mu, \Sigma^{ab}, \bar{\Sigma}_{ab}\}$ is the set of quantum fields, $w_\mu(x)$ is the vector field defining the diffeomorphism and $\lambda^{ab}(x)$ is the parameter of the $\sf{O}(4)_{\rm loc}$-transformation. Thus, for both the ``B''- and the ``G''-type transformations, $\delta(w,\lambda)A=-{\cal W}_{\rm D} (w) A-{\cal W}_{\rm L}(\lambda)A$.

We can verify that the background type operators satisfy the algebra
\begin{equation}\label{ward-B-algebra}
\begin{array}{lcl}\displaystyle [{\cal W}^\Br_{~\Dr} (w_1), {\cal W}^\Br_{~\Dr} (w_2)] &=& {\cal W}^\Br_{~\Dr} ([w_1, w_2]) \\
\displaystyle [{\cal W}^\Br_{~\Lr} (\lambda_1), {\cal W}^\Br_{~\Lr} (\lambda_2)] &=& {\cal W}^\Br_{~\Lr} ([\lambda_1, \lambda_2]) \\
\displaystyle [{\cal W}^\Br_{~\Dr} (w), {\cal W}^\Br_{~\Lr} (\lambda)] &=& {\cal W}^\Br_{~\Lr} ({\cal L}_w \lambda)\:, \end{array}
\end{equation}
while the ``true'' ones obey the relations
\begin{equation}\label{ward-G-algebra}
\begin{array}{lcl}\displaystyle [{\cal W}^\Gr_{~\Dr} (w_1), {\cal W}^\Gr_{~\Dr} (w_2)] &=& {\cal W}^\Gr_{~\Dr} ([w_1, w_2]) \\
\displaystyle [{\cal W}^\Gr_{~\Lr} (\lambda_1), {\cal W}^\Gr_{~\Lr} (\lambda_2)] &=& {\cal W}^\Gr_{~\Lr} ([\lambda_1, \lambda_2]) \\
\displaystyle [{\cal W}^\Gr_{~\Dr} (w), {\cal W}^\Gr_{~\Lr} (\lambda)] &=& {\cal W}^\Gr_{~\Lr} ({\cal L}_w \lambda)\:, \end{array}
\end{equation}
Here, the brackets $[\cdot,\cdot]$ on the RHS denote the Lie bracket of the vector fields $w_{1,2}$ and the commutator of the matrices $\lambda_{1,2}$, respectively, while ${\cal L}_w$ stands for the Lie derivative w.\,r.\,t. to the vector field $w$. From \eqref{ward-B-algebra} and \eqref{ward-G-algebra} we infer the direct product structure already mentioned before: ${\bf G}= {\sf Diff}({\cal M})\ltimes {\sf O}(4)_{\rm loc}$.

Like their precursors before the background split, the commutation relations \eqref{ward-G-algebra} are {\it not ${\sf O}(4)_{\rm loc}$ covariant} since the Lie derivative ${\cal L}_w$ contains ordinary partial derivatives $\partial_\mu$ rather then ${\sf O}(4)_{\rm loc}$ covariant ones, $\nabla_\mu$.\footnote{See ref. \cite{eomega-corfu} for the transformations $\widetilde{\delta_{\rm D}}$ and their Ward operators $\widetilde{\cal W_{\rm D}}$ which apply prior to the background split.}

In order to deal with this situation, within the background field setting, we define modified diffeomorphisms which include a certain ${\sf O}(4)_{\rm loc}$ transformation \cite{jackiw1, jackiw2}:
\begin{align}
\widetilde{\widetilde{\delta^\Br_{\:\Dr}}} (w) \equiv \delta^\Br_{\:\Dr} (w) + \delta^\Br_{\:\,\Lr} (w\cdot \bar{\omega})\:, \\
\widetilde{\widetilde{\delta^\Gr_{\:\Dr}}} (w) \equiv \delta^\Gr_{\:\Dr} (w) + \delta^\Gr_{\:\,\Lr} (w\cdot \bar{\omega})\:,
\end{align}
with $(w\cdot\bar{\omega})^{ab}\equiv w^{\mu} \bar{\omega}^{ab}{}_\mu$. In terms of their Ward operators, the modified ``background'' diffeomorphisms satisfy the commutation relations
\begin{equation}\label{ward-algebra-back-new}
\begin{array}{lcl}\displaystyle [\widetilde{\widetilde{{\cal W}^\Br_{~\Dr}}} (w_1), \widetilde{\widetilde{{\cal W}^\Br_{~\Dr}}} (w_2)] &=& \widetilde{\widetilde{{\cal W}^\Br_{~\Dr}}} ([w_1, w_2]) - {\cal W}^\Br_{~\Lr} (w_1 w_2 \cdot \bar{F}) \\
\displaystyle [{\cal W}^\Br_{~\Lr} (\lambda_1), {\cal W}^\Br_{~\Lr} (\lambda_2)] &=& {\cal W}^\Br_{~\Lr} ([\lambda_1, \lambda_2]) \\
\displaystyle [\widetilde{\widetilde{{\cal W}^\Br_{~\Dr}}} (w), {\cal W}^\Br_{~\Lr} (\lambda)] &=& 0 \end{array}
\end{equation}
while their ``gauge'' counterparts have the algebra
\begin{equation}\label{ward-algebra-true-new}
\begin{array}{lcl}\displaystyle [\widetilde{\widetilde{{\cal W}^\Gr_{~\Dr}}} (w_1), \widetilde{\widetilde{{\cal W}^\Gr_{~\Dr}}} (w_2)] &=& \widetilde{\widetilde{{\cal W}^\Gr_{~\Dr}}} ([w_1, w_2]) + {\cal W}^\Gr_{~\Lr} (w_1 w_2 \cdot \bar{F}) \\
\displaystyle [{\cal W}^\Gr_{~\Lr} (\lambda_1), {\cal W}^\Gr_{~\Lr} (\lambda_2)] &=& {\cal W}^\Gr_{~\Lr} ([\lambda_1, \lambda_2]) \\
\displaystyle [\widetilde{\widetilde{{\cal W}^\Gr_{~\Dr}}} (w), {\cal W}^\Gr_{~\Lr} (\lambda)] &=& {\cal W}^\Gr_{~\Lr} (w \cdot \bar{\nabla} \lambda)\:, \end{array}
\end{equation}
where $(w_1w_2\cdot \bar{F})^{ab}\equiv w_1^\mu w_2^\nu \bar{F}^{ab}{}_{\mu\nu}$ and $(w\cdot \bar{\nabla}\lambda)^{ab}=w^{\mu}\bar{\nabla}_\mu\lambda^{ab}$. Note that the modified transformations enjoy fully ${\sf O}(4)_{\rm loc}$ covariant Lie algebra relations.

Both algebras, \eqref{ward-algebra-back-new} and \eqref{ward-algebra-true-new}, respectively, are going to become important in a moment: The ``background'' transformations and their commutators will determine the theory space on which the RG flow is taking place, while the algebra of the ``gauge'' transformations determines the ghost action \cite{JEUM}.

\noindent{\bf (B)} We choose the gauge fixing conditions ${\cal F}_\mu$ and ${\cal G}^{ab}$ to be linear in $\varepsilon^a_{~\mu}$ and independent of $\tau^{ab}_{~~\mu}$ \cite{percacci1}. Concretely, we shall employ the following family of functions:
\begin{subequations}\label{gc}
\begin{equation}
{\cal F}_\mu = \bar{e}_a^{~\nu} \big[\bar{D}_\nu \varepsilon^a_{~\mu} + \beta_{\rm D} \bar{D}_\mu \varepsilon^a_{~\nu}\big]\:,\label{gc-diff}
\end{equation}
\begin{equation}
{\cal G}^{ab} = \frac{1}{2}\bar{g}^{\mu\nu}\big[\varepsilon^a_{~\mu} \bar{e}^b_{~\nu} - \varepsilon^b_{~\nu} \bar{e}^a_{~\nu}\big] \equiv \varepsilon^{[ab]}\label{gc-o4}
\end{equation}
\end{subequations}
Thus, in total, there are three gauge fixing parameters: $\alpha_{\rm D}$, $\alpha_{\rm L}$ and $\beta_{\rm D}$\footnote{As can be inferred from \eqref{gf} and \eqref{gc}, the diffeomorphism gauge parameter $\alpha_{\rm D}$ is dimensionless whereas the Lorentz-gauge parameter $\alpha_{\rm L}$ is of mass dimension $-4$. Therefore, it has to be rescaled properly. We perform this rescaling by means of the mass parameter $\bar{\mu}$ that will be introduced in a moment. Within the approximations used, no scale derivatives of dimensionless couplings appear on the right-hand side of the flow equation. Therefore, including an additional factor of $g$ into $\alpha_{\rm L}$ will not lead to additional contributions.}. Using \eqref{gc-diff}, \eqref{gc-o4} in \eqref{gf} we can verify that the resulting gauge fixing action $S_{\rm gf} [\varepsilon, \tau; \bar{e}, \bar{\omega}]$ is indeed background gauge invariant:
\begin{equation}
{\cal W}^\Br_{~\Lr} \, S_{\rm gf} = 0 = \widetilde{\widetilde{{\cal W}^\Br_{~\Dr}}} \, S_{\rm gf} ~~ \Leftrightarrow ~~ {\cal W}^\Br_{~\Lr} \, S_{\rm gf} = 0 = {\cal W}^\Br_{~\Dr} \, S_{\rm gf}\:.
\end{equation}

\noindent{\bf (C)} The ghost sector requires some care, and this is indeed the reason for introducing the modified diffeomorphisms. We would like the ghost action $S_{\rm gh} [\varepsilon, \tau, {\cal C}, \bar{{\cal C}}, \Sigma, \bar{\Sigma}; \bar{e}, \bar{\omega}]$ to be background gauge invariant, too. However, straightforwardly applying the Faddeev-Popov procedure to the original transformations
\begin{align}
\delta^\Gr(w,\lambda) = \left( \begin{array}{c} \delta^\Gr_{\:\Dr} (w) \\ \delta^\Gr_{\:\,\Lr} (\lambda) \end{array} \right)
\end{align} 
we obtain, in the $\bar{\Sigma}-{\cal C}$-sector, the ghost action\footnote{We consider the case $\varepsilon^a_{~\mu} = 0$ which is all we need here \cite{eomega-corfu}.} 
\begin{align}
S_{\rm gf}^{\bar{\Sigma}-{\cal C}} [{\cal C}, \bar{\Sigma}; \bar{e}, \bar{\omega}] = - \int\dr^4x\,\bar{e}\left.\left(\bar{\Sigma}_{ab} \frac{\partial {\cal G}^{ab}}{\partial \varepsilon^c_{~\nu}} \delta^\Gr_{\:\Dr} ({\cal C}) \varepsilon^c_{~\nu}\right)\right|_{\varepsilon = 0} 
\end{align}
which, with \eqref{gc-o4}, evaluates to
\begin{align}\label{sgf}
S_{\rm gf}^{\bar{\Sigma}-{\cal C}} [{\cal C}, \bar{\Sigma}; \bar{e}, \bar{\omega}] = - \int\dr^4x\,\bar{e}\:\bar{\Sigma}_{ab} \:\bar{e}^{b\mu} {\cal L}_{\cal C} \bar{e}^a_{~\mu}\:. 
\end{align}
While this functional is invariant under background diffeomorphisms, it fails to be invariant under the ${\sf O (4)}_{\rm loc}$ transformations $\delta_{\:\,\Lr}^\Br (\lambda)$, the reason being that the Lie derivative of an ${\sf O (4)}$ tensor does not define an ${\sf O (4)}$ tensor. Rather, we have ${\cal L}_{\cal C} (\lambda^a_{~b} \, \bar{e}^b_{~\mu}) \neq \lambda^a_{~b} \, {\cal L}_{\cal C} \bar{e}^b_{~\mu}$, since $\lambda^a_{~b} (x)$ is a spacetime scalar which transforms non-trivially under diffeomorphisms. Stated differently, ${\sf O (4)}_{\rm loc}$ transformations and (ordinary) diffeomorphisms do not commute, and this is exactly what the above Lie algebra relations express.\footnote{An analogous complication arises in QEG coupled to ${\sf SU}(N)$ Yang-Mills fields where the group of gauge transformations, ${\sf Diff}({\cal M})\ltimes {\sf SU}(N)_{\rm loc}$, has a similar semidirect product structure as in the pure gravity Einstein-Cartan case, where ${\bf G}={\sf Diff}({\cal M})\ltimes {\sf O}(4)_{\rm loc}$; see \cite{JEUM} for a detailed discussion.}

The way out consists in applying the Faddeev-Popov procedure to the modified, that is, ${\sf O (4)}_{\sf loc}$-covariantized (true) gauge transformations:
\begin{align}
\widetilde{\widetilde{\delta^\Gr}} = \left( \begin{array}{c} \widetilde{\widetilde{\delta^\Gr_{\:\Dr}}} (w) \\ \delta_{\:\,\Lr}^\Gr (\lambda) \end{array} \right) \:.
\end{align}  
They are broken by the ten gauge fixing conditions 
\begin{align}
\left( \begin{array}{c} {\cal F}_\mu \\ {\cal G}^{ab} \end{array} \right) \equiv \big( {\cal Q}^I \big)
\end{align}
for which we use a uniform notation now where $\big({\cal Q}^I\big) \equiv \big({\cal F}_\mu\big)$ for $I = 1, \cdots, 4$ and $\big({\cal Q}^I\big) \equiv \big({\cal G}^{ab}\big)$ for $I = 5, \cdots, 10$. Denoting, in the same fashion, the ten parameters of the gauge transformations as $\big( \Lambda^I \big)  = \big(w^\mu, \lambda^a_{~b}\big)$, the Faddeev-Popov determinant reads 
\begin{align}
{\rm det} \left. \left(\frac{\delta {\cal Q}^I (x)}{\delta \Lambda^J (y)}\right)\right|_{\Lambda = 0}\:,
\end{align}
and exponentiating it we obtain a ghost action which has the structure
\begin{align}\label{gh-action-exp}
- \int\dr^4 x \:\bar{e} \left( \begin{array}{c} \bar{{\cal C}_\mu} \\ \bar{\Sigma}_{ab} \end{array} \right)^{\rm T} \left(\begin{array}{cc} \Omega^\mu_{~\nu} & \Omega^\mu_{~cd} \\ \Omega^{ab}_{~~\nu} & \Omega^{ab}_{~~cd} \end{array} \right) \left( \begin{array}{c} {\cal C}^\nu \\ \Sigma^{cd} \end{array} \right) 
\end{align}
The Faddeev-Popov operator $\Omega$ is rather complicated; here we must refer to \cite{eomega-corfu} for its explicit form. Suffice it to say that one can now verify explicitly that the ghost action \eqref{gh-action-exp} is indeed invariant under background gauge transformations:
\begin{equation}
{\cal W}^\Br_{~\Lr} \, S_{\rm gh} = 0 = \widetilde{\widetilde{{\cal W}^\Br_{~\Dr}}} \, S_{\rm gh} ~~ \Leftrightarrow ~~ {\cal W}^\Br_{~\Lr} \, S_{\rm gh} = 0 = {\cal W}^\Br_{~\Dr} \, S_{\rm gh}\:.
\end{equation}
This property is the main prerequisite for arriving at a background gauge invariant effective average action.

\section{A 3-Parameter Truncation:\\ Outline of the Computation}
In this section we introduce the 3-parameter ``Holst truncation'' and briefly outline our basic strategy for commuting its RG flow. This section may serve as a first overview that will guide the reader through the technical details of the computation given in Sections 6 and 7. Moreover, the reader who is mainly interested in the results can skip those details and proceed directly to Section \ref{Results} which contains a discussion of the main results.

\subsection{The truncation ansatz}
Our truncation ansatz is of the form of the Euclidean Holst action \eqref{Holst-action} supplemented by the gauge fixing and the ghost terms associated to the gauge conditions ${\cal F}_\mu(\bar{\varepsilon};\bar{e},\bar{\omega})$ and ${\cal G}^{ab}(\bar{\varepsilon};\bar{e})$ discussed in the previous section:
\begin{equation}\label{Truncation}
 \Gamma_k= \Gamma_{k \: \text{grav}}+\Gamma_{k \: \text{gh}} \quad \text{with}\quad \Gamma_{k \: \text{grav}}=\Gamma_{k \: \text{Ho}}+\Gamma_{k \: \text{gf}}\:.
\end{equation}
Explicitly, the three action functionals are given by the ``running'' Holst action
\begin{equation}
 \Gamma_{k\:{\rm Ho}}[e, \omega] = - \frac{1}{16 \pi G_k} \int\:\dr^4 x \:e \Big\{e_a^{~\mu} e_b^{~\nu}  \Big(F^{ab}_{~~\mu\nu} - \frac{1}{2 \gamma_k}\varepsilon^{ab}_{~~cd}F^{cd}_{~~\mu\nu}\Big) - 2 \Lambda_k\Big\}\label{Holst-truncation}\:,
\end{equation}
the gauge fixing part
\begin{equation}
 \Gamma_{k\:{\rm gf}}[e, \bar{e}, \bar{\omega}] = \frac{1}{2 \alpha_\Dr}\cdot\frac{1}{16 \pi G_k} \int\:\dr^4 x \:\bar{e}\:\bar{g}^{\mu\nu}{\cal F}_\mu {\cal F}_\nu + \frac{1}{2 \alpha_\Lr} \int\:\dr^4 x \:\bar{e}\:{\cal G}^{ab}{\cal G}_{ab} \label{gf-trunk}
\end{equation}
and the ghost contribution
\begin{align}
  \Gamma_{k\:{\rm gh}} [\bar{e}, \bar{\omega}, \xi, \bar{\xi}, \Upsilon, \bar{\Upsilon}]&= - \Big\{\int\!\dr^4 x\:\bar{e}\:\bar{\xi}_\mu \big[(\bar{g}^{\mu\alpha}\partial_\nu \!+\! \beta_\Dr \delta^\alpha_{~\nu} \partial^\mu) \partial_\alpha \!+\! \bar{e}^{l\mu} (\partial_k \bar{\omega}^k_{~l\nu}) \!+\! \bar{\Gamma}^\sigma_{\sigma\nu}\partial^\mu \!+\! {\cal O} (\bar{\omega}^2) \big]\xi^\nu \nonumber \\
& ~ + \int\!\dr^4 x \:\bar{e}\:\bar{\xi}_\mu \big[\bar{e}_d^{~\mu}\partial_c + \bar{e}_d^{~\mu}\bar{e}_a^{~\nu}\bar{\omega}^a_{~c \nu} + \bar{e}_m^{~\mu}\bar{e}_c^{~\nu} \bar{\omega}^{m}_{~~d\nu}\big]\Upsilon^{cd} \nonumber\\
& ~ + \int\!\dr^4 x \:\bar{e}\:\bar{\Upsilon}_{ab}\big[\bar{e}^a_{~\nu} \partial^b + \bar{\omega}^{ab}_{~~\nu}\big] \xi^\nu + \int\!\dr^4 x\:\bar{e}\:\bar{\Upsilon}_{ab} \Upsilon^{ab}\Big\}\:.\label{gh-trunk}
\end{align}
Therein, we have substituted the quantum fields $\hat{e}=\bar{e}+\varepsilon,  \hat{\omega}=\bar{\omega}+\tau,\cdots$ by their expectation values $e=\bar{e}+\bar{\varepsilon}, \omega=\bar\omega+\bar\tau,\cdots$. The couplings $G$, $\Lambda$, $\gamma$ appearing in \eqref{Holst-action} are now allowed to acquire a scale dependence: However, the gauge fixing parameters $\alpha_D$ and $f$ are still approximated to be $k$-independent. In addition the ghost sector is treated classically, i.\,e. we neglect all renormalization effects in the ghost field couplings. This approximation has been proven reliable in QEG \cite{Groh:2010ta,Eichhorn:2010tb}, and in analogy to most QEG investigations we also apply it here.

\subsection{The background configuration}\label{bgrndconf}
In order to derive the RG equations of the running couplings $G_k, \Lambda_k, \gamma_k$ we insert the truncation ansatz into the FRGE \eqref{FRGE} and, after having performed the functional derivatives implicit in $\Gamma_k^{(2)}$, we set the fluctuations $\bar\varepsilon, \bar\tau$ to zero. We are then left with the problem of projecting the supertrace appearing on the RHS of the FRGE onto the 3-dimensional theory space spanned by the field monomials present in the running Holst action.

In order to ``project out'' its three invariants, associated with the couplings $G_k$, $\Lambda_k$ and $\gamma_k$, we insert appropriate background configurations $(\bar{e},\bar{\omega})$ on both sides of the FRGE. For the background vielbein we choose
\begin{equation}\label{backgrnd_e} 
 \bar{e}^a_{\ \mu} = \kappa\, \delta^a_{\ \mu}= \text{const}
\end{equation}
with $\kappa$ being a constant real number. For the background metric, this implies
\begin{equation}
 \bar{g}_{\mu\nu}=\kappa^2 \eta_{\mu\nu},
\end{equation}
i.\,e. this metric is conformally flat with constant conformal factor $\kappa^2$.

The spacetime volume $\int \ddx \bar{e}$, the invariant associated with the cosmological constant, constitutes the only invariant that is independent of $\bar{\omega}$. Therefore it is projected out of the trace on the RHS of \eqref{proper-time-equation} by inserting the $\bar{e}$-configurations \eqref{backgrnd_e} along with $\bar{\omega}^{ab}{}_\mu=0$.

In order to distinguish the two terms in \eqref{Holst-truncation} that are linear in $\bar{F}^{ab}_{\ \ \mu\nu}$ we exploit the fact that the Immirzi term contains the ${\sf O}(4)$ dualized field strength, $\star \bar{F}^{ab}_{\ \ \mu\nu}$. Therefore its sign changes when we switch from a selfdual ($\bar{\omega}^{(+)}$) to an anti-selfdual background spin connection ($\bar{\omega}^{(-)}$) whereas that of the Hilbert-Palatini term remains unaffected.\footnote{Here we make use of the identity $F\big(\omega^{(\pm)}\big)^{ab}_{\ \ \mu\nu}=\big[F\big(\omega\big)^{(\pm)}\big]^{ab}_{\ \ \mu\nu}$ with $\omega^{(\pm)}$ ($F^{(\pm)}$) denoting the (anti-) selfdual projection of $\omega$ ($F$).}

Concretely, we choose $\bar{\omega}^{(\pm)}$ to be ``quasi-Abelean'',
\begin{equation}
 \bar{\omega}^{(\pm)\:ab}_{\ \ \ \ \ \ \mu}(x)=\frac{1}{2}\, n^{(\pm)\:ab} \,v_{\mu}(x),
\end{equation}
with $n^{(\pm)\:ab}$ denoting a constant (anti-)selfdual ${\sf O}(4)$ tensor, i.\,e.
\begin{equation}
 \frac{1}{2}\, \varepsilon^{ab}_{\ \ cd}\,n^{(\pm)\: cd} = \pm \,n^{(\pm)\:ab}\:.
\end{equation}
The vector field $v_\mu$(x) is completely arbitrary at this point. The associated field strength is obtained as
\begin{equation}
 \bar{F}^{(\pm)\:ab}_{\ \ \ \ \ \ \mu\nu}=n^{(\pm)\:ab} \partial_{[\mu}\, v_{\nu]}\:.
\end{equation}
In summary, the three invariants are unambiguously identified on the RHS of the FRGE by inserting the three background configurations $(\kappa\, \delta,0)$, $(\kappa\,\delta, \bar{\omega}^{(+)})$ and $(\kappa \,\delta, \bar{\omega}^{(-)})$ into the supertrace $\text{STr}[\cdots]$.

To this end, we have to expand $\text{STr}[\cdots]$ up to terms of order ${\cal O}(\partial \bar{\omega}^{(\pm)})$ inclusively, i.\,e. we account for all terms that are independent of $\bar{\omega}^{(\pm)}$ or that are of first order in $\bar{\omega}^{(\pm)}$ and contain at most one partial derivative. Among the latter terms, we finally distinguish the terms whose signs change when switching from $\bar{\omega}^{(+)}$ to $\bar{\omega}^{(-)}$ from those whose signs do not. All higher powers and all higher derivatives of $\bar{\omega}^{(\pm)}$ can be neglected since they do not contribute to the desired order ${\cal O}(\bar{F}^{(\pm)})$.

\subsection{The functional flow equation}
Suppressing all indices for the sake of notational simplicity, the operator $\Gamma^{(2)}_k$ evaluated on the background configurations will be decomposed according to 
\begin{equation}\label{M1}
\Gamma^{(2)}_k\equiv H + V \equiv H_0 + \frac{1}{\gamma_k} \bar{H} + V 
\end{equation}
in the graviton as well as in the ghost sector.\footnote{In the following, we will distinguish the graviton and ghost blocks by attaching a superscript ``grav'' and ``gh'' to the respective operator.} Therein, $H$ denotes the free part of order ${\cal O}(\bar{\omega}^{0})$, which is further decomposed according to 
\begin{equation}
 H=H_0 + \frac{1}{\gamma_k} \bar{H} 
\end{equation}
with $H_0$ and $\bar{H}$ being independent of $\gamma_k$. Since the ghost sector is independent of $\gamma_k$ as a whole, we have $\bar{H}^{\text{gh}}=0$. The interaction part $V$ contains all contributions of the order ${\cal O}(\bar{\omega})$ and ${\cal O}(\partial \bar{\omega})$; in particular, it contains matrix elements of the orders $\left(\frac{1}{\gamma_k}\right)^0$ as well as $\left(\frac{1}{\gamma_k}\right)^1$. 

In order to write down the proper-time FRGE for our concrete truncation it is most convenient to take a step backward and start from the representation of $\Gamma_k$ as an RG-improved 1-loop-determinant \cite{avact}:
\begin{equation}
\Gamma_k=S+\frac{1}{2}{\rm STr}\, \ln \Gamma^{(2)}_k\:.
\end{equation}
We will obtain the following equation for $\Gamma_k$ from it:
\begin{equation}\label{FEraw}
\begin{split}
\Gamma_k  = S +& \frac{1}{4} \,{\rm Tr}\,{\rm ln}\bigg(\frac{(H^{\rm grav}_0)^2}{\bar{\mu}^6}\bigg) + \frac{1}{2} \, {\rm Tr}\,{\rm ln}\bigg(\openone + \frac{1}{\gamma_k} {\cal M} \bigg)\\
& + \frac{1}{2} {\rm Tr}\bigg[(H^{\rm grav}_0)^2 (H^{\rm grav})^{-1} V^{\rm grav}(H^{\rm grav}_0)^{-2}\bigg] \\
& - 2 \bigg\{ \frac{1}{4} {\rm Tr}\,{\rm ln}\bigg(\frac{(H^{\rm gh}_0)^2}{\bar{\mu}^2}\bigg) + \frac{1}{2} {\rm Tr}\bigg((H^{\rm gh}_0) V^{\rm gh}(H^{\rm gh}_0)^{-2}\bigg) \bigg\} \:
\end{split}
\end{equation} 
Here we also introduced the operator ${\cal M} \equiv  (H_0)^{-1} \bar{H}$.

In eq. \eqref{FEraw} a parameter with the dimension of a mass appears: $[\bar{\mu}]=1$. It has to be introduced in order to rescale the fluctuations $\bar{\varepsilon}^a_{\ \mu}$ and $\bar{\tau}^{ab}_{\ \ \mu}$ for dimensional reasons; here, it renders the arguments of the logarithms dimensionless.

Expressing the above logarithms in terms of proper-time integrals leads to
\begin{equation}\label{PTIntegrals}
\begin{split}
\Gamma_k  = S &- \frac{1}{4} \:\ddashint \frac{\dr s}{s}\Big( {\rm Tr}\,\eu^{-s(H^{\rm grav}_0)^2} - {\rm Tr}\,\eu^{-s\,\bar{\mu}^6}\Big) - \frac{1}{4}\:\ddashint \frac{\dr s}{s}\,{\rm Tr}\,\eu^{-s\,{\cal N}^2}\\
& + \frac{1}{2}\:\ddashint \frac{\dr s}{s} {\rm Tr}\Big((H^{\rm grav}_0)^2 (H^{\rm grav})^{-1} V^{\rm grav}\eu^{-s(H^{\rm grav}_0)^{2}}\Big) \\
& - 2 \Big\{ - \frac{1}{4}\:\ddashint \frac{\dr s}{s}\Big( {\rm Tr}\,\eu^{-s (H^{\rm gh}_0)^2} - \eu^{-s\,\bar{\mu}^2}\Big)+ \frac{1}{2}\:\ddashint \frac{\dr s}{s} {\rm Tr}\Big((H^{\rm gh}_0) V^{\rm gh}\eu^{-s(H^{\rm gh}_0)^{2}}\Big) \Big\}
\end{split}\raisetag{2.2cm}
\end{equation}
with the operator ${\cal N} \equiv  \openone + \frac{1}{\gamma_k} {\cal M}$. The symbol $\ddashint$ denotes a proper-time integral that demands regularization, giving rise to a $k$-dependence of \eqref{PTIntegrals}. Taking its derivative w.\,r.\,t. the scale $k$ leads to the desired functional RG equation for $\Gamma_k$ then. It is equivalent to \eqref{proper-time-equation} and will define the flow on the theory space ${\cal T}_{\text{QECG}}$ that we are going to analyze.

Since we will be able to determine the spectrum of $H_0$ analytically, our strategy is to compute the traces in \eqref{PTIntegrals} in the eigenbasis of $H_0$. The projection discussed in the previous subsection \ref{bgrndconf} then allows for an expansion of the RHS of \eqref{PTIntegrals} in terms of the invariants contained in the truncation \eqref{Holst-truncation} leading to the desired system of beta functions for the 3 running couplings.

\section{The Structure of the Hessian Operator $\boldsymbol\Gamma^{(2)}_k$}
In this section we begin with the calculational program outlined in the previous section by setting up the matrix blocks which constitute the Hessian operator $\Gamma^{(2)}_k$. Expanding the truncation ansatz \eqref{Truncation} around vanishing fluctuations $(\bar{\varepsilon},\bar{\tau})$, and restricting ourselves to the second order contribution of the graviton sector, leads to the quadratic form
\begin{equation}\label{GammaQuadGrav}
\Gamma^{\rm quad}_{k\:{\rm grav}}[\bar{\varepsilon}, \bar{\tau}; \bar{e}, \bar{\omega}] = \frac{1}{2} \cdot\frac{1}{16 \pi G_k} \int \dr^4 x \left(\!\!\begin{array}{c} \bar{\varepsilon}^m_{~\:\beta} \\ \tau^{mn}_{~~\:\beta}\end{array}\!\! \right)^{\!\!\!\!\rm T} \left( \begin{array}{cc} T_{m~k}^{~~\beta~\alpha} & T_{m~kl}^{~~\beta~~\alpha} \\ T_{mn~k}^{~~~\beta~\alpha} & T_{mn~kl}^{~~~\beta~~\alpha} \end{array} \right) \left(\!\!\begin{array}{c} \bar{\varepsilon}^k_{~\alpha} \\  \tau^{kl}_{~~\alpha}\!\!\end{array} \right)
\end{equation}
with the block matrices
\begin{align}
\begin{split}
T_{m~k}^{~~\beta~\alpha} = &- \frac{1}{2}\big(K^{(\gamma_k)}_{abmk} \bar{F}^{ab}_{~~\mu\nu} - 2 \Lambda_k \varepsilon_{abmk} \bar{e}^a_{~\mu} \bar{e}^b_{~\nu}\big)\varepsilon^{\mu\nu\beta\alpha}\label{eps-eps} + \frac{1}{\alpha^\prime_\Lr}\:\frac{1}{2} \bar{e} \big(\eta_{mk}\bar{g}^{\beta\alpha} - \bar{e}_m^{~\alpha} \bar{e}_k^{~\beta}\big)\\
&- \frac{1}{\alpha_\Dr} \big(\bar{D}^{~\:~~\beta b}_{\sigma\:m~~\mu} + \beta_\Dr \bar{D}^{~\:~~\beta b}_{\mu\:m~~\sigma}\big) \:\bar{e}\: \bar{g}^{\mu\nu}\bar{e}_b^{~\sigma} \bar{e}_a^{~\rho}\big(\bar{D}^{~\:a~~\alpha}_{\rho\:~\nu k} + \beta_\Dr \bar{D}^{~\:a~~\alpha}_{\nu\:~\rho k}\big)\:,
\end{split}\\ \label{eps-tau}
T_{m~kl}^{~~\beta~~\alpha} = &- K^{(\gamma_k)}_{abcm}\bar{e}^c_{~\nu} \bar{\nabla}^{~\:ab}_{\mu~~\:kl} \varepsilon^{\mu\nu\beta\alpha}\:,\\
\label{tau-eps}
T_{mn~k}^{~~~\beta~\alpha} =  &- K^{(\gamma_k)}_{abck}\bar{\nabla}^{~\:~~~ab}_{\mu\:mn~~}\bar{e}^c_{~\nu} \varepsilon^{\mu\nu\beta\alpha}\:,\\
\label{tau-tau} 
T_{mn~kl}^{~~~\beta~~\alpha} = &- K^{(\gamma_k)}_{nkab}\bar{e}^a_{~\mu} \bar{e}^b_{~\nu} \eta_{ml} \varepsilon^{\mu\nu\beta\alpha}
\end{align}
and the tensor 
\begin{equation}
K^{(\gamma_k)}_{abcd} \equiv \varepsilon_{abcd} - \frac{1}{\gamma_k} (\eta_{ac}\eta_{bd} - \eta_{ad}\eta_{bc}). 
\end{equation}
Here we also set $\alpha^\prime_\Lr \equiv \alpha_\Lr/(16 \pi G_k)$ and inserted the gauge conditions discussed in Section \ref{GFandGhosts}. Moreover, the operator $\bar{D}^{\ a\ \ \:\beta}_{\mu\ \alpha b}$ is defined such that $\bar{D}^{\ a\ \ \:\beta}_{\mu\ \alpha b}v^b_{\ \beta}=\big(\bar{D}_\mu v\big)^a_{\ \alpha}$ for an arbitrary tensor $v^a_{\ \alpha}$, i.\,e. $\bar{D}^{\ a\ \ \:\beta}_{\mu\ \alpha b}\equiv \delta^{\ \beta}_\alpha \delta^a_{\ b} \partial_\mu + \delta^{\ \beta}_\alpha \bar{\omega}^a_{\ b\mu} - \delta^a_{\ b} \bar{\Gamma}^\beta_{\mu \alpha}$; the generalization to tensors of higher rank is straightforward.

In contrast to the graviton sector, the ghost sector is by construction quadratic in the ghost fields already at the level of \eqref{Truncation}. Since we will evaluate $\Gamma^{(2)}_k$ for vanishing fluctuations and ghost fields, the classical fields $(e,\omega)$ appearing in the ghost contribution \eqref{gh-trunk} can be substituted by the background configurations already {\it before} taking the functional derivatives:\footnote{For this reason, there are no mixed graviton-ghost components in $\Gamma^{(2)}_k$.}
\begin{align}\label{GammaQuadGh}
  \Gamma^{\text{quad}}_{k\:{\rm gh}} [\bar{e}, \bar{\omega}, \xi, \bar{\xi}, \Upsilon, \bar{\Upsilon}]= &- \Big\{\!\!\int\!\dr^4 x\:\bar{e}\:\bar{\xi}_\mu \big[(\bar{g}^{\mu\alpha}\partial_\nu \!+\! \beta_\Dr \delta^\alpha_{~\nu} \partial^\mu) \partial_\alpha \!+\! \bar{e}^{l\mu} (\partial_k \bar{\omega}^k_{~l\nu}) \!+\! \bar{\Gamma}^\sigma_{\sigma\nu}\partial^\mu \!+\! {\cal O} (\bar{\omega}^2) \big]\xi^\nu \nonumber \\
&+ \int\!\dr^4 x \:\bar{e}\:\bar{\xi}_\mu \big[\bar{e}_d^{~\mu}\partial_c + \bar{e}_d^{~\mu}\bar{e}_a^{~\nu}\bar{\omega}^a_{~c \nu} + \bar{e}_m^{~\mu}\bar{e}_c^{~\nu} \bar{\omega}^{m}_{~~d\nu}\big]\Upsilon^{cd} \nonumber\\
&+ \int\!\dr^4 x \:\bar{e}\:\bar{\Upsilon}_{ab}\big[\bar{e}^a_{~\nu} \partial^b + \bar{\omega}^{ab}_{~~\nu}\big] \xi^\nu + \int\!\dr^4 x\:\bar{e}\:\bar{\Upsilon}_{ab} \Upsilon^{ab}\Big\}\:.
\end{align}

Returning to the graviton sector, we notice that the quadratic form in  \eqref{GammaQuadGrav} does not yet define an {\it operator} ``$\Gamma^{(2)}_{k\:\text{grav}}$'' with a well-defined spectrum but only an {\it integral kernel}. This is due to the fact that the mass dimension of the two types of fluctuations does not coincide: We chose the vielbein to be dimensionless, $[\hat{e}^a_{\ \mu}]=0$, whereas $\hat{\omega}^{ab}_{\ \ \mu}$ as a connection has the dimension of a mass, $[\hat{\omega}^{ab}_{\ \ \mu}]=1$. This obstacle is overcome by an appropriate rescaling of the fluctuations,
\begin{equation}
\bar{\varepsilon}^a_{\ \mu} \rightarrow \bar{\mu}^{1/2}\:\bar{\varepsilon}^a_{\ \mu}, \qquad \bar{\tau}^{ab}_{\ \ \mu}\rightarrow \bar{\mu}^{-1/2} \:\bar{\tau}^{ab}_{\ \ \mu}\:,
\end{equation}
with $\bar{\mu}$ having the dimension of a mass, $[\bar{\mu}]=1$. The above quadratic form is now
\begin{equation}
\begin{split}
\Gamma^{\rm quad}_{k\:{\rm grav}}&[\bar{\varepsilon}, \bar{\tau}; \bar{e}, \bar{\omega}]=
\\
&= \frac{1}{2} \cdot\frac{1}{16 \pi G_k} \int \!\dr^4 x \left(\!\!\begin{array}{c} \bar{\varepsilon}^m_{~\:\beta} \\ \bar{\tau}^{mn}_{~~\:\beta}\end{array}\!\! \right)^{\!\!\!\!\rm T} \left( \begin{array}{cc} T_{m~k}^{~~\beta~\alpha} & T_{m~kl}^{~~\beta~~\alpha} \\ T_{mn~k}^{~~~\beta~\alpha} & T_{mn~kl}^{~~~\beta~~\alpha} \end{array} \right) \left(\!\!\begin{array}{c} \bar{\varepsilon}^k_{~\alpha} \\ \bar{\tau}^{kl}_{~~\alpha}\end{array}\!\! \right)\\
&= \frac{1}{2} \cdot\frac{1}{16 \pi G_k} \int\! \dr^4 x \left(\!\!\begin{array}{c} \bar{\mu}^{\frac{1}{2}}\:\bar{\varepsilon}^m_{~\:\beta} \\ \bar{\mu}^{ - \frac{1}{2}}\:\bar{\tau}^{mn}_{~~\:\beta}\end{array}\!\! \right)^{\!\!\!\!\rm T} \left( \begin{array}{cc} \bar{\mu}^{-1}\:T_{m~k}^{~~\beta~\alpha} & T_{m~kl}^{~~\beta~~\alpha} \\ T_{mn~k}^{~~~\beta~\alpha} & \bar{\mu}\:T_{mn~kl}^{~~~\beta~~\alpha} \end{array} \right) \left(\!\!\begin{array}{c} \bar{\mu}^{\frac{1}{2}}\:\bar{\varepsilon}^k_{~\alpha} \\ \bar{\mu}^{ - \frac{1}{2}}\:\bar{\tau}^{kl}_{~~\alpha}\end{array} \!\!\right).
\end{split}
\end{equation}
Its kernel defines the desired operator $\Gamma^{(2)}_{k\;\text{grav}}$ with uniform mass dimension 1 in all block matrices (up to the global prefactor). 

The same phenomenon is encountered in the ghost sector \eqref{GammaQuadGh}. Here it is cured by a rescaling of the form $\xi^\mu \to \bar{\mu}^{1/2}\:\xi^\mu,\ \bar{\xi}_\mu \to \bar{\mu}^{1/2}\:\bar{\xi}_\mu$ and $\Upsilon^{ab} \to \bar{\mu}^{ - 1/2}\:\Upsilon^{ab},\ \bar{\Upsilon}_{ab} \to \bar{\mu}^{ - 1/2}\:\bar{\Upsilon}_{ab}\:.$

The necessity of introducing a parameter with the dimension of a mass has already been encountered previously in the literature, for instance in \cite{kreimer, zanelli}. We will comment on its physical significance below, discussing in particular the way it enters the RG equations. Like the concrete choice of the decomposition of the fluctuations and the ghosts, the $\bar{\mu}$-parameter parametrizes a family of representations of $\Gamma^{(2)}_k$, that is, it distinguishes different bases in field space. In general the flow will depend on the specific parametrization. Accidentally, However, in a single-field setting \cite{elisa2} the LHS of the FRGE is independent of the chosen representation.

\subsection{The decomposition of the fluctuation fields}
\noindent{\bf (A)} In order to analyze the spectrum of the operators $\Gamma^{(2)}_{k\:\text{grav}}$ and $\Gamma^{(2)}_{k\:\text{gh}}$ and to partially diagonalize them, the rescaled fluctuations are now decomposed according to 
\begin{equation}\label{vielbein-dec}
\bar{\mu}^{\frac{1}{2}}\:\bar{\varepsilon}^a_{~\mu} (x) = \frac{\partial^a}{\sqrt{ - \Box}} \partial_\mu a (x) + \frac{\partial_\mu}{\sqrt{ - \Box}} b^a (x) + \frac{\partial^a}{\sqrt{ - \Box}} c_\mu (x) + d^a_{~\mu} (x) 
\end{equation}
and 
\begin{equation}\label{connection-dec}
\bar{\mu}^{ - \frac{1}{2}} \bar{\tau}^{ab}_{~\mu} (x) = \frac{\partial_\mu \partial^{[a}}{\sqrt{ - \Box}^2}A^{b]} (x) + \frac{\partial^{[a}}{\sqrt{ - \Box}} B^{b]}_{~~\mu} (x) + \frac{1}{2}\varepsilon^{ab}_{~~cd}\frac{\partial_\mu \partial^c}{\sqrt{ - \Box}^2} C^d (x) + \frac{1}{2}\varepsilon^{ab}_{~~cd} \frac{\partial^c}{\sqrt{ - \Box}} D^d_{~\mu} (x)
\end{equation}
with all tensor valued component fields being transverse:
\begin{equation}
\begin{split}
\partial_a b^a = 0 = \partial^\mu c_\mu\:,~~ \partial_a d^a_{~\mu} = 0\:,~~&\partial^\mu d^a_{~\mu} = 0,\\
\partial_a A^a = 0 = \partial_a C^a\:,~~\partial_a B^a_{~\mu} = 0 = \partial_a D^a_{~\mu}\:,~~ &\partial^\mu B^a_{~\mu} = 0 = \partial^\mu D^a_{~\mu}\:.
\end{split}
\end{equation}

\noindent{\bf (B)} In a second step the tensors of second rank are further decomposed into a trace, a symmetric-traceless tensor, and an antisymmetric tensor. With $\Phi\in\{d, B, D\}$ we write
\begin{equation}\label{2ndDec}
\Phi^a_{~\mu} = \stackrel{\hspace{-0.3cm}{\rm t}}{{\Phi}^a_{~\mu}} + \stackrel{\hspace{-0.3cm}{\rm st}}{\Phi^a_{~\mu}} + \stackrel{\hspace{-0.3cm}{\rm a}}{\Phi^a_{~\mu}}
\end{equation}
with all tensors on the RHS of \eqref{2ndDec} being transverse. Moreover we have $\bar{e}^{\ \mu}_a \stackrel{\hspace{-0.3cm}{\rm st}}{\Phi^a_{~\mu}}=0$.

Expressing the trace and the antisymmetric tensor by
\begin{equation}
\stackrel{\hspace{-0.3cm}{\rm t}}{\Phi^a_{~\mu}} = \frac{(\bar{e}^a_{~\mu} \Box - \partial^a \partial_\mu)}{\sqrt{ - \Box}^2} \Phi \qquad \text{and} \qquad \stackrel{\hspace{-0.3cm}{\rm a}}{\Phi^a_{~\mu}} = \varepsilon^a_{~bcd} \bar{e}^b_{~\mu} \frac{\partial^c}{\sqrt{ - \Box}} \Phi^d\:,
\end{equation}
respectively, we are left with the symmetric-traceless part $\stackrel{\hspace{-0.3cm}{\rm st}}{\Phi^a_{~\mu}}$ as the only tensor of second rank. Therefore, after inserting this decomposition for all tensor-valued components we will drop the superscript denoting the symmetric-traceless tensors.

\noindent{\bf (C)} The ghost fields are similarly decomposed according to
\begin{equation}\label{ghost-dec}
\bar{\mu}^{\frac{1}{2}}\:\xi^\mu = \frac{\partial^\mu}{\sqrt{ - \Box}} f + g^\mu,\qquad \bar{\mu}^{\frac{1}{2}}\:\bar{\xi}_\mu = \frac{\partial_\mu}{\sqrt{ - \Box}} \bar{f} + \bar{g}_\mu 
\end{equation}
and 
\begin{equation}\label{ghost-dec2}
\bar{\mu}^{ - \frac{1}{2}}\:\Upsilon^{ab} = \frac{\partial^{[a}}{\sqrt{ - \Box}} F^{b]} + \frac{1}{2}\varepsilon^{ab}_{~~cd}\frac{\partial^{c}}{\sqrt{ - \Box}} G^{d},\quad \bar{\mu}^{ - \frac{1}{2}}\:\bar{\Upsilon}_{ab} = \frac{\partial_{[a}}{\sqrt{ - \Box}} \bar{F}_{b]} + \frac{1}{2}\varepsilon_{ab}^{~~cd}\frac{\partial_{c}}{\sqrt{ - \Box}} \bar{G}_{d}
\end{equation}
with again all vector-valued component fields being transverse.

\noindent{\bf (D)} Now, several comments are in order: 

\noindent{\bf (i)} All partial derivatives appearing in the above decompositions are rescaled by an appropriate power of $\sqrt{-\Box}$ with $\Box = \bar{g}^{\mu\nu} \partial_\mu \partial_\nu$ being a positive-definite operator in the Euclidean case. Therefore, the mass dimension of the components coincides with the mass dimension of the rescaled fluctuations and the rescaled ghost fields, respectively. 

\noindent{\bf (ii)} Small letters $\{a,b,\cdots\}$ denote the component fields of the vielbein fluctuation and the diffeomorphism ghost fields, whereas capital letters $\{A,B,\cdots\}$ have been assigned to the spin connection fluctuation and to the ${\sf O}(4)$ ghost fields. 

\noindent{\bf (iii)} The appearance of the ordinary partial derivative has the status of an approximation; it would be desirable to replace all partial derivatives $\partial_\mu$ by the full covariant background derivative $\bar{D}_\mu$ including the d'Alembertians in the denominators.

\subsection{The matrix operator $\boldsymbol\Gamma^{(2)}_k$ in the component field basis}
By inserting the above decompositions \eqref{vielbein-dec}, \eqref{connection-dec} and \eqref{ghost-dec}, \eqref{ghost-dec2} into the quadratic forms \eqref{GammaQuadGrav} and \eqref{GammaQuadGh}, respectively, we arrive at a representation of $\Gamma^{(2)}_k$ in terms of the component fields. As far as the interaction part of order ${\cal O}(\bar{\omega})$ is concerned, we do not perform the rescaling due to the inverse $\sqrt{-\Box}$-operators at this stage of the computation. Rather, along with indices $i_{\text{S}}$, $i_{\text{V}}$, $i_{\text{T}}$ labeling the scalars, the vectors and the traceless-symmetric tensors that appear in the above decompositions, we introduce numbers $a^{\text{S}}_{i_{\text{S}}}$, $b^{\text{S}}_{i_{\text{S}}}$ that denote the powers of the $\bar{\mu}$- and the $1/\sqrt{-\Box}$-rescaling for the scalar field $i_{\text{S}}$; for vectors and tensors, we proceed in complete analogy.

Instead of computing $\Gamma^{(2)}_k$ itself we focus on the so-called {\it reduced inverse propagator}, denoted $\tilde{\Gamma}^{(2)}_k$, that is connected to the former by the defining relations\pagebreak[1]
\begin{align}
\big(\Gamma^{(2)}_k\big)^{\rm S}_{{j_{\rm S}}{i_{\rm S}}} &\equiv \bar{\mu}^{a^{\rm S}_{j_{\rm S}}} \Big(\frac{1}{\sqrt{- \Box}}\Big)^{b^{\rm S}_{j_{\rm S}}}\: \big(\tilde{\Gamma}^{(2)}_k\big)^{\rm S}_{{j_{\rm S}}{i_{\rm S}}}\ \bar{\mu}^{a^{\rm S}_{{i_{\rm S}}}} \Big(\frac{1}{\sqrt{- \Box}}\Big)^{b^{\rm S}_{{i_{\rm S}}}}\:,\label{s-red}\\
\big(\Gamma^{(2)}_k\big)^{\rm V}_{{j_{\rm V}}{i_{\rm V}}\:mk} &\equiv \bar{\mu}^{a^{\rm V}_{j_{\rm V}}} \Big(\frac{1}{\sqrt{- \Box}}\Big)^{b^{\rm V}_{j_{\rm V}}}\: \big(\tilde{\Gamma}^{(2)}_k\big)^{\rm V}_{{j_{\rm V}}{i_{\rm V}}\:mk}\ \bar{\mu}^{a^{\rm V}_{{i_{\rm V}}}} \Big(\frac{1}{\sqrt{- \Box}}\Big)^{b^{\rm V}_{{i_{\rm V}}}}\label{v-red}\:,\\
\big(\Gamma^{(2)}_k\big)^{\rm T}_{{j_{\rm T}}{i_{\rm T}}\:mnkl} &\equiv \bar{\mu}^{a^{\rm T}_{j_{\rm T}}} \Big(\frac{1}{\sqrt{- \Box}}\Big)^{b^{\rm T}_{j_{\rm T}}} \:\big(\tilde{\Gamma}^{(2)}_k\big)^{\rm T}_{{j_{\rm T}}{i_{\rm T}}\:mnkl}\ \bar{\mu}^{a^{\rm T}_{{i_{\rm T}}}} \Big(\frac{1}{\sqrt{- \Box}}\Big)^{b^{\rm T}_{{i_{\rm T}}}}\label{t-red}\:.
\end{align}
Here, we have converted all spacetime indices to frame indices by means of the (inverse) background vielbein.

For the free part $\Gamma^{\text{quad}\:(0)}_k$ of order ${\cal O}(\bar{\omega}^0)$, however, the rescaling can be performed immediately, and we end up with, in the graviton sector,
\begin{equation}\label{FreeGravitonSector}
\begin{split}
&\Gamma^{{\rm quad}\:(0)}_{k\:{\rm grav}} [\bar{\varepsilon}, \bar{\tau}; \bar{e}, \bar{\omega}]_{\bar{e}={\rm const}, \bar{\omega}=0} = \frac{1}{2} \cdot \frac{1}{16 \pi G_k} \int \dr^4 x \:\bar{e} \times\\
& \times \left\{\left(\!\! \begin{array}{c} \Ss{a} \\ \Ss{d} \\ \Ss{B} \\ \Ss{D} \end{array} \!\!\right)^{\!\!\!\!\rm T} \left( \!\!\begin{array}{cccc} \Ss{- \frac{(1 + \beta_\Dr)^2}{\alpha_\Dr}\Box \bar{\mu}^{-1}} & \Ss{3\big(2 \Lambda_k  - \frac{\beta_\Dr(1+\beta_\Dr)}{\alpha_\Dr}\Box\big)\bar{\mu}^{-1}} & \Ss{0} & \Ss{0} \\ \Ss{3\big(2 \Lambda_k  - \frac{\beta_\Dr(1+\beta_\Dr)}{\alpha_\Dr}\Box\big)\bar{\mu}^{-1}} & \Ss{3\big(4\Lambda_k - 3 \frac{\beta_\Dr^2}{\alpha_\Dr}\Box\big)\bar{\mu}^{-1}} &  \Ss{6 \sqrt{ - \Box}} & \Ss{- \frac{6}{\gamma_k} \sqrt{ - \Box}} \\ \Ss{0} & \Ss{6 \sqrt{ - \Box}} & \Ss{3 \bar{\mu}} & \Ss{- \frac{3}{\gamma_k} \bar{\mu}} \\ \Ss{0} & \Ss{- \frac{6}{\gamma_k} \sqrt{ - \Box}} & \Ss{- \frac{3}{\gamma_k} \bar{\mu}} & \Ss{3 \bar{\mu}} \end{array}\!\!\right) \left( \!\!\begin{array}{c} \Ss{a} \\ \Ss{d} \\ \Ss{B} \\ \Ss{D} \end{array} \!\!\right) \right.\\
&+ \left( \!\!\begin{array}{c} \Ss{b^m} \\ \Ss{c^m} \\ \Ss{A^m} \\ \Ss{D^m} \\ \Ss{d^m} \\ \Ss{B^m} \\ \Ss{C^m} \end{array} \!\!\right)^{\!\!\!\!\rm T} \left(\!\! \begin{array}{cc@{\extracolsep{-1mm}}c@{\extracolsep{-1mm}}cc@{\extracolsep{-1mm}}c@{\extracolsep{-1mm}}c}  \Ss{\frac{1}{2\alpha_\Lr^\prime}\bar{\mu}^{-1}} & \Ss{- \big(2 \Lambda_k + \frac{1}{2\alpha_\Lr^\prime}\big)\bar{\mu}^{-1}} & \Ss{0} & \Ss{0} & \Ss{0} & \Ss{0} & \Ss{0} \\ \Ss{- \big(2 \Lambda_k + \frac{1}{2\alpha_\Lr^\prime}\big)\bar{\mu}^{-1}} & \Ss{- \big(\frac{\Box}{\alpha_\Dr} - \frac{1}{2\alpha_\Lr^\prime}\big)\bar{\mu}^{-1}} & \Ss{0} & \Ss{2 \sqrt{ - \Box}} & \Ss{0} & \Ss{- \frac{2}{\gamma_k} \sqrt{ - \Box}} & \Ss{0} \\  \Ss{0} & \Ss{0} & \Ss{0} & \Ss{\bar{\mu}} & \Ss{0} &\Ss{ - \frac{1}{\gamma_k} \bar{\mu}} & \Ss{0} \\ \Ss{0} & \Ss{2 \sqrt{ - \Box}} & \Ss{\bar{\mu}} & \Ss{\bar{\mu}} & \Ss{- \frac{2}{\gamma_k} \sqrt{ - \Box}} & \Ss{-\frac{1}{\gamma_k} \bar{\mu}} & \Ss{- \frac{1}{\gamma_k} \bar{\mu}} \\  \Ss{0} & \Ss{0} & \Ss{0} & \Ss{- \frac{2}{\gamma_k} \sqrt{ - \Box}} & \Ss{2\big(2 \Lambda_k + \frac{1}{\alpha_\Lr^\prime}\big)\bar{\mu}^{-1}} & \Ss{2 \sqrt{ - \Box}} & \Ss{0} \\ \Ss{0} & \Ss{- \frac{2}{\gamma_k} \sqrt{ - \Box}} & \Ss{- \frac{1}{\gamma_k} \bar{\mu}} & \Ss{- \frac{1}{\gamma_k} \bar{\mu}} & \Ss{2 \sqrt{ - \Box}} & \Ss{\bar{\mu}} & \Ss{\bar{\mu}} \\ \Ss{0} & \Ss{0} & \Ss{0} & \Ss{- \frac{1}{\gamma_k} \bar{\mu}} & \Ss{0} & \Ss{\bar{\mu}} & \Ss{0} \end{array}\!\!\right) \left(\!\! \begin{array}{c} \Ss{b_m} \\ \Ss{c_m} \\ \Ss{A_m} \\ \Ss{D_m} \\ \Ss{d_m} \\ \Ss{B_m} \\ \Ss{C_m} \end{array}\!\! \right) \\
&+ \left. \left(\!\! \begin{array}{c} \Ss{d^{mn}} \\ \Ss{B^{mn}} \\ \Ss{D^{mn}} \end{array}\!\! \right)^{\!\!\!\!\rm T} \left(\!\! \begin{array}{ccc} \Ss{- 2\Lambda_k \bar{\mu}^{-1}} & \Ss{- \sqrt{ - \Box}} & \Ss{\frac{1}{\gamma_k} \sqrt{ - \Box}} \\ \Ss{- \sqrt{ - \Box}} & \Ss{- \frac{\bar{\mu}}{2}} & \Ss{\frac{\bar{\mu}}{2 \gamma_k}} \\ \Ss{\frac{1}{\gamma_k} \sqrt{ - \Box}} & \Ss{\frac{\bar{\mu}}{2 \gamma_k}} & \Ss{- \frac{\bar{\mu}}{2}} \end{array}\!\!\right) \left(\!\! \begin{array}{c} \Ss{d_{mn}} \\ \Ss{B_{mn}} \\ \Ss{D_{mn}} \end{array} \!\!\right)\right\}\:. \\
\end{split}\raisetag{1.7cm}
\end{equation}
Similarly we find in the ghost sector:
\begin{multline}\label{FreeGhostSector}
\Gamma^{(0)}_{k\:{\rm gh}} [\xi, \bar{\xi}, \Upsilon, \bar{\Upsilon}; \bar{e}]\\
=  - \int \dr^4 x\: \bar{e} \left(\!\! \begin{array}{c} \bar{f} \\ \bar{g}_m \\ \bar{F}_m \\ \bar{G}_m \end{array} \!\!\right)^{\!\!\!\!\rm T} \left( \begin{array}{cccc} (1+\beta_\Dr)\Box\bar{\mu}^{-1} & 0 & 0 & 0 \\ 0 & 0 & - \frac{\sqrt{ - \Box}}{2} & 0 \\ 0 & - \frac{\sqrt{ - \Box}}{2} & \frac{\bar{\mu}}{2} & 0 \\ 0 & 0 & 0 & \frac{\bar{\mu}}{2} \end{array}\!\!\right)\left(\!\! \begin{array}{c} f \\ g^m \\ F^m \\ G^m \end{array}\!\! \right)\:. 
\end{multline}
\vspace{0.3cm}

\noindent For real $\gamma_k$ the kernel of the graviton sector \eqref{FreeGravitonSector} is Hermitean with first-order derivatives appearing in the off-diagonal elements. Up to a global prefactor its mass dimension is 1 like the mass dimension of the ghost operator. The latter becomes singular for $\beta_D=-1$ as in this limit the condition \eqref{gc-diff} does not fix the gauge completely. 

In \eqref{FreeGravitonSector}, \eqref{FreeGhostSector} we have ordered the fields according to their parity: The true scalars, vectors and tensors, respectively, whose definition involves no or an even number of Levi-Civita tensors, are followed by their pseudo-tensor counterparts with an odd number of Levi-Civita tensors entering their definition. The off-diagonal elements then couple a true tensor with a pseudo-tensor giving rise to a pseudo-scalar; all these elements are proportional to $\frac{1}{\gamma_k}$ since the Immirzi term constitutes the only pseudo-scalar in the truncation considered. 

Finally, the free part decomposes into a scalar, a vector and a tensor block. This is due to the fact that the only objects that could couple tensors of different rank at order ${\cal O}(\bar{\omega}^0)$ are the partial derivative $\partial_m$ and the Levi-Civita tensor $\varepsilon^{mkab}$. However, since all component fields are transverse and symmetric, all these contributions vanish. 

This last observation has a far reaching consequence: Since the trace of the product of a diagonal matrix and an arbitrary matrix does not change when the off-diagonal blocks are removed from the latter, it is sufficient to compute separately the scalar, the vector and the tensor block of the interaction part as well. Matrix elements that couple tensors of different rank will not contribute to the trace. Schematically we have:
\vspace{0.5cm}
\begin{equation}\label{Trace_Arg}
\begin{split}
&{\rm Tr}\left\{\left(\begin{array}{cc} a & b \\ c & d \end{array}\right)\left(\begin{array}{cc} A & 0 \\ 0 & D \end{array}\right)\right\} = {\rm Tr} \left(\begin{array}{cc} a\,A & b\,D \\ c\,A & d\,D \end{array}\right)\\[0.3cm]
&= {\rm Tr}\big(a\,A\big) + {\rm Tr}\big(d\,D\big)\\[0.3cm]
&= {\rm Tr}\left(\begin{array}{cc} a\,A & 0 \\ 0 & d\,D \end{array}\right)= {\rm Tr}\left\{\left(\begin{array}{cc} a & 0 \\ 0 & d \end{array}\right)\left(\begin{array}{cc} A & 0 \\ 0 & D \end{array}\right)\right\}\:.
\end{split}
\end{equation}
\vspace{0.3cm}

\noindent Moreover, this argument can be transferred to the sub-block structure of the vector block in the ghost sector as well. Therefore, we can use the structure of the free part in order to constrain the set of contributions of the interaction part that has to be computed.

\subsection{The interaction part}
After the computation of the free part of the quadratic form \eqref{GammaQuadGrav}, \eqref{GammaQuadGh} in terms of the component fields, $\Gamma^{\text{quad}\:(0)}_k$, we now will turn to the interaction part, $\Gamma^{\text{quad}\:(1)}_k$, that contains the contributions of the orders ${\cal O}(\bar{\omega})\equiv {\cal O}(v)$ and ${\cal O}(\partial\bar{\omega})\equiv {\cal O}(\partial v)$. Here, the $\frac{1}{\sqrt{-\Box}}$-operators, that are due to the rescalings, will not be integrated by parts; rather, they still act (on the fields) towards the left. As it turns out, this will be advantageous at a later stage of the computation.

Since we cannot explicitly write down the complete first-order part of the quadratic form here due to its sheer size, we will explain the general procedure by means of an example. Looking at \eqref{GammaQuadGrav} and neglecting the global prefactor $\frac{1}{2}\cdot\frac{1}{16 \pi G_k}$ as well as the $\frac{1}{\sqrt{-\Box}}$-rescalings, the $b^m$-$A^k$ component of the $\bar\varepsilon$-$\bar\tau$ block is given by 
\begin{multline}
\int\dr^4 x\big(\partial_\beta b^m\big) \big[ - \varepsilon^{\mu\nu\beta\alpha} K^{(\gamma_k)}_{abcm} \bar{e}^c_{~\nu} n^{(\pm)\:b}_{~~~~\:l} v_\mu\big]\partial_\alpha \partial^{[a} A^{l]}\\
= \int\dr^4 x\: b^m \big[ \varepsilon^{\mu\nu\beta\alpha} K^{(\gamma_k)}_{abcm} \bar{e}^c_{~\nu} n^{(\pm)\:b}_{~~~~\:l} (\partial_\beta v_\mu) \big]\partial_\alpha \partial^{[a} A^{l]}\:.
\end{multline}
Neglecting surface terms, we obtain for $\bar{e}^a_{\ \mu}\equiv \text{const}$:
\begin{multline}
\int\dr^4 x\: b^m \Big[\frac{1}{2} \varepsilon^{\mu\nu\beta\alpha} \big(K^{(\gamma_k)}_{abcm} n^{(\pm)\:b}_{~~~~\:l} - K^{(\gamma_k)}_{lbcm} n^{(\pm)\:b}_{~~~~\:a}\big) \bar{e}^c_{~\nu} (\partial_\beta v_\mu) \Big]\partial_\alpha \partial^{a} A^{l}\\
=\int \dr^4 x\:\frac{\bar{e}}{2} \: b^m \Big\{n^{(\pm) \:b}_{~~~~\:l}\big[\big((\partial_b v_m) - (\partial_m v_b) \big)\Box + \big((\partial_m v_a) - (\partial_a v_m) \big)\partial^a\partial_b\big]\\
\label{bAComponent}\hspace{2.3cm} - \frac{1}{\gamma_k}\big(n^{(\pm)}_{~~~\:ma} \varepsilon_{~~~l}^{prs} + \eta_{ml}n^{(\pm) \:b}_{~~~~\:a}\varepsilon_b^{~prs}\big)\big(\partial_p v_r\big) \partial^a \partial_s\Big\} A^l\:.
\end{multline}
Thereby, we have exploited the relations $\varepsilon^{\mu\nu\rho\sigma}=\bar{e}\bar{e}_{a}{}^{\mu}\bar{e}_{b}{}^{\nu}\bar{e}_{c}{}^{\rho}\bar{e}_{d}{}^{\sigma}$ and $\varepsilon_{abcd}\varepsilon^{efgh}=4!\delta^{[e}_{a}\delta^{f}_{b}\delta^{g}_{c}\delta^{h]}_{d}$ for the Levi-Civita density. Moreover, now all tensor indices have been converted into ${\sf O}(4)$ frame indices.

The final expression exhibits the following general structure: Since both $b^m$ and $A^k$ constitute true vectors, the part of the quadratic form that is independent of $\gamma_k$ forms a true scalar that does not contain a Levi-Civita tensor. In contrast, the part $\propto \frac{1}{\gamma_k}$ does so, and therefore we obtain a pseudo-scalar. Expressed in the parity-ordered representation of \eqref{FreeGravitonSector} the part $\propto \Big(\frac{1}{\gamma_k}\Big)^0$ in the diagonal blocks of the interaction part coupling tensors with tensors and pseudo-tensors with pseudo-tensors, respectively, either contains no Levi-Civita tensor or two Levi-Civita tensors; in contrast, the part $\propto \Big(\frac{1}{\gamma_k}\Big)^1$ always contains exactly one Levi-Civita tensor.

In the off-diagonal blocks that couple tensors with pseudo-tensors or vice versa the structure of the $\Big(\frac{1}{\gamma_k}\Big)^0$- and the $\Big(\frac{1}{\gamma_k}\Big)^1$-contributions is just reversed. Therefore, all contributions $\propto \Big(\frac{1}{\gamma_k}\Big)^0$ constitute scalars whereas all contributions $\propto \Big(\frac{1}{\gamma_k}\Big)^1$ are pseudo-scalars.

Furthermore, when discussing \eqref{bAComponent} a further subtlety should be pointed out: Since all fields are transverse, a zero direction of the form
\begin{equation}
\int\dr^4 x\:\bar{e}\:b^m \big[\cdots \big((\partial_m v_a) + v_a \partial_m\big) \big] A^k = \int\dr^4 x\:\bar{e} (\partial_m b^m) \big[ - \cdots  v_a \big] A^k = 0\:.
\end{equation}
can be added to the quadratic form. Stated differently: At the level of the quadratic form, the expressions $\int\dr^4 x\:\bar{e}\:b^m \big[\cdots (\partial_m v_a) \big] A^k$ and $\int\dr^4 x\:\bar{e}\:b^m \big[\cdots - v_a \partial_m \big] A^k$ are equivalent up to a partial integration. If such an ambiguity arises in one component, there is always a corresponding ambiguity in the component at the transposed position, as well.

In order to define a Hermitean operator $\Gamma^{(2)}_k$, we have to fix a convention about how to deal with such ambiguities: We agree on always writing them in the first way, i.\,e. such that the partial derivative always acts on the vector field $v_\mu$.\footnote{By inserting transverse projectors to the left and to the right of the kernel, this equivalence could be transformed immediately to the level of the components of the associated operator.}

We have checked the hermiticity of the resulting quadratic form for each component up to ${\cal O}(\partial v)$. Therefore, we now have a self-adjoint operator $\Gamma^{(2)}_k$ at our disposal, that will enter the RHS of the functional flow equation \eqref{PTIntegrals}. The explicit implementation will be explained in the next chapter.

\vspace{1cm}
\section{Block Structure of the Flow Equation}

\subsection{A proper-time equation for first-order gravity}

Suppressing all indices, the inverse (effective) propagator can be decomposed as
\begin{equation}
 \Gamma^{(2)}_k = H + V = H_0 + \frac{1}{\gamma_k} \bar{H}+V
\end{equation}
with $H_0$ denoting the $\gamma_k$-independent, parity-even, block-diagonal part of the free (${\cal O}(v^0)$) propagator 
\begin{equation}
H\equiv H_0 + \frac{1}{\gamma_k} \bar{H}\:.
\end{equation}
The parity-odd part is given by $\bar{H}$. It contains only entries in the off-diagonal blocks, and since all pseudo-scalar contributions that occur in the above quadratic form \eqref{FreeGravitonSector} are proportional to $\frac{1}{\gamma_k}$, $\bar{H}$ is independent of $\gamma_k$, as well.

The gauge fixing term is a scalar, and therefore it only contributes to $H_0$. Furthermore, since the ghost sector is block-diagonal in the parity-ordered representation, too, we have $\bar{H}^{\text{gh}}\equiv 0$. The interaction part $V$ contains all contributions of ${\cal O}(v)$ and ${\cal O}(\partial v)$. They are independent of $\gamma_k$ or linear in $\frac{1}{\gamma_k}$ with parity-even and parity-odd couplings appearing in both the diagonal and the off-diagonal blocks.

Now, $H_0$ can be further decomposed into blocks according to 
\begin{equation}
 H_0 = H_0^\text{S} \oplus H_0^\text{V\;I}\oplus H_0^\text{V\;II}\oplus H_0^\text{T}\oplus H_0^\text{gh\;S}\oplus H_0^\text{gh\;V}
\end{equation}
with the labels S, V and T meaning that the respective block acts on scalar-, vector-, or tensor-valued fields. Here, we have already indicated that the graviton vector block decouples into two separate sub-blocks V\;I and V\;II; the Grassmann-valued part of the field space that is embodied by the ghost fields is treated separately. The respective numbers of independent field components are given by $d_\text{S}\equiv d_\text{gh\;S}=1$, $d_\text{V}\equiv d_\text{gh\;V}=3$ and $d_\text{T}=5$, since all tensors are transverse, and in addition the ones of rank two are symmetric and traceless.

After the $\bar \mu$-rescaling, the graviton fluctuations have mass-dimension $\frac{1}{2}$, whereas the rescaled ghost fields have mass dimension $\frac{3}{2}$. Therefore, $\Gamma^{(2)}_{k\,\text{grav}}$ has mass-dimension 3, and $\Gamma^{(2)}_{k\;\text{ghost}}$ has mass-dimension 1 so that an appropriate rescaling has to be incorporated into the trace and the RHS of \eqref{proper-time-equation}.\footnote{The delta distribution $\delta^{(4)} (x-y)$ that occurs due to the functional derivatives and that has mass-dimension 4 according to our convention is not counted, since the trace Tr contains a compensating space-time integration.} In order to perform this rescaling, we again employ the mass parameter $\bar\mu$, leading to 
\begin{equation}
 \Gamma_k - S= \frac{1}{2} \Big\{{\rm Tr}\,{\rm ln}\Big(\frac{\Gamma^{(2)}_{k\:{\rm grav}}}{\bar{\mu}^3}\Big) - 2\,{\rm Tr}\,{\rm ln}\Big(\frac{\Gamma^{(2)}_{k\:{\rm gh}}}{\bar{\mu}}\Big)\Big\}\:. \label{1loop-impr}
\end{equation}
We expand the graviton trace of \eqref{1loop-impr} according to
\begin{equation}
 \begin{split}
  &\frac{1}{2} {\rm Tr}\,{\rm ln}\Big(\frac{H^{\rm grav} + V^{\rm grav}}{\bar{\mu}^3}\Big) = \frac{1}{2} {\rm Tr}\,{\rm ln}\Big[\frac{H^{\rm grav}}{\bar{\mu}^3}\Big(\openone + (H^{\rm grav})^{-1} V^{\rm grav}\Big)\Big] \nonumber\\
 &=  \frac{1}{2} {\rm Tr}\,{\rm ln}\Big(\frac{H^{\rm grav}}{\bar{\mu}^3}\Big) + \frac{1}{2} {\rm Tr}\,{\rm ln}\Big[\openone + (H^{\rm grav})^{-1} V^{\rm grav}\Big]\hspace{7cm}\phantom{x}
 \end{split}
\end{equation}
 
\begin{equation}\label{1loop_expanded}
 \begin{split}
&= \frac{1}{2} {\rm Tr}\,{\rm ln}\Big[\frac{H^{\rm grav}_0}{\bar{\mu}^3}\Big\{\openone + \frac{1}{\gamma_k} (H^{\rm grav}_0)^{-1} \bar{H}^{\rm grav}\Big\}\Big] + \frac{1}{2} {\rm Tr}\Big( (H^{\rm grav})^{-1} V^{\rm grav}\Big) + {\cal O}(\bar{\omega}^2)\\
&= \frac{1}{2} {\rm Tr}\,{\rm ln}\Big(\frac{H^{\rm grav}_0}{\bar{\mu}^3}\Big) + \frac{1}{2} {\rm Tr}\,{\rm ln}\Big(\openone + \frac{1}{\gamma_k} {\cal M} \Big) + \frac{1}{2} {\rm Tr}\Big( (H^{\rm grav})^{-1} V^{\rm grav}\Big) + {\cal O}(\bar{\omega}^2)\\
&= \frac{1}{4} {\rm Tr}\,{\rm ln}\Big(\frac{(H^{\rm grav}_0)^2}{\bar{\mu}^6}\Big)\! +\! \frac{1}{2} {\rm Tr}\,{\rm ln}\Big(\!\openone\! +\! \frac{1}{\gamma_k} {\cal M} \!\Big)\! +\! \frac{1}{2} {\rm Tr}\Big[(H^{\rm grav}_0)^2 (H^{\rm grav})^{-1} V^{\rm grav}(H^{\rm grav}_0)^{-2}\Big]\! +\! {\cal O}(\bar{\omega}^2).
 \end{split}\raisetag{2.3cm}
\end{equation}
Here we see that the dependence on the Immirzi parameter is controlled by the operator
\begin{equation}
 {\cal M} \equiv  (H^{\rm grav}_0)^{-1} \bar{H}^{\rm grav}\nonumber\:.
\end{equation}
(The fact that the traces in the first step of \eqref{1loop_expanded} can be separated even for non-commuting operators follows from $ \delta \Tr \ln A = \Tr ( A^{-1} \delta A )$
which holds true even if $[A,\delta a]\neq0$.) Note that we have introduced a factor of $(H_0^{\rm grav})^2$ in the last term of the bottom line of \eqref{1loop_expanded}: This is needed as we are going to regularize each of the traces by a proper-time cutoff w.\,r.\,t. the spectrum of this positive definite operator (for more details see below).

For the ghost part we proceed analogously. With $H^{\text{gh}}\equiv H_0^{\text{gh}}$ we obtain
\begin{equation}
\begin{split}
 \Gamma_k  - S &= \frac{1}{4} {\rm Tr}\,{\rm ln}\bigg(\frac{(H^{\rm grav}_0)^2}{\bar{\mu}^6}\bigg) + \frac{1}{2} {\rm Tr}\,{\rm ln}\Big(\openone + \frac{1}{\gamma_k} {\cal M} \Big)\\&
+ \frac{1}{2} {\rm Tr}\Big((H^{\rm grav}_0)^2 (H^{\rm grav})^{-1} V^{\rm grav}(H^{\rm grav}_0)^{-2}\Big) \\
\displaystyle & - 2 \bigg\{ \frac{1}{4} {\rm Tr}\,{\rm ln}\bigg(\frac{(H^{\rm gh}_0)^2}{\bar{\mu}^2}\bigg) + \frac{1}{2} {\rm Tr}\Big((H^{\rm gh}_0) V^{\rm gh}(H^{\rm gh}_0)^{-2}\Big) \bigg\}
\end{split}\label{unser-fluss}
\end{equation}
This final expression demands for a regularization of these otherwise not well-defined traces. In a next step, the scale derivative of the regularized RHS of \eqref{unser-fluss} is used to define the RHS of the desired flow equation on ${\cal T}_{\text{QECG}}$, which will be analyzed in the next chapter. 

Since we want to compute all traces in eq. \eqref{unser-fluss} (except for the second one) in the eigenbasis of $H_0$, we first have to determine the spectrum of this operator. This is the topic of the next subsection.

\subsection{The block structure of $H_0$ and its spectrum}
Due to the threefold index structure of $\Gamma_k^{(2)}$, $H_0$ is obtained as the tensor product of three components:

\vspace{0.5cm}
\noindent{\bf (i)} $H_0$ is a matrix operator in field space with entries depending on $\sqrt{-\Box}$. The corresponding indices $i_\text{S}$, $i_{\text{V\,I}}$, $i_{\text{V\,II}}$, $i_{\text{T}}$, $i_{\text{gh\,S}}$ and $i_{\text{gh\,V}}$ enumerate all types of fields in the corresponding sector. 

\vspace{0.5cm}
\noindent{\bf (ii)} Concerning its ${\sf O}(4)$ index structure, $H_0$ is diagonal as can be seen from \eqref{FreeGravitonSector}. However, since all vector- and tensor-valued fields are transverse, the identity matrix on these subspaces is not given by the full Kronecker delta but rather by the corresponding transverse projector. Stated differently: The transversality of the fields in the quadratic form has to be taken into account by inserting appropriate transverse projectors. We denote them by $P_{\text{S}}$, $P_{\text{V}}$ and $P_{\text{T}}$ for the respective sector; their explicit form will be introduced below. 

\vspace{0.5cm}
\noindent{\bf (iii)} Finally, as was already mentioned, a tensorial delta function arises. 

\vspace{0.5cm}
The blocks constituting $H_0$ are then obtained as follows. For the graviton, they read as follows in the scalar, first and second vector, and the tensor sector, respectively.

\vspace{0.5cm}
\noindent{\bf Scalar Sector:}
\begin{equation}
\begin{split}
 H_0^{\rm S} &\stackrel{\wedge}{=} \big(H_0^{\rm S}\big)^{i_{\rm S}}_{~j_{\rm S}} (- \Box_x)\:P_{\rm S}\:\frac{\delta^{(4)} (x - y)}{\bar{e}}\\
&= {\frac{1}{16 \pi G_k}}\left({\setlength\arraycolsep{0.1em} \begin{array}{cccc} \Ss{- \frac{(1 + \beta_\Dr)^2}{\alpha_\Dr}\Box_x \bar{\mu}^{-1}} & \Ss{3\big(2 \Lambda_k  - \frac{\beta_\Dr(1+\beta_\Dr)}{\alpha_\Dr}\Box_x \big)\bar{\mu}^{-1}} & \Ss{0} & \Ss{0} \\\Ss{3\big(2 \Lambda_k  - \frac{\beta_\Dr(1+\beta_\Dr)}{\alpha_\Dr}\Box_x \big)\bar{\mu}^{-1}} & \Ss{3\big(4\Lambda_k - 3 \frac{\beta_\Dr^2}{\alpha_\Dr}\Box_x \big)\bar{\mu}^{-1}} &  \Ss{6 \sqrt{ - \Box_x}} & \Ss{0}  \\ \Ss{0} & \Ss{6 \sqrt{ - \Box_x}} & \Ss{3 \bar{\mu}} & \Ss{0} \\ \Ss{0} & \Ss{0} & \Ss{0} & \Ss{3 \bar{\mu}} \end{array}}\right)P_{\rm S}\:\frac{\delta^{(4)} (x - y)}{\bar{e}}
\end{split}
\end{equation}

\noindent{\bf First Vector Sector:}
\begin{equation}
\begin{split}
 H_0^{{\rm V\;I}} &\stackrel{\wedge}{=} \big(H_0^{{\rm V\;I}}\big)^{i_{{\rm V\;I}}}_{~j_{{\rm V\;I}}} (- \Box_x)\:P_{{\rm V}~k}^{~m}\:\frac{\delta^{(4)} (x - y)}{\bar{e}}\\
&= \frac{1}{16 \pi G_k} \left( \begin{array}{cccc}  \frac{1}{2\alpha_\Lr^\prime}\bar{\mu}^{-1} & - \big(2 \Lambda_k + \frac{1}{2\alpha_\Lr^\prime}\big)\bar{\mu}^{-1} & 0 & 0 \\ - \big(2 \Lambda_k + \frac{1}{2\alpha_\Lr^\prime}\big)\bar{\mu}^{-1} & - \big(\frac{\Box_x}{\alpha_\Dr} - \frac{1}{2\alpha_\Lr^\prime}\big)\bar{\mu}^{-1} & 0 & 2 \sqrt{ - \Box_x} \\  0 & 0 & 0 & \bar{\mu} \\ 0 & 2 \sqrt{ - \Box_x} & \bar{\mu} & \bar{\mu} \end{array}\right)P_{{\rm V}~k}^{~m}\:\frac{\delta^{(4)} (x - y)}{\bar{e}}
\end{split}\label{VBI}\raisetag{0.8cm}
\end{equation}

\vspace{0.5cm}
\noindent{\bf Second Vector Sector:}
\begin{equation}
\begin{split}
 H_0^{{\rm V\;II}} &\stackrel{\wedge}{=} \big(H_0^{{\rm V\;II}}\big)^{i_{{\rm V\;II}}}_{~j_{{\rm V\;II}}} (- \Box_x)\:P_{{\rm V}~k}^{~m}\:\frac{\delta^{(4)} (x - y)}{\bar{e}}\\
&= \frac{1}{16 \pi G_k} \left( \begin{array}{ccc}   2\big(2 \Lambda_k + \frac{1}{\alpha_\Lr^\prime}\big)\bar{\mu}^{-1} & 2 \sqrt{ - \Box_x} & 0 \\ 2 \sqrt{ - \Box_x} & \bar{\mu} & \bar{\mu} \\ 0 & \bar{\mu} & 0 \end{array}\right)P_{{\rm V}~k}^{~m}\:\frac{\delta^{(4)} (x - y)}{\bar{e}}
\end{split}\label{VBII}
\end{equation}

\vspace{0.5cm}
\noindent{\bf Tensor Sector:}
\begin{equation}
\begin{split}
 H_0^{{\rm T}} &\stackrel{\wedge}{=} \big(H_0^{{\rm T}}\big)^{i_{{\rm T}}}_{~j_{{\rm T}}} (- \Box_x)\:P_{{\rm V}~~kl}^{~mn}\:\frac{\delta^{(4)} (x - y)}{\bar{e}}\\
&= \frac{1}{16 \pi G_k} \left( \begin{array}{ccc} - 2\Lambda_k \bar{\mu}^{-1}& - \sqrt{ - \Box_x} & 0 \\ - \sqrt{ - \Box_x} & - \frac{\bar{\mu}}{2} & 0 \\ 0 & 0 & - \frac{\bar{\mu}}{2} \end{array}\right)P_{{\rm T}~~kl}^{~mn}\:\frac{\delta^{(4)} (x - y)}{\bar{e}}
\end{split}
\end{equation}

\vspace{0.5cm}
\noindent For the ghosts the result in the scalar and vector sector, respectively, is as follows.

\vspace{0.5cm}
\noindent{\bf Scalar Sector:}
\begin{equation}
\begin{split}
H_0^{{\rm gh}\,{\rm S}} &\stackrel{\wedge}{=} \big(H_{0}^{{\rm gh}\,{\rm S}}\big)^{i_{{\rm gh}\,{\rm S}}}_{~j_{{\rm gh}\,{\rm S}}} (- \Box_x)\:P_{\rm S}\:\frac{\delta^{(4)} (x - y)}{\bar{e}}= - (1+\beta_\Dr)\Box_x \, \bar{\mu}^{-1}\:P_{\rm S}\:\frac{\delta^{(4)} (x - y)}{\bar{e}}
\end{split}
\end{equation}

\noindent{\bf Vector Sector:}
\begin{equation}
\begin{split}
H_{0}^{{\rm gh}\,{\rm V}} &\stackrel{\wedge}{=} \big(H_{0}^{{\rm gh}\,{\rm V}}\big)^{i_{{\rm gh}\,{\rm V}}}_{~j_{{\rm gh}\,{\rm V}}} (- \Box_x)\:P_{{\rm V}~k}^{~m}\:\frac{\delta^{(4)} (x - y)}{\bar{e}}\\
&= - \left( \begin{array}{ccc} 0 & - \frac{\sqrt{ - \Box_x}}{2} & 0 \\ - \frac{\sqrt{ - \Box_x}}{2} & \frac{\bar{\mu}}{2} & 0 \\ 0 & 0 & \frac{\bar{\mu}}{2} \end{array}\right) P_{{\rm V}~k}^{~m}\:\frac{\delta^{(4)} (x - y)}{\bar{e}}
\end{split}
\end{equation}

This block structure has a representation-theoretic background \cite{kuhfussnitsch,vN}. The eigenvectors of $H_0$ will, of course, exhibit an analogous threefold tensor product structure. Viewed as identity operators on the subspaces perpendicular to the longitudinal direction, the transverse  projectors will also appear in the inverse operator $(H_0)^{-1}$ that appears on the RHS of \eqref{proper-time-equation}. 

\vspace{1cm}
\subsection{The spectrum of $H_0$}
With $\Box = \bar{g}^{\mu\nu}\partial_\mu \partial_\nu= \eta^{ab}\partial_a\partial_b$ and $\bar{e}^a_{\ \mu}=\kappa \delta^a_{\ \mu}= \text{const}$ so that $\det(e^a_{\ \mu}) = \kappa^4$, $g_{\mu\nu}=\kappa^2 \eta_{\mu\nu}$ the eigenfunctions of $\Box$ are given by the (normalized) plane waves $\frac{1}{(2 \pi)^2 }e^{i \kappa p x}$. Therein, $p x\equiv p^\mu x_\mu$ denotes the standard Minkowski inner product (without any further factors of $\kappa$!). Then the associated eigenvalue of $-\Box$ is $p^2= p^\mu p^\nu \eta_{\mu\nu}$. This can be inferred by carefully keeping track of the powers of $\kappa$:

With $\partial_\mu\equiv \bar{\partial}_\mu=\frac{\partial}{\partial x^\mu}$, let $\{i \bar{p}_\mu, i \bar{p}^\mu, i \bar{p}_a, i\bar{p}^a\}$ denote the eigenvalues of the operators $\{\bar{\partial}_\mu,\bar{\partial}^\mu,\bar{\partial}_a,\bar{\partial}^a\}$, whereby the last three ones are obtained by acting with $\bar{g}^{\mu\nu}$, $\bar{e}^a_{\ \mu}$, and $\bar{e}_a^{\ \mu}$ on $\bar{\partial}_\mu$. Then we have $\bar{p}_\mu= \kappa p_\mu$, $\bar{p}^\mu = \bar{g}^{\mu\nu}\bar{p}_{\mu}=\kappa^{-1} p^\mu$, $\bar{p}_a=\bar{e}_a^{\ \mu}\bar{p}_\mu=p_a$ and $\bar{p}^a=\bar{e}^a_{\ \mu} \bar{p}^{\mu}=p^a$ with $p^\mu=\eta^{\mu\nu} p_{\nu}$, $p_a= \delta_a^{\ \mu} p_\mu$, and $p^a=\delta^{a}_{\ \mu} p^\mu = \eta^{ab} p_b$. The eigenvalue of $\Box$ is therefore obtained as $-\bar{p}^\mu \bar{p}_\mu = - \bar{p}^a \bar{p}_a= - p^\mu p_\mu=-p^a p_a$.

The completeness relation of these eigenfunctions is given by
\begin{equation}
 \label{ebeneWellen-Vollst}
\int\frac{\dr^4 p}{(2 \pi)^4}\:\eu^{i \kappa p (x - y)} = \frac{1}{\kappa^4}\int\:\frac{\dr^4 p}{(2 \pi)^4}\:\eu^{i p (x-y)} = \frac{\delta^{(4)}(x-y)}{\bar{e}}\:.
\end{equation}

All derivatives that occurred in our computations up to now correspond to the set $\{\bar{\partial}_\mu,\bar{\partial}^\mu,\bar{\partial}_a,\bar{\partial}^a\}$. Since we have converted all indices to Lorentz indices, we are in particular allowed to replace $\bar{\partial}_a \rightarrow i p_a$, $\bar{\partial}^a\rightarrow i p^a$. From now on, we will, of course, again drop the bars on the operators and their eigenvalues.

\subsubsection{The spectrum of the algebraic part of $H_0$}
Acting with the matrices  $(H_0^{\rm S})^{i_{\rm S}}_{~j_{\rm S}}$, $(H_0^{{\rm V}I})^{i_{{\rm V}I}}_{~j_{{\rm V}I}}$, $(H_0^{{\rm V}II})^{i_{{\rm V}II}}_{~j_{{\rm V}II}}$, $(H_0^{\rm T})^{i_{\rm T}}_{~j_{\rm T}}$, $(H_{0}^{{\rm gh}\,{\rm S}})^{i_{{\rm gh}\,{\rm S}}}_{~j_{{\rm gh}\,{\rm S}}}$ and $(H_{0}^{{\rm gh}\,{\rm V}})^{i_{{\rm gh}\,{\rm V}}}_{~j_{{\rm gh}\,{\rm V}}}$ that contain Laplacians on the plane waves just introduced leads to the substitution $-\Box \rightarrow p^2$, $\sqrt{-\Box}\rightarrow p \equiv \sqrt{p^2}$.\footnote{In contrast to Lorentzian signature, the operator $-\Box$ is positive-semidefinite in the Euclidean setting considered here. Therefore, in a Lorentzian treatment the proper-time equation would have to be based on a Fourier representation instead of a Laplace representation leading to Fresnel integrals instead of Gauss integrals. However, in the Lorentzian case there are no real selfdual or anti-selfdual fields of the kind of the chosen background configuration $\bar{\omega}^{(\pm)}$.}

The determination of the spectrum of $H_0$ amounts to a purely algebraic problem that can be solved analytically. We obtain the eigenvalues $\bar{\lambda}_{\alpha_{\rm S}}^{\rm S} (p^2)$, $\bar{\lambda}_{\alpha_{\rm V}}^{\rm V} (p^2)$, $\bar{\lambda}_{\alpha_{\rm T}}^{\rm T} (p^2)$, $\bar{\lambda}_{\alpha_{{\rm gh}\,{\rm S}}}^{{\rm gh}\,{\rm S}} (p^2)$ and $\bar{\lambda}_{\alpha_{{\rm gh}\,{\rm V}}}^{{\rm gh}\,{\rm V}} (p^2)$ that have (again up to a global prefactor $\frac{1}{16 \pi G_k}$ in the graviton sector) mass dimension 1. Thereby, the indices ${\alpha_{\rm S}}$, ${\alpha_{\rm V}}$, ${\alpha_{\rm T}}$, $\alpha_{{\rm gh}\,{\rm S}}$ and $\alpha_{{\rm gh}\,{\rm V}}$ enumerate the different eigenvalues within each block; their range is identical to the one of the indices $i_{\rm S},\cdots, i_{\rm gh\,V}$ denoted by  $n_{\rm S} = 4$, $n_{\rm V} = 7$, $n_{\rm T} = 3$, $n_{{\rm gh}\,{\rm S}} = 1$ and $n_{{\rm gh}\,{\rm V}} = 3$. Finally, we introduce a symbolic index $\chi$ for the different types of blocks, i.\,e. $\chi \in \{{\rm S}, {\rm V}, {\rm T}, {{\rm gh}\,{\rm S}}, {{\rm gh}\,{\rm V}}\}$.

Knowing the eigenvalues, the determination of the eigenvectors amounts to the solution of a system of linear equations. The corresponding eigenvectors are denoted by $v_{~\:{\alpha_\chi}}^\chi$ with components $\big(v_{~\:{\alpha_\chi}}^\chi)^{i_\chi} (p^2)$ and $i_\chi, \alpha_\chi \in \{1, \cdots, n_\chi\}$. The blockwise eigenvalue equation reads
\begin{equation}
 \big(H^\chi_0\big)^{i_\chi}_{~j_\chi} (p^2)\times \big(v_{~\:\alpha_\chi}^\chi)^{j_\chi} (p^2) = \bar{\lambda}_{\alpha_\chi}^\chi (p^2)\times \big(v_{~\:\alpha_\chi}^\chi)^{i_\chi} (p^2)\:,\hspace{0.5cm}\alpha_\chi \in \{1, \cdots, n_\chi\}\:.
\end{equation}

For each block, the eigenvectors form a basis of $\mathbbm{R}^{n_\chi}$. The associated completeness and orthogonality relations are given by
\begin{equation}
 \sum_{\alpha_\chi = 1}^{n_\chi}\big(v^{\chi\:{\alpha_\chi}}\big)_{i_\chi} \big(v^\chi_{~\:{\alpha_\chi}}\big)^{j_\chi} = \delta_{i_\chi}^{~j_\chi} 
\qquad \text{and}\qquad 
\sum_{i_\chi = 1}^{n_\chi}\big(v^{\chi\:{\alpha_\chi}}\big)_{i_\chi} \big(v^\chi_{~\:{\beta_\chi}}\big)^{i_\chi} = \delta^{\alpha_\chi}_{~{\beta_\chi}}
\end{equation}
with $\{\delta_{i_\chi}^{~j_\chi}\} \equiv \openone_{n_\chi \times n_\chi}$. Moreover, we have introduced the tensor $\eta^{{\alpha_\chi}{\beta_\chi}}$ with $\{\eta^{{\alpha_\chi}{\beta_\chi}}\} = {\rm diag} (\underbrace{+1, \cdots, +1}_{n_\chi})$ in order to define the dual vector $v^{\alpha_\chi} \equiv  \eta^{{\alpha_\chi}{\beta_\chi}}\, v_{\beta_\chi}$. The matrix elements of the transition map from the base to the dual base and vice versa are then given by
\begin{equation}
 \big(v^\chi_{~\:{\alpha_\chi}}\big)^{i_\chi} \equiv \langle i_\chi\,|\,\alpha_\chi \rangle \qquad\text{and}\qquad \big(v^{\chi\:{\alpha_\chi}}\big)_{i_\chi} \equiv \langle \alpha_\chi\,|\,i_\chi \rangle\:.
\end{equation}

\subsubsection{The transverse projectors and their eigenvectors}
In this subsection we determine the projectors $P_S$, $P_V$, and $P_T$ needed to block-diagonalize $H_0$.

Starting with the scalar sector, we note that the dimension of the scalar subspace is $d_{\rm S}=1$, so we (trivially) have $P_{\rm S}=1$.

The transverse projector $P_{\rm V}$ in the vector sector fulfills
\begin{equation}
 \bigg[\frac{\partial^m \partial_k}{\Box} + P_{{\rm V}~k}^{~m} \big(-i \hat{\partial}\big)\bigg]\frac{\delta^{(4)} (x - y)}{\bar{e}} = \delta^m_{~k} \frac{\delta^{(4)} (x - y)}{\bar{e}}
\end{equation}
with $\hat{\partial}=\frac{ \partial}{\sqrt{ - \Box}}$ being the rescaled partial derivative. Inserting the completeness relation \eqref{ebeneWellen-Vollst} for the tensorial delta distribution leads to the momentum representation of the transverse projector:
\begin{equation}
 P_{{\rm V}~k}^{~m} \big(\!\!-\!i \hat{\partial_x}\big)\!\! \int\!\! \frac{\dr^4 p}{(2 \pi)^4} \eu^{i \kappa p (x - y)} = \!\!\int\!\!\frac{\dr^4 p}{(2 \pi)^4} \Big(\delta^m_{~k} \!-\! \frac{\partial^m \partial_k}{\Box}\Big) \eu^{i \kappa p (x - y)}
= \!\!\int\!\!\frac{\dr^4 p}{(2 \pi)^4} \Big(\delta^m_{~k}\! -\! \frac{p^m p_k}{p^2}\Big) \eu^{i \kappa p (x - y)}\nonumber
\end{equation}
Thus the momentum representation of the transverse projector reads
\begin{equation}
 P_{{\rm V}~k}^{~m} (\hat{p}) \equiv \delta^m_{~k} - \frac{p^m p_k}{p^2}
\end{equation}
with $\hat{p}^m\equiv  p^m/p$. Its eigenvectors spanning the subspace of transverse vectors with dimension $d_{\rm V}=3$ are denoted by $t_I^{~m} (\hat{p})$, $I \in \{1, 2, 3\}$; like polarization vectors, they can be assumed to be real without loss of generality. Their completeness relation then can be written as the sum of a longitudinal and a transverse part:
\begin{equation}
 \delta^m_{~k} = \hat{p}^m \hat{p}_k + t_I^{~m} (\hat{p}) t^{I}_{\ k} (\hat{p})\:.
\end{equation}
On the transverse subspace, we introduce tensors $\eta_{IJ}$, $\eta^{IJ}$, $\delta^I_{~J}$, $\delta_I^{~J}$ all being numerically equal to diag$(+1,+1,+1)$ here.

Finally, in the momentum representation the projector onto transverse-traceless, symmetric tensors $(d_{\rm T}=5)$ is given by 
\begin{equation}
P_{{\rm T}~~kl}^{~mn}\equiv  \frac{1}{2}\Big(P_{{\rm V}~k}^{~m}P_{{\rm V}~l}^{~n} + P_{{\rm V}~l}^{~m}P_{{\rm V}~k}^{~n} - \frac{2}{3}P_{\rm V}^{~mn}P_{{\rm V}kl}\Big)
\end{equation}
with eigenvectors
\begin{equation}
 \frac{1}{2} \Big(t_I^{~m} t_J^{~n} + t_I^{~n} t_J^{~m} - \frac{2}{3} \,t^{Km} t_K^{~n} \eta_{IJ}\Big)\:.
\end{equation}

\vspace{1cm}
\subsection{Generalized position- and momentum-representation}
Having analyzed the spectrum of $H_0$ in all graviton and ghost sectors, we have a transition map from a generalized position space to a generalized momentum space at our disposal. For example, the graviton vector block $H_0^{\rm V}$ in \eqref{VBI}, \eqref{VBII} was expressed in terms of the position representation according to 
\vspace{0.3cm}
\begin{equation}
 \langle x\, m\, i_{\rm V}| H_0^{\rm V} |y\, k\, j_{\rm V} \rangle = \big(H_0^{\rm V}\big)^{i_{\rm V}}_{~j_{\rm V}} (- \Box_x)\: P_{{\rm V}~k}^{~m} \big(- i \hat{\partial}_x\big) \frac{\delta^{(4)}(x-y)}{\bar{e}}\:.
\end{equation}
\vspace{0.05cm}

\noindent More generally, the Hilbert space on which $H_0$ operates has a tensor product structure spanned by the basis kets $\big\{|\,x \,i_{\rm S}\rangle,\:|\,x \,m\,i_{\rm V}\rangle,\:|\,x \,m\,n\,i_{\rm T}\rangle,\:|\,x \,i_{{\rm gh}\,{\rm S}}\rangle,\:|\,x \,m \,i_{{\rm gh}\,{\rm V}}\rangle \big\}$. The quantum numbers that classify the $H_0$ eigenvectors define another basis $\{p, \alpha_{\rm S}\}$, $\{p, I, \alpha_{\rm V}\}$, $\{p, I, J, \alpha_{\rm T}\}$, $\{p, \alpha_{{\rm gh}\,{\rm S}}\}$ and $\{p, I, \alpha_{{\rm gh}\,{\rm V}}\}$. The transition matrix elements are given by the following projections (the ``wavefunctions'')\footnote{The occurrence of complex plane waves does not affect our goal to express everything in terms of real fields. After having exploited their property of being eigenfunctions of $\Box$ and their completeness, they do not explicitly enter our computation anymore.}:
\begin{equation}
 \begin{split}
  \langle x\, i_{\rm S} | p \,\alpha_{\rm S} \rangle &= \big(v^{\rm S}_{~\:\alpha_{\rm S}}\big)^{i_{\rm S}}(p^2) \,\frac{\eu^{i \kappa p x}}{(2 \pi)^2}\;,\\
  \langle x\, m\, i_{\rm V} | p\,I\,\alpha_{\rm V} \rangle &= \big(v^{\rm V}_{~\:\alpha_{\rm V}}\big)^{i_{\rm V}} (p^2) t_I^{~m}(\hat{p})\frac{\eu^{i \kappa p x}}{(2 \pi)^2}\;,\\
  \langle x\, m\,n\, i_{\rm T} | p\,I\,J\,\alpha_{\rm T} \rangle &= \big(v^{\rm T}_{~\:\alpha_{\rm T}}\big)^{i_{\rm T}} (p^2) \frac{1}{2}\Big(t_I^{~m}(\hat{p}) t_J^{~n}(\hat{p})\! +\! t_I^{~n}(\hat{p}) t_J^{~m}(\hat{p})\! -\! \frac{2}{3}P_{\rm V}^{~mn}(\hat{p})\eta_{IJ}\!\Big)\frac{\eu^{i \kappa p x}}{(2 \pi)^2}\;,\\
  \langle x\, i_{{\rm gh}\,{\rm S}} | p \,\alpha_{{\rm gh}\,{\rm S}} \rangle &= \big(v^{{\rm gh}\,{\rm S}}_{~~~~\alpha_{{\rm gh}\,{\rm S}}}\big)^{i_{{\rm gh}\,{\rm S}}} (p^2)\,\frac{\eu^{i \kappa p x}}{(2 \pi)^2}\;,\\
  \langle x\, m \, i_{{\rm gh}\,{\rm V}} | p \,I\,\alpha_{{\rm gh}\,{\rm V}} \rangle &= \big(v^{{\rm gh}\,{\rm V}}_{~~~~\alpha_{{\rm gh}\,{\rm V}}}\big)^{i_{{\rm gh}\,{\rm V}}}(p^2)\,t_I^{~m}(\hat{p})\frac{\eu^{i \kappa p x}}{(2 \pi)^2}\;.
 \end{split}\raisetag{1cm}
\end{equation}
The corresponding completeness relations of the momentum and position eigenvectors are given in Appendix \ref{App:Completeness}.

As an example, let us consider the the graviton vector block of the eigenvalue equation. It can then be written as
\begin{align}
 \langle x\,m\,i_{\rm V}\,|H_0^{\rm V}\,&|\,y\,k\,j_{\rm V}\rangle \langle y\,k\,j_{\rm V}\,|\,p\,I\,\alpha_{\rm V}\rangle\nonumber\\
&\equiv \int\dr^4 y\:\bar{e}\sum_{k=1}^{4}\sum_{j_{\rm V}=1}^7 \Big[ \big(H_0^{\rm V}\big)^{i_{\rm V}}_{~j_{\rm V}} (-\Box_x) P_{{\rm V}~k}^{~m} \Big(\frac{- i \partial_x}{\sqrt{ - \Box_x}}\Big) \frac{\delta^{(4)}(x-y)}{\bar{e}}\Big]\nonumber\\
&\hspace{3cm} \big(v^{\rm V}_{~\:\alpha_{\rm V}}\big)^{j_{\rm V}}(p^2) \frac{1}{(2 \pi)^2}\eu^{i \kappa p y} \:t_I^{~k} (\hat{p})\nonumber\\  
&= \sum_{k=1}^{4}\sum_{j_{\rm V}=1}^7  \big(H_0^{\rm V}\big)^{i_{\rm V}}_{~j_{\rm V}} (p^2) \: P_{{\rm V}~k}^{~m} (\hat{p})\big(v^{\rm V}_{~\:\alpha_{\rm V}}\big)^{j_{\rm V}}(p^2)\frac{1}{(2 \pi)^2}\eu^{i \kappa p x} \: t_I^{~k} (\hat{p})\nonumber\\  
&= \bar{\lambda}^{\rm V}_{\alpha_{\rm V}} (p^2) \times \big(v^{\rm V}_{~\:\alpha_{\rm V}}\big)^{i_{\rm V}}(p^2)\:t_I^{~m} (\hat{p})\frac{1}{(2 \pi)^2}\eu^{i \kappa p x}\nonumber\\
&= \bar{\lambda}^{\rm V}_{\alpha_{\rm V}} (p^2) \times \langle x\,m\,i_{\rm V}\,|\,p\,I\,\alpha_{\rm V}\rangle\:.\label{ew-gl-kompakt}
\end{align}
In the first line of \eqref{ew-gl-kompakt}, as always, summation andintegration over repeated indices $(y, k, j_{\rm V})$ is understood.

The occurrence of the the transverse projectors in $H_0$ gives rise to a $d_\chi$-fold degeneracy of all the eigenvalues. Since the tensor structure of $H_0$ w.\,r.\,t. the ${\sf O}(4)$ indices and the spacetime coordinates is given by a transverse projector and tensorial delta distribution, respectively, the only non-trivial part in the determination of the spectrum consists in the computation of the eigenvalues of the matrix part of the corresponding $H_0$ block.

\vspace{1cm}
\subsection{Computation of the trace in the eigenbasis of $\boldsymbol{H_0}$}
The starting point of the proper-time equation is the RG improved 1-loop equation
\vspace{0.5cm}
\begin{equation}
\begin{split}
 \Gamma_k  - S &= \frac{1}{4} {\rm Tr}\,{\rm ln}\Big(\frac{(H^{\rm grav}_0)^2}{\bar{\mu}^6}\Big) + \frac{1}{2} {\rm Tr}\,{\rm ln}\Big(\openone + \frac{1}{\gamma_k} {\cal M} \Big)\\[0.2cm]
& + \frac{1}{2} {\rm Tr}\Big((H^{\rm grav}_0)^2 (H^{\rm grav})^{-1} V^{\rm grav}(H^{\rm grav}_0)^{-2}\Big) \\[0.2cm]
& - 2 \Big\{ \frac{1}{4} {\rm Tr}\,{\rm ln}\Big(\frac{(H^{\rm gh}_0)^2}{\bar{\mu}^2}\Big) + \frac{1}{2} {\rm Tr}\Big((H^{\rm gh}_0) V^{\rm gh}(H^{\rm gh}_0)^{-2}\Big) \Big\} \:.\label{Gamma-S}
\end{split}
\end{equation}
\vspace{0.1cm}

\noindent Except for the trace containing ${\cal M}$, we are now going to compute all the traces {\it in the eigenbasis of $H_0$} that we have just constructed in the previous subsection.

Afterwards, we will in each trace apply the proper-time representation to a positive~(!) operator and implement the necessary cutoff in the proper-time integral. We shall symbolically denote the regularized trace-log combinations as Tr\,ln$[\cdots]_{\sf REG}$ and the proper-time integrals that demand for regularization by the symbol $\ddashint$. At a later stage we shall replace it by $\int_0^\infty \dr s\, C(s)$ with some appropriate regulator function $C(s)$.

\subsubsection{The free part}
Returning to the example of the graviton vector block, we obtain
\vspace{0.4cm}
\begin{align}
 {\rm Tr}\big\{f(H_0^{\rm V})\big\} &= \sum_{p\,I\,{\alpha}_{\rm V}} \langle p\,I\,\alpha_{\rm V}\,|f(H_0^{\rm V})|\,p\,I\,\alpha_{\rm V} \rangle\nonumber\\
&= \sum_{p\,I\,{\alpha}_{\rm V}} \sum_{x\,m\,i_{\rm V}} \sum_{y\,k\,j_{\rm V}} \langle p\,I\,\alpha_{\rm V}\,|\,x\,m\,i_{\rm V}\rangle  \langle x\,m\,i_{\rm V}\,|f(H_0^{\rm V})|\,y\,k\,j_{\rm V}\rangle \langle y\,k\,j_{\rm V}\,|\,p\,I\,\alpha_{\rm V} \rangle\nonumber\\
&= \sum_{p\,I\,{\alpha}_{\rm V}} \sum_{x\,m\,i_{\rm V}} f\big(\bar{\lambda}^{\rm V}_{\alpha_{\rm V}}(p^2)\big)\,\langle p\,I\,\alpha_{\rm V}\,|\,x\,m\,i_{\rm V}\rangle  \langle x\,m\,i_{\rm V}\,|\,p\,I\,\alpha_{\rm V} \rangle\nonumber\\
&= \bigg(\!\int \dr^4 x\:\bar{e}\bigg)\:d_{\rm V} \int\frac{\dr^4 p}{(2 \pi)^4}\sum_{\alpha_{\rm V} = 1}^{n_{\rm V}}f\big(\bar{\lambda}^{\rm V}_{\alpha_{\rm V}}(p^2)\big)\label{spur-funktion-h0} 
\end{align}
\vspace{0.0cm}

\noindent with $\sum_p \equiv \int \dr^4 p$ and $\sum_x \equiv \int \dr^4 x\:\bar{e}$. 

This structure generalizes all other blocks, labeled by $\chi\in \{$S, V, T, gh\,S, gh\,V$\}$,
\vspace{0.5cm}
\begin{equation}
 \Tr\Big\{f\big(H_0^{\chi}\big)\Big\}=\bigg(\!\int \dr^4 x\:\bar{e}\bigg)\:d_{\rm \chi} \int\frac{\dr^4 p}{(2 \pi)^4}\sum_{\alpha_{\rm \chi} = 1}^{n_{\rm \chi}}f\big(\bar{\lambda}^{\rm \chi}_{\alpha_{\rm \chi}}(p^2)\big)
\end{equation}
\vspace{0.1cm}

\noindent with the factor $d_\chi$ being due to the degeneracy originating from the occurrence of the transverse projectors mentioned above.

In order to take into account the $\bar{\mu}$ rescaling in the argument of the logarithm we further need
\vspace{0.5cm}
\begin{equation}
 {\rm Tr}(\openone^\chi) =  d_\chi \cdot n_\chi\: \bigg(\!\int\! \dr^4 x\:\bar{e}\bigg)\: \bigg(\!\int \!\!\frac{\dr^4 p}{(2 \pi)^4}\bigg)\:.
\end{equation}
\vspace{0.1cm}

\noindent With $d_\chi\cdot n_\chi$ being the number of field components of the corresponding sector, this trace amounts to the (infinite) number of field degrees of freedom multiplied by the (infinite) sum over all momentum modes. 
\pagebreak

If we now apply the basic proper-time integral identity
\vspace{0.5cm}
\begin{equation}
 {\rm ln}\bigg(\frac{A}{B}\bigg) = - \int_0^\infty \frac{\dr s}{s}\Big\{\eu^{-s\,A} - \eu^{-s\,B}\Big\}
\end{equation}
\vspace{0.1cm}

\noindent we obtain for the free part
\vspace{0.5cm}
\begin{equation}
\begin{split}
\frac{1}{4}\Big\{{\rm Tr}\,&{\rm ln}\Big[\frac{\big(H^{\rm grav}_0\big)^2}{\bar{\mu}^6}\Big]_{\sf REG} - 2 {\rm Tr}\,{\rm ln}\Big[\frac{\big(H^{\rm gh}_0\big)^2}{\bar{\mu}^2}\Big]_{\sf REG}\Big\}=\\ &- \frac{1}{4}\int\! \dr^4 x\:\bar{e}\int\frac{\dr^4 p}{(2 \pi)^4}\bigg\{\ddashint\frac{\dr s}{s}\Big(\sum_{\chi \in\{{\rm S}, {\rm V}, {\rm T}\}} d_\chi \sum_{\alpha_\chi = 1}^{n_\chi}\eu^{ - s \,\big(\bar{\lambda}^\chi_{\alpha_\chi}(p^2)\big)^2} - 40\:\eu^{ - s \,\bar{\mu}^6}\Big)\nonumber\\
&\hspace{3cm} - 2\:\ddashint\frac{\dr s^\prime}{s^\prime}\Big(\sum_{\chi \in\{{{\rm gh}\,{\rm S}}, {{\rm gh}\,{\rm V}}\}} d_\chi \sum_{\alpha_\chi = 1}^{n_\chi}\eu^{ - s^\prime \,\big(\bar{\lambda}^\chi_{\alpha_\chi}(p^2)\big)^2} - 10\:\eu^{ - s^\prime \,\bar{\mu}^2}\Big)\bigg\}
\end{split}
\end{equation}
\vspace{0.1cm}

\noindent Therein, $\sum_{\chi \in\{{\rm S}, {\rm V}, {\rm T}\}} d_\chi\,n_\chi = 40$ and $\sum_{\chi \in\{{{\rm gh}\,{\rm S}}, {{\rm gh}\,{\rm V}}\}} d_\chi\,n_\chi = 10$ amount to the number of field components in the graviton and the ghost sector, respectively. Finally, one ought to note that the proper-time variables $s$ and $s'$ in the graviton and the ghost sector, respectively, have different mass dimensions.

\vspace{0.5cm}
\subsubsection{The interaction part}
In contrast to the free part, the operators that make up the interaction part are not simply proportional to the transverse projectors; instead they carry a rather complicated inherent ${\sf O}(4)$ index structure. So returning to the above example of the graviton vector sector, we have in the generalized position representation
\begin{equation}
 \langle x\,m\,i_{\rm V}\,|V^{\rm V}|\,y\,k\,j_{\rm V}\rangle = \big(V^{\rm V}\big)^{i_{\rm V}~\:m}_{~\:j_{\rm V}~\:k} (-i \hat\partial)\:\frac{\delta^{(4)}(x-y)}{\bar{e}}\nonumber\:.
\end{equation}
However, under the trace this operator is multiplied by the inverse of the free part, $\big(H^{\rm V}\big)^{-1}$. Combined with the $t_{\rm I}$'s contained in the eigenvectors, this leads to 
\begin{equation}
 \begin{split}
  & \frac{1}{2}{\rm Tr}\big\{\big(H^{\rm V}\big)^{-1} V^{\rm V} \big(H^{\rm V}_0\big)^2\big[\big(H^{\rm V}_0\big)^{-2}\big]_{\sf REG}\}\\
&= \frac{1}{2}\int \dr^4 x\:\bar{e}\int\dr^4 y\:\bar{e}\int\dr^4 z\:\bar{e}\int\dr^4 p\sum_{i_{\rm V}, j_{\rm V}, k_{\rm V} = 1}^{n_{\rm V}}\sum_{\alpha_{\rm V} = 1}^{n_{\rm V}}\sum_{k, m, n = 1}^{4}\sum_{I=1}^{3}\\
&\hspace{0.9cm}\frac{1}{(2 \pi)^4}\:\eu^{- i \kappa p x} t^I_{~m}(\hat{p})\big(v^{{\rm V}\alpha_{\rm V}}\big)_{i_{\rm V}}(p^2)\times\\
&\hspace{0.5cm} \times \Big[\big((H^{\rm V})^{-1}\big)^{i_{\rm V}}_{~j_{\rm V}}(-\Box_x) P_{{\rm V}~k}^{~m} \big(- i \hat\partial\big)\frac{\delta^{(4)}(x-y)}{\bar{e}}\Big]\Big[\big(V^{\rm V}\big)^{j_{\rm V}~\:k}_{~\:k_{\rm V}~n}(-i \partial_a^y) \frac{\delta^{(4)}(y-z)}{\bar{e}}\Big]\times\\
&\hspace{0.5cm} \times\,\big(\bar{\lambda}^{\rm V}_{\alpha_{\rm V}}(p^2)\big)^2\Big(\ddashint\dr s\:\eu^{-s\,(\bar{\lambda}^{\rm V}_{\alpha_{\rm V}}(p^2))^2}\Big) \eu^{i \kappa p z} t_I^{~n}(\hat{p})\big(v^{\rm V}_{~\alpha_{\rm V}}\big)^{k_{\rm V}}(p^2)\\
&= \frac{1}{2}\int \dr^4 x\:\bar{e}\int\frac{\dr^4 p}{(2 \pi)^4} \sum_{i_{\rm V}, j_{\rm V} = 1}^{n_{\rm V}}\sum_{\alpha_{\rm V} = 1}^{n_{\rm V}}\sum_{k, n = 1}^{4}\big(\bar{\lambda}^{\rm V}_{\alpha_{\rm V}}(p^2)\big)^2\Big(\ddashint\dr s\:\eu^{-s\,(\bar{\lambda}^{\rm V}_{\alpha_{\rm V}}(p^2))^2}\Big)\times\\
&\hspace{2cm}\times\,\big(v^{\rm V \alpha_{\rm V}}\big)_{i_{\rm V}}(p^2) \Big[\big((H^{\rm V})^{-1}\big)^{i_{\rm V}}_{~j_{\rm V}}(p^2) \big(V^{\rm V}\big)^{j_{\rm V}~\:k}_{~\:k_{\rm V}~n}(p_a) P_{{\rm V}k}^{~~n} (\hat{p})\Big]\big(v^{\rm V}_{~\alpha_{\rm V}}\big)^{k_{\rm V}}(p^2)\:.
 \end{split}\raisetag{6.3cm}
\end{equation}
The operator under the trace, like $H^{\rm V}$, is proportional to $P_{\rm V}$. Therefore, we have to compute the contraction of $\big(H^{-1}\big)^\chi$ with $P_\chi$ for $\chi \in \{ {\rm S,V,T}\}$. In fact, when computing $\big(H^{-1}\big)^\chi$, we only have to invert the matrix part, which can be easily performed by a computer algebra program.

Before we proceed, two remarks are in order: First of all, at this point the $\frac{1}{\sqrt{-\Box}}$-rescalings to the left of $V^\chi$ can be performed much more easily than by partial integration. We simply let the $\frac{1}{\sqrt{-\Box}}$ operators act on the dual bra-vectors to their left according to $\langle p\,I\,\alpha_{\rm V}\,|\frac{1}{\sqrt{ - \Box}} = \frac{1}{p}\,\langle p\,I\,\alpha_{\rm V}\,|$ and thereby exploiting the fact that they are eigenfunctions of $\frac{1}{\sqrt{-\Box}} \equiv \frac{1}{\sqrt{\pi}} \int_0^\infty {\rm d} t\  t^{1/2} e^{t \Box}$ with eigenvalue $\frac{1}{p}$.

Moreover, only terms with an even number of $p_a$'s will contribute to the $\int {\rm d}^4 p$-integral. Therefore, we can drop all terms that contain an odd number of partial derivatives. It turns out that these are exactly the terms where no derivative acts on $v_a\equiv \bar{e}_a^{\ \mu} v_\mu$; in contrast, $(\partial_a v_b)$ is always followed by an even number of partial derivatives.

Making use of the symmetric integration $\int {\rm d}^4 p f(p^2) p_a p_b = \frac{1}{4}\eta_{ab} \int {\rm d}^4 p f(p^2)p^2$ after the contraction with the transverse projectors, we first of all observe that all non-vanishing contributions to the trace stemming from the interaction part are proportional to $\int \ddx\! \bar{e}\, n^{(\pm)ab} f_{ab}$, as expected. The terms whose signs are independent of the chosen background configuration have to be associated with the invariant $\int \ddx\! \bar{e}\, \bar{F}^{(\pm)ab}_{~~~~~\mu\nu}\bar{e}_a^{\ \mu}\bar{e}_b^{\ \nu}= \int \ddx\! \bar{e}\, n^{(\pm)ab} f_{ab}$, whereas those terms whose signs change when switching from $\bar{\omega}^{(+)}$ to $\bar{\omega}^{(-)}$ contribute to the Immirzi term $\int \ddx \bar{e} \star\bar{F}^{(\pm)ab}_{~~~~~\mu\nu}\bar{e}_a^{\ \mu}\bar{e}_b^{\ \nu}=\pm \int \ddx \bar{e}\; n^{(\pm)ab} f_{ab}$.

After the contraction, the ${\sf O}(4)$ index structure of the objects $\big(V^\chi\big)^{i_\chi~~m}_{~~j_\chi~~k}$, $\chi \in \{ {\rm S,V,T}\}$ disappears; instead, we obtain $n_\chi\times n_\chi$ matrices $\big(V^\chi_{\rm contr}\big)^{i_\chi}_{~~j_\chi}$ which carry the labels $i_\chi,j_\chi=1,\cdots,(n_\chi)$ of the fields in the corresponding sector.

These matrices exhibit the expected parity structure. Separating the contributions that change their sign under $\bar{\omega}^{(+)} \mapsto  \bar{\omega}^{(-)}$ from those that do not, according to $\big(V^\chi_{\rm contr}\big)^{i_\chi}_{~~j_\chi}\equiv \big(V^{\chi (+)}_{\rm contr}\big)^{i_\chi}_{~~j_\chi}\mp\big(V^{\chi (-)}_{\rm contr}\big)^{i_\chi}_{~~j_\chi} $, the schematic structure of these matrices is given by
\begin{align}
 \big(V^{\chi\,(+)}_{\rm contr}\big)&\simeq \left( \begin{array}{cc} (V^{\chi\,(+)}_{\rm contr})_{\rm TT} & \frac{1}{\gamma_k} (V^{\chi\,(+)}_{\rm contr})_{\rm TP} \\ \frac{1}{\gamma_k} (V^{\chi\,(+)}_{\rm contr})_{\rm PT} & (V^{\chi\,(+)}_{\rm contr})_{\rm PP} \end{array} \right)\\ \intertext{and}
\big(V^{\chi\,(-)}_{\rm contr}\big)&\simeq \left( \begin{array}{cc} \frac{1}{\gamma_k} (V^{\chi\,(-)}_{\rm contr})_{\rm TT} & (V^{\chi\,(-)}_{\rm contr})_{\rm TP} \\ (V^{\chi\,(-)}_{\rm contr})_{\rm PT} & \frac{1}{\gamma_k} (V^{\chi\,(-)}_{\rm contr})_{\rm PP} \end{array} \right)\:.
\end{align}
Thereby, we have again made use of the parity-ordered notation, i.\,e. ordering the fields of the corresponding sector such that the parity-even ones are followed by the parity-odd ones. The subscripts 'T' and 'P' refer to the combination of tensors and pseudo-tensors that the corresponding matrix element couples. We see that a scalar $\big(V^{\chi (+)}_{\rm contr}\big)$ can be either formed by coupling two tensors or two pseudo-tensors $\Big({\cal O}\big( (\frac{1}{\gamma_k})^0\big)\Big)$ or by coupling a tensor and a pseudo-tensor via the parity-violation ${\sf O}(4)$ duality operator stemming from the Immirzi tensor $\Big({\cal O}\big( (\frac{1}{\gamma_k})^1\big)\Big)$. For the pseudo-scalar contributions $\big(V^{\chi (-)}_{\rm contr}\big)$, the situation is reversed.

Since $\big( H^\chi\big)^{-1}$ respects the block structure of $H^\chi = H_0^\chi+ \frac{1}{\gamma_k} \bar{H}_0^\chi$, its structure is given by 
\vspace{0.5cm}
\begin{equation}
 \big(H^{\chi}\big)^{-1}\simeq \left( \begin{array}{cc} A^{\chi\,{\rm even}}_{\rm TT}(\gamma_k) & A^{\chi\,{\rm odd}}_{\rm TP}(\gamma_k) \\ A^{\chi\,{\rm odd}}_{\rm PT}(\gamma_k) & A^{\chi\,{\rm even}}_{\rm PP}(\gamma_k) \end{array} \right)
\end{equation}
\vspace{0.1cm}

\noindent with $A^{\chi\, {\rm even}}(\gamma_k)$ and $A^{\chi\, {\rm odd}}(\gamma_k)$ denoting even and odd functions of $\gamma_k$, respectively. Therefore we obtain 
\vspace{0.5cm}
\begin{equation}
 \big(H^{\chi}\big)^{-1}\big\{\big(V^{\chi\,(+)}_{\rm contr}\big) \mp \big(V^{\chi\,(-)}_{\rm contr}\big)\big\}\simeq \left( \begin{array}{cc} B^{\chi\,{\rm even}}_{\rm TT}(\gamma_k) \mp B^{\chi\,{\rm odd}}_{\rm TT}(\gamma_k) & B^{\chi\,{\rm odd}}_{\rm TP}(\gamma_k) \mp B^{\chi\,{\rm even}}_{\rm TP}(\gamma_k) \\ B^{\chi\,{\rm odd}}_{\rm PT}(\gamma_k) \mp B^{\chi\,{\rm even}}_{\rm PT}(\gamma_k) & B^{\chi\,{\rm even}}_{\rm PP}(\gamma_k) \mp B^{\chi\,{\rm odd}}_{\rm PP}(\gamma_k)\end{array} \right)
\end{equation}
\vspace{0.1cm}

\noindent wherein the components are given by different even (odd) functions $B^{\chi\,{\rm even}}$ $(B^{\chi\,{\rm odd}})$ of $\gamma_k$. Since all other ingredients to the trace are independent of $\gamma_k$ and block diagonal in the case of $H_0^\chi$, it follows that the $\gamma_k$-dependence of the trace will be of the form $f^{\chi\,{\rm even}}(\gamma_k)\mp f^{\chi\,{\rm odd}}(\gamma_k)$ with some even (odd) function $f^{\chi\,{\rm even}}$ $(f^{\chi\,{\rm odd}})$ of $\gamma_k$ yet to be determined.

Since the ghost sector is completely block-diagonal in the parity-ordered representation and therefore parity-even, it does not contribute to the Immirzi invariant. Explicitly, the matrices after contraction are obtained as follows. 

\vspace{0.5cm}
\noindent{\bf Scalars in the graviton sector ($a$, $d$, $B$, $D$):}
\begin{equation}
\begin{split}
 &\big(V^{{\rm S}\,(+)}_{\rm contr}\big)^{i_{\rm S}}_{~j_{\rm S}} \mp \big(V^{{\rm S}\,(-)}_{\rm contr}\big)^{i_{\rm S}}_{~j_{\rm S}} = \frac{1}{16 \pi G_k}\: n^{(\pm)\:ab} f_{ab}\: \times\\
&\times\left(\!\!\!\begin{array}{cccc} \Ss{-\frac{(1+\beta_\Dr)}{2 \alpha_\Dr}\bar{\mu}^{-1}} & \Ss{\Big[-\frac{1}{2}\Big(1 \mp \frac{1}{\gamma_k}\Big) - \frac{1}{2 \alpha_\Dr}\Big(5\beta_\Dr + \frac{1}{2}\Big)\Big]\bar{\mu}^{-1}} & \Ss{\big(1 \mp \frac{1}{\gamma_k}\Big)p^{-1}} & \Ss{\Big(\pm 1 - \frac{1}{\gamma_k}\Big)p^{-1}} \\ \Ss{\Big[-\frac{1}{2}\Big(1 \mp \frac{1}{\gamma_k}\Big) + \frac{3}{2 \alpha_\Dr}\Big(\beta_\Dr + \frac{1}{2}\Big)\Big]\bar{\mu}^{-1}} & \Ss{\Big[-\Big(1 \mp \frac{1}{\gamma_k}\Big) + \frac{3 \beta_\Dr}{2 \alpha_\Dr}\Big]\bar{\mu}^{-1}} & \Ss{0} & \Ss{0} \\ \Ss{- \Big(1 \mp \frac{1}{\gamma_k}\Big)p^{-1}} & \Ss{-2 \Big(1 \mp \frac{1}{\gamma_k}\Big)p^{-1}} & \Ss{0} & \Ss{0} \\ \Ss{\Big(\mp 1 + \frac{1}{\gamma_k}\Big)p^{-1}} & \Ss{2\Big(\mp 1 + \frac{1}{\gamma_k}\Big) p^{-1}} & \Ss{0} & \Ss{0} \end{array}\!\! \right)
\end{split}\raisetag{4.5cm}
 \end{equation}

\begin{sideways}  
\parbox{\textheight}{\vspace{4cm} {\bf Vectors in the graviton sector ($b^m$, $c^m$, $A^m$, $D^m$, $d^m$, $B^m$, $C^m$):} 
\begin{equation}
\begin{split}
&\big(V^{{\rm V}\,(+)}_{\rm contr}\big)^{i_{\rm V}}_{~j_{\rm V}} \mp \big(V^{{\rm V}\,(-)}_{\rm contr}\big)^{i_{\rm V}}_{~j_{\rm V}} = \frac{1}{16 \pi G_k}\: n^{(\pm)\:ab} f_{ab}\: \times\\
&\times\left(\!\!\!{\setlength\arraycolsep{0.1em}
\begin{array}{ccccccc} \Sss{0} & \Sss{\Big[\frac{1}{2}\Big(1 \mp \frac{1}{\gamma_k}\Big) - \frac{3}{4 \alpha_\Dr}\Big]\bar{\mu}^{-1}} & \Sss{- \frac{1}{2}\Big(1 \mp \frac{1}{\gamma_k}\Big) p^{-1}} & \Sss{- \frac{1}{2}\Big(\frac{5}{4} \mp \frac{2}{\gamma_k}\Big) p^{-1}} & \Sss{\Big(\mp 1 + \frac{1}{\gamma_k}\Big) \bar{\mu}^{-1}} & \Sss{\Big(\mp 1 + \frac{1}{\gamma_k}\cdot\frac{5}{8}\Big)p^{-1}} & \Sss{\frac{1}{2}\Big(\mp 1 + \frac{1}{\gamma_k}\Big)p^{-1}} \\ \Sss{\Big[\frac{1}{2}\Big(1\mp\frac{1}{\gamma_k}\Big)-\frac{1}{8 \alpha_\Dr}\Big]\bar{\mu}^{-1}} & \Sss{- \frac{3}{4 \alpha_\Dr} \bar{\mu}^{-1}} & \Sss{- \frac{1}{2} \Big(1 \mp \frac{1}{\gamma_k}\Big)p^{-1}} & \Sss{- \frac{1}{2}\Big(\frac{5}{4} \mp \frac{2}{\gamma_k}\Big)p^{-1}} & \Sss{\Big[\Big(\pm 1 - \frac{1}{\gamma_k}\Big)\pm\frac{1}{2\alpha_\Dr}\Big]\bar{\mu}^{-1}} & \Sss{\Big(\mp \frac{3}{4} + \frac{1}{\gamma_k}\cdot\frac{5}{8}\Big)p^{-1}} & \Sss{\Big(\mp \frac{1}{4} + \frac{1}{\gamma_k}\cdot\frac{1}{2}\Big)p^{-1}} \\ \Sss{- \frac{1}{2} \Big(1 \mp \frac{1}{\gamma_k}\Big)p^{-1}} & \Sss{\frac{3}{4} \Big(\frac{3}{2} \mp \frac{1}{\gamma_k}\Big)p^{-1}} & \Sss{0} & \Sss{0} & \Sss{\Big(\pm \frac{5}{4} - \frac{1}{\gamma_k} \cdot \frac{13}{8}\Big)p^{-1}} & \Sss{0} & \Sss{0} \\ \Sss{\frac{1}{2}\Big(\frac{5}{4} \mp \frac{2}{\gamma_k}\Big)p^{-1}} & \Sss{- \frac{1}{2}\Big(\frac{5}{4} \mp \frac{2}{\gamma_k}\Big) p^{-1}} & \Sss{0} & \Sss{0} & \Sss{\frac{5}{4}\Big(\mp 1 + \frac{1}{\gamma_k}\Big) p^{-1}} & \Sss{0} & \Sss{0} \\ \Sss{(\mp 1 + \frac{1}{\gamma_k}\Big)\bar{\mu}^{-1}} & \Sss{\Big[\Big(\pm 1 - \frac{1}{\gamma_k}\Big) \mp \frac{3}{4 \alpha_\Dr}\Big]\bar{\mu}^{-1}} & \Sss{0} & \Sss{0} & \Sss{- \Big(1 \mp \frac{1}{\gamma_k}\Big) \bar{\mu}^{-1}} & \Sss{\pm \frac{1}{4}\cdot\frac{1}{\gamma_k} p^{-1}} & \Sss{\mp \frac{1}{4}\cdot\frac{1}{\gamma_k} p^{-1}} \\ \Sss{\Big(\pm \frac{1}{4} - \frac{1}{\gamma_k}\cdot\frac{5}{8}\Big)p^{-1}} & \Sss{\Big(\mp \frac{3}{4} + \frac{1}{\gamma_k} \cdot\frac{5}{8}\Big) p^{-1}} & \Sss{0} & \Sss{0} & \Sss{- \frac{1}{2} \Big(\frac{5}{2} \mp \frac{3}{\gamma_k}\Big)p^{-1}} & \Sss{0} & \Sss{0} \\ \Sss{\frac{1}{2} \Big(\mp 1 + \frac{1}{\gamma_k} \Big)p^{-1}} & \Sss{\Big(\pm 1 - \frac{1}{\gamma_k} \cdot \frac{9}{8}\Big)p^{-1}} & \Sss{0} & \Sss{0} & \Sss{\frac{1}{2}\Big(\frac{13}{4} \mp \frac{3}{\gamma_k}\Big) p^{-1}} & \Sss{0} & \Sss{0} \end{array}}\!\!\!\right)
\end{split}
\end{equation}
}
\end{sideways}

\noindent{\bf Tensors in the graviton sector ($d^{mn}$, $B^{mn}$, $D^{mn}$):}
\begin{equation}
 \begin{split}
  \big(V^{{\rm T}\,(+)}_{\rm contr}\big)^{i_{\rm T}}_{~j_{\rm T}} \mp \big(&V^{{\rm T}\,(-)}_{\rm contr}\big)^{i_{\rm T}}_{~j_{\rm T}} = \frac{1}{16 \pi G_k}\: n^{(\pm)\:ab} \:f_{ab} \:\times\\
&\times\left(\begin{array}{ccc} \frac{5}{6}\Big(1 \mp \frac{1}{\gamma_k}\Big) \bar{\mu}^{-1} & - \frac{1}{2} \Big(\frac{5}{4} \pm \frac{1}{\gamma_k}\cdot \frac{5}{6}\Big)p^{-1} & \frac{1}{\gamma_k} \cdot \frac{5}{24} p^{-1} \\ \frac{1}{2}\Big(\frac{5}{4} \mp \frac{1}{\gamma_k}\cdot\frac{5}{3}\Big) p^{-1} & 0 & 0 \\ \Big(\pm \frac{5}{6} - \frac{1}{\gamma_k} \cdot \frac{5}{24}\Big) p^{-1} & 0 & 0 \end{array}\right)
 \end{split}
\end{equation}

\vspace{0.5cm}
\noindent{\bf Scalars in the ghost sector ($\bar{f}$, $f$):}
\begin{equation}
 \big(V^{{\rm gh}\,{\rm S}}_{\rm contr}\big)^{i_{{\rm gh}\,{\rm S}}}_{~j_{{\rm gh}\,{\rm S}}} = n^{(\pm)\:ab}\: f_{ab} \times\Big(0\Big)\equiv0
\end{equation}

\vspace{0.5cm}
\noindent{\bf Vectors in the ghost sector ($\bar{g}_m$, $\bar{F}_m$, $\bar{G}_m$, $g^m$, $F^m$, $G^m$):}
\begin{equation}
 \big(V^{{\rm gh}\,{\rm V}}_{\rm contr}\big)^{i_{{\rm gh}\,{\rm V}}}_{~j_{{\rm gh}\,{\rm V}}} = n^{(\pm)\:ab}\: f_{ab} \times \left(\begin{array}{ccc} -\frac{3}{4} \bar{\mu}^{-1} & \frac{3}{16} p^{-1} & 0 \\ \frac{3}{8} p^{-1} & 0 & 0 \\ 0 & 0 & 0 \end{array} \right)
\end{equation}
\vspace{0.5cm}

Up to the global prefactor in the graviton sector, all the above matrices have mass dimension 1; when multiplied with $\big( H^\text{grav}\big)^{-1}$, this prefactor cancels. These objects are independent of the cosmological constant and of the ${\sf O}(4)$ gauge fixing parameter that both enter only the free part. Moreover, no $\bar{\tau}$-$\bar{\tau}$-components contribute to the interaction part. The off-diagonal $p^{-1}$-terms originate directly from the first-order structure of the truncation ansatz.

Finally, after having performed a partial trace in form of the momentum integration that makes several terms vanish, the matrices are no longer Hermitean w.\,r.\,t. the remaining inner product.
\pagebreak 

Two remarks are in order here.

\noindent{\bf (A)} Let us assume for a moment that the Immirzi term was not present in the original truncation ansatz. Then we have $H\equiv H_0$ not only in the ghost, but also in the graviton sector. In the matrices $V^{\chi\,(+)}_{\text{contr}}$ and $V^{\chi\,(-)}_{\text{contr}}$ all contributions $\propto \frac{1}{\gamma_k}$ vanish, and $V^{\chi\,(+)}_{\text{contr}}$ is block-diagonal whereas $V^{\chi\,(-)}_{\text{contr}}$ contains only off-diagonal elements. The trace argument \eqref{Trace_Arg} that we applied sectorwise to the interaction part can now be used blockwise: Since both $H\equiv H_0$ and $V^{\chi\,(+)}_{\text{contr}}$ are block-diagonal in the parity-ordered representation, $V^{\chi\,(-)}_{\text{contr}}$ will not contribute to the trace. Therefore, if only the invariants $\int \ddx \bar{e}$ and $\int \ddx \bar{e} \bar{F}^{ab}_{~~\mu\nu} \bar{e}_a^{~\mu}\bar{e}_b^{\ \nu}$ were present in the truncation, the relevant interaction terms would be obtained by sending $\gamma_k\rightarrow \infty$ in $V^{\chi\,(+)}_{\text{contr}}$ and by setting $V^{\chi\,(-)}_{\text{contr}}\rightarrow 0.$
\vspace{0.5cm}

\noindent{\bf (B)} Finally, one might wonder whether choosing a constant (anti-)selfdual background spin connection and expanding to ${\cal O}\Big(\big(\bar{\omega}^{(\pm)}\big)^2\Big)$ would have been more advantageous. We decided against this possibility for three main reasons: First, both invariants containing $\bar{F}^{(\pm)}(\bar{\omega})$ would manifest themselves as surface contributions which one had to carefully keep track of. Second, the expansion up to second order in the (constant) spin connection generates an at least a comparable number of terms as the first order expansion in the more general connection. Third, there are other field monomial like $T^a\wedge T_a$ stemming from the Nieh-Yan invariant, and other torsion squared invariants that require an inverted vielbein for their construction, that contain second order spin connection terms. Therefore we would not be able to unambiguously identify the contributions to the flow of the couplings in our truncation.
\vspace{0.5cm}

In order to present the final result of this subsection it is useful to introduce the reduced matrices $\check{V}^{\chi}_{\rm contr}$ by separating off $n^{(\pm)\:ab} f_{ab}$ from $V^{\chi}_{\rm contr}$:
\begin{equation}
 \big(V^{\chi}_{\rm contr}\big)^{i_\chi}_{~j_\chi} \equiv\big(n^{(\pm)\:ab} f_{ab}\big)\big(\check{V}^{\chi}_{\rm contr}\big)^{i_\chi}_{~j_\chi}~~\forall\:\chi \in\{{\rm S}, {\rm V}, {\rm T}, {{\rm gh}\,{\rm S}}, {{\rm gh}\,{\rm S}}\}\:.
\end{equation}
In terms of the $\big(\check{V}^{\chi}_{\rm contr}\big)$-matrix elements, we thus obtain the following representation of the interaction contributions to the supertrace:
\begin{equation}
 \begin{split}
  & \frac{1}{2}\Big\{{\rm Tr}\Big(\big(H^{\rm grav}\big)^{-1} V^{\rm grav}\big(H^{\rm grav}_0\big)^2  \big[\big(H^{\rm grav}_0\big)^{-2}\big]_{\sf REG}\Big) - 2\,{\rm Tr}\Big(\big(H^{\rm gh}_0\big)V^{\rm gh} \big[\big(H^{\rm gh}_0\big)^{-2}\big]_{\sf REG}\Big)\Big\}\nonumber\\
&= \frac{1}{2}\big(\int\dr^4\:\bar{e}\:n^{(\pm)\:ab} f_{ab}\big)\int\frac{\dr^4 p}{(2 \pi)^4}\bigg[ \sum_{\chi \in \{{\rm S}, {\rm V}, {\rm T}\}}\sum_{\alpha_\chi = 1}^{n_\chi}\sum_{i_\chi, j_\chi, k_\chi = 1}^{n_\chi}\ddashint\dr s\:\eu^{ - s(\bar{\lambda}^\chi_{~\alpha_\chi}(p^2))^2} \nonumber\\
&\hspace{3.5cm} \big(v^{\chi \alpha_\chi}\big)_{i_\chi} (p^2) \big(\big(H^\chi\big)^{-1}\big)^{i_\chi}_{~j_\chi} (p^2) \big(\check{V}^\chi_{\rm contr}\big)^{j_\chi}_{~k_\chi} (p^2) \big(v^\chi_{~\alpha_\chi}\big)^{k_\chi} (p^2) \big(\bar{\lambda}^\chi_{\alpha_\chi} (p^2)\big)^2\nonumber\\
&~~~~ -\, 2 \sum_{\chi \in \{{{\rm gh}\,{\rm S}}, {{\rm gh}\,{\rm V}}\}}\sum_{\alpha_\chi = 1}^{n_\chi}\sum_{i_\chi, j_\chi, k_\chi = 1}^{n_\chi}\ddashint\dr s^\prime\:\eu^{ - s^\prime(\bar{\lambda}^\chi_{~\alpha_\chi}(p^2))^2}\nonumber\\
&\hspace{6cm} \big(v^{\chi \alpha_\chi}\big)_{i_\chi} (p^2) \big(H^\chi_0\big)^{i_\chi}_{~j_\chi} (p^2) \big(\check{V}^\chi_{\rm contr}\big)^{j_\chi}_{~k_\chi} (p^2) \big(v^\chi_{~\alpha_\chi}\big)^{k_\chi} (p^2)  \bigg]\nonumber\\
&= \frac{1}{2}\big(\int\dr^4\:\bar{e}\:n^{(\pm)\:ab} f_{ab}\big)\int\frac{\dr^4 p}{(2 \pi)^4}\bigg[ \sum_{\chi \in \{{\rm S}, {\rm V}, {\rm T}\}}\sum_{\alpha_\chi = 1}^{n_\chi}\sum_{i_\chi, j_\chi, k_\chi = 1}^{n_\chi}\ddashint\dr s\:\eu^{ - s(\bar{\lambda}^\chi_{~\alpha_\chi}(p^2))^2}\nonumber\\
&\hspace{3.3cm} \big(v^{\chi \alpha_\chi}\big)_{i_\chi} (p^2) \big(\big(H^\chi\big)^{-1}\big)^{i_\chi}_{~j_\chi} (p^2) \big(\check{V}^\chi_{\rm contr}\big)^{j_\chi}_{~k_\chi} (p^2) \big(v^\chi_{~\alpha_\chi}\big)^{k_\chi} (p^2) \big(\bar{\lambda}^\chi_{\alpha_\chi} (p^2)\big)^2 \nonumber\\
&~~~ -\, 2 \sum_{\chi \in \{{{\rm gh}\,{\rm S}}, {{\rm gh}\,{\rm V}}\}}\sum_{\alpha_\chi = 1}^{n_\chi}\sum_{j_\chi, k_\chi = 1}^{n_\chi}\ddashint\dr s^\prime\:\eu^{ - s^\prime(\bar{\lambda}^\chi_{~\alpha_\chi}(p^2))^2}\nonumber\\
&\hspace{6.6cm} \bar{\lambda}^\chi_{~\alpha_\chi}(p^2) \big(v^{\chi \alpha_\chi}\big)_{j_\chi} (p^2) \big(\check{V}^\chi_{\rm contr}\big)^{j_\chi}_{~k_\chi} (p^2) \big(v^\chi_{~\alpha_\chi}\big)^{k_\chi} (p^2) \bigg]\:.\label{V-Spur}
 \end{split}
\end{equation}
We shall further evaluate this representation in Section \ref{Section76}.

\subsubsection{Algebraic properties of ${\cal M}$ and evaluation of $\frac{1}{2} \Tr \text{ln} \big\{\openone + \frac{1}{\gamma_k} {\cal M}\big\}$}\label{AlgebraicFormM}
After the by now well-known manipulations, we obtain in the graviton sector
\begin{equation}
 \frac{1}{2}{\rm Tr}\,{\rm ln}\Big\{\openone^\chi + \frac{1}{\gamma_k}{\cal M}^\chi\Big\}
= \frac{1}{2}\!\int\!\!\big(\dr^4 x\:\bar{e}\big)d_\chi\int\!\!\frac{\dr^4 p}{(2 \pi)^4} \sum_{i_\chi = 1}^{n_\chi}\Big({\rm ln}\Big\{\openone^\chi + \frac{1}{\gamma_k}{\cal M}^\chi(p^2)\Big\}\Big)^{i_\chi}_{~i_\chi}\:,
\end{equation}
with $\chi\in\{S,V,T\}$. Here, ${\cal M}^\chi$ denotes the corresponding block of the full (block-diagonal) matrix ${\cal M}=(H_0)^{-1}\bar{H}$, cf. \eqref{M1}.

Therefore, we have to compute the algebraic trace of the $n_\chi\times n_\chi$ matrices in field space, $\Big({\rm ln}\Big\{\openone^\chi + \frac{1}{\gamma_k}{\cal M}^\chi(p^2)\Big\}\Big)^{i_\chi}_{~i_\chi}$, in a regularized manner. In doing so, several quite remarkable algebraic properties of the matrices ${\cal M}^\chi$ can be exploited.

\paragraph{(A) Algebraic properties of the matrices ${\cal M}^\chi$.}

The matrices ${\cal M}^\chi$ are sectorwise given in Appendix \ref{Appendix:MMatrix} . They only depend on the ratio $\frac{p}{\bar{\mu}}$, and are independent of $\Lambda_k$, $\alpha_D$, and $\alpha_L'$, respectively.

For any $\chi\in\{S,V,T\}$ they can be written as ${\cal M}^{{\chi}} = \Pi^{{\chi}}_{+} - \Pi^{{\chi}}_{-}$ where $\Pi^{{\chi}}_{+}$ and $\Pi^{{\chi}}_{-}$ are orthogonal projectors: $\Pi^{{\chi}}_{+}\cdot\Pi^{{\chi}}_{-} = 0$. This implies that for any $n=1,2,3 \cdots$
\begin{equation}
\big({\cal M}^{{\chi}}\big)^2 = \Pi^{{\chi}}_{+} + \Pi^{{\chi}}_{-}, \quad \big({\cal M}^{{\chi}}\big)^{2n - 1} = {\cal M}^{{\chi}}\quad \text{and} \quad \big({\cal M}^{{\chi}}\big)^{2n} = \big({\cal M}^{{\chi}}\big)^2. 
\end{equation}
Furthermore, defining $\Pi^{{\chi}}_{0} \equiv  \openone^{{\chi}} - \Pi^{{\chi}}_{+} - \Pi^{{\chi}}_{-}$ we end up with a set of three orthogonal projectors:
\begin{equation}
 \Pi^{{\chi}}_{i}\cdot\Pi^{{\chi}}_{j} = \delta_{ij} \,\Pi^{{\chi}}_{j}\:,\quad \forall\, i,\,j\in\{+, -, 0\}.
\end{equation}

Exploiting in particular that $\big({\cal M}^{{\chi}}\big)^2=\openone^{{\chi}} - \Pi^{{\chi}}_{0}$ and that $\Pi^{{\chi}}_{0}$ is orthogonal to ${\cal M}^\chi$ we obtain the remarkable relation
\begin{equation}
({\cal M}^{{\chi}})^3 = {\cal M}^{{\chi}} 
\end{equation}
which makes it obvious that the spectrum of ${\cal M}^\chi$ consists of $0$ and $\pm1$.

Now we introduce the quantity $\nu_\chi$ that denotes the numbers of scalars, vectors, and tensors in the decomposition of the spin connection fluctuation $\bar{\tau}^{kl}_{~~\alpha}$. Since the fields $\{A,B\}$ and $\{C,D\}$ are connected by an ${\sf O}(4)$ dualization, $\nu_\chi$ is an even number; explicitly, we have $\nu_{\rm S}=2$, $\nu_{\rm V}=4$, and $\nu_{\rm T}=2$, summing up to the expected $\sum_{\chi \in \{{\rm S,V,T}\}} \nu_\chi d_\chi=24$ independent components of the spin connection. 

The spectrum of each of the orthogonal projectors $\Pi^\chi_+$ and $\Pi^\chi_-$ contains the eigenvalue +1 with $\nu_\chi/2$-fold degeneracy and the eigenvalue 0 with $(n_\chi-\nu_\chi/2)$-fold degeneracy. Therefore, in the matrix ${\cal M}^\chi$ each of the eigenvalues +1 and -1 occurs with a degeneracy $\nu_\chi/2$, whereas 0 occurs with $(n_\chi-\nu_\chi)$-fold degeneracy. In the spectrum of $\big({\cal M}^\chi\big)^2$, the degeneracy of +1 is given by $\nu_\chi$, and the degeneracy of 0 is given by $n_\chi-\nu_\chi$. The 0-eigenvalues originate from the component fields of the vielbein fluctuations, i.\,e. from the $\bar{\varepsilon}^k_{~\alpha}$-decomposition. Therefore, the algebraic trace over the indices $i_\chi$ and $j_\chi$ of these matrices is given by
\begin{equation}
 {\rm Tr}_{\rm alg} \big\{{\cal M}^{\chi}\big\} = 0 \hspace{0.3cm}\mbox{and}\hspace{0.3cm}{\rm Tr}_{\rm alg} \big\{\big({\cal M}^{\chi}\big)^2\big\} = \nu_{\chi}\:.
\end{equation}

\vspace{0.5cm}
The spectrum of the operator ${\cal N}^\chi\equiv \openone_{n_\chi\times n_\chi}+ \frac{1}{\gamma_k} {\cal M}^\chi$ consists of the eigenvalues $1+\frac{1}{\gamma_k}$ and  $1-\frac{1}{\gamma_k}$ (each with degeneracy $\nu_\chi/2$), and 1 (with degeneracy $n_\chi-\nu_\chi$). This spectrum is denoted by ${\rm spec}({\cal N}^{\chi}) = \big\{\lambda^{\chi}_{{\cal N}{\alpha_{\chi}}}\}$ where the eigenvalues $\lambda^\chi_{{\cal N}\,{\alpha_\chi}}=1,1\pm \frac{1}{\gamma_k}$ are distinguished by an index $\alpha_{\chi} \in \{1, \cdots, n_{\chi}\}$.

\paragraph{(B) Evaluation of $\frac{1}{2} \Tr \text{ln} \big\{\openone + \frac{1}{\gamma_k} {\cal M}\big\}$.} Consistent with the other contributions to the supertrace computed earlier, this logarithm is represented as a proper-time integral after squaring its argument in order to exponentiate a positive operator; with ${\rm Tr}_{\rm alg}\big\{{\rm ln}\:{\cal N}^{\chi}\big\} = \frac{1}{2} {\rm Tr}_{\rm alg}\big\{{\rm ln}\big({\cal N}^{\chi}\big)^2\big\}$ we obtain
\begin{equation}
 \frac{1}{2} {\rm Tr}_{\rm alg}\,\big[{\rm ln}\big({\cal N}^{\chi}\big)^2\big]_{\sf REG}\big\} = - \frac{1}{2}\ddashint\frac{\dr s}{s}{\rm Tr}_{\rm alg} \Big(\eu^{-s\,({\cal N}^{\chi})^2}\Big)\ = - \frac{1}{2}\sum_{\alpha_{\chi}=1}^{n_{\chi}} \ddashint\frac{\dr s}{s} \Big(\eu^{-s\,\big(\lambda^{\chi}_{{\cal N}{\alpha_{\chi}}}\big)^2}\Big)
\end{equation}
Since the spectrum of ${\cal N}^\chi$ is invariant w.\,r.\,t. $\gamma_k \rightarrow - \gamma_k$, this trace, that will only contribute to the beta function of the cosmological constant, constitutes an even function of $\gamma_k$, as well.

Due to the above algebraic relations this algebraic trace can be computed exactly. In order to give a first idea concerning the $\gamma_k$-dependence of the unregularized trace, we state the result:
\begin{equation}\label{First_Idea}
 {\rm Tr}_{\rm alg}\,{\rm ln}\Big\{\openone_{n_{\chi}\times n_{\chi}} + \frac{1}{\gamma_k}{\cal M}^{\chi}\Big\}= \frac{1}{2}\,\nu_{\chi}\, {\rm ln}\Big(\frac{{\gamma_k}^2 - 1}{{\gamma_k}^2}\Big)\:.
\end{equation}
We observe that \eqref{First_Idea} becomes singular for $\gamma_k\rightarrow \pm 1$ and $\gamma_k \rightarrow0$.

\subsection{The supertrace in terms of eigenvalues and interaction\\matrix elements}\label{Section76}

Combining the results of the previous subsection we can now write down a representation of the functional traces constituting $\Gamma_k$ in which the linear algebra part is evaluated at the formal level to a maximal extent. It involves the matrix elements and eigenvalues of the free part of the Hessian, as well as the reduced matrix elements of its interaction part:\begin{equation}
 \begin{split}
  \Gamma_k &= S - \frac{1}{4}\int \dr^4 x\:\bar{e}\int\frac{\dr^4 p}{(2 \pi)^4}\bigg[\ddashint\frac{\dr s}{s}\Big(\sum_{\chi \in\{{\rm S}, {\rm V}, {\rm T}\}} d_\chi \sum_{\alpha_\chi = 1}^{n_\chi}\eu^{ - s \,\big(\bar{\lambda}^\chi_{\alpha_\chi}(p^2)\big)^2} - 40\:\eu^{ - s \,\bar{\mu}^6}\Big)\\
&\hspace{3.7cm} - 2\ddashint\frac{\dr s^\prime}{s^\prime}\Big(\sum_{\chi \in\{{{\rm gh}\,{\rm S}}, {{\rm gh}\,{\rm V}}\}} d_\chi \sum_{\alpha_\chi = 1}^{n_\chi}\eu^{ - s^\prime \,\big(\bar{\lambda}^\chi_{\alpha_\chi}(p^2)\big)^2} - 10\:\eu^{ - s^\prime \,\bar{\mu}^2}\Big)\\
&\hspace{7cm} + \sum_{\chi \in\{{\rm S}, {\rm V}, {\rm T}\}} d_{\chi}\sum_{\alpha_{\chi}=1}^{n_{\chi}} \ddashint\frac{\dr s}{s} \Big(\eu^{-s\,\big(\lambda^{\chi}_{{\cal N}{\alpha_{\chi}}}\big)^2}\Big)\bigg]\\
& + \frac{1}{2}\big(\int\dr^4 x\:\bar{e}\:n^{(\pm)\:ab} f_{ab}\big)\int\frac{\dr^4 p}{(2 \pi)^4}\bigg[ \sum_{\chi \in \{{\rm S}, {\rm V}, {\rm T}\}}\sum_{\alpha_\chi = 1}^{n_\chi}\:\sum_{i_\chi, j_\chi, k_\chi = 1}^{n_\chi} \ddashint\dr s\:\eu^{ - s(\bar{\lambda}^\chi_{~\alpha_\chi}(p^2))^2}\\
&\hspace{2.3cm}\big(v^{\chi \alpha_\chi}\big)_{i_\chi} (p^2) \big(\big(H^\chi\big)^{-1}\big)^{i_\chi}_{~j_\chi} (p^2) \big(\check{V}^\chi_{\rm contr}\big)^{j_\chi}_{~k_\chi} (p^2) \big(v^\chi_{~\alpha_\chi}\big)^{k_\chi} (p^2) \big(\bar{\lambda}^\chi_{\alpha_\chi} (p^2)\big)^2 \\
&~ -\, 2 \sum_{\chi \in \{{{\rm gh}\,{\rm S}}, {{\rm gh}\,{\rm V}}\}}\sum_{\alpha_\chi = 1}^{n_\chi}\:\sum_{i_\chi, j_\chi, k_\chi = 1}^{n_\chi}\ddashint\dr s^\prime\:\eu^{ - s^\prime(\bar{\lambda}^\chi_{~\alpha_\chi}(p^2))^2}\\
&\hspace{4.5cm} \big(v^{\chi \alpha_\chi}\big)_{i_\chi} (p^2) \big(H^\chi_0\big)^{i_\chi}_{~j_\chi} (p^2) \big(\check{V}^\chi_{\rm contr}\big)^{j_\chi}_{~k_\chi} (p^2) \big(v^\chi_{~\alpha_\chi}\big)^{k_\chi} (p^2) \bigg]\:.
 \end{split}\label{ToBeRegularized}
\end{equation}

In deriving the representation \eqref{ToBeRegularized} we have done as much as is possible in order to break down the formidable task of evaluating the supertrace to a set of manageable smaller problems. In particular, the linear algebra on the 40 dimensional field space of the spin connection and the vielbein got reduced to the analytically tractable problem of diagonalizing various ``small'' algebraic matrix blocks.

\subsection{The regularization scheme}
The intermediate result \eqref{ToBeRegularized} is still formal in the sense that the details of its regularization need to be specified. Besides the proper-time regularization, an additional damping of the momentum integration $\int {\rm d}^4 p$ will be needed; its explicit implementation constitutes the issue of the following subsection. Together with the proper-time regularization, it defines the renormalization scheme we employ and is therefore, in a sense, equivalent to picking a certain ${\cal R}_k$ in the standard case.

First let us introduce the familiar dimensionless couplings in cutoff units
\begin{equation}
g_k\equiv G_k k^2,\qquad\lambda_k \equiv \Lambda_k k^{-2} 
\end{equation}
and the dimensionless quantities
\begin{equation}
 \begin{aligned}
  \mu &\equiv \bar{\mu}\; k^{-1},& y &\equiv p / k,\\
  \lambda^\chi_{\alpha_\chi}(y^2) &\equiv 16 \pi G_k\:\bar{\lambda}^\chi_{\alpha_\chi}(p^2)\:k^{-1},& \chi &\in\{{\rm S}, {\rm V}, {\rm T}\},\\
  \lambda^\chi_{\alpha_\chi} (y^2) &\equiv \bar{\lambda}^\chi_{\alpha_\chi}(p^2)\:k^{-1}, & \chi &\in\{{{\rm gh}\,{\rm S}}, {{\rm gh}\,{\rm V}}\}\:.
 \end{aligned}
\end{equation}

The dimensionful Lorentz gauge parameter $\alpha_L'$ can be rescaled either by means of $k$ or by means of $\bar{\mu}$. We decide for the latter and define $\alpha_L'\equiv f \bar{\mu}^{-2} \mu^{-2}= f \mu^{-4} k^{-2}$ which implies for the original Lorentz gauge parameter $\alpha_L=\frac{\alpha_L'}{16 \pi G_k}$:
\begin{equation}
 \alpha_L= 16 \pi g_k f \bar{\mu}^{-4}
\end{equation}
Obviously, a redefinition according to $f\rightarrow \mu^4$ amounts to a rescaling by means of the cutoff scale $k$. Therewith, all objects on the RHS of \eqref{ToBeRegularized} can be substituted by their dimensionless counterparts.

\subsubsection{The $\hat{\boldsymbol{p}}^{2n}$-regularization}
The positive operators that are exponentiated by means of the proper-time integrals will be subject to another kind of regularization over and above the proper-time cutoff. The reason is that some of the eigenvalues of $\big(H_0^{\text{grav}}\big)^2$ and $\big(H_0^{\text{gh}}\big)^2$, rather than growing $\propto p^2$, approach {\it constant} values for $p^2\rightarrow \infty$. The additional regularization cures (numerical) instabilities due to the otherwise necessary cancellation of large contributions.

We define the operator ${\bf \hat{p}}^2$, whose generalized position space representation in the graviton vector block is given by
\begin{equation}
 \langle x\,m\,i_{\rm V}\,|{\bf \hat{p}}^2|\,y\,k\,j_{\rm V}\rangle = \big(-\Box_x\big) \delta^{i_{\rm V}}_{~j_{\rm V}} P_{{\rm V}~k}^{~m} \Big(\frac{-i\partial_x}{\sqrt{- \Box_x}}\Big) \frac{\delta^{(4)}(x - y)}{\bar{e}}\:,
\end{equation}
i.\,e. ${\bf \hat{p}}^2$ exhibits a simple product structure consisting of $-\Box$ times the identity w.\,r.\,t. the field labels of the considered block times the transverse projector times a tensorial delta distribution. This operator is positive; it commutes with $H_0^\chi$ and
${\cal N}^\chi$, but not with $V^\chi$.

In the case of the free logarithms the regularization acts as follows: Let $\Omega$ be some positive operator with mass dimension $\hat{n}$, $[\Omega]=\hat n$. Then we typically encounter
\begin{equation}
 \Tr \ln \bigg(\frac{\Omega}{\bar{\mu}^{\hat{n}}}\bigg) \equiv \Tr \ln \big(\Omega {\bf \hat{p}}^{2n}{\bf \hat{p}}^{-2n}\bar{\mu}^{-\hat{n}}\big)\equiv \Tr \ln \big(\Omega {\bf \hat{p}}^{2n}\big)- \Tr \ln \big(\bar{\mu}^{\hat{n}} {\bf \hat{p}}^{2n}\big),
\end{equation}
and we regularize according to the prescription
\begin{equation}
 {\rm Tr}\,\Big[{\rm ln}\Big(\frac{\Omega}{\bar{\mu}^{\hat{n}}}\Big)\Big]_{\sf REG} \equiv  - \ddashint \frac{\dr s}{s}\Big({\rm Tr}\,\eu^{-s\,\Omega\,{\bf \hat{p}}^{2n}} - {\rm Tr}\,\eu^{-s\,\mu^{\hat{n}}\,{\bf \hat{p}}^{2n}}\Big)\label{reg1}
\end{equation}
Therein, $n=1,2,\cdots$ is an integer that (partially) specifies the regularization scheme, and $\ddashint$ denotes the proper-time integral with a (sharp or smooth) cutoff. 

The traces that contain the interaction contributions have the structure
\begin{equation}
 \Tr \Big( A \Omega^{-1}\Big)=\Tr \Big( A {\bf \hat{p}}^{2n}{\bf \hat{p}}^{-2n}\Omega^{-1}\Big)=\Tr \Big( A {\bf \hat{p}}^{2n}(\Omega{\bf \hat{p}}^{2n})^{-1}\Big)
\end{equation}
with some operator $A$ which does not commute with ${\bf \hat{p}}^2$ in general. In these cases the regularization is implemented by the rule
\begin{equation}
 {\rm Tr}\,\big\{A\,\big[\Omega^{-1}\big]_{\sf REG}\big\} \equiv  {\rm Tr}\Big\{A\,{\bf \hat{p}}^{2n} \ddashint \dr s\,\eu^{-s\,\Omega\,{\bf \hat{p}}^{2n}}\Big\}\:.\label{reg2}
\end{equation}
As we said already, this ``${\bf \hat{p}}^{2n}$-regularization'' is needed since some of the eigenvalues of $\big(H_0^{\text{grav}}\big)^2$ and $\big(H_0^{\text{gh}}\big)^2$ approach constant values for large values of $p^2$.

\subsubsection{The proper-time cutoff}
The proper-time regularization is implemented using standard regulator functions \cite{bonannozappala}. We will employ three different schemes: The sharp cutoff, the $C^m_k$-regularization and the $f^m_k$-regularization.

\noindent{\bf (i)} In the case of the {\it sharp cutoff} the proper-time integral of a positive operator with mass dimension $\hat{n}$ as in the previous subsection is cut off according to 
\begin{equation}
 \ddashint {\rm d}s\equiv \int_{\Lambda_{\text{UV}}^{-\hat{n}}}^{k^{-\hat{n}}} {\rm d}s,
\end{equation}
$k$ and $\Lambda_{\text{UV}}$ denoting IR and UV cutoff scales, respectively. The IR flow equation we are going to derive is not sensitive to the UV regularization; we may therefore formally send $\Lambda_{\text{UV}}$ to infinity after having performed the scale derivative, $\Lambda_{\text{UV}}\rightarrow\infty$.

\noindent{\bf (ii)} The {\it $C^m_k$-scheme} regularizes the proper-time integrals according to
\begin{equation}
 \ddashint \dr s \equiv \int_0^\infty \dr s\:C^m_k (s) 
\end{equation}
with regulator functions
\begin{equation}
 C^m_k (s) \equiv  \frac{\Gamma(m+1, sk^{\hat{n}}) - \Gamma(m+1, s{\Lambda_{\rm UV}}^{\hat{n}})}{\Gamma (m+1)}
\end{equation}
and $m \ge 0$, $s \ge 0$, $k \ge 0$; here $\Gamma(\alpha,x)$ denotes the incomplete gamma function 
\begin{equation}
 \Gamma(\alpha,x)\equiv \int_x^\infty {\rm d} r\, r^{\alpha-1} e^{-r}.
\end{equation}

\noindent{\bf (iii)} The {\it $f^m_k$-regularization} is obtained by substituting $s\rightarrow zs$ in the functions $C^m_k$:
\begin{equation}
 \ddashint {\rm d}s\equiv \int_0^\infty {\rm d}s f^m_k(s)
\end{equation}
with $f^m_k(s)\equiv C^m_k(zs)=\frac{\Gamma(m+1, zsk^{\hat{n}}) - \Gamma(m+1, zs{\Lambda_{\rm UV}}^{\hat{n}})}{\Gamma (m+1)}$.
In order to reproduce the sharp cut off in the limit $m\rightarrow \infty$, we have to choose $z=m$ for the free logarithms and $z= m+1$ for the traces containing interaction contributions; in these cases, the cutoff scale is given by $(m)^{1/\hat{n}} k$ and $(m+1)^{1/\hat{n}} k$, respectively. 

In the following we will combine the formulae pertaining to the $C^m_k$- and the $f^m_k$-scheme by introducing two book keeping variables: $\rho_H$ for the free part and $\rho_V$ for the interaction contributions. Then we can easily switch from one scheme to the other by setting
\begin{equation}
\begin{aligned}
 \rho_H&=1& \rho_V&=1\qquad & &\text{for the $C^m_k$-scheme,}\\
\rho_H&=m& \rho_V&=m+1 \qquad& &\text{for the $f^m_k$-scheme.}
\end{aligned}
\nonumber 
\end{equation}

In the graviton sector there are two proper-time integrals containing the exponential of the eigenvalues of $\big(H_0^{\text{grav}}\big)^2$. In order to implement the IR cutoff w.\,r.\,t. an operator that is ``as close as possible'' to the usual $k$-independent Laplacian $\Box$, the global prefactor $(16 \pi g_k)^2$ is defined into the IR cutoff scale of the proper-time integral. This amounts to an approximate implementation of the ``$Z=\zeta$'' rule that is usually employed in the context of QEG. Note however that while we cut off the spectrum of $g_k$-independent operators they still depend on the running $\lambda_k$.
\vspace{0.5cm}

The following remarks are in order here.

\noindent{\bf (A)} The transition from the exact FRGE to its lowest order proper-time approximation considered here is accomplished by identifying $\Gamma_k^{(2)}$ with the argument of ${\cal R}_k$, but in the following neglecting the scale dependence of $\Gamma_k^{(2)}$. Therefore, no scale derivatives of dimensionless couplings occur on the RHS of the proper-time equation, and we can safely redefine the Lorentz gauge parameter $\alpha_L'$ by means of an additional factor of $g_k$. The inclusion of $g_k$ into the IR cutoff scale of the proper-time integral does not lead to any scale derivatives of $g_k$, either.
\vspace{0.5cm}

\noindent{\bf (B)} For the same reason, the assumption of a constant {\it dimensionless} mass parameter $\mu$ does not imply any restrictions. Moreover, only for the choice $\mu= \text{const}$ the system of partial differential equations, $\partial_t u_\alpha= \beta_\alpha(u_1,u_2,\cdots)$, that we are about to derive will be autonomous. Stated differently: Only for $\mu=\text{const}$ the vector field $\vec{\beta}\equiv(\beta_\lambda,\beta_\gamma,\beta_g)$ and the fixed point structure of the flow it generates will not exhibit an explicit $k$-dependence.

However, if $\mu$ was a function of the other dimensionless couplings, we obtain an autonomous system of flow equations, as well. A natural choice is given by identifying $\bar{\mu}$ with the running Planck mass according to $\bar{\mu}\equiv {G_k}^{-1/2}\Rightarrow \mu_k={g_k}^{-1/2}$. We will also comment on the results obtained with this second choice in the next section.
\vspace{0.5cm}

\noindent{\bf (C)} The parameter $\bar{\mu}$ results from our freedom in parametrizing the fluctuation fields and therefore of choosing a representation of $\Gamma_k^{(2)}$. Since within the truncation considered the LHS of the FRGE is evaluated for vanishing fluctuations, it happens to be independent of the chosen representation. As a consequence we are not led to a flow equation for $\mu$. This is due to the ``single-metric'' character of the present truncation ansatz. In a more advanced ``bi-metric'' treatment the situation would be different, similar to those already performed in metric QEG \cite{MRS}. There, the LHS of the flow equations, too, depends on the fluctuations and therefore on the chosen representation of $\Gamma_k^{(2)}$. Thus, within such an approach one might be able to derive a flow equation $\partial_t \mu_k=\cdots$ for $\mu$ as well, providing us with a now 4-component vector field $\vec{\beta}\equiv (\beta_\lambda,\beta_\gamma,\beta_g,\beta_\mu)$ with no explicit $k$ dependence.
\vspace{0.5cm}

\noindent{\bf (D)} In setting up the above simplified flow equation we relaxed the requirement of {\it background independence} to some extent since several steps of its derivation are not covariant under background gauge transformations: In the transverse-traceless decompositions of the fluctuation and ghost fields partial rather than covariant derivatives are used, the parameter $k$ is a cutoff in the spectrum of $\bar{g}^{\mu\nu} \partial_\mu\partial_\nu\equiv \Box$, and correspondingly the $\hat{p}^{2n}$  regularization refers to this operator involving partial derivatives. As a result, the domain of validity of the simplified flow equation, in the space of metrics, is restricted to a vicinity of flat space. Ultimately we would like to go beyond this approximation, of course, but given its much higher calculational complexity it seems sensible to embark on the general case where $\bar{g}_{\mu\nu}$ is allowed to be ``far away'' from $\eta_{\mu\nu}$ only after having gained some first insights and technical experience.

\subsection{The $\beta$-functions for $g$, $\gamma$ and $\lambda$}
At this point we finally are in the position to write down the desired proper-time approximation to the FRGE. We start out from the representation \eqref{ToBeRegularized} for the functional traces of $\Gamma_k$, insert the two regularization prescriptions outlined in the previous subsection, and then we take a derivative w.\,r.\,t. the RG scale $t\equiv \ln k$. Making its dependence on the three invariants and on $k^2$ manifest, the result for $\partial_t\Gamma_k$ has the structure:

\begin{equation}
 \begin{split}
  \partial_t\,\Gamma_k &= \big(\int \dr^4 x\:\bar{e}\big)\:k^4\:\Big[I^{\rm grav}_{\rm F} (\lambda_k, \mu) + I^{\rm grav}_{\cal N} (\gamma_k) - 2\,I^{\rm gh}_{\rm F} (\mu)\Big]\\
& + \big(\int \dr^4 x\:\bar{e}\:n^{(\pm)\:ab} f_{ab}\big)\:k^2\:\Big[I^{\rm grav}_{{\rm V}(+)} (\lambda_k, \gamma_k, \mu) - 2\,I^{\rm gh}_{\rm V} (\mu)\Big]\\
& \mp \big(\int \dr^4 x\:\bar{e}\:n^{(\pm)\:ab} f_{ab}\big)\:k^2\:I^{\rm grav}_{{\rm V}(-)} (\lambda_k, \gamma_k, \mu)\label{fluss-einheitlich}
 \end{split}
\end{equation}
Here the coefficient functions $I^{\rm grav}_{\rm F}, I^{\rm grav}_{\cal N}, \cdots$ are dimensionless, and each one of the three invariants is multiplied by a power of $k$ that amounts precisely to its inverse mass dimension. The notation for the $I$'s is self-explaining: the superscript distinguishes graviton and ghost sector, while the subscript indicates the term in \eqref{ToBeRegularized} the respective coefficient function originates from. For the $C^m_k$/$f^m_k$-regularization they read explicitly
\begin{equation}
 \begin{aligned}\nonumber
  I^{\rm grav}_{\rm F}(\lambda_k,\mu)&=\frac{1}{8\pi^2}\frac{n+3}{2}\!\int_0^\infty\!\!\!\dr y\:y^3\bigg[\sum_{\genfrac{}{}{0pt}{2}{\chi \in}{\{{\rm S}, {\rm V}, {\rm T}\}}} \!\!d_\chi \sum_{{\alpha_\chi} = 1}^{n_\chi}\Big[\frac{\zeta_{\rm H}}{\zeta_{\rm H} + y^{2n}\big(\lambda^\chi_{\alpha_\chi} (y^2)\big)^2}\Big]^{m+1}
 \\&\hspace{7cm}- 40 \Big[\frac{\zeta_{\rm H}}{\zeta_{\rm H} + y^{2n}\mu^6}\Big]^{m+1}\bigg]\\
  I^{\rm grav}_{\cal N}(\gamma_k)&=\frac{(\zeta_{\rm H})^{\frac{2}{n}}}{8\pi^2}\frac{3}{m!}\Gamma\Big(\frac{2}{n}\Big)\Gamma\Big(m+1-\frac{2}{n}\Big) \bigg[\Big(1+\frac{1}{\gamma_k}\Big)^{-\frac{4}{n}} + \Big(1-\frac{1}{\gamma_k}\Big)^{-\frac{4}{n}} - 2\bigg]\\
    I^{\rm grav}_{\rm V(\pm)}(\lambda_k,\gamma_k,\mu)&=- \frac{n\!+\!3}{8 \pi^2}\frac{m\!+\!1}{\zeta_{\rm V}}\!\int_0^\infty\!\!\!\dr y\,y^{3+2n}\bigg[ \sum_{\genfrac{}{}{0pt}{2}{\chi \in}{\{{\rm S}, {\rm V}, {\rm T}\}}}\sum_{{\alpha_\chi}=1}^{n_\chi}\sum_{i_\chi, j_\chi = 1}^{n_\chi} \Big[\frac{\zeta_{\rm V}}{y^{2n} \big(\lambda^\chi_{\alpha_\chi}(y^2)\big)^2\! +\! \zeta_{\rm V}}\Big]^{m+2}\\
  &\hspace{2cm}\big(\lambda^\chi_{\alpha_\chi}(y^2)\big)^2 \big(v^{\chi{\alpha_\chi}}\big)_{i_\chi}\big(\big(H^\chi\big)^{-1}\big)^{i_\chi}_{~j_\chi}(y^2)\big(\check{V}^{\chi(\pm)}_{\rm contr}\big)^{j_\chi}_{~k_\chi}(y^2)\big(v^\chi_{~{\alpha_\chi}}\big)^{k_\chi} \bigg]
  \end{aligned}
  \end{equation}
  
\begin{equation}
 \begin{aligned}\label{SmoothCutoffCoeffients}
  I^{\rm gh}_{\rm F}(\mu)&=\frac{1}{8\pi^2}\frac{n+1}{2}\!\int_0^\infty\!\!\!\dr y\:y^3\bigg[\!\!\sum_{\genfrac{}{}{0pt}{2}{\chi \in}{\{{{\rm gh}\,{\rm S}}, {{\rm gh}\,{\rm V}}\}}}\!\!\!\! d_\chi \sum_{{\alpha_\chi} = 1}^{n_\chi}\Big[\frac{\zeta_{\rm H}}{\zeta_{\rm H} + y^{2n}\big(\lambda^\chi_{\alpha_\chi} (y^2)\big)^2}\Big]^{m+1}\\
&\hspace{8cm}- 10 \Big[\frac{\zeta_{\rm H}}{\zeta_{\rm H} + y^{2n}\mu^2}\Big]^{m+1}\bigg]\\
  I^{\rm gh}_{\rm V}(\mu)&=-\frac{n\!+\!1}{8\pi^2}\frac{m\!+\!1}{\zeta_{\rm V}}\!\int_0^\infty\!\!\!\!\dr y\:y^{3+2n}\bigg[\!\!\!\sum_{\genfrac{}{}{0pt}{2}{\chi \in}{\{{{\rm gh}\,{\rm S}}, {{\rm gh}\,{\rm V}}\}}}\!\sum_{{\alpha_\chi}=1}^{n_\chi}\sum_{i_\chi, j_\chi = 1}^{n_\chi}\Big[\frac{\zeta_{\rm V}}{y^{2n} (\lambda^\chi_{\alpha_\chi}(y^2))^2 \!+\! \zeta_{\rm V}}\Big]^{m+2}\\
  &\hspace{4.5cm} \lambda^\chi_{\alpha_\chi} (y^2) \big(v^{\chi{\alpha_\chi}}\big)_{i_\chi}\big(\check{V}^\chi_{\rm contr}\big)^{i_\chi}_{~j_\chi}(y^2)\big(v^\chi_{~{\alpha_\chi}}\big)^{j_\chi}\bigg]\\[-0.7cm]
\end{aligned}
\end{equation}

\vspace*{.7cm}
\noindent Here, $\rho_H=1=\rho_V$ for the $C^m_k$ and $\rho_H=m$, $\rho_V=m+1$ for the $f^m_k$ regularization.

The analogous results for the sharp proper-time cutoff are listed in Appendix \ref{Appendix:SharpCutoff}.

In the formulae \eqref{SmoothCutoffCoeffients} for the coefficient functions the integration is over the dimensionless momentum variable $y\equiv p/k$. As the angular part of the momentum integration is trivial it appeared upon replacing $\int \frac{{\rm d}^4 p}{(2\pi)^4}$ with $\frac{k^4}{8 \pi^2}\int_0^\infty \dr y y^3$.

In writing down \eqref{SmoothCutoffCoeffients} we also employed the ``$(\pm)$ notation'' for the reduced matrix elements $\check{V}^{\rm grav}_{\rm contr}$ and decomposed accordingly $\check{V}^{\rm grav}_{\rm contr} \equiv \check{V}^{{\rm grav}(+)}_{\rm contr} \mp \check{V}^{{\rm grav}(-)}_{\rm contr}$ into terms that stay invariant under $\bar{\omega}^{(+)}\mapsto\bar{\omega}^{(-)}$ and those that change their sign.

The integrands of the $y$-integrals depend on $\lambda_k$, $\gamma_k$, $\mu$, $\alpha_D$, $\beta_D$ and $f$, whereby $\gamma_k$ enters via the spectrum of ${\cal N}$ and the products $(H^{\text{grav}})^{-1}V_{\text{contr}}^{\text{grav} (+)}$ and $(H^{\text{grav}})^{-1}V_{\text{contr}}^{\text{grav} (-)}$. The spectrum of the free operators $H_0^{\text{grav}}$ and $H_0^{\text{gh}}$ only depends on $\lambda_k$, $\mu$, $\alpha_D$, $\beta_D$ and $f$. As a consequence of the rough implementation of the $z=\zeta$-rule the RHS of the flow equations does not depend on $g_k$ any more.

In the arguments of $I^{\rm grav}_{\rm F}(\lambda_k,\mu),\cdots$ we have suppressed the dependence on the gauge parameters. Choosing a value $\beta_D\neq -1$ singles out a concrete gauge condition whose implementation is fixed by the gauge parameters $\alpha_D$ and $f$. The ``family parameters'' $n$ and $(n,m)$ determine the explicit form of the sharp and the $C^m_k$/$f^m_k$-proper-time regularization, respectively. Finally, a value of the dimensionless mass parameter $\mu$ has to be fixed. However, as we have already argued, it is consistent and by no means restricting to assume this value to be $k$ independent within the present approximation. Even though $\mu$ is not a coupling since it is not associated with an invariant, it is included as an argument of the functions on the RHS of \eqref{fluss-einheitlich}.

At this point it remains to insert the spectra and matrix elements which we obtained analytically into $I^{\rm grav}_{\rm F}, I^{\rm grav}_{\cal N},\cdots$ and to perform the various finite sums in \eqref{SmoothCutoffCoeffients}; the resulting expressions are extremely complicated and lengthy. They would fill many pages and, hence, cannot be displayed here. These expressions (besides $\mu$ and the gauge parameters) still contain the dimensionless momentum variable $y$, and the last computational step consists in a numerical integration over $y$.

Finally, comparing the coefficients in front of the three invariants, the system of beta functions is obtained as
\begin{equation}\label{vollst-exakt-fluss}
\boxed{\begin{aligned}
\partial_t\,g_k &= \beta_g = \Big[2 + 16 \pi \Big(I^{\rm grav}_{{\rm V}(+)} (\lambda_k, \gamma_k, \mu) - 2\,I^{\rm gh}_{\rm V} (\mu)\Big)g_k\Big]g_k \\  
\partial_t\,\gamma_k &= \beta_\gamma = 16 \pi g_k\,\gamma_k\Big[\gamma_k\, I^{\rm grav}_{{\rm V}(-)} (\lambda_k, \gamma_k, \mu) - \Big(I^{\rm grav}_{{\rm V}(+)} (\lambda_k, \gamma_k, \mu) - 2\,I^{\rm gh}_{\rm V} (\mu)\Big)\Big] \\
\displaystyle \partial_t\,\lambda_k &= \beta_\lambda = 8\pi g_k\Big[\Big(I^{\rm grav}_{\rm F} (\lambda_k, \mu) + I^{\rm grav}_{\cal N} (\gamma_k) - 2\,I^{\rm gh}_{\rm F} (\mu)\Big) \\
& \hspace{3cm} + 2\Big(I^{\rm grav}_{{\rm V}(+)} (\lambda_k, \gamma_k, \mu) - 2\,I^{\rm gh}_{\rm V} (\mu)\Big)\lambda_k\Big] - 2\,\lambda_k\:.
\end{aligned}}
\end{equation}
This system of equations constitutes our central result. It describes the RG flow on the 3-dimensional truncated theory space coordinatized by the dimensionless Newton constant $g_k\equiv k^2 G_k$, cosmological constant $\lambda_k\equiv \Lambda_k/k^2$, and the Immirzi parameter $\gamma_k$.

The $I$-functions on the RHS of \eqref{vollst-exakt-fluss} are much too complicated to easily deduce anything about the schematic structure of the beta functions or about their dependence on the couplings and parameters, in particular. The next section will be devoted to the analysis of the physical content of \eqref{vollst-exakt-fluss}.

Nevertheless, one particular property of the flow can already be inferred at this stage: From our previous considerations we can deduce that $I^{\rm grav}_{{\rm V}(-)} (\lambda_k, \gamma_k, \mu)$ is odd and $I^{\rm grav}_{\cal N} (\gamma_k)$ is invariant under the interchange $\gamma_k\leftrightarrow -\gamma_k$. {\it Therefore, the flow defined by \eqref{vollst-exakt-fluss} is mapped onto itself for $\gamma_k\rightarrow -\gamma_k$, i.\,e. it is symmetric w.\,r.\,t. the $(\gamma_k=0)$-plane.}

\section{Analysis of the RG Flow}\label{Results}

In this section we will analyze and interpret the physical content of the system of differential equations \eqref{vollst-exakt-fluss} of the three running couplings $\lambda$, $\gamma$, and $g$ in some detail.

Unless otherwise stated, we will choose the parameter of the ${\bf \hat{p}}^{2n}$-regularization to be $n=3$. This is its minimal value that ensures the convergence of all momentum integrals present in our calculation. Larger values of $n$ tend to result in less pronounced characteristics of the RG flow. The scheme dependence of our results will be tested by employing the three regularization schemes, sharp proper-time cutoff, $C^m_k$- and $f^m_k$-regularization, introduced in the previous section, and a variation of the cutoff parameter $m$ therein.

Investigating the gauge dependence in the three dimensional space of gauge parameters $f$, $\alpha_D$ and $\beta_D$ in full generality is an extremely tedious task. For that reason we restrict ourselves to the following discrete subset of gauge parameters: We always choose the Lorentz gauge parameter $f=1$. For the diffeomorphism gauge, the parameter $\beta_D$ takes on the values $0$ or $1$, and most of the discussion will concern one of the values $\alpha_D\in\{0.1,1,10\}$. Due to numerical instabilities, the (presumptive) fixed point value $\alpha_D=0$ cannot be realized. In addition to their gauge fixing dependence we shall also analyze the qualitative dependence of our results on the mass parameter $\mu$.

This section is divided into three subsections, each of which is devoted to a specific truncation. First, we analyze the RG flow in the two dimensional $(\lambda,g)$ coupling space; this amounts to discussing a truncation of the form of the Hilbert-Palatini action. Even though the couplings present in the Einstein-Hilbert truncation of metric gravity are denoted by the same symbols, we should refrain from expecting any further similarities concerning the results. Due to the different theory space under consideration, the results of this subsection are conceptually independent from the well-known QEG results.

In the second subsection, we include the Immirzi term, but concentrate on the $(\gamma,g)$-subspace, setting $\lambda=0$, while in the last subsection we discuss the full three dimensional $(\lambda, \gamma, g)$ coupling space.

\subsection{The 2-dimensional $(\lambda,g)$ subspace}
The system of flow equations on the $(\lambda,g)$ subspace can be derived from the full system \eqref{vollst-exakt-fluss} by omitting all contributions from the supertrace that are pseudo-scalars (all contributing to the running of the Immirzi parameter) and all contributions to the scalar part that are due to the Immirzi term. This corresponds to neglecting $\hat{V}^{{\rm grav}(-)}_{\rm contr}$ completely, while taking into account $\hat{V}^{{\rm grav}(+)}_{\rm contr}$ in the limit $\gamma_k \to \infty$. Using the notation $I^{\rm grav}_{\rm V} (\lambda, \mu)\equiv \lim_{\gamma \rightarrow \infty} I^{\rm grav}_{{\rm V}(+)} (\lambda, \gamma, \mu)$ and exploiting that $\lim_{\gamma \rightarrow \infty} I^{\rm grav}_{\cal N} (\gamma_k)=0$ we thus obtain the following system of flow equations:
\begin{subequations}\label{g-lambda-betas}
\begin{align} \partial_t\,g_k = \beta_g (g_k, \lambda_k) &= \Big[2 +\eta_{\rm N}(\lambda_k, g_k, \mu) \Big]g_k \label{g-beta-EH}\\  
 \partial_t\,\lambda_k = \beta_\lambda (g_k, \lambda_k) &= 8\pi g_k\Big[I^{\rm grav}_{\rm F} (\lambda_k, \mu) - 2\,I^{\rm gh}_{\rm F} (\mu)\Big] +\big[\eta_{\rm N}(\lambda_k, g_k, \mu)-2\big] \lambda_k\label{lambda-beta-EH}
\end{align}
\end{subequations}
Here $\eta_{\rm N}$ denotes the anomalous dimension of Newton's constant which reads explicitly
\begin{equation}
\eta_{\rm N} (\lambda, g, \mu) = 16 \pi \Big(I^{\rm grav}_{{\rm V}} (\lambda, \mu) - 2\,I^{\rm gh}_{\rm V} (\mu)\Big)g.
\end{equation} 

Before we turn to the numerical analysis we stress again that this system of RG equations cannot be seen as being analogous to the Einstein-Hilbert truncation of metric gravity. As the theory space analyzed here allows for torsion and the fluctuations of the torsion tensor are {\it not suppressed at all} in the limit $\gamma\rightarrow \infty$ in the path integral, this limit should rather describe an RG flow ``most distant'' to the one from metric gravity.

\paragraph{(A) The NGFP conditions.} The above system \eqref{g-lambda-betas} clearly allows for a Gaussian fixed point (GFP) at vanishing couplings. A non-Gaussian fixed point (NGFP) can only exist if $\eta_N=-2$ at that point, implying $\beta_g=0$. As $\eta_N$ is linear in $g$, we obtain from that condition a simple relation for the fixed point coordinates $(g^\ast,\lambda^\ast)$:
\begin{equation}
 g^\ast(\lambda^\ast) = - \Big[8 \pi\big(I^{\rm grav}_{{\rm V}} (\lambda^\ast, \mu) - 2\,I^{\rm gh}_{\rm V} (\mu)\big)\Big]^{-1}\:.\label{fp-bdg-1-EH}
\end{equation}
Inserting \eqref{fp-bdg-1-EH} into the second condition $\beta_\lambda (\lambda^\ast, g^\ast) = 0$ we are led to the function
\begin{equation}
 \beta_\lambda (g^\ast(\lambda), \lambda)=- \frac{I^{\rm grav}_{\rm F} (\lambda, \mu) - 2\,I^{\rm gh}_{\rm F} (\mu)}{I^{\rm grav}_{{\rm V}} (\lambda, \mu) - 2\,I^{\rm gh}_{\rm V} (\mu)} - 4\,\lambda\:, \label{fp-bdg-2-EH}
\end{equation}
whose zeros correspond to all non-Gaussian FP-values $\lambda^\ast$ that are possible: $\beta_\lambda(g^\ast(\kern-1pt\lambda^\ast\kern-1pt),\lambda^\ast\kern-1pt)\!=0$. The corresponding fixed point value of Newton's constant can then be obtained by \eqref{fp-bdg-1-EH}.

\paragraph{(B) Optimized choices for $\mu$ and $\alpha_{\rm D}$.}
\begin{figure}
 \includegraphics[width=\textwidth]{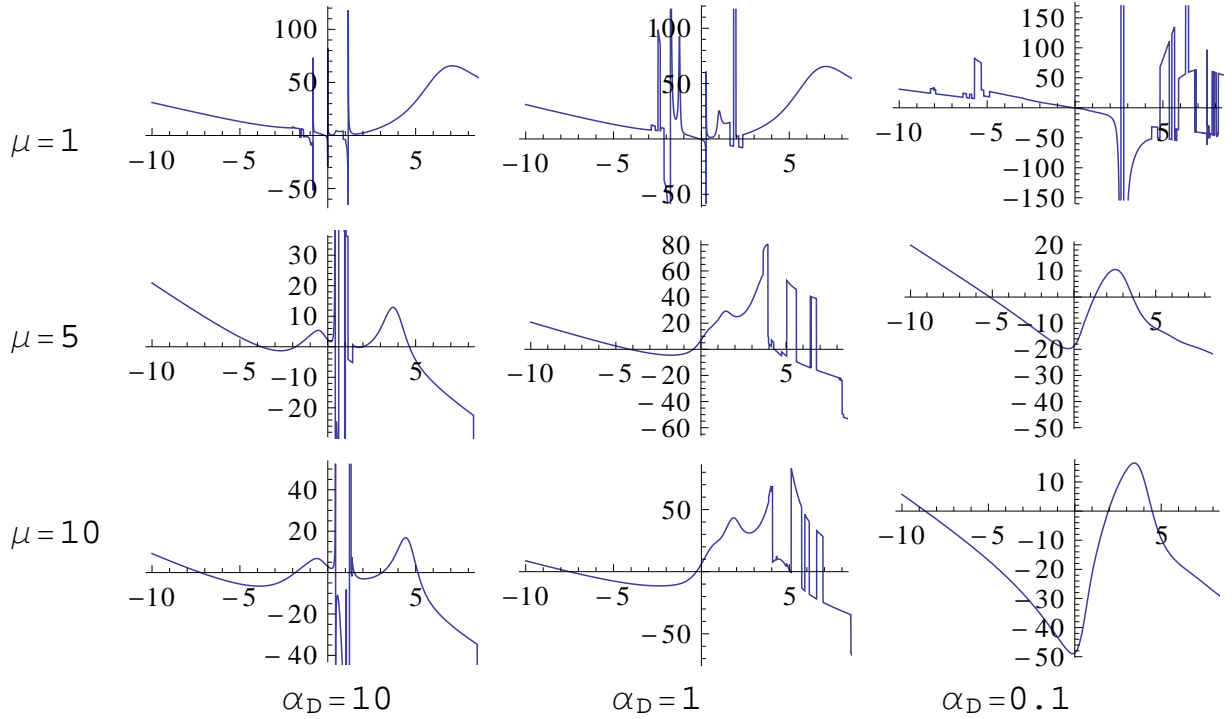}
\caption{The function $\beta_\lambda(\lambda,g^\ast(\lambda))$ for $\beta_{\rm D}=0$ and various choices for $\alpha_{\rm D}$ and $\mu$. Generically the function is plagued by divergences and numerical instabilities. For small values of $\alpha_{\rm D}$ and not too small $\mu$ values (lower right corner) the function smoothens considerably and three zeros emerge in a stable way.}
\label{BetaLambdaTable}
\end{figure}

Before determining the actual fixed point values $\lambda^\ast$ numerically, let us have a look at the global behavior of the function \eqref{fp-bdg-2-EH}. In Fig. \ref{BetaLambdaTable} we have plotted $\beta_\lambda(g^\ast(\lambda),\lambda)$ as a function of $\lambda$ for $\beta_{\rm D}=0$ and various values of $\mu$ and $\alpha_{\rm D}$. We observe that for small $\mu\approx 1$ the evaluation of the function is plagued by numerical instabilities leading to discontinuous jumps and sharp peaks that render a sensible numerical analysis of its zeros impossible. For $\mu=2$ the situation improves, but we do not find zeros in \eqref{fp-bdg-2-EH} except for those due to nearby singularities of the function. For all larger $\mu$ we begin to observe how a numerically reliable systematics emerges: 

{\noindent \bf (i)} The linear part $-4\lambda$ of the function \eqref{fp-bdg-2-EH} is dominant in the asymptotic regions of large arguments. This global behavior causes the occurrence of one zero at negative $\lambda$, while we typically find a ``bump'' at small positive $\lambda$ that causes two additional zeros of \eqref{fp-bdg-2-EH}; one that may occur at small $\lambda$ on both sides of the origin, and a second one occurring at about $\lambda\approx 5$. 

{\noindent \bf (ii)} While the first two of those zeros are found in intervals where the function is completely regular, the third is visible only for small and large values of $\alpha_{\rm D}$, whereas for $\alpha_{\rm D}=1$ it lies in a region that is numerically not accessible (cf. Fig. \ref{BetaLambdaTable}). We will thus discuss the properties of the corresponding fixed point only for the former $\alpha_{\rm D}$ values. 

{\noindent \bf (iii)} We may thus conclude that the three fixed points the zeros of \eqref{fp-bdg-2-EH} give rise to seem to be a generic property of the RG equations on a large portion of  parameter space. In the following we shall refer to them as the {\bf NGFP at large negative $\boldsymbol{\lambda}$}, the {\bf NGFP at small $\boldsymbol{\lambda}$}, and as the {\bf NGFP at large positive $\boldsymbol{\lambda}$}, respectively.

It is further encouraging that for all values of $\mu>2$ the function \eqref{fp-bdg-2-EH} smoothens considerably to small $\alpha_{\rm D}$, say $\alpha_{\rm D}\lesssim0.1$, in particular in the limit of $\alpha_{\rm D} \rightarrow 0$. As we expect zero to be a fixed point in the flow of this gauge parameter the values obtained in this limit are at the same time numerically and physically most credible. As the limiting value $\alpha_{\rm D}=0$ was not directly accessible in our numerical treatment, we could only approximate it by choosing small values as e.g. $\alpha_{\rm D} = 0.1$ and $\alpha_{\rm D}=0.001$.

\paragraph{(C) Fixed point properties.} In order to examine the properties of these non-Gaussian fixed points and their parameter dependence we analyzed the fixed points for all combinations of the three parameters $\mu\in \{5,10,15\}$, $\alpha_D\in\{0.1,1,10\}$ and $\beta_D\in\{0,1\}$. Furthermore, we regularized the flow for each of these combinations in 7 different manners, choosing the sharp proper-time cutoff, the $C^m_k$- or $f^m_k$-regularization for the three values of the cutoff parameter $m\in\{1,10,50\}$.

As a first general result it was found that the smooth $C^m_k$- and $f^m_k$-regularizations are continuously related to the sharp proper-time cutoff. In particular, in the case of $f^m_k$-regularization for large $m=50$ the expected agreement of the results with the sharp cutoff was almost perfect. 

In general, employing the smooth cutoff functions fully confirmed the picture obtained by the sharp cutoff, qualitatively as well as quantitatively. For this reason we only give the exact figures obtained by the sharp cutoff in the following tables, while the qualitative discussion refers to all regularization schemes, unless otherwise stated.

Let us now discuss the numerical results on the properties of the three fixed points.

\begin{table}\renewcommand*{\arraystretch}{0.75}
\centering
\begin{tabular}{|c|c|c|c|c|c|c|c|}\hline
 \multicolumn{3}{|c|}{} & $\lambda^\ast$ & $g^\ast$ & $\lambda^\ast g^\ast$ & $\Theta_1$ & $\Theta_3$ \\ \hline\hline
  & $\alpha_\Dr = 0.1$ & $\beta_\Dr = 0$                                         & -5.16 & 3.86 & -19.89 &  3.32  & 2.55 \\ \cline{2-8}
  & & $\beta_\Dr = 1$                                                                                       & -3.69 & 3.29 & -12.17 & 3.00   & 1.85 \\ \cline{3-8}
 $\mu=5$& \raisebox{0.45cm}[-0.5ex]{$\alpha_\Dr = 1$} & $\beta_\Dr = 0$          & -4.18 & 3.30 & -13.79 & 3.22   & 1.81 \\ \cline{2-8}
 &  & $\beta_\Dr = 1$                                                                                       & -2.93 & 2.87 & -8.41  & 2.99   & 0.58 \\ \cline{3-8}
 & \raisebox{0.45cm}[-0.5ex]{$\alpha_\Dr = 10$} & $\beta_\Dr = 0$                                           & -3.78 & 3.08 & -11.66 & 3.22   & 1.28 \\ \hline

  & $\alpha_\Dr = 0.1$ & $\beta_\Dr = 0$                                         & -8.65 & 3.39 & -29.34 &  3.55  & 2.43 \\ \cline{2-8}
  & & $\beta_\Dr = 1$                                                                                       & -6.82 & 3.01 & -20.50 & 3.40   & 1.90 \\ \cline{3-8}
 $\mu=10$& \raisebox{0.45cm}[-0.5ex]{$\alpha_\Dr = 1$} & $\beta_\Dr = 0$         & -7.51 & 3.02 & -22.67 & 3.51   & 1.90 \\ \cline{2-8}
 &  & $\beta_\Dr = 1$                                                                                       & -6.48 & 2.89 & -18.72 & 3.41   & 1.62 \\ \cline{3-8}
 & \raisebox{0.45cm}[-0.5ex]{$\alpha_\Dr = 10$} & $\beta_\Dr = 0$                                           & -7.24 & 2.93 & -21.22 & 3.51   & 1.72 \\ \hline

  & $\alpha_\Dr = 0.1$ & $\beta_\Dr = 0$                                         &-10.79 & 3.02 & -32.60 &  3.59  & 2.41 \\ \cline{2-8}
  & & $\beta_\Dr = 1$                                                                                       & -8.64 & 2.70 & -23.29 &  3.47  & 1.91 \\ \cline{3-8}
 $\mu=15$ & \raisebox{0.45cm}[-0.5ex]{$\alpha_\Dr = 1$} & $\beta_\Dr = 0$         & -9.48 & 2.71 & -25.66 &  3.56  & 1.92 \\ \cline{2-8}
 &  & $\beta_\Dr = 1$                                                                                       & -8.29 & 2.61 & -21.63 &  3.47  & 1.69 \\ \cline{3-8}
 & \raisebox{0.45cm}[-0.5ex]{$\alpha_\Dr = 10$} & $\beta_\Dr = 0$                                           & -9.20 & 2.64 & -24.27 &  3.56  & 1.76 \\ \hline
 \end{tabular}
\caption{Properties of the {\bf NGFP at large negative $\boldsymbol{\lambda}$} for various parameter choices.}
 \label{g-lambda-1st-FP}
\end{table}

\begin{table}\renewcommand*{\arraystretch}{0.75}
\centering
\begin{tabular}{|c|c|c|c|c|c|c|c|}\hline
 \multicolumn{3}{|c|}{} & $\lambda^\ast$ & $g^\ast$ & $\lambda^\ast g^\ast$ & $\Theta_1$ & $\Theta_3$ \\ \hline\hline
  & & $\beta_\Dr = 1$                                                                                       &  1.10 & 6.82 &  7.48 & -18.75 & 1.99  \\ \cline{3-8}
  & \raisebox{0.45cm}[-0.5ex]{$\alpha_\Dr = 0.1$} & $\beta_\Dr = 0$                                         &  1.25 & 5.07 &  6.35 & -14.66 & 1.89  \\ \cline{2-8}
  & & $\beta_\Dr = 1$                                                                                       & -0.48 & 3.13 & -1.52 & -8.41  & 2.02  \\ \cline{3-8}
 \raisebox{0.45cm}[-0.5ex]{$\mu=5$}& \raisebox{0.45cm}[-0.5ex]{$\alpha_\Dr = 1$} & $\beta_\Dr = 0$          & -0.46 & 3.34 & -1.54 & -10.56 & 2.00  \\ \cline{2-8}
 &  & $\beta_\Dr = 1$                                                                                       & -2.22 & 2.54 & -5.63 & -0.71  & 2.79 \\ \cline{3-8}
 & \raisebox{0.45cm}[-0.5ex]{$\alpha_\Dr = 10$} & $\beta_\Dr = 0$                                           & -1.83 & 2.18 & -3.98 & -2.41  & 2.76  \\ \hline

  & & $\beta_\Dr = 1$                                                                                       &   1.72 & 7.86 & 13.48 & -32.35 & 2.02  \\ \cline{3-8}
  & \raisebox{0.45cm}[-0.5ex]{$\alpha_\Dr = 0.1$} & $\beta_\Dr = 0$                                         &   1.94 & 4.70 &  9.12 & -19.23 & 1.88    \\ \cline{2-8}
  & & $\beta_\Dr = 1$                                                                                       &  -0.28 & 2.79 & -0.78 & -14.75 & 2.02  \\ \cline{3-8}
 \raisebox{0.45cm}[-0.5ex]{$\mu=10$}& \raisebox{0.45cm}[-0.5ex]{$\alpha_\Dr = 1$} & $\beta_\Dr = 0$         &  -0.28 & 3.31 & -0.94 & -19.14 & 2.00   \\ \cline{2-8}
 &  & $\beta_\Dr = 1$                                                                                       &  -1.81 & 1.65 & -2.99 & -3.80  & 2.47   \\ \cline{3-8}
 & \raisebox{0.45cm}[-0.5ex]{$\alpha_\Dr = 10$} & $\beta_\Dr = 0$                                           &  -1.85 & 1.48 & -2.73 & -5.03  & 2.60   \\ \hline

  & & $\beta_\Dr = 1$                                                                                       &  2.07 & 7.84 & 16.20 & -38.71 & 2.02  \\ \cline{3-8}
  & \raisebox{0.45cm}[-0.5ex]{$\alpha_\Dr = 0.1$} & $\beta_\Dr = 0$                                         &  2.34 & 4.31 & 10.08 & -20.94 & 1.88  \\ \cline{2-8}
  & & $\beta_\Dr = 1$                                                                                       & -0.25 & 2.49 & -0.63 & -16.41 & 2.02  \\ \cline{3-8}
 \raisebox{0.45cm}[-0.5ex]{$\mu=15$}& \raisebox{0.45cm}[-0.5ex]{$\alpha_\Dr = 1$} & $\beta_\Dr = 0$         & -0.27 & 3.05 & -0.81 & -21.86 & 2.00  \\ \cline{2-8}
 &  & $\beta_\Dr = 1$                                                                                       & -1.98 & 1.40 & -2.76 & -4.34  & 2.42  \\ \cline{3-8}
 & \raisebox{0.45cm}[-0.5ex]{$\alpha_\Dr = 10$} & $\beta_\Dr = 0$                                           & -2.08 & 1.24 & -2.59 & -5.53  & 2.56  \\ \hline
 \end{tabular}
\caption{Properties of the {\bf NGFP at small $\boldsymbol\lambda$} for various parameter choices.}
 \label{g-lambda-2nd-FP}
\end{table}

\begin{table}\renewcommand*{\arraystretch}{0.75}
\centering
\begin{tabular}{|c|c|c|c|c|c|c|c|}\hline
 \multicolumn{2}{|c|}{} & $\lambda^\ast$ & $g^\ast$ & $\lambda^\ast g^\ast$ & $\Theta_1$ & $\Theta_3$ \\ \hline\hline
  & $\alpha_\Dr = 0.001$             & 5.00 & 5.20 & 26.00  &  9.19  & 2.66  \\ \cline{2-7}
 $\mu=5$& $\alpha_\Dr = 0.1$         & 3.59 & 1.74 &  6.24  &  8.51  & 3.48  \\ \cline{2-7}
 & $\alpha_\Dr = 10$                 & 4.64 & 4.39 & 20.40  & 11.15  & 2.61  \\ \hline

  & $\alpha_\Dr = 0.001$             & 5.33  & 7.94 & 42.27 & 28.29 & 2.61    \\ \cline{2-7}
  $\mu=10$& $\alpha_\Dr = 0.1$       &  4.48 & 1.94 & 8.71  & 13.94 & 3.21   \\ \cline{2-7}
  & $\alpha_\Dr = 10$                &  5.13 & 5.85 & 29.99 & 24.39 & 2.49   \\ \hline
\end{tabular}
\caption{Properties of the {\bf NGFP at large positive $\boldsymbol\lambda$} for $\beta_{\rm D}=0$ and different $(\alpha_{\rm D},\mu)$.}
 \label{g-lambda-3rd-FP}
\end{table}

\paragraph{(i) NGFP at large negative $\boldsymbol{\lambda}$.} In Table \ref{g-lambda-1st-FP} we  list the coordinates of the fixed point, their product, and its critical exponents for various values of the parameters $(\alpha_{\rm D}, \beta_{\rm D}, \mu)$. The fixed point always occurs at negative values of $\lambda^\ast$, but its exact position strongly depends on the mass parameter $\mu$; we find that $\lambda^\ast$ decreases monotonically as $\mu$ is increased. To a lesser extent, the position is also gauge dependent: We observe that $\lambda^\ast$ increases with $\alpha_{\rm D}$. The dependence on $\beta_{\rm D}$ turns out relatively weak; however, for $\alpha_{\rm D}=0.1$ the existence of the FP could not be verified for $\beta_{\rm D}=1$.

The corresponding $g^\ast$ coordinate is, in comparison, remarkably stable. It is positive, lies in a range of 2.9 to 3.4 and does not show significant dependence on any of the parameters. Due to this fact, we find that, unlike in QEG, there is no compensation of gauge- and scheme-dependence in the product $\lambda^\ast g^\ast$. The relative variability of this product is hence similar to the one of $\lambda^\ast$ itself.

At this FP the critical exponent $\Theta_1$ can be related to the coupling $\lambda$, as the corresponding eigenvector of the stability matrix points in this direction for almost all parameter choices. We find that $\lambda$ is a relevant direction and the critical exponent $\Theta_1$ is remarkably gauge- and scheme-independent in a small range of about $3.2$ to $3.5$. The second critical exponent $\Theta_3$ cannot be directly associated with a single coupling.\footnote{The unconventional enumeration of the exponents is owed to our convention that in the full three dimensional coupling space the order of the couplings is fixed by $(\lambda, \gamma, g)$ forming a right handed triad.} It gives rise to a second relevant direction, such that the fixed point has a two dimensional critical hypersurface $\mathscr{S}_{\rm UV}$ in this truncation. The numerical values for $\Theta_3$ are not quite as robust as for $\Theta_1$ and show a variability in the range $0.58$ to $2.43$.

\paragraph{(ii) NGFP at small $\boldsymbol{\lambda}$.} The numerical values of the characteristic properties of this fixed point are listed in Table \ref{g-lambda-2nd-FP}. The position of this FP is strongly dependent on the gauge parameter $\alpha_{\rm D}$: Both coordinates, $\lambda^\ast$ and $g^\ast$, decrease with increasing $\alpha_{\rm D}$ such that, again, no compensation of this behavior is found in the product $g^\ast \lambda^\ast$. However, while $\lambda^\ast$ may change its sign, $g^\ast$ is always found positive. Comparatively, the dependence of the FP position on $\mu$ and $\beta_{\rm D}$ is small.

The critical exponents $\Theta_1$ and $\Theta_3$ can be associated approximately with the $\lambda$- and $g$-direction, respectively. While the FP is UV attractive in the $g$-direction, $\lambda$ turns out an irrelevant coupling. However, the corresponding critical exponent $\Theta_1$, first, takes on remarkably large values and, second, shows a severe gauge- and mass-parameter dependence and may be regarded less reliable, while $\Theta_3$ varies to a similar extent as at the first FP considered above.

\paragraph{(iii) NGFP at large positive $\boldsymbol{\lambda}$.} The numerical values of the characteristic properties of this fixed point are listed in Table \ref{g-lambda-3rd-FP}. The picture we find here is similar to the last FP concerning the gauge dependence of its properties. Again, the $\lambda$-direction has a considerably large critical exponent with a high variability, only this time the FP is UV attractive also in this direction.

\paragraph{(iv) Summary.} Taken together we conclude that all three fixed points seem suitable for the asymptotic safety construction; they all have at least one UV attractive direction. Moreover, asymptotically safe theories constructed at any of these fixed points show an anti-screening behavior due to the positive value of $g^\ast$, as is known from QEG. In principle, only an experiment can reveal which RG trajectory is realized in nature and which of the fixed points serves as its UV limit. 

For theoretical reasons, the second fixed point (NGFP at small $\lambda$) could be preferred due its higher predictivity, but the first shows a higher degree of robustness in its properties, such that it is most likely an inherent feature of the theory space rather than an artifact of our truncation. Asymptotically safe theories w.\,r.\,t. this FP show a negative $\lambda^\ast$, while the sign of the UV cosmological constant at the second FP is scheme dependent within this truncation.

The possibility of a negative $\lambda^\ast$-value is a new feature of QECG compared to the QEG results. A second main difference compared to QEG results is that here the critical exponents of all three fixed points are found to be real and do not form complex conjugated pairs.

\paragraph{(D) The phase portrait.}

\begin{figure}[htp]
\centering
{\small
\begin{psfrags}
\input{gLambda_PhasePortrait_0001-psfrag.tex}
 \subfigure[]{
\includegraphics[width=0.45\linewidth]{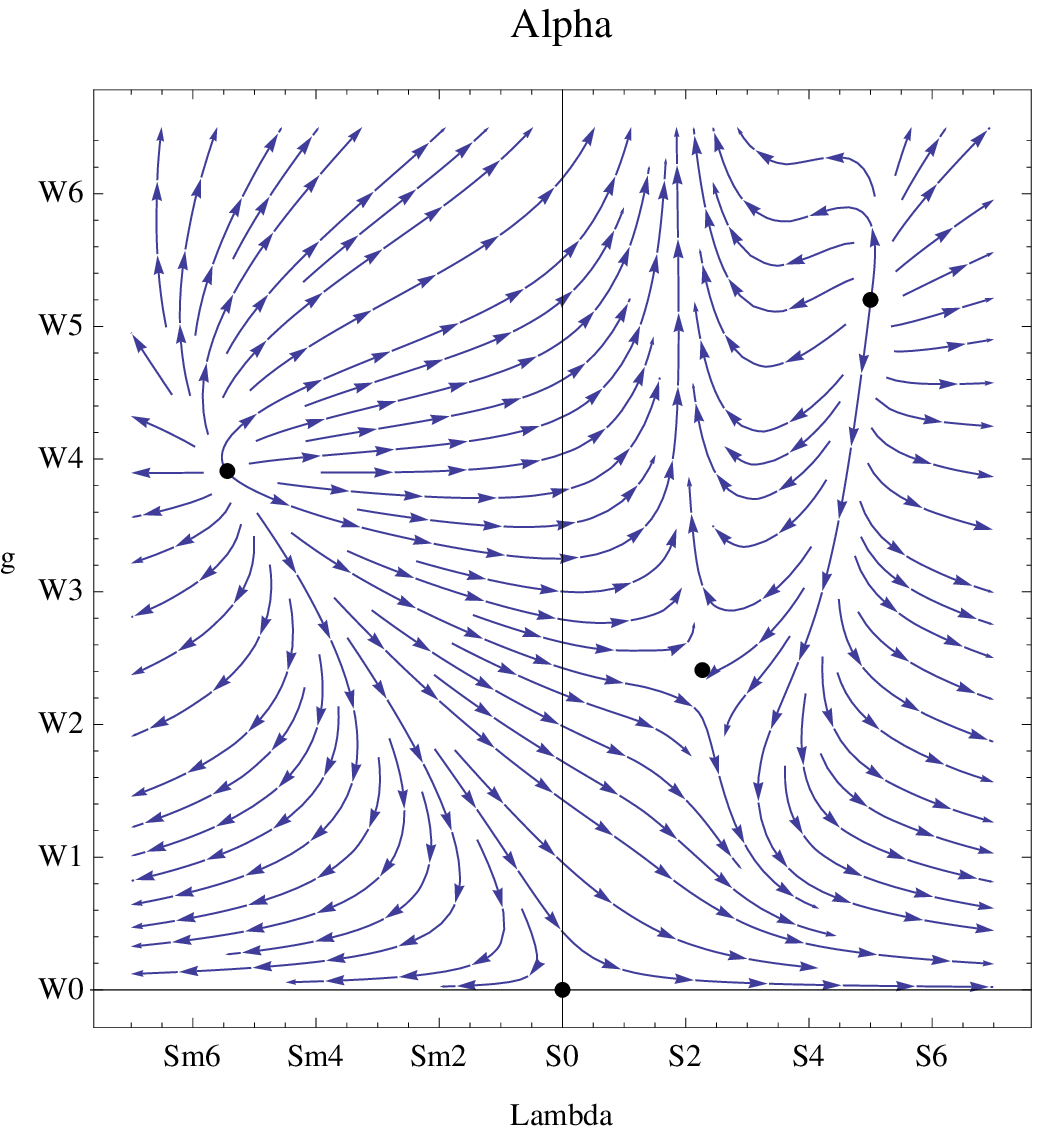}
\label{gLambda_0001}
}
\end{psfrags}\quad
\begin{psfrags}
\input{gLambda_PhasePortrait_01-psfrag.tex}
\subfigure[]{
\includegraphics[width=0.45\linewidth]{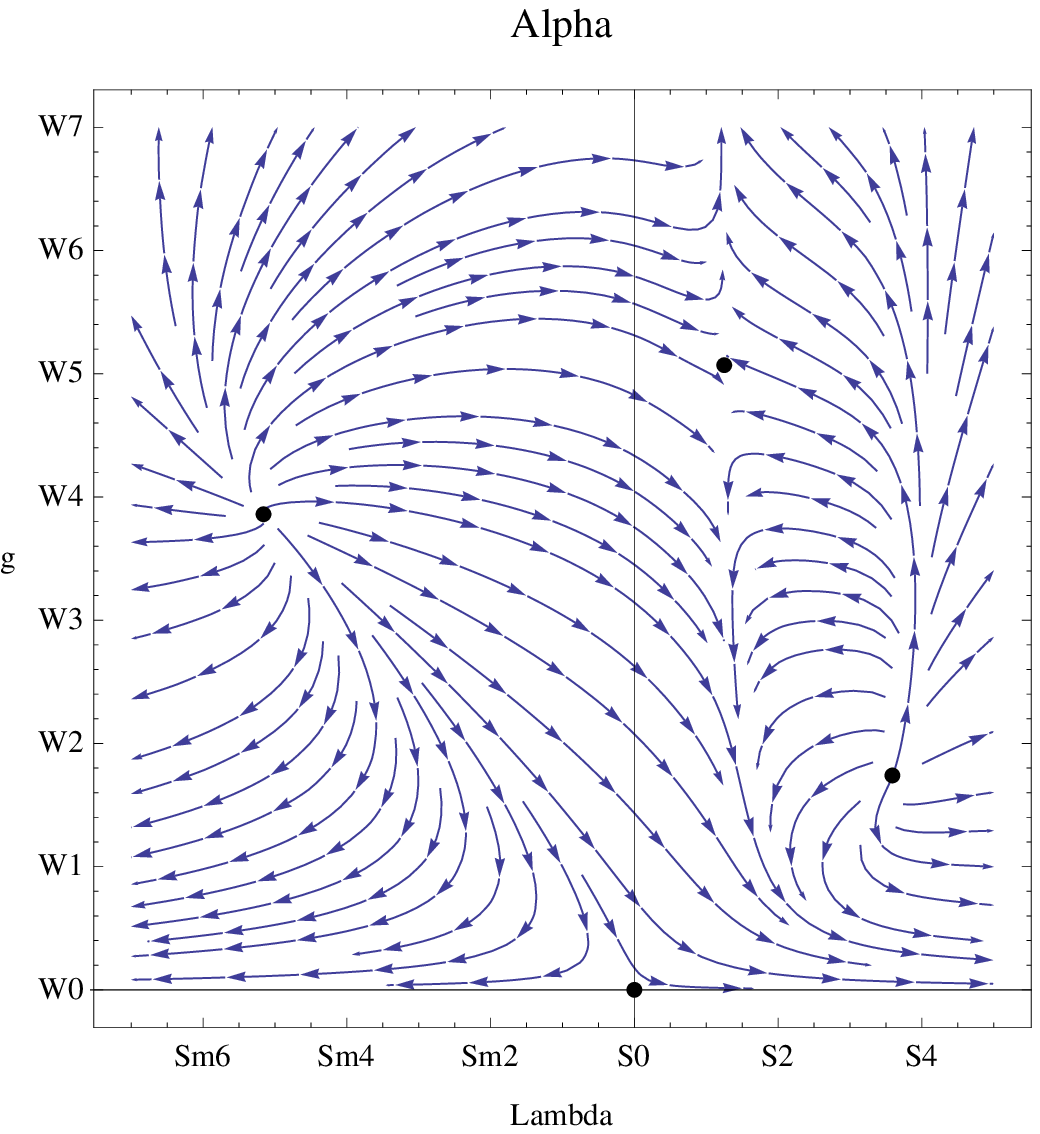}
\label{gLambda_01}
}
\end{psfrags}\\
\begin{psfrags}
\input{gLambda_PhasePortrait_1-psfrag.tex}  
\subfigure[]{
\includegraphics[width=0.45\linewidth]{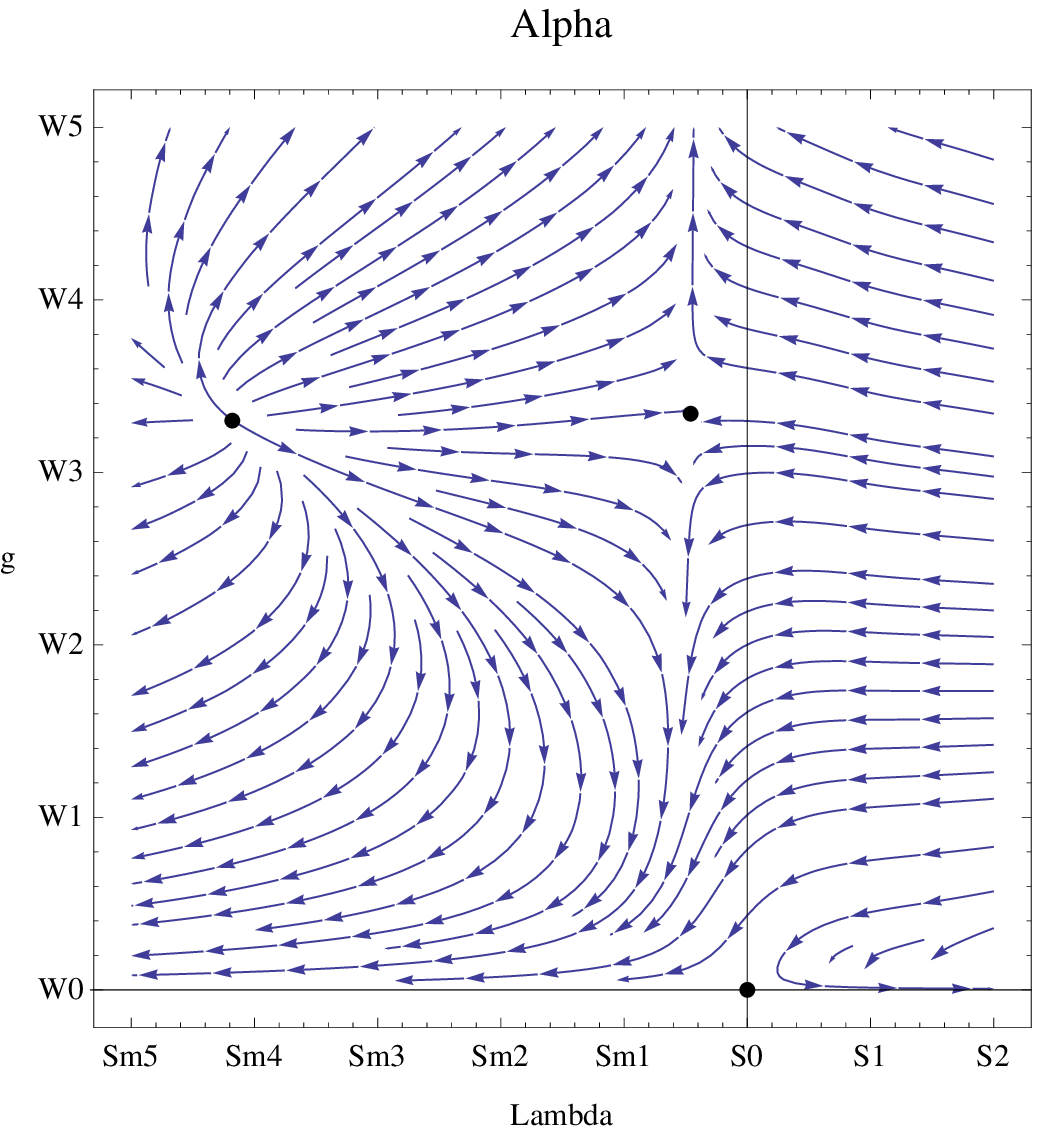}
\label{gLambda_1}
}
\end{psfrags}}
\caption{Phase portrait of the $g$-$\lambda$-truncation obtained for the sharp cutoff with the parameter values $\mu=5$, $\beta_{\rm D}=0$ and different values for $\alpha_{\rm D}$. The arrows point in the direction of decreasing $k$.}
\label{gLambda_PhasePortrait}
\end{figure}

In Fig. \ref{gLambda_PhasePortrait} we have plotted the phase portrait of the RG flow in the $g$-$\lambda$-plane resulting from the system \eqref{g-lambda-betas}.\footnote{Here and in all following phase portraits the arrows point in the direction of decreasing $k$-values.} To obtain the flow diagrams we employed the sharp cutoff with the parameter values $\mu=5$ and $\beta_{\rm D}=0$. The three diagrams differ in the parameter $\alpha_{\rm D}$ chosen from $\alpha_{\rm D}\in \{0.001, 0.1, 1\}$. The fixed point structure is similar in all three cases, only for $\alpha_{\rm D}=1$ the NGFP to the right is absent, which is most probably due to numerical difficulties in evaluating the $\beta$-function (cf. Fig. \ref{BetaLambdaTable}). A second main difference between the cases $\alpha_{\rm D}\in\{0.001, 0.1\}$ and $\alpha_{\rm D}=1$ is that in the latter case the second NGFP occurs at negative $\lambda$. Its critical hypersurface $\mathscr{S}_{\rm UV}$ starts off nearly vertically and is then, close to the GFP, bent to negative $\lambda$, resulting in a barrier separating the NGFP at large negative $\lambda$ from the GFP. For smaller $\alpha_{\rm D}$ this critical surface bents to positive $\lambda$ such that there exists a trajectory connecting the first NGFP with the GFP. In this case, if we start from a positive cosmological constant $\lambda_{\rm IR}$ in the IR, the flow may run to any of the three fixed points in the UV depending on the precise initial conditions.\footnote{We stress already here that from a phenomenological point of view a negative effective cosmological constant at high scales (and a fixed point value $\lambda^\ast < 0 $) is perfectly acceptable if the trajectory reaches positive $\lambda$ values in the IR.}

\noindent{\bf (E) The choice $\boldsymbol{\mu=1/\sqrt{g_k}}$.} To end this section, let us address the issue raised earlier concerning the assumed running of the $\mu$ parameter. While we opted for $\bar{\mu} = \mu k$, a choice of $\bar{\mu}= 1/\sqrt{G_k}\ \Rightarrow \ \mu=1/\sqrt{g_k}\,$ seemed equally plausible. We performed an analysis of the RG flow in the $(\lambda,g)$ system for this choice as well, leading to an even 
more severe scheme dependence. In this case even the existence of the NGFP was scheme dependent. For this reason we restricted all subsequent investigations to the choice $\mu=$const.

\subsection{The 2-dimensional $(\gamma,g)$ subspace}

In this section we consider the running of the Immirzi parameter coupled to the running Newton's constant, i.\,e. we analyze the Holst truncation without the cosmological term. Thus our truncated theory space is the $(\gamma, g)$-plane. We obtain the RG equations from the full system \eqref{vollst-exakt-fluss} by setting $\lambda_k=0$ in the $\beta$-functions for $g_k$ and $\gamma_k$:
\begin{subequations}
\label{g-gamma-betas}
\begin{align}
\partial_t\,g_k = \beta_g (g_k, \gamma_k) &= \Big[2 + 16 \pi g_k\,f^{(+)}(0, \gamma_k, \mu)\Big]g_k\label{g-beta-g-gamma}\\
\partial_t\,\gamma_k = \beta_\gamma (g_k, \gamma_k) &= 16 \pi g_k\,\gamma_k\Big[f^{(-)}(0, \gamma_k, \mu) - f^{(+)}(0, \gamma_k, \mu)\Big]\label{gamma-beta-g-gamma}  
\end{align}
\end{subequations}
with
\begin{align}
 f^{(+)}(\lambda, \gamma, \mu)&\equiv\frac{\eta_{\rm N} (\lambda, g, \mu)}{16 \pi g}=I^{\rm grav}_{{\rm V}(+)} (\lambda, \gamma, \mu) - 2\,I^{\rm gh}_{\rm V} (\mu)\\
f^{(-)}(\lambda, \gamma, \mu)&\equiv\gamma\, I^{\rm grav}_{{\rm V}(-)} (\lambda, \gamma, \mu)\:.
\end{align}
The flow diagrams obtained from \eqref{g-gamma-betas} show a reflection symmetry $\gamma\mapsto-\gamma$. Obviously, $g=0$ is a fixed line of the flow, such that, in particular, a GFP at $(\gamma^\ast,g^\ast)=(0,0)$ exists.

Analyzing the $(\gamma,g)$ truncation it was found that, as for the $(\lambda,g)$ system, the results using the sharp proper-time cutoff lie in line with those employing the smooth regularization functions. For that reason we restrict the discussion of the numerical results in this section mainly to this regularization scheme.

\subsubsection{The pseudo fixed points ${\bf NGFP'_{\pm}}$}\label{PseudoFPs}
{\noindent \bf (A)} The $\gamma^\ast$ coordinate of a possibly existent NGFP at a finite, non-zero $\gamma$ has to satisfy
\begin{equation}
f^{(-)}(0, \gamma^\ast, \mu) - f^{(+)}(0, \gamma^\ast, \mu)=0
\label{fp-bdg-1-g-gamma},
\end{equation}
which follows from the condition $\beta_\gamma (g^\ast, \gamma^\ast) = 0$ if $\gamma^\ast \neq 0$. Provided a solution to this condition is found, the corresponding $g^\ast$ value is then implied by the relation $\eta_{\rm N}^\ast = -2$: 
\begin{equation}
 g^\ast = - \Big[8 \pi f^{(+)}(0, \gamma^\ast, \mu)\Big]^{-1}\:.\label{fp-bdg-2-g-gamma}
\end{equation}
The linearized flow near such a NGFP is governed by a triangular stability matrix:
\begin{equation}
 {\cal B} \equiv \big\{ {\cal B}_{ij} \big\} =\left. \left( \begin{array}{cc} \frac{\partial \beta_\gamma}{\partial \gamma} & \frac{\partial \beta_\gamma}{\partial g} \\ \frac{\partial \beta_g}{\partial \gamma} & \frac{\partial \beta_g}{\partial g} \end{array} \right) \right|_{\gamma = \gamma^\ast, g = g^\ast} = \left. \left( \begin{array}{cc} \frac{\partial \beta_\gamma}{\partial \gamma} & 0 \\ \frac{\partial \beta_g}{\partial \gamma} & -2 \end{array} \right) \right|_{\gamma = \gamma^\ast, g = g^\ast}\nonumber\:.
\end{equation}
Therefore, the critical exponents, the negative eigenvalues of ${\cal B}$ are simply given by
\begin{equation}
 \Theta_2 = - \left. \frac{\partial \beta_\gamma}{\partial \gamma} \right|_{\gamma = \gamma^\ast, g = g^\ast}\hspace{0.5cm}\mbox{and}\hspace{0.5cm} \Theta_3 = 2\:.
\end{equation}
As the eigenvector $V^3$ corresponding to the eigenvalue $-\Theta_3$ is given by $V^3=(0,1)^{\rm T}$, it is possible to associate this critical exponent with the Newton constant $g$.

\begin{table}
\renewcommand*{\arraystretch}{0.75}
\centering
\begin{tabular}{|c|c|c|c|c|c|}\hline
\multicolumn{2}{|c|}{} &  $\gamma^\ast$ & $g^\ast$ & $\Theta_2$ & $\Theta_3$ \\ \hline\hline
&$\alpha_\Dr = 1$ & 1.054 & -0.026 & 4.05 &2 \\ \cline{2-6}
$\mu=1$ &$\alpha_\Dr = 10$ & 1.007 & 1.252 & 5598.6 & 2 \\ \cline{2-6}
&$\alpha_\Dr = 0.1$ & 1.054 & -0.026 & 4.93 & 2 \\ \hline\hline 

&$\alpha_\Dr = 1$ & 1.138 & 1.722 & 7.14 & 2 \\ \cline{2-6}
$\mu=2$&$\alpha_\Dr = 10$ & 1.017 & 0.042 & 66.92 & 2 \\ \cline{2-6}
&$\alpha_\Dr = 0.1$ & 0.981 & -16.370 & -794.2 & 2 \\ \hline\hline 

&$\alpha_\Dr = 1$ & 1.061 & 2.118 & 18.56 & 2 \\ \cline{2-6}
$\mu=5$ & $\alpha_\Dr = 10$ & 1.009 & 0.057 & 116.13 & 2 \\ \cline{2-6}
&$\alpha_\Dr = 0.1$ & 0.979 & -8.625 & -269.8 & 2 \\ \hline\hline 
\end{tabular}
\caption{Properties of the pseudo fixed points {\bf $\mbox{NGFP}^\prime_\pm$} in the $(\gamma, g)$ truncation for various gauge parameters using the sharp proper-time cutoff and $\beta_{\rm D}=0$.}
\label{g-gamma-pseudo-tab}
\end{table}

For $\mu\ge1$ up to the maximal value of $\mu=50$ that was analyzed we find a reflexion symmetric pair of such fixed points, one in each half-space $\gamma>0$ and $\gamma<0$, that lies very close to $\gamma=\pm 1$. We will denote this pair of fixed points by ${\bf NGFP'_{\pm}}$. 

In Table \ref{g-gamma-pseudo-tab} we give its coordinates together with its critical exponents for various values of the parameters $\mu$ and $\alpha_{\rm D}$. We thereby restrict ourselves to $\beta_{\rm D}=0$ as we did not find any notable differences in the case $\beta_{\rm D}=1$.

For all mass parameters $\mu>1$ that were studied the numerical data show the following systematics: For $\alpha_{\rm D}=1$ and $\alpha_{\rm D}=10$ we always find the absolute value of the fixed point coordinate $|\gamma^\ast|>1$ together with a positive $g^\ast$ coordinate and a positive critical exponent $\Theta_2$. The deviation from $\gamma^\ast=\pm 1$ is approximately one order of magnitude less for $\alpha_{\rm D}=10$ compared to $\alpha_{\rm D}=1$ for a fixed value of $\mu$. For $\alpha_{\rm D}=0.1$ we always find $|\gamma^\ast|<1$, with $g^\ast<0$ and $\Theta_2<0$.
Increasing $\mu$ results in $|\gamma^\ast|$ approaching 1. In general we find that the closer $|\gamma^\ast|$ is to 1, the larger the critical exponent $\Theta_2$.

For smaller values $\mu\lesssim1$ (we analyzed the cases $\mu\in\{0.2,0.5,0.8,1\}$) this general systematics is lost.

\vspace{0.5cm}
\noindent {\bf (B)} The strong gauge dependence of a critical exponent as is found here usually hints at an insufficiency of the approximation. In the present case we can understand this behavior on a deeper level, and this will shed light on the nature of these fixed points. 

\begin{figure}[tbp]
\subfigure[]{
\includegraphics[width=0.45\textwidth]{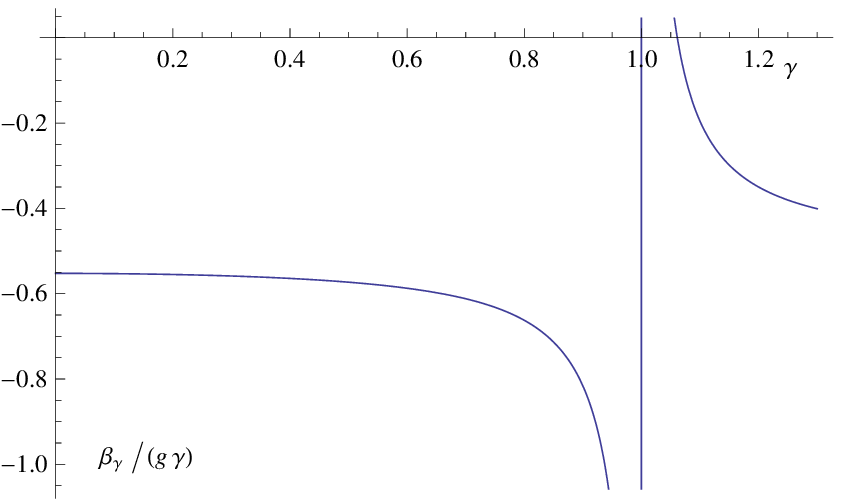}
\label{beta-gamma-pol1-fig}
}
\subfigure[]{
\includegraphics[width=0.45\textwidth]{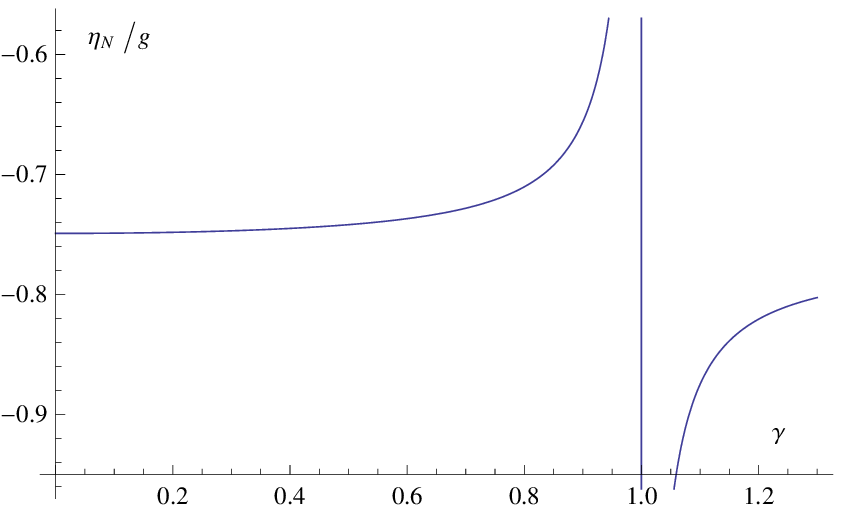}
\label{etaN-pol1-fig}
}
\caption{The functions $\gamma \mapsto \beta_\gamma/(g\,\gamma)$ and $\gamma \mapsto \eta_{\rm N}/g$ in the vicinity of their poles at $\gamma=1$. Note that only in a very narrow range in $\gamma$ the functions deviate considerably from a constant value and that this is the reason why their zeros are as close to the singularities. (The example employs the sharp cutoff and the parameters $(\mu, \alpha_{\rm D},\beta_{\rm D})=(5,1,0).$)}
\end{figure}

Let us consider the following two functions depending on the Immirzi parameter only:
\begin{equation}\label{betagammaoverggamma}
 \gamma\mapsto\frac{\beta_\gamma (g, \gamma)}{g \gamma} =  16 \pi \Big[f^{(-)}(0, \gamma, \mu) - f^{(+)}(0, \gamma, \mu)\Big]\:,
\end{equation}
\begin{equation}\label{etaNoverg}
 \gamma\mapsto\frac{\eta_{\rm N} (g, \gamma)}{g} = 16 \pi f^{(+)}(0, \gamma, \mu)\:.
\end{equation}
Their graphs are shown in Fig. \ref{beta-gamma-pol1-fig} and \ref{etaN-pol1-fig} for $\alpha_{\rm D}=1$, $\beta_{\rm D}=0$, and $\mu=5$. Both figures reveal a qualitatively similar behavior: The functions have a pole at $\gamma=\pm1$ that gives rise to a sign change. As we move away from these poles, the functions very quickly approach a horizontal asymptote, to a good approximation. The main differences between the two functions are their asymptotic value and their slopes: While $\frac{\beta_\gamma (g, \gamma)}{g \gamma}$ has a negative slope everywhere, $\frac{\eta_{\rm N} (g, \gamma)}{g}$ is a monotonically increasing function for $\gamma\neq \pm 1$. 

It is a very generic observation in the numerical data that a significant deviation of the functions \eqref{betagammaoverggamma}, \eqref{etaNoverg} from their horizontal asymptote is restricted to a very narrow region close to $\gamma=\pm 1$. Outside this region, $\beta_\gamma/(g\gamma)$ and $\eta_{\rm N}/g$ are virtually constant; by virtue of \eqref{betagammaoverggamma}, \eqref{etaNoverg} this is equivalent to saying that with good accuracy
\begin{equation}\label{fsconstant}
\begin{aligned}
f^{\pm}(0,\gamma,\mu)&\approx \text{const w.\,r.\,t. } \gamma \qquad \text{ for } |\gamma|>1+\delta \\
f^{\pm}(0,\gamma,\mu)&\approx \text{const w.\,r.\,t. } \gamma \qquad \text{ for } |\gamma|<1-\delta
\end{aligned}
\end{equation}
for some small $\delta<<1$ and with $\mu$ fixed. The constants in the first and second line of \eqref{fsconstant} can be different a priori. Up to a factor of $16 \pi$ we shall denote them $b_0^{(\pm)}$ and $b_\infty^{(\pm)}$, respectively:
\begin{equation}\label{bsdef}
\begin{aligned}
b_0^{(\pm)}\equiv b_0^{(\pm)}(\mu)&\equiv \lim_{\gamma \rightarrow 0} 16 \pi f^{\pm}(0,\gamma,\mu),\\
b_\infty^{(\pm)}\equiv b_\infty^{(\pm)}(\mu)&\equiv \lim_{\gamma \rightarrow \infty} 16 \pi f^{\pm}(0,\gamma,\mu).
\end{aligned}
\end{equation}

According to the numerical analysis the statement \eqref{fsconstant} holds with a remarkable precision for all gauge parameters and regularization schemes studied; a variation of the gauge parameters merely results in a vertical translation of the functions.

\vspace{0.5cm}
{\noindent \bf (C)} The origin of the singularities at $\gamma=\pm1$ is easy to understand. As in this limit the action only depends on one of the two chiral components of $\omega^{ab}_{~~~\mu}$, but the path integral is performed over the space of {\it all} spin connections, it diverges due to the contributions from the unsuppressed modes that do not appear in the integrand. An analogous remark applies to the functional trace of the FRGE.

\vspace{0.5cm}
{\noindent \bf (D)} The function $\frac{\beta_\gamma (g, \gamma)}{g \gamma}$ contains most of the relevant information on the fixed point properties. Its zeros correspond to the $\gamma^\ast$-values of the ${\bf NGFP'_\pm}$. If the horizontal asymptote has a negative value, we find $|\gamma^\ast|>1$, while for a positive value $|\gamma^\ast|<1$ holds. The critical exponent $\Theta_2$ is given by
\begin{equation}\label{theta2-struct}
 \Theta_2=-\partial_\gamma\beta_\gamma\Big|_\ast=g^\ast \gamma^\ast\, \partial_\gamma\Big(\frac{\beta_\gamma}{g\gamma}\Big)\bigg|_\ast
\end{equation}
and is thus proportional to the derivative of the function at its zero. From this we can understand that the absolute value of $\Theta_2$ must diverge when the fixed point approaches $\gamma^\ast\rightarrow\pm 1$, as the derivative necessarily diverges at the poles. Also it becomes clear that $\Theta_2$ changes its sign if and only if $g^\ast$ changes sign, as the other factors in \eqref{theta2-struct} have a fixed sign.

\vspace{0.5cm}
{\noindent \bf (E)} Taken together these observations show that not only the variability of the critical exponent $\Theta_2$, but also the very existence of this pair of fixed points, stems from the continuous interpolation between two distinct regimes of the functions $f^{(\pm)}$: the constant regime for $\gamma\not\approx \pm 1$ and the divergent behavior near the singular point $\gamma=\pm 1$, whose origin is deeply rooted in the very construction of the theory. We conclude that {\it both the poles and the nearby zeros are most probably an artifact of the approximation} and that the constant regime (horizontal asymptote) is likely to actually apply for {\it all} values of $\gamma\neq\pm1$. 

For this reason we refer to the fixed points ${\bf NGFP'_\pm}$ as {\it pseudo fixed points}: We expect them to disappear in an exact treatment of the problem.

In Section \ref{Hypothesis} a comparative study will be carried out, comparing numerically the results of our calculation with a simple ``effective'' model in which the two functions $f^{(\pm)}(0,\gamma,\mu)$ are replaced by {\it constants}. Besides the fact that the pseudo fixed points disappear, we find an astonishingly good agreement of the resulting $\beta$-functions with the ``exact'' ones in the whole $(\gamma,g)$-plane, except extremely close to $\gamma=\pm1$.

\subsubsection{Coordinate charts for theory space}

\begin{figure}[ht]
\begin{center}
{\small
\psfrag{I}[tc][tc]{{\bf $\mbox{NGFP}^\prime_0$}}
\psfrag{III}[bc][bc]{{\bf $\mbox{NGFP}^\prime_\infty$}}
\psfrag{I+}[br][br]{{\bf $\mbox{NGFP}^\prime_+$}}
\psfrag{I-}[bl][bl]{{\bf $\mbox{NGFP}^\prime_-$}}
\psfrag{KI}[bl][bl]{$\gamma$ chart}
\psfrag{KII}[bl][bl]{{$\hat{\gamma}$ chart}}
\includegraphics[width=0.5\textwidth]{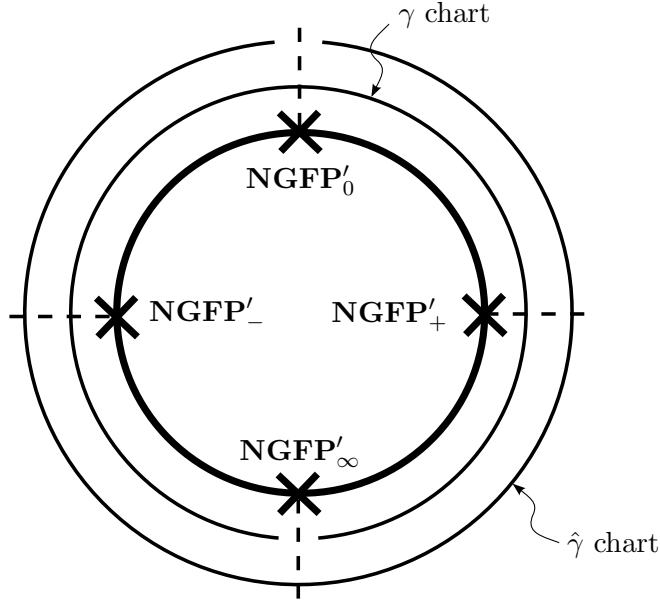}
}
\caption{The 1-dimensional subspace of theory space related to the Immirzi parameter (fat circle) is covered by two coordinate charts: the $\gamma$-chart which applies everywhere except in a vicinity of ${\bf NGFP'_\infty}$, and the $\hat\gamma$-chart which can be used everywhere except near ${\bf NGFP'_0}$. The pseudo fixed points ${\bf NGFP'_\pm}$ are at $\gamma=\hat\gamma=\pm 1$.}
\label{chart-fig}
\end{center}
\end{figure}

In order to search for fixed points in the complete $(\gamma,g)$ theory space we have to include the limits $\gamma\rightarrow\pm\infty$ in a well defined way. To this end, we cover its 1-dimensional $\gamma$-subspace by two coordinate charts. Away from ``$\gamma=\pm\infty$'' we use a chart on which the standard Immirzi parameter serves as the coordinate. Close to ``$\gamma=\pm\infty$'' we introduce a second chart with a new coordinate $\hat\gamma$, however. On their overlap the couplings and $\beta$-functions are related by
\begin{equation}
\hat\gamma=\gamma^{-1} \quad \text{and} \quad \beta_{\hat \gamma}(g,\hat\gamma) = -{\hat\gamma}^2 \beta_\gamma(g, {\hat\gamma}^{-1})\;,
\end{equation}
respectively, whereby the ``transition function'' follows from the relation $\partial_t\big({\gamma_k}^{-1}\big)= - {\gamma_k}^{-2}(\partial_t \gamma_k)$. Thus, ``$\gamma=\pm\infty$'' amounts to the regular coordinate value $\hat\gamma=0$. Similar to a stereographic projection of a sphere both charts apply to the whole truncated theory space, except for one point. This situation is depicted in Fig. \ref{chart-fig}.

If we now rewrite the beta functions \eqref{g-gamma-betas} in terms of the new coupling $\hat{\gamma}$, we find the following RG equations on the second chart:
\begin{subequations}\label{g-gammahat-betas}
\begin{align}
\partial_t g_k=\beta_g(g_k,\gamma_k)&=\beta_g(g_k,\hat{\gamma}_k^{-1})=\Big[2+16 \pi g_k \,f^{(+)}(0,\hat{\gamma}_k^{-1},\mu)\Big]g_k\\
 \partial_t\hat{\gamma}_k=\beta_{\hat\gamma}(g_k,\hat{\gamma}_k)&=-\hat{\gamma}_k^2\beta_\gamma(g_k,\hat{\gamma}_k^{-1})=-16 \pi g \hat{\gamma}\Big[f^{(-)}(0,\hat{\gamma}_k^{-1},\mu)-f^{(+)}(0,\hat{\gamma}_k^{-1},\mu)\Big].
\end{align}
\end{subequations}
Note the similarity of these equations with those on the $\gamma$-chart, eqs. \eqref{g-gamma-betas}.

\subsubsection{The physical fixed points ${\bf NGFP'_0}$ and ${\bf NGFP'_\infty}$}

From now on we analyze the combined set of the flow equations \eqref{g-gamma-betas} and \eqref{g-gammahat-betas} on the $\gamma$- and $\hat\gamma$-charts, respectively. In this way our analysis is unbiased towards the ultimate physical relevance of either $\gamma=0$ or ``$\gamma=\pm\infty$'', respectively. We start by searching for nontrivial fixed points.

\paragraph{(A) Position of the fixed points.} From subsection \ref{PseudoFPs} we already know that the limits $\gamma\rightarrow0$ and $\gamma\rightarrow \infty$ of each of the functions $f^{(\pm)}(0,\gamma,\mu)$ exist and are almost equal. Therefore we find two additional fixed points at $\gamma^\ast=0$ and $\hat \gamma^\ast=0$ with the respective $g^\ast$ coordinates
\begin{equation}
 g^\ast_0= - \lim_{\gamma\rightarrow0}\Big[8 \pi f^{(+)}(0,\gamma,\mu)\Big]^{-1}\quad \text{and}\quad g^\ast_\infty= - \lim_{\hat\gamma\rightarrow0}\Big[8 \pi f^{(+)}(0,\hat{\gamma}^{-1},\mu)\Big]^{-1}\:.
\end{equation}
We shall denote these fixed points by {\bf $\mbox{NGFP}^\prime_0$} $\equiv (\gamma^\ast, g^\ast_0)$ and {\bf $\mbox{NGFP}^\prime_\infty$} $\equiv (\hat{\gamma}^\ast, g^\ast_\infty)$.

\paragraph{(B) Critical exponents.} Also for the fixed points {\bf $\mbox{NGFP}^\prime_{0,\infty}$} we find the stability matrix triangular as $\partial_g\beta_\gamma|_{\gamma=0}=0=\partial_g \beta_{\hat{\gamma}}|_{\hat{\gamma}=0}$. Moreover, it even turns out diagonal as also $\partial_\gamma\beta_g$ vanishes, for $\gamma\rightarrow\infty$ due to the horizontal asymptote of $f^{(+)}$, and for $\gamma=0$ due to the reflexion symmetry under $\gamma\mapsto -\gamma$. Thus, the critical exponents are given by 
\begin{equation}\Theta_3=2,\quad\Theta_2=-\partial_\gamma\beta_\gamma\big|_{\gamma=0,g=g^\ast_0}=-c_0 g^\ast_0,\quad \Theta_{\hat 2}=-\partial_{\hat\gamma}\beta_{\hat\gamma}\big|_{\hat\gamma=0,g=g^\ast_\infty}=c_\infty g^\ast_\infty\:.
\end{equation}
Thereby, $c_0$ and $c_\infty$ are defined as the ($\mu$ dependent) limits
\begin{equation}
\begin{aligned}
 c_0(\mu) &\equiv \lim_{\gamma \to 0}\Big(\frac{\beta_\gamma}{g \gamma}\Big) = \lim_{\gamma \to 0} 16 \pi \Big[f^{(-)}(0,\gamma,\mu)-f^{(+)}(0,\gamma,\mu)\Big],\\
c_\infty(\mu) &\equiv \lim_{\gamma \to \pm \infty}\Big(\frac{\beta_\gamma}{g \gamma}\Big) = \lim_{\gamma \to \pm \infty} 16 \pi \Big[f^{(-)}(0,\gamma,\mu)-f^{(+)}(0,\gamma,\mu)\Big]\:.
\end{aligned}
\end{equation}
As for a diagonal stability matrix the eigenvectors point into the directions of the coupling axes, we can associate the critical exponents to the different couplings according to $\Theta_2 \equiv \Theta_\gamma$, $\Theta_{\hat{2}} \equiv \Theta_{\hat{\gamma}}$ and $\Theta_3 \equiv \Theta_g$.

\begin{table}[ht]
\renewcommand*{\arraystretch}{0.75}
\centering
\begin{tabular}{|c|c|c|c|c|c|c|c|c|}\hline
\multicolumn{2}{|c|}{} &  $c_0$ & $c_\infty$ & $g^\ast_0$ & $g^\ast_\infty$ & $\Theta_\gamma$ & $\Theta_{\hat{\gamma}}$ & $\Theta_g$ \\ \hline\hline
&$\alpha_\Dr = 1$ & 8.52 & 7.67 & -0.26 & -0.26 & 0.22 & -0.20 & 2 \\ \cline{2-9}
$\mu=1$&$\alpha_\Dr = 10$ & -29.78 & -25.21 & 4.74 & 4.42 & 141.1 & -129.9 & 2 \\ \cline{2-9}
&$\alpha_\Dr = 0.1$ & 10.46 & 9.39 & -0.025 & -0.025 & 0.27 & -0.24 & 2 \\ \hline\hline 
&$\alpha_\Dr = 1$ & -0.69 & -0.55 & 2.09 & 2.03 & 1.46 & -1.12 & 2 \\ \cline{2-9}
$\mu=2$&$\alpha_\Dr = 10$ & -25.21 & -24.34 & 0.04 & 0.04 & 1.05 & -1.01 & 2 \\ \cline{2-9}
&$\alpha_\Dr = 0.1$ & 0.61 & 0.65 & 5.51 & 5.10 & -3.39 & 3.31 & 2 \\ \hline\hline
&$\alpha_\Dr = 1$ &  -0.55 & -0.49 & 2.67 & 2.59 & 1.48 & -1.27 & 2 \\ \cline{2-9}
$\mu=5$&$\alpha_\Dr = 10$ & -17.85 & -17.55 & 0.06 & 0.06 & 1.05 & -1.03 & 2 \\ \cline{2-9}
&$\alpha_\Dr = 0.1$ & 0.62 & 0.65 & 6.75 & 6.29 & -4.19 & 4.07 & 2 \\ \hline\hline 
&$\alpha_\Dr = 1$ & -1.34 & -1.33 & 1.57 & 1.56 & 2.09 & -2.08 & 2 \\ \cline{2-9}
$\mu=50$&$\alpha_\Dr = 10$ & -31.95 & -31.94 & 0.033 & 0.033 & 1.063 & -1.063 & 2 \\ \cline{2-9}
&$\alpha_\Dr = 0.1$ & 1.18 & 1.18 & 4.97 & 4.94 & -5.85 & 5.83 & 2 \\ \hline\hline 
\end{tabular}
\caption{Properties of the fixed points {\bf $\mbox{NGFP}^\prime_0$} and {\bf $\mbox{NGFP}^\prime_\infty$} on the $(\gamma,g)$ theory space using the sharp proper-time cutoff and $\beta_{\rm D}=0$.}
\label{duality-tab}
\end{table}

\paragraph{(C) Numerical results.}
Table \ref{duality-tab} shows the values of the quantities $c_0$, $c_\infty$, $g^\ast_0$, $g^\ast_\infty$, $\Theta_\gamma$, $\Theta_{\hat{\gamma}}$ and $\Theta_g$ obtained for different choices of the parameters $\mu$ and $\alpha_{\rm D}$. Here, again, we restricted ourselves to the sharp cutoff regularization scheme and to the case $\beta_{\rm D}=0$, because the results do not change significantly, neither qualitatively nor quantitatively, for other choices of the proper-time cutoff or the $\beta_{\rm D}$ value.\pagebreak

The most remarkable property of the numerical values in the table are the approximate equalities $c_0\approx c_\infty$ and $g^\ast_0 \approx g^\ast_\infty$ that were already asserted in the last subsection. For the critical exponents they imply $\Theta_\gamma\approx -\Theta_{\hat{\gamma}}$. According to their very definition, the approximate equality of $c_0$ and $c_\infty$ entails that
\begin{equation}\label{bEqualities}
\lim_{\gamma\rightarrow 0} f^{(\pm)}(0,\gamma,\mu) \approx \lim_{\gamma\rightarrow\infty} f^{(\pm)}(0,\gamma,\mu), \quad \text{i.\,e.} \quad b^{(\pm)}_0(\mu)\approx b^{(\pm)}_\infty(\mu)
\end{equation}
for every fixed $\mu$.

For $\mu\ge 2$ we again find a stable systematics in Table \ref{duality-tab}: First of all, we find for all regularization schemes and all parameter choices a positive value for $g^\ast_0 \approx g^\ast_\infty > 0$, i.\,e. the fixed points show the usual anti-screening behavior. Moreover, for $\alpha_{\rm D} =1$ and $\alpha_{\rm D}=10$ we always find $\Theta_\gamma>0$, while for $\alpha_{\rm D}=0.1$, $\Theta_\gamma$ is negative. (For $\Theta_{\hat{\gamma}}$ we find the converse statement.) 

Thus we find for the ``physical'' fixed points ${\bf NGFP_0'}$ and ${\bf NGFP_\infty'}$ that with remarkable accuracy 
\begin{equation}\label{ApproxConjecture}
 c_0(\mu) \approx c_\infty(\mu),\qquad b^{(\pm)}_0(\mu)\approx b^{(\pm)}_\infty(\mu), \qquad g_0^\ast \approx g^\ast_\infty, \qquad \Theta_\gamma \approx - \Theta_{\hat\gamma}.
\end{equation}
These approximate equalities get increasingly better for larger values of $\mu$. It is plausible to speculate that they become exact in the limit $\mu \rightarrow \infty$.

\begin{figure}[htp]
\centering
{\small
\begin{psfrags}
 \input{asymptotes-psfrag.tex}
 \includegraphics[width=0.6\linewidth]{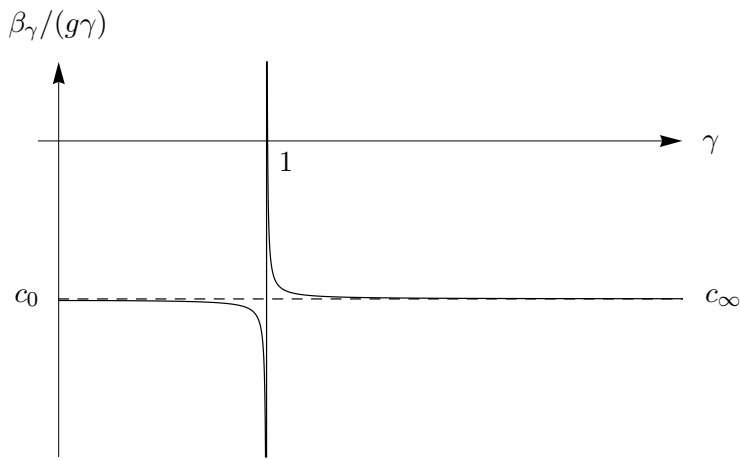}
\end{psfrags}
}
\caption{Schematic sketch of the function $\gamma\mapsto\beta_\gamma/(g\gamma)$ when $c_0=c_\infty$. In this case the function is almost perfectly constant, except in a narrow vicinity of $\gamma=\pm1$ in which it has a pole which entails a zero. The truncation is insufficient there presumably. We conjecture that the exact result is instead the everywhere defined constant function corresponding to the dashed line.}
\label{asymptotes}
\end{figure}

\paragraph{(D) A conjecture concerning the exact flow.} When we combine the result \eqref{bEqualities} with our earlier observation \eqref{fsconstant} it follows that the two constants ``const'' in \eqref{fsconstant} are actually equal $(b^{(\pm)}_0=b^{(\pm)}_\infty)$. For any fixed $\mu$, the functions $f^{(\pm)}(0,\gamma,\mu)$ assume the same $\gamma$-independent value for all $\gamma$ to the left ($\gamma<1-\delta$) and to the right ($\gamma>1+\delta$) of the narrow interval containing the singularity. In Fig. \ref{asymptotes} we sketched this situation for the related function $\beta_\gamma/(g\gamma)$.

This important result motivates the conjecture that in a more general truncation, or in an exact treatment, we would actually yield functions which assume the same constant value {\it everywhere}, including the interval $(1-\delta,1+\delta)$:
\begin{equation}\label{fsconstant2}
f^{(\pm)}(0,\gamma,\mu)\equiv \frac{b^{(\pm)}(\mu)}{16\pi}=\text{const w.\,r.\,t. } \gamma\qquad \text{for all }\gamma\in(-\infty,+\infty)
\end{equation}
With the notation of \eqref{bsdef}, the conjecture amounts to $b^{(\pm)}\equiv b^{(\pm)}_0\stackrel{!}{=} b^{(\pm)}_\infty$.

If this conjecture is true, the above approximate equalities \eqref{ApproxConjecture} become exact, and 
\begin{equation}\label{exactduality}
c_0=c_\infty,\quad g_0^\ast=g_\infty^\ast,\quad \Theta_\gamma=-\Theta_{\hat\gamma}
\end{equation}
holds true for {\it all} values of $\mu$, not only in the limit $\mu\rightarrow \infty$.

Stated differently, we conjecture that in the exact treatment there is neither a pole at $\gamma=+1$ (and, by reflection symmetry, $\gamma=-1$), nor is there the concomitant zero of $\beta_\gamma/(g\gamma)$ that gave rise to the pseudo fixed points ${\bf NGFP'_\pm}$. In a sense, the pole and the zero would annihilate one another, cf. Fig. \ref{asymptotes}.

If the conjecture is true the two fixed points ${\bf NGFP'_0}$ and ${\bf NGFP'_\infty}$ are at the same value $g^\ast$, but one of them has two and the other has only one UV attractive directions; those directions are parallel to the direction of the coupling axes. The fixed points are mapped onto each other and switch their roles by the ``duality operation'' $\gamma \leftrightarrow 1/\gamma$.

The ${\bf NGFP'_\infty}$ corresponds to a theory with freely fluctuating torsion, that is thus ``maximally different'' from metric gravity. At the ${\bf NGFP'_0}$ some components of the torsion tensor are suppressed completely, while others remain fluctuating. Hence, also an asymptotically safe theory defined at this fixed point does not directly correspond to metric gravity.

\subsubsection{The phase portrait of the $(\gamma,g)$ truncation}
\begin{figure}[tbp]
\begin{center}
{\small
\psfrag{g}[bl][bl]{$g$}
\psfrag{G}[bl][bl]{$\gamma$}
\psfrag{H}[bl][bl]{$\hat{\gamma}$}
\psfrag{N}[tl][tl]{${\bf NGFP'_0}$}
\psfrag{M}[tr][tr]{${\bf NGFP'_\infty}$}
\psfrag{P}[tl][tl]{${\bf NGFP'_+}$}
\psfrag{I}[bc][bc]{$\infty$}
\psfrag{0}[bc][bc]{0}
\psfrag{1}[bc][bc]{1}
\psfrag{2}[bc][bc]{2}
\psfrag{3}[bc][bc]{3}
\psfrag{4}[bc][bc]{4}
\psfrag{A}[bc][bc]{0.2}
\psfrag{B}[bc][bc]{0.5}
\psfrag{5}[bc][bc]{5}
\includegraphics[width=0.95\textwidth]{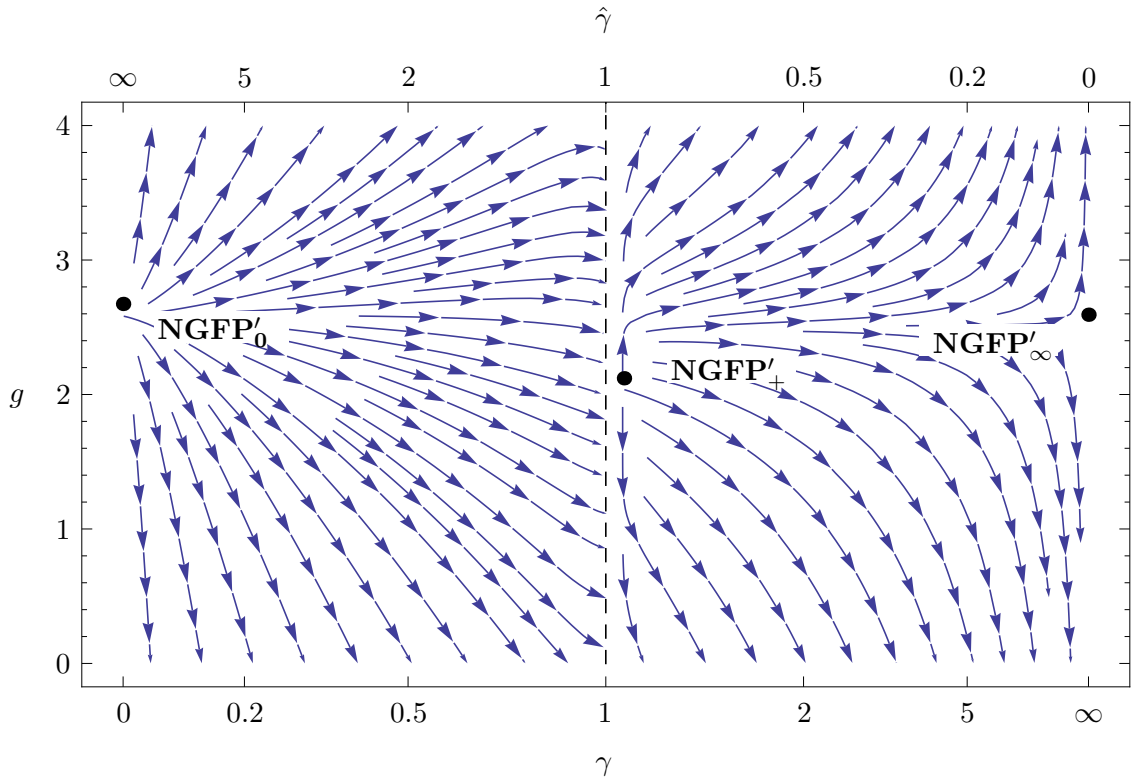}
}
\caption{The RG flow in the $(\gamma, g)$-plane for $(\alpha_{\rm D},\beta_{\rm D}, \mu)=(1,0,5)$ and the sharp cutoff.}
\label{gamma-g-fluss-fig}
\end{center}
\end{figure}

In Fig. \ref{gamma-g-fluss-fig} we have plotted the phase portrait of the RG flow of the $(\gamma,g)$ theory space, for $(\alpha_{\rm D},\beta_{\rm D}, \mu)=(1,0,5)$. It amounts to a typical example of the RG flows resulting from the $\beta$-functions \eqref{g-gamma-betas} for various choices of gauge parameters. In this plot we have compactified the $\gamma$-coordinate by an {\it arctangent}-rescaling; this explains the highly nonlinear scale along the $\gamma$ and $\hat{\gamma}$ axis.

The RG flow is governed by the two physical fixed points ${\bf NGFP'_0}$ and ${\bf NGFP'_\infty}$, both of which are UV attractive in the $g$ direction. The fact that always one of the two is UV attractive and the other UV repulsive in the $\gamma$ direction reaches out to almost the whole $(\gamma, g)$ space and gives rise to a preferred direction of the flow with respect to $\gamma$. For $\alpha_{\rm D}=10$ and $\alpha_{\rm D}=1$, the Immirzi parameter flows towards increasing values of $\gamma$ when we decrease $k$, as ${\bf NGFP'_0}$ is UV attractive in both directions; for $\alpha_{\rm D}=0.1$ $\gamma$ decreases when going to the IR. 

Only in the narrow vertical strips between $\gamma=\pm 1$ and $\gamma=\gamma^\ast$, the location of the pseudo fixed points ${\bf NGFP'_\pm}$, the flow in $\gamma$ direction is reversed. 

In Fig. \ref{gamma-g-fluss-fig} we also find an additional confirmation of our assertion that the pseudo fixed points and the divergences at $\gamma=\pm1$ should mutually annihilate in an exact treatment. The flow within the vertical strips is completely decoupled from the rest of the $(\gamma,g)$ plane. At $\gamma=\pm1$ the flow simply stops, as both $\beta$-functions diverge, but their ratio stays finite. On the other hand, at $\gamma=\gamma^\ast$ the beta function $\beta_\gamma$ vanishes for all $g$. Thus no trajectory crosses this line, but all are bent into the fixed point. 

Looking at the $(\gamma,g)$-plane as a whole, one is tempted, therefore, to cut out the strips and to simply connect the trajectories on both sides. This is in fact exactly what amounts to the conjecture and leads to the ``effective'' model RG equations that we want to discuss next.

\subsubsection{Testing the conjecture concerning the exact flow (2-dimensional case)} \label{Hypothesis}

\paragraph{(A)} Above we found strong evidence for the simple $\gamma$-independent form \eqref{fsconstant2} of the two functions $f^{(\pm)}(0,\gamma,\mu)\equiv b^{(\pm)}/16\pi$. (We keep $\mu$ fixed.) If this conjecture is correct, the RG equations \eqref{g-gamma-betas} boil down to the following simplified system:
\begin{subequations}\label{HYP-2-betas}
\begin{align} \partial_t\,g_k = \beta_g (g_k, \gamma_k) &= \Big[2 + b^{(+)} g_k\Big]g_k\label{beta-g-HYP-2}\\
\partial_t\,\gamma_k = \beta_\gamma (g_k, \gamma_k) &= \Big[b^{(-)} - b^{(+)}\Big]g_k\,\gamma_k\:. \label{beta-gamma-HYP-2} 
\end{align}
\end{subequations}
Here, $b^{(\pm)}$ are gauge parameter and $\mu$-dependent constants. If the conjecture applies we may identify them with either $b^{(\pm)}_0 \equiv {\displaystyle \lim_{\gamma \to 0}}\, 16 \pi f^{(\pm)} (0,\gamma,\mu)$ or $b^{(\pm)}_\infty \equiv {\displaystyle\lim_{\gamma \to \pm \infty}}\, 16 \pi f^{(\pm)} (0,\gamma,\mu)$. 

As we have already discussed, using the numerical data these limits do indeed coincide to a good approximation, corroborating our hypothesis. 

\paragraph{(B)} According to the conjectured RG equations \eqref{HYP-2-betas} the fixed points ${\bf NGFP'_0}$ and ${\bf NGFP'_\infty}$ possess the same $g^\ast$-coordinate $g^\ast=-2/b^{(+)}$ and have the critical exponents $\Theta_\gamma=-2\big(1- b^{(-)}/b^{(+)}\big)$ and $\Theta_\gamma=2\big(1- b^{(-)}/b^{(+)}\big)$, respectively, as well as $\Theta_g=2$.

\begin{figure}[p!]
\begin{center}
{\small
\psfrag{g}[bl][bl]{$g$}
\psfrag{G}[bl][bl]{$\gamma$}
\psfrag{H}[bl][bl]{$\hat{\gamma}$}
\psfrag{N}[tl][tl]{${\bf NGFP'_0}$}
\psfrag{M}[tr][tr]{${\bf NGFP'_\infty}$}
\psfrag{P}[tl][tl]{${\bf NGFP'_+}$}
\psfrag{I}[bc][bc]{$\infty$}
\psfrag{0}[bc][bc]{0}
\psfrag{1}[bc][bc]{1}
\psfrag{2}[bc][bc]{2}
\psfrag{3}[bc][bc]{3}
\psfrag{4}[bc][bc]{4}
\psfrag{A}[bc][bc]{0.2}
\psfrag{B}[bc][bc]{0.5}
\psfrag{5}[bc][bc]{5}
\includegraphics[width=0.95\textwidth]{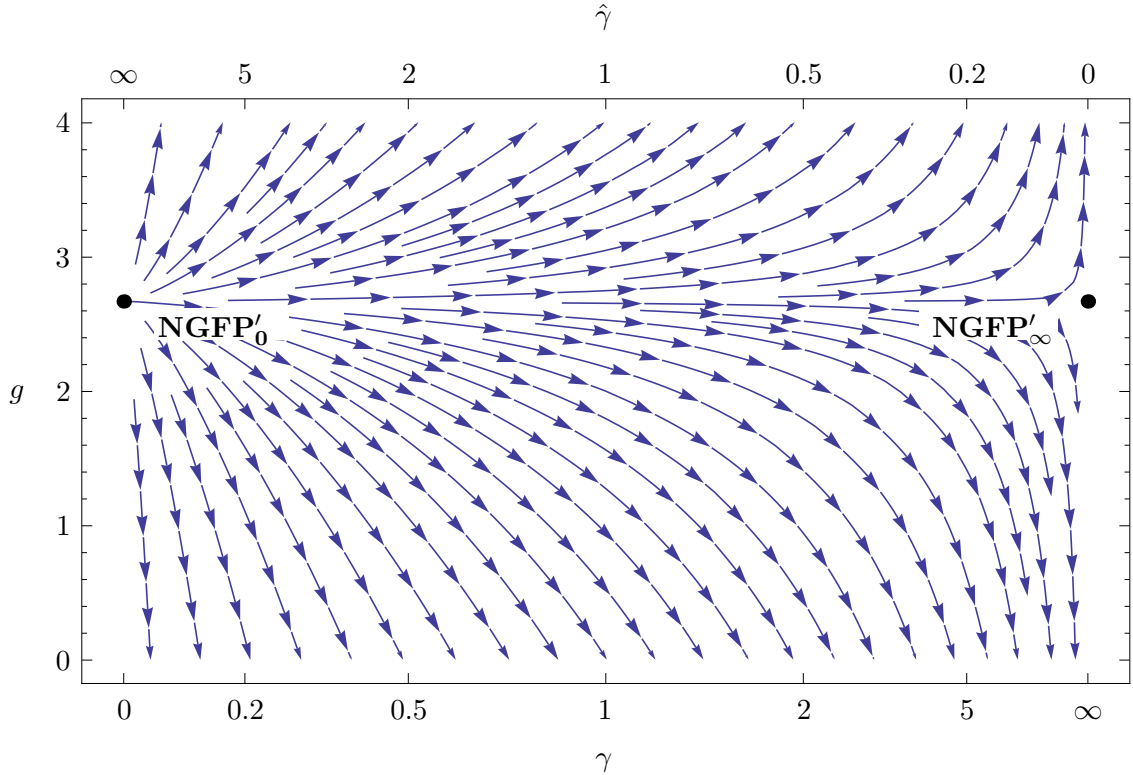}
}
\caption{The RG flow in the $(\gamma, g)$-plane for the simplified RG equations \eqref{HYP-2-betas} with $b^{(+)}=-0.75$ and $b^{(-)}=-1.305$. These values of $b^{(\pm)}$ correspond to the $(\alpha_{\rm D},\beta_{\rm D}, \mu)$ values also used in Fig. \ref{gamma-g-fluss-fig}.}
\label{gamma-g-fluss-approx}
\end{center}
\end{figure}

\begin{table}[p!]\renewcommand*{\arraystretch}{0.75}
\centering
\begin{tabular}{|c|c|c|c|c|}\hline
\multicolumn{2}{|c|}{} & $\alpha_\Dr = 1$ & $\alpha_\Dr = 10$ & $\alpha_\Dr = 0.1$ \\ \hline\hline
 & $\big(\frac{b^{(-)}}{b^{(+)}}\big)_0$ & 1.11 & 71.54 & 1.13 \\ \cline{2-5}
\raisebox{0.5cm}[-0.5ex]{$\mu = 1$} & $\big(\frac{b^{(-)}}{b^{(+)}}\big)_\infty$ & 1.10 & 65.94 & 1.12 \\ \hline
& $\big(\frac{b^{(-)}}{b^{(+)}}\big)_0$ & 1.74 & 1.53 & -1.10 \\ \cline{2-5}
\raisebox{0.5cm}[-0.5ex]{$\mu = 5$} & $\big(\frac{b^{(-)}}{b^{(+)}}\big)_\infty$ & 1.64 & 1.52 & -1.03 \\ \hline
& $\big(\frac{b^{(-)}}{b^{(+)}}\big)_0$ & 2.05 & 1.53 & -1.92 \\ \cline{2-5}
\raisebox{0.5cm}[-0.5ex]{$\mu = 50$} & $\big(\frac{b^{(-)}}{b^{(+)}}\big)_\infty$ & 2.04 & 1.53 & -1.91 \\ \hline 
\end{tabular}
\caption{The ratios $\big(\frac{b^{(-)}}{b^{(+)}}\big)_0$ and $\big(\frac{b^{(-)}}{b^{(+)}}\big)_\infty$ for $\beta_{\rm D}=0$, the sharp proper-time cutoff and various values of $\mu$- and the gauge parameter $\alpha_{\rm D}$.}
\label{hyp-quot-tab}
\end{table}

\paragraph{(C)} The system \eqref{HYP-2-betas} is simple enough to be solved analytically. It gives rise to RG trajectories of the form
\begin{equation}
 g_k = \frac{G_0\,k^2}{1 - \frac{1}{2} b^{(+)} G_0 k^2},\qquad
\gamma_k = \gamma_0 \Big[1 - \frac{1}{2}\,b^{(+)} G_0 k^2\Big]^{1-\frac{b^{(-)}}{b^{(+)}}}\:,\label{hypothese-g-gamma}
\end{equation}
where $G_0\equiv \lim_{k\rightarrow 0}(g_k/k^2)$ and $\gamma_0=\lim_{k\rightarrow0} \gamma_k$ are constants of integration.

\paragraph{(D)} Fig. \ref{gamma-g-fluss-approx} shows the phase portrait of the simplified system \eqref{HYP-2-betas} where we have, again, compactified the $\gamma$-coordinate. It is adapted to the parameter values $(\alpha_{\rm D},\beta_{\rm D}, \mu)=(1,0,5)$ giving rise to the coefficients $b^{(+)}=-0.75$ and $b^{(-)}=-1.305$. Comparing this phase portrait to the one in Fig. \ref{gamma-g-fluss-fig} that employed the same $(\alpha_{\rm D},\beta_{\rm D}, \mu)$ values we find that, apart from the region close to $\gamma=\pm1$, the two RG flows show a remarkably good qualitative and quantitative agreement. This agreement constitutes the central argument in favor of our conjecture.

\paragraph{(E)} If this conjecture is true, the ratio $b^{(-)}/b^{(+)}$ obviously plays a central role for the properties of the resulting flow. We have examined this ratio for various parameter configurations and part of the resulting figures is given in Table \ref{hyp-quot-tab}. 

First of all, we note the good agreement of the pairs of values obtained for the same parameters in the two limits $\gamma\rightarrow0$ and $\gamma\rightarrow\infty$. 

The second striking result is that the absolute ratio stays close to unity for almost all parameter values although numerator and denominator separately vary over {\it several orders of magnitude} for the different $(\alpha_{\rm D},\beta_{\rm D}, \mu)$ configurations. We may conclude that {\it a highly nontrivial and effective compensation of scheme dependences is at work here}.

\paragraph{(F)} The value 1 for the ratio $b^{(-)}/b^{(+)}$ is special, as for this value, by \eqref{beta-gamma-HYP-2}, the Immirzi parameter is seen to have {\it no running with $k$ at all}.

In order to assess how likely it is that an exact treatment of the flow indeed might yield $\partial_t\gamma_k=0$ we recall that a ratio of 1 of these to quantities corresponds to {\it equal scalar and pseudo scalar} contributions to the total supertrace. While the original Holst action for a vanishing cosmological constant $\lambda$ treats scalars and pseudo scalars on the same footing, it may well be that our choice of ${\sf O}(4)$ gauge fixing action breaks this possibly existent symmetry as it only contains scalar and no pseudo scalar terms. However, we did not find a simple generalization of the gauge fixing condition that respects this symmetry.

If this symmetry could be established to hold as a property of the flow, computed with a better (but as to yet, unknown) gauge fixing which does not break it, the action
\begin{equation}
\frac{1}{G_k} \int\dr^4 x\:\bar{e}\Big(F^{ab}_{~~\mu\nu} - \frac{1}{\gamma_k}\star F^{ab}_{~~\mu\nu}\Big)e_a^{~\mu} e_b^{~\nu}
\end{equation}
is renormalized ``as a whole'' only, and an arbitrary fixed value $\gamma_k=$const can be assigned to the Immirzi parameter.

However, whether or not this ``perfect'' gauge fixing exists, as soon as a non-zero cosmological constant is present, the scalar terms in the action stand out as no corresponding pseudo scalar monomial exists. Thus the symmetry is unavoidably violated and we expect a running of the Immirzi parameter that is driven by the cosmological constant. We will come back to this issue when discussing the full $(\lambda,\gamma,g)$-system in the next subsection.

\subsection{The full 3-dimensional $(\lambda,g,\gamma)$ truncation}
In this section we analyze the full $(\lambda,g,\gamma)$ system. The section is divided into three parts that are devoted to the following main points: First, we will formulate and test the 3-dimensional generalization of the conjecture concerning the exact flow that was introduced in the previous subsection. Second, we will analyze the fixed point structure of this simplified set of RG equations and discuss how they carry over to the case of the full $\beta$-functions we actually obtained. In the third part we discuss qualitatively the RG flow in the $(\lambda,\gamma,g)$ theory space. 

Unless otherwise stated, all explicit numerical examples in this section were obtained for the parameter values $(\alpha_{\rm D}, \beta_{\rm D}, \mu)=(1,0,5)$, as it was confirmed that it may serve as a typical case, also in the three dimensional coupling space.

\subsubsection{Testing the conjecture concerning the exact flow (3-dimensional case)}
In the two dimensional $(\gamma,g)$ truncation we found that the functions\footnote{Here and in the following we suppress the argument $\mu$ which is considered fixed.} $16 \pi f^{(\pm)}(\lambda,\gamma)$ could be very well approximated by constants $b^{(\pm)}$ for the case of a fixed $\lambda=0$. Moreover, our results indicated that the ratio $b^{(-)}/b^{(+)} = 1$, expressing a symmetry between scalar and pseudo scalar contributions, could possibly be realized, if not a symmetry breaking gauge fixing term had been used. 

\vspace{0.5cm}
\noindent{\bf (A)} Switching on the cosmological constant now, the simplest generalization one could think of would be that the functions $f^{(\pm) }(\lambda,\gamma)$ turn out $\lambda$-independent as well. However, this is not the case and, moreover, is not expected. In fact, the inclusion of a cosmological constant term breaks the symmetry of scalar and pseudo scalar invariants already at the level of the truncation and therefore should drive the value of the ratio $b^{(-)}/b^{(+)}$ away from unity.

So the natural generalization of our conjecture \eqref{fsconstant2} therefore is that the former constants $b^{(\pm)}$ w.\,r.\,t. $\gamma$ will depend on $\lambda$ now: 
\begin{equation}\label{GenConj}
f^{(\pm)}(\lambda,\gamma)\equiv \frac{b^{(\pm)}(\lambda)}{16\pi}=\text{const w.\,r.\,t. } \gamma \qquad \text{for all }\gamma\in(-\infty,\infty). 
\end{equation}

If this generalized conjecture is true the limits
\begin{equation}
B^{(\pm)}_0(\lambda)\equiv\lim_{\gamma\rightarrow0} 16 \pi f^{(\pm)}(\lambda,\gamma)\quad \text{and}\quad B^{(\pm)}_\infty(\lambda)\equiv\lim_{\gamma\rightarrow\infty} 16 \pi f^{(\pm)}(\lambda,\gamma),
\end{equation}
should coincide: $B_0^{(\pm)}(\lambda)=B_\infty^{(\pm)}(\lambda)$ for all $\lambda$. 

\vspace{0.5cm}
\noindent{\bf (B)} We have tested this hypothesis choosing any parameter combination of $\alpha_{\rm D}\kern-2pt\in\kern-3pt \{0.\kern-0.6pt 1,\kern-2pt 1,\kern-2pt 1\kern-0.6pt 0\}$\kern-1pt, $\mu\in \{1,5,50\}$ and $\beta_{\rm D}=0$ for the sharp proper-time cutoff, we found that the functions $B^{(\pm)}_0(\lambda)$ and $B^{(\pm)}_\infty(\lambda)$ indeed do show a remarkably good agreement, although their actual shape is severely parameter dependent. 

In general, the functions $B^{(\pm)}_{0,\infty}(\lambda)$ start off near zero for large negative $\lambda$, show a pronounced dependence on $\lambda$ close to zero up to $\lambda\approx4$ before getting numerically unstable. As the relevant fixed points mostly occur in the domain of (relative) numerical stability, these instabilities do not have to concern us too much.

The most profound and strongest argument in favor of our generalized hypothesis therefore is that the coincidence of the functions $B_0$ and $B_\infty$ is largely gauge- and $\mu$-independent.

\begin{figure}[htb]
\begin{center}
{\small
\psfrag{Y}[br][br]{$\left(\frac{B^{(-)}(\lambda)}{B^{(+)}(\lambda)}\right)_0$}
\psfrag{L}[bc][bc]{$\lambda$}
\includegraphics[width=0.75\textwidth]{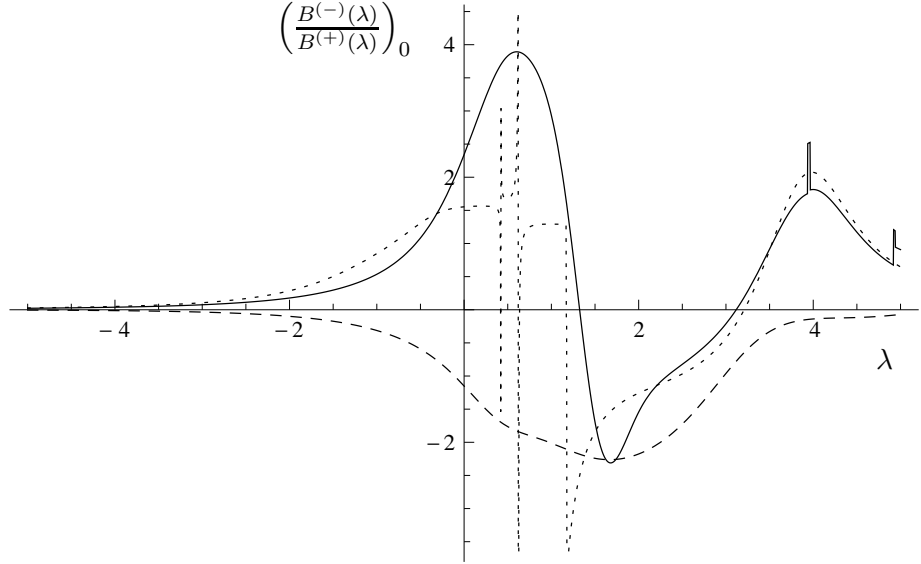}
}
\caption{The ratio $\left(\frac{B^{(-)}(\lambda)}{B^{(+)}(\lambda)}\right)_0$ for $\mu=5$, $\beta_{\rm D}=0$ and the sharp cutoff. All three functions ($\alpha_{\rm D}=1$ solid, $\alpha_{\rm D}=10$ dotted, $\alpha_{\rm D}=0.1$ dashed) decay for large negative $\lambda$.}
\label{B_ratios}
\end{center}
\end{figure}

\noindent{\bf (C)} Let us move on and discuss the ratio $B^{(-)}(\lambda)/B^{(+)}(\lambda)$ that is of crucial importance for the running of the Immirzi parameter. In Fig. \ref{B_ratios} we plot it for three different choices of the gauge parameter $\alpha_{\rm D}$. We find that, while its behavior for positive $\lambda$ is quite gauge dependent, for negative $\lambda$ the absolute value of the ratio decreases monotonically, starting from the value at $\lambda=0$ that was already obtained in the last subsection. Hence, in the domain of large negative $\lambda$, the scalar contributions ($B^{(+)}$) outweigh the pseudo scalar ones ($B^{(-)}$) by far.

We therefore conclude that there do not exist any extended intervals of $\lambda$-values for which $B^{+}(\lambda)=B^{(-)}(\lambda)$, not even approximately. The important consequence is that {\it the Immirzi parameter has a non-zero beta function and is subject to a nontrivial RG running therefore}. (See for instance, eq. \eqref{beta-gamma-3d} below.)

\subsubsection{The physical fixed points ${\bf NGFP_0}$ and ${\bf NGFP_\infty}$}
Assuming now the generalized conjecture \eqref{GenConj} is true, the exact RG equations on the 3-dimensional theory space simplify to
\begin{subequations}\label{3d-fluss}
\begin{align}
\partial_t\,g_k = \beta_g (\lambda_k,\gamma_k,g_k) &= \Big[2 + B^{(+)} (\lambda_k) g_k\Big] g_k\label{beta-g-3d} \\ 
\partial_t\,\gamma_k = \beta_\gamma (\lambda_k,\gamma_k,g_k) &= \Big[B^{(-)}(\lambda_k) - B^{(+)}(\lambda_k)\Big] g_k\,\gamma_k\label{beta-gamma-3d}\\
\partial_t\,\lambda_k = \beta_\lambda (\lambda_k,\gamma_k,g_k) &= 8\pi g_k \Big[I^{\rm grav}_{\rm F} (\lambda_k, \mu) + I^{\rm grav}_{\cal N} (\gamma_k) - 2\,I^{\rm gh}_{\rm F} (\mu)\Big]\nonumber\\
&\ \  \ + g_k\,\lambda_k\,B^{(+)} (\lambda_k) - 2\,\lambda_k\:.\label{beta-lambda-3d}
\end{align}
\end{subequations}
In view of the results in the previous subsections these RG equations should be a good approximation to the exact system \eqref{vollst-exakt-fluss}. Note, that in the limits $\gamma=0$ and $\gamma\rightarrow\infty$ this approximation is best as $\partial_\gamma f^{(\pm)}=0$ in these limits, due to symmetry in $\gamma$ and a constant asymptote, respectively. As the decay to the constant asymptote is, in addition, faster than $1/\gamma$, fixed points lying in the $\gamma=0$-plane and the $\gamma\rightarrow\infty$-plane and all their properties are the same for the hypothetical flow employing $B=B_0$ and $B=B_\infty$, respectively, and the actually calculated flow.

\paragraph{(A)} Since $\beta_\gamma=0$ and $\beta_{\hat\gamma}=0$ for $\gamma=0$ and $\hat{\gamma}=0$, respectively, those two planes of constant $\gamma$ are invariant under the RG flow. Searching for fixed points on these planes we can therefore ignore the $\beta$-function of the Immirzi parameter. We solve $\beta_g=0$  in the non-Gaussian case for $g^\ast(\lambda) = -2/B^{(+)} \big(\lambda \big)$ and insert this into $\beta_\lambda=0$, giving a condition for $\lambda\equiv \lambda^\ast$,
\begin{equation}\label{3d-FP-cond}
  \frac{16 \pi}{B^{(+)} \big(\lambda^\ast\big)} \Big[I^{\rm grav}_{\rm F} \big(\lambda^\ast, \mu\big) + I^{\rm grav}_{\cal N} \big(\gamma^\ast\big) - 2\,I^{\rm gh}_{\rm F} \big(\mu\big)\Big] + 4\lambda^\ast = 0\:,
\end{equation}
that can only be solved numerically for $\lambda^\ast$.

We denote the resulting fixed points by {\bf $\mbox{NGFP}_0$} $\equiv (\lambda^\ast_0, \gamma^\ast, g^\ast_0)$ and {\bf $\mbox{NGFP}_\infty$} $\equiv (\lambda^\ast_\infty, \hat{\gamma}^\ast, g^\ast_\infty)$, with $\gamma^\ast=0$ and $\hat{\gamma}^\ast=0$, respectively. As $\lim_{\hat{\gamma}\rightarrow0}I^{\rm grav}_{\cal N} \big(\gamma^\ast\big)=0$ the equation \eqref{3d-FP-cond} reduces to \eqref{fp-bdg-2-EH} such that the three FPs found in the 2-dimensional $(g,\lambda)$-truncation carry over to the plane $\hat{\gamma}=0$. For $\gamma\rightarrow 0$ we will again find three different fixed points that, however, have different $(\lambda,g)$-coordinates.

Besides these non-Gaussian fixed points, the simplified system \eqref{3d-fluss} allows for a {\it fixed line}, namely $(\lambda=0,\gamma, g=0)$ with $\gamma\in(-\infty,\infty)$. It contains the Gaussian fixed point at $\lambda^\ast=\gamma^\ast=g^\ast=0$.

\paragraph{(B)} Let us come back to the non-Gaussian fixed points. First of all, we note that {\it the duality of the two fixed points under $\gamma\rightarrow\frac{1}{\gamma}$, that was present in the 2-dimensional $(\gamma, g)$ truncation, does not carry over to the three dimensional flow}. As $\lim_{\gamma\rightarrow0} I^{\rm grav}_{\cal N} \big(\gamma^\ast\big)\neq0$, the fixed point condition \eqref{3d-FP-cond} differs for the two $\gamma$-planes even if our hypothesis $B^{(\pm)}_0(\lambda)=B^{(\pm)}_\infty(\lambda)$ is satisfied, thus giving rise to {\it different fixed point coordinates} $\lambda^\ast_0$ and $\lambda^\ast_\infty$. This also implies different values for $g^\ast_0$ and $g^\ast_\infty$.

\paragraph{(C)} Now we can move on and discuss the stability properties of the fixed points. The stability matrices at the fixed points {\bf $\mbox{NGFP}_0$} and {\bf $\mbox{NGFP}_\infty$}, respectively, are of the form
\begin{equation}\label{StabMatrices}
 {\cal B}_0 \equiv \left. \left( \begin{array}{ccc} \frac{\partial \beta_\lambda}{\partial \lambda} & 0 &\frac{\partial \beta_\lambda}{\partial g} \\ 0 &\frac{\partial \beta_\gamma}{\partial \gamma} & 0 \\ \frac{\partial \beta_g}{\partial \lambda} & 0 & -2 \end{array} \right)\right|_{\lambda^\ast_0, \gamma^\ast, g^\ast_0}
\quad\text{and}\qquad
 {\cal B}_\infty \equiv \left. \left( \begin{array}{ccc} \frac{\partial \beta_\lambda}{\partial \lambda} & 0 &\frac{\partial \beta_\lambda}{\partial g} \\ 0 &\frac{\partial \beta_{\hat{\gamma}}}{\partial \hat{\gamma}} & 0 \\ \frac{\partial \beta_g}{\partial \lambda} & 0 & -2 \end{array} \right)\right|_{\lambda^\ast_\infty, \hat{\gamma}^\ast, g^\ast_\infty}
\end{equation}
This is due to the fact that the derivatives $\partial_\lambda B^{(\pm)}_{0,\infty}$ stay finite at $\lambda^\ast_{0,\infty}$, such that
\begin{equation}
 \left. \frac{\partial \beta_\gamma}{\partial g} \right|_{\lambda^\ast_0, \gamma^\ast, g^\ast_0} = 0 = \left. \frac{\partial \beta_\gamma}{\partial \lambda} \right|_{\lambda^\ast_0, \gamma^\ast, g^\ast_0}\:.
\end{equation}
Moreover, because of their symmetry in $\gamma$, and because of their rapid approach of a constant asymptote, we find that the $\gamma$- and $\hat\gamma$-derivatives of $\beta_g$ and $\beta_\lambda$ vanish at the respective fixed points.

From the form of the stability matrices \eqref{StabMatrices} we can directly infer that one critical exponent of each fixed point corresponds to the $\gamma$-direction. It can be easily expressed in terms of the $B^{(\pm)}$-functions according to
\begin{equation}
\Theta_\gamma=-2 \Big[1 - \Big(\frac{B^{(-)}}{B^{(+)}}\Big)_0 \big(\lambda^\ast_0\big)\Big]\qquad\text{and}\qquad \Theta_{\hat{\gamma}} = 2 \Big[1 - \Big(\frac{B^{(-)}}{B^{(+)}}\Big)_\infty \big(\lambda^\ast_\infty\big)\Big]\:.
\end{equation}
Since $\lambda^\ast_0\neq \lambda^\ast_\infty$, even if the conjecture is satisfied, this does {\it not} imply $\Theta_\gamma=-\Theta_{\hat\gamma}$. We observe that the critical exponent depends only on the ratio of pseudo scalar to scalar contributions; in particular, it is close to $-2$ or $2$ if the scalars outweigh the pseudo scalars, as is the case for large negative $\lambda^\ast$-values (cf. Fig. \ref{B_ratios}).

As already found for the $(\lambda,g)$ truncation, the other critical exponents $\Theta_1$ and $\Theta_3$ can be associated with the couplings $\lambda$ and $g$, respectively, but only approximately.

\begin{table}\renewcommand*{\arraystretch}{0.75}
\centering
\subtable[]{\begin{tabular}{|c|c|c|c|c|c|c|c|}\hline
 {\bf $\mbox{NGFP}^1_0$}& $\lambda^\ast_0$ & $g^\ast_0$ & $\lambda^\ast_0g^\ast_0$ & $\Theta_1$ & $\Theta_3$ & $\Theta_\gamma$ & $\big(\frac{B^{(-)}}{B^{(+)}}\big)_0$ \\ \hline \hline
$\alpha_\Dr = 0.1$ & -7.42& 3.65& -27.09 & 3.73& 2.28& -2.00& 0.00015\\ \hline
$\alpha_\Dr = 1$ & -6.78 & 3.37 & -22.86 & 3.71 & 1.94 & -1.98 & 0.01\\ \hline  
$\alpha_\Dr = 10$ & -6.66 & 3.32 & -22.09 & 3.70 & 1.85& -1.97& 0.013 \\ \hline
{\bf $\mbox{NGFP}^1_\infty$}& $\lambda^\ast_\infty$ & $g^\ast_\infty$ & $\lambda^\ast_\infty g^\ast_\infty$ & $\Theta_1$ & $\Theta_3$ & $\Theta_{\hat{\gamma}}$ & $\big(\frac{B^{(-)}}{B^{(+)}}\big)_\infty$ \\ \hline \hline
$\alpha_\Dr = 0.1$ & -5.16& 3.86& -19.89& 3.32& 2.55& 2.01& -0.005\\ \hline
$\alpha_\Dr = 1$ & -4.18& 3.30& -13.79 & 3.22& 1.81& 1.94& 0.03\\ \hline  
$\alpha_\Dr = 10$ & -3.78& 3.08& -11.66 & 3.22& 1.28 & 1.90& 0.05\\ \hline
\end{tabular}
}
\subtable[]{\begin{tabular}{|c|c|c|c|c|c|c|c|}\hline
 {\bf $\mbox{NGFP}^2_0$}& $\lambda^\ast_0$ & $g^\ast_0$ & $\lambda^\ast_0g^\ast_0$ & $\Theta_1$ & $\Theta_3$ & $\Theta_\gamma$ & $\big(\frac{B^{(-)}}{B^{(+)}}\big)_0$ \\ \hline \hline
$\alpha_\Dr = 0.1$ & 2.08& 4.04& 8.38 & -13.20& 1.81& -6.26& -2.13\\ \hline
$\alpha_\Dr = 1$ & 0.07 & 3.25 & 0.21 & -23.40 & 2.00 & 3.13 & 2.56\\ \hline  
$\alpha_\Dr = 10$ & -1.08 & 1.30 & -1.41 & 6.64 & 2.45& -0.43& 0.78 \\ \hline
{\bf $\mbox{NGFP}^2_\infty$}& $\lambda^\ast_\infty$ & $g^\ast_\infty$ & $\lambda^\ast_\infty g^\ast_\infty$ & $\Theta_1$ & $\Theta_3$ & $\Theta_{\hat{\gamma}}$ & $\big(\frac{B^{(-)}}{B^{(+)}}\big)_\infty$ \\ \hline \hline
$\alpha_\Dr = 0.1$ & 1.25 & 5.06& 6.35& -14.66& 1.89 & 5.99& -1.99\\ \hline
$\alpha_\Dr = 1$  & -0.46& 3.34& -1.54 & -10.56& 2.00 & -0.14& 1.07\\ \hline  
$\alpha_\Dr = 10$ & -1.83& 2.18& -3.98 & -2.41& 2.76& 1.34& 0.33\\ \hline
\end{tabular}
}
\subtable[]{\begin{tabular}{|c|c|c|c|c|c|c|c|}\hline
 {\bf $\mbox{NGFP}^3_0$}& $\lambda^\ast_0$ & $g^\ast_0$ & $\lambda^\ast_0g^\ast_0$ & $\Theta_1$ & $\Theta_3$ & $\Theta_\gamma$ & $\big(\frac{B^{(-)}}{B^{(+)}}\big)_0$ \\ \hline \hline
$\alpha_\Dr = 0.001$ & 4.52 & 10.60& 47.92 & 24.42 & 2.78& 2.25& 2.13\\ \hline
$\alpha_\Dr = 0.1$ & 3.37 & 2.35 & 7.92 & 6.14 & 4.25 & -2.89 & -0.44\\ \hline  
$\alpha_\Dr = 10$ & -4.26 & 7.91 & 33.71 & 21.16 & 2.54& 1.47& 1.73 \\ \hline
{\bf $\mbox{NGFP}^3_\infty$}& $\lambda^\ast_\infty$ & $g^\ast_\infty$ & $\lambda^\ast_\infty g^\ast_\infty$ & $\Theta_1$ & $\Theta_3$ & $\Theta_{\hat{\gamma}}$ & $\big(\frac{B^{(-)}}{B^{(+)}}\big)_\infty$ \\ \hline \hline
$\alpha_\Dr = 0.001$ & 5.00 & 5.20 & 26.00& 9.19& 2.66 & 0.60 & 0.70\\ \hline
$\alpha_\Dr = 0.1$ & 3.59 & 1.73& 6.24 & 8.51 & 3.48& 2.48 & -0.24\\ \hline  
$\alpha_\Dr = 10$ & 4.64 & 4.39& 20.40 & 11.15 & 2.61 & 0.56& 0.72\\ \hline
\end{tabular}
}
\caption{Properties of the three pairs of (``physical'') fixed points {\bf $\mbox{NGFP}_0$} and {\bf $\mbox{NGFP}_\infty$} in the full 3-dimensional $(\lambda, \gamma, g)$ truncation. The values were obtained for the sharp cutoff and the choice $(\mu, \beta_{\rm D})=(5,0)$.}
\label{3d-FP-tab}
\end{table}

\paragraph{(D)} In Table \ref{3d-FP-tab}, subtable (a), we have listed the fixed point properties for the parameter values $\mu=5$, $\beta_{\rm D}=0$ and $\alpha_{\rm D}\in\{0.1,1,10\}$. We find indeed a pair of non-Gaussian fixed points {\bf $\mbox{NGFP}_0^1$} and {\bf $\mbox{NGFP}_\infty^1$} that both occur at large negative $\lambda^\ast$ and positive $g^\ast$, for all three values of $\alpha_{\rm D}$. As the ratio $B^{(-)}/B^{(+)}$ is tiny at those $\lambda$-values we find in addition $\Theta_\gamma\approx\Theta_{\hat\gamma}\approx -2$. The other critical exponents $\Theta_1$ and $\Theta_3$ are in the range already known from the $(\lambda,g)$-truncation; for the $\gamma\rightarrow\infty$ case the data carries over exactly.

\paragraph{(E)} The subtables (b) and (c) of Table \ref{3d-FP-tab} show results obtained for the 3-dimensional ``lifts'' {\bf $\mbox{NGFP}_{0,\infty}^2$} and {\bf $\mbox{NGFP}_{0,\infty}^3$} the other two fixed points that were found in the $(g,\lambda)$-truncation. Here, the critical exponents turn out far less stable w.\,r.\,t. a variation of the gauge parameter, confirming the picture obtained already in the two dimensional truncation. In the three dimensional case we find in addition that the critical exponent corresponding to the $\gamma$-direction may even change sign, such that the attractivity properties of the fixed point change even qualitatively.

\paragraph{(F)} Table \ref{3d-FP-tab} gives a rather typical insight into the fixed point structure of the truncation, also for other choices of the parameters $\mu$ and $\beta_{\rm}$. As a generic feature the pair of fixed points at large negative $\lambda$, {\bf $\mbox{NGFP}_0^1$} and {\bf $\mbox{NGFP}_\infty^1$}, is most stable and occurs at a {\it positive} $g^\ast$ coordinate, such that the typical {\it anti-screening} behavior is also found in the three dimensional truncation. Moreover, at these negative $\lambda$-values the $\gamma=0$-plane is found UV repulsive, while the $\hat\gamma=0$-plane is UV attractive, such that the fixed point in the $\gamma=0$-plane leads to a more predictive theory. The corresponding critical exponents satisfy the approximate equality $\Theta_\gamma\approx-\Theta_{\hat\gamma}\approx -2$.

The other two pairs of fixed points, {\bf $\mbox{NGFP}_{0}^{2,3}$} and {\bf $\mbox{NGFP}_{\infty}^{2,3}$}, lead to far less robust results.

\paragraph{(G)} As a last remark, we want to mention that in the original system of flow equations \eqref{vollst-exakt-fluss} also pseudo fixed points  ${\bf NGFP_\pm}$ close to $\gamma=\pm1$ are likely to occur. In complete analogy to the discussion in the 2-dimensional $(\gamma,g)$ truncation we discard them as unphysical and do not study them in more detail.

\subsubsection{Is there a fixed point with a non-zero and finite Immirzi parameter?}
When analyzing the system of flow equations \eqref{3d-fluss} it becomes obvious that there is another possible mechanism that could lead to a FP: At those points $\lambda^\ast$ where the ratio $B^{(-)}/B^{(+)}$ equals 1 we have $\beta_\gamma=0$. Inserting this value $\lambda^\ast$ into $\beta_g=0$ we can solve for $g^\ast(\lambda^\ast)$ and substitute the two values ($\lambda^\ast$, $g^\ast$) into $\beta_\lambda=0$ in order to find a corresponding $\gamma^\ast$-coordinate. While the first two conditions are typically solvable (following our conjecture about the exact flow in the $(g,\gamma)$-truncation, the cosmological constant could be given by $\lambda^\ast=0$) it is not clear whether the last condition has a solution. We tested this scenario for different gauge parameters and found that, due to the last condition, the existence of such a FP is strongly gauge dependent in our truncation. For this reason we did not study its properties in more detail.

\subsubsection{Comparison with a perturbative calculation}
In ref. \cite{benedetti-speziale} a perturbative one-loop calculation in the $(\gamma, g)$ sector of the Holst theory has been reported. Its result is not easily compared to ours as a different gauge fixing is used there, the ${\sf O}(4)$ ghost contribution is neglected completely and, most importantly, the cosmological constant is set to zero throughout. Nevertheless, specializing our beta functions correspondingly we obtain a flow equation which has the same general structure as the perturbative one, with the same sign of the anomalous dimension. In view of the strong $\lambda$-dependence of $\beta_\gamma$ which we observe in our 2-dimensional truncation a direct comparison of the sign of $\beta_\gamma$ seems not meaningful, however.

\section{Summary}
In this paper we have explored the functional RG flow of quantum gravity on the Einstein-Cartan theory space which, by definition, consists of all action functionals that depend on the spin connection and the vielbein and that are invariant under the semidirect product group of spacetime diffeomorphisms and local Lorentz transformations. In the first part of the paper we developed general methods and calculational tools which are needed to evaluate the functional RG equation for the corresponding effectice average action. These techniques and partial results were presented in such a way that the various items in our ``tool kit'' can be applied to many different truncations which go beyond the one studied here, namely the ``Holst truncation'' with its 3 running parameters, Newton's constant, the cosmological constant, and, most interestingly, the Immirzi parameter. Since a priori no value of the scale dependent prefactor of the Immirzi term in $\Gamma_k$ may be excluded, in particular not the one which amounts to ``$\gamma=\pm \infty$'' in the usual parametrization, we were led to cover the ``Immirzi direction'' of theory space by two coordinate charts. In this way we found, for instance, an intriguing (small $\gamma) \ \leftrightarrow$ (large $\gamma$) duality in the 2-dimensional $(\gamma, g)$-truncation with $\lambda_k\equiv 0$, the flow being invariant under $\gamma\mapsto 1/\gamma$.

Let us now summarize our main results obtained with the complete truncation ansatz. Putting all the details together, the following picture for the typical, most reliable RG flow on the three-dimensional $(\lambda,g,\gamma)$ theory space emerged 

\noindent{\bf (i)} In the $\gamma=0$ and the $\gamma=\infty$ planes, respectively, we have non-Gaussian fixed points {\bf $\mbox{NGFP}_{0}^{1}$} and {\bf $\mbox{NGFP}_{\infty}^{1}$} with a positive $g^\ast$ and a negative $\lambda^\ast$ coordinate. 

\noindent{\bf (ii)} Besides this pair of fixed points, there may exist two other pairs, \kern-1pt(${\bf NGFP^{2}_{0}}$, \kern-1pt${\bf NGFP^{2}_{\infty}}$\kern-1pt) and (${\bf NGFP^{3}_{0}}$, ${\bf NGFP^{3}_{\infty}}$), whose properties turned out less robust under variations of the gauge fixing and $\mu$ parameters in our present approximation.

\noindent{\bf (iii)} In contrast to the conjectured exact flow in the $(\gamma,g)$-truncation, the $(\lambda^\ast,g^\ast)$-co\-or\-di\-nates do not coincide for each pair of fixed points, such that the duality under $\gamma\mapsto 1/\gamma$ which is visible in the 2-dimensional $(\gamma, g)$-truncation breaks down in the 3-dimensional truncation. 

\noindent{\bf (iv)} Both fixed points ${\bf NGFP^1_{0,\infty}}$ are UV attractive in the two directions of the $(\lambda,g)$-plane. They are, at least in the small $\alpha_{\rm D}$ limit, in an interplay with the respective Gaussian fixed point in each of the planes $\gamma=0,\infty$. This allows for the existence of a separatrix in those planes that separates trajectories with positive or negative IR cosmological constant, respectively. This feature is shared with metric gravity.

\noindent{\bf (v)} The $\beta$-function of the Immirzi parameter vanishes in the $\gamma=0$ and $\gamma=\infty$ planes. In between the $\gamma$-flow shows a preferred direction. This direction, i.\,e. the sign of $\beta_\gamma$ only depends on the $\lambda$-coordinate. At large negative $\lambda$ it corresponds to the stability properties of the non-Gaussian fixed points: While ${\bf NGFP^1_\infty}$ is UV attractive, ${\bf NGFP^1_0}$ is repulsive, generically, leading to a tendency of the RG flow towards smaller absolute values of the Immirzi parameter in the IR. For some larger or even positive $\lambda$, however, the $\gamma$-flow often changes sign  and thus reverses its direction.

\noindent{\bf (vi)} In principle, both other pairs of fixed points (${\bf NGFP^{2}_{0}}$, ${\bf NGFP^{2}_{\infty}}$) and (${\bf NGFP^{3}_{0}}$, ${\bf NGFP^{3}_{\infty}}$) allow for the asymptotic safety construction, too, but due to their pronounced gauge dependence let us focus here on the most promising first pair, (${\bf NGFP^{1}_{0}}$ and ${\bf NGFP^{1}_{\infty}}$).

\noindent{\bf (vii)} In this picture the set of all asymptotically safe trajectories emanating from ${\bf NGFP^{1}_{0}}$ or ${\bf NGFP^{1}_{\infty}}$ can be divided into three classes. They differ with respect to the running of the Immirzi parameter they imply: 

\noindent{\bf (a)} All RG trajectories that are asymptotically safe w.\,r.\,t. ${\bf NGFP^1_0}$ lie completely in the $\gamma=0$-plane, such that in this case the Immirzi parameter does not run at all and certain components of the torsion fluctuations are completely suppressed.

\noindent{\bf (b)} There exist trajectories which are asymptotically safe w.\,r.\,t. ${\bf NGFP^1_\infty}$ that are confined to the $\hat\gamma=0$-plane. Those trajectories do not show a running of $\gamma$ either, but in this case all torsion components fluctuate freely.

\noindent{\bf (c)} All other trajectories that are asymptotically safe w.\,r.\,t. ${\bf NGFP^1_\infty}$ run to an arbitrary value of $\gamma$ in the IR. It can be speculated that in the exact flow a complete RG trajectory exists connecting the two non-Gaussian fixed points, running from ${\bf NGFP^1_\infty}$ in the UV to ${\bf NGFP^1_0}$ in the IR. In our approximation, however, such a trajectory cannot be continued beyond the singularity at $\gamma=\pm1$.

The singularities of the beta functions at $\gamma=\pm1$ are due to the fact that for these values of the Immirzi parameter the (anti-)selfdual projection of $\omega^{ab}{}_\mu$ drops out from the Holst action. However, in the functional integral related to the flow equations one still performs an integration over the projection which decouples. This entails a singularity of the functional integral which is mirrored by the FRGE. In fact, if one wants to study {\it chiral gravity} where only one or the other of the two chiral projections is quantized one must modify the FRGE computation from the outset in such a way that it corresponds to a functional integral over one chiral component only \cite{uli-chiral}.

The picture summarized above is the outcome of detailed reliability checks. In particular we tested the scheme and gauge fixing dependence of the results, and the dependence on the $\mu$ parameter one must introduce for dimensional reasons. Our overall assessment is that while the features listed above are generic the results are somewhat less robust than for a comparable truncation in metric gravity. Our preliminary explanation is that this might be due to the larger gauge sector in Einstein-Cartan theory.\footnote{During the course of writing up this paper further evidence for this interpretation was found \cite{e-only, robertofermions}.}

It might also be that, in the latter theory, the observables are less directly related to the running couplings than in QEG so that then there is more room for the compensation of gauge dependencies on the way from the RG flow to observable quantities. In this context we also should emphasize that, strictly speaking, for the Asymptotic Safety construction to work it is by no means necessary that the fixed point of a projected flow exists {\it for all} values of the gauge parameters and $\mu$. Exact Asymptotic Safety is perfectly consistent with a NGFP in a projected or truncated flow existing only for some, or perhaps only a single value of the couplings kept fixed. An instructive lesson is provided by the $\alpha_{\rm D}$ parameter in QEG. It is believed to approach $\alpha_{\rm D}^\ast=0$ for $k\rightarrow\infty$ in the exact theory. So, if $\alpha_{\rm D}$ is kept fixed in some truncation, and the pertinent NGFP happens to disappear if one makes $\alpha_{\rm D}$ very large, i.\,e. very different from its exact NGFP value, then we have no reason to question the reliability of this truncation; rather, we would expect this to happen generically.

Thus the overall conclusion is that, like metric gravity, Einstein-Cartan gravity has very good chances of turning out nonperturbatively renormalizable in the end. Clearly more work, in particular on more general truncations will be needed in order to further substantiate this conjecture, as this has been done in QEG. In particular it will be important to see how the asymptotically safe $e$-$\omega$-theory relates to its metric counterpart. At present the relatively large uncertainties (of critical exponents, etc.) make it hard to judge about their possible (in-)equivalence.

\vspace{1cm}
{\noindent \bf Acknowledgement:} We would like to thank U. Harst for many discussions and for his help in the numerical analysis and in preparing the manuscript. 

\vspace{2cm}
\appendix
\noindent{\Large\bf Appendix}
\vspace{-0.6cm}
\section{Completeness relations of the generalized \\position- and momentum-bases}\label{App:Completeness}
 
The completeness relations of the generalized momentum and position eigenvectors introduced in the main text are, sectorwise, given in the following table:
\begin{equation}
 \begin{split}
  \int\dr^4 p\sum_{\alpha_{\rm V} = 1}^{n_{\rm V}}\sum_{I = 1}^{3}&\langle x\,m\,i_{\rm V}\,|\,p\,I\,\alpha_{\rm V} \rangle \langle p\,I\,\alpha_{\rm V}\,|\,y\,k\,j_{\rm V} \rangle \\
&= \int\dr^4 p\big(v_{~\:\alpha_{\rm V}}^{\rm V}\big)^{i_{\rm V}}(p^2)\big(v^{{\rm V}\:\alpha_{\rm V}}\big)_{j_{\rm V}}(p^2) t_I^{~m} (\hat{p})t^I_{~k} (\hat{p}) \frac{1}{(2 \pi)^4} \eu^{i\kappa p (x-y)}\\
&= \int \frac{\dr^4 p}{(2 \pi)^4}\delta^{i_{\rm V}}_{~{j_{\rm V}}}P_{{\rm V}~k}^{~m}(\hat{p})\:\eu^{i\kappa p (x-y)}\\
&= \delta^{i_{\rm V}}_{~{j_{\rm V}}}P_{{\rm V}~k}^{~m}\big(- i \hat\partial_x\big)\frac{\delta^{(4)} (x - y)}{\bar{e}}\:,
 \end{split}
\end{equation}
\begin{align}
&\int\dr^4 p\sum_{\alpha_{\rm S} = 1}^{n_{\rm S}}\langle x\,i_{\rm S}\,|\,p\,\alpha_{\rm S} \rangle \langle p\,\alpha_{\rm S}\,|\,y\,j_{\rm S} \rangle = \delta^{i_{\rm S}}_{~{j_{\rm S}}}\frac{\delta^{(4)} (x - y)}{\bar{e}},\\
&\int\!\dr^4 p\!\sum_{\alpha_{\rm T} = 1}^{n_{\rm T}}\sum_{I, J = 1}^{3}\langle x\,m\,n\,i_{\rm T}\,|\,p\,I\,J\,\alpha_{\rm T} \rangle \langle p\,I\,J\,\alpha_{\rm T}\,|\,y\,k\,l\,j_{\rm T} \rangle = \delta^{i_{\rm T}}_{~{j_{\rm T}}}P_{{\rm V}~~\:kl}^{~mn}\big(\!-\! i \hat\partial_x\big)\frac{\delta^{(4)} (x\! -\! y)}{\bar{e}},\\
&\int\dr^4 p\sum_{\alpha_{{\rm gh}\,{\rm S}} = 1}^{n_{{\rm gh}\,{\rm S}}}\langle x\,i_{{\rm gh}\,{\rm S}}\,|\,p\,\alpha_{{\rm gh}\,{\rm S}} \rangle \langle p\,\alpha_{{\rm gh}\,{\rm S}}\,|\,y\,j_{{\rm gh}\,{\rm S}} \rangle = \delta^{i_{{\rm gh}\,{\rm S}}}_{~~~~{j_{{\rm gh}\,{\rm S}}}}\frac{\delta^{(4)} (x - y)}{\bar{e}},\\
&\int\!\dr^4 p\!\sum_{\alpha_{{\rm gh}\,{\rm V}} = 1}^{n_{{\rm gh}\,{\rm V}}}\langle x\,m\,i_{{\rm gh}\,{\rm V}}\,|\,p\,I\,\alpha_{{\rm gh}\,{\rm V}} \rangle \langle p\,I\,\alpha_{{\rm gh}\,{\rm V}}\,|\,y\,k\,j_{{\rm gh}\,{\rm V}} \rangle = \delta^{i_{{\rm gh}\,{\rm V}}}_{~~~~{j_{{\rm gh}\,{\rm V}}}}P_{{\rm V}~k}^{~m}\big(\!-\! i \hat\partial_x\big)\frac{\delta^{(4)} (x\! -\! y)}{\bar{e}},
\end{align}

\begin{equation}
\begin{split}
\int \dr^4 x\:\bar{e}\sum_{m=1}^4 \sum_{i_{\rm V} = 1}^{n_{\rm V}}&\langle p^\prime\,J\,\beta_{\rm V}\,|\,x\,m\,i_{\rm V}\rangle \langle x\,m\,i_{\rm V}|\,p\,I\,\alpha_{\rm V} \rangle\\
&= \int\dr^4 x\:\bar{e}\:\frac{1}{(2 \pi)^4}\eu^{i \kappa(p - p^\prime)x}\big(v^{{\rm V}\:{\beta_{\rm V}}}\big)_{i_{\rm V}}({p^\prime}^2)\big(v^{\rm V}_{~\:{\alpha_{\rm V}}}\big)^{i_{\rm V}}(p^2) t^J_{~m} (\hat{p}^\prime)t_I^{~m} (\hat{p})\\
&= \frac{\bar{e}}{\kappa^4}\delta^{(4)} (p - p^\prime)\big(v^{{\rm V}\:{\beta_{\rm V}}}\big)_{i_{\rm V}}({p^\prime}^2)\big(v^{\rm V}_{~\:\alpha_{\rm V}}\big)^{i_{\rm V}}(p^2) t^J_{~m} (\hat{p}^\prime)t_I^{~m} (\hat{p})\\ 
&=\label{ortsraum-vollst-grav-vektor} \delta^{(4)} (p - p^\prime)\big(v^{{\rm V}\:{\beta_{\rm V}}}\big)_{i_{\rm V}}({p^\prime}^2)\big(v^{\rm V}_{~\:\alpha_{\rm V}}\big)^{i_{\rm V}}(p^2) t^J_{~m} (\hat{p}^\prime)t_I^{~m} (\hat{p}),
 \end{split}
\end{equation}

\begin{equation}\shoveleft{
 \int \dr^4 x\:\bar{e} \sum_{i_{\rm S} = 1}^{n_{\rm S}}\langle p^\prime\,\beta_{\rm S}\,|\,x\,i_{\rm S}\rangle \langle x\,i_{\rm S}|\,p\,\alpha_{\rm S}\rangle =  \delta^{(4)} (p - p^\prime)\big(v^{{\rm S}\:{\beta_{\rm S}}}\big)_{i_{\rm S}}({p^\prime}^2)\big(v^{\rm S}_{~\:\alpha_{\rm S}}\big)^{i_{\rm S}}(p^2)},
\end{equation}

\begin{multline}
\int \dr^4 x\:\bar{e}\sum_{m, n = 1}^4 \sum_{i_{\rm T} = 1}^{n_{\rm T}}\langle p^\prime\,K\,L\,\beta_{\rm T}\,|\,x\,m\,n\,i_{\rm T}\rangle \langle x\,m\,n\,i_{\rm T}|\,p\,I\,J\,\alpha_{\rm T}\rangle\\
= \delta^{(4)} (p - p^\prime)\big(v^{{\rm T}\:{\beta_{\rm T}}}\big)_{i_{\rm T}}({p^\prime}^2)\big(v^{\rm T}_{~\:\alpha_{\rm T}}\big)^{i_{\rm T}}(p^2)\frac{1}{4}\Big(t^K_{~m}(\hat{p}^\prime)t^L_{~n}(\hat{p}^\prime) + t^K_{~n}(\hat{p}^\prime)t^L_{~m}(\hat{p}^\prime) - \frac{2}{3}P_{{\rm V}mn}(\hat{p}^\prime)\eta^{KL}\Big)\\
\times\Big(t_I^{~m}(\hat{p})t_J^{~n}(\hat{p}) + t_I^{~n}(\hat{p})t_J^{~m}(\hat{p}) - \frac{2}{3}P_{\rm V}^{~mn}(\hat{p})\eta_{IJ}\Big),
\end{multline}

\begin{multline}
 \int \dr^4 x\:\bar{e} \sum_{i_{{\rm gh}\,{\rm S}} = 1}^{n_{{\rm gh}\,{\rm S}}}\langle p^\prime\,\beta_{{\rm gh}\,{\rm S}}\,|\,x\,i_{{\rm gh}\,{\rm S}}\rangle \langle x\,i_{{\rm gh}\,{\rm S}}|\,p\,\alpha_{{\rm gh}\,{\rm S}}\rangle =\\  \delta^{(4)} (p - p^\prime)\big(v^{{{\rm gh}\,{\rm S}}\:{\beta_{{\rm gh}\,{\rm S}}}}\big)_{i_{{\rm gh}\,{\rm S}}}({p^\prime}^2)\big(v^{{\rm gh}\,{\rm S}}_{~~~~\alpha_{{\rm gh}\,{\rm S}}}\big)^{i_{{\rm gh}\,{\rm S}}}(p^2), 
\end{multline}

\begin{multline}
  \int \dr^4 x\:\bar{e} \sum_{m=1}^4 \sum_{i_{{\rm gh}\,{\rm V}} = 1}^{n_{{\rm gh}\,{\rm V}}}\langle p^\prime\,J\,\beta_{{\rm gh}\,{\rm V}}\,|\,x\,m\,i_{{\rm gh}\,{\rm V}}\rangle \langle x\,m\,i_{{\rm gh}\,{\rm V}}|\,p\,I\,\alpha_{{\rm gh}\,{\rm V}}\rangle\\
=  \delta^{(4)} (p - p^\prime)\big(v^{{{\rm gh}\,{\rm V}}\:{\beta_{{\rm gh}\,{\rm V}}}}\big)_{i_{{\rm gh}\,{\rm V}}}({p^\prime}^2)\big(v^{{\rm gh}\,{\rm V}}_{~~~~{\alpha_{{\rm gh}\,{\rm V}}}}\big)^{i_{{\rm gh}\,{\rm V}}}(p^2) t^J_{~m} (\hat{p}^\prime)t_I^{~m} (\hat{p}).
\end{multline}
Note that while the measure of the inner product on $x$-space contains a volume factor $\bar{e}$, the inner product on $p$-space does not. This asymmetry originates from the fact that we have explicitly introduced an additional factor of $\kappa$ in the exponent of the plane waves \eqref{ebeneWellen-Vollst}.

\newpage
\section{Explicit form of the matrix ${\cal M}$}\label{Appendix:MMatrix}
The matrix ${\cal M}=(H_0^{\rm grav})^{-1}\bar{H}^{\rm grav}$ whose algebraic properties are discussed in detail in Section \ref{AlgebraicFormM} consists of the following block matrices: 

\noindent {\bf Scalars in the graviton sector ($a$, $d$, $B$, $D$):}
\begin{equation}\label{ScalarM}
\big({\cal M}^{\rm S}\big)^{i_{\rm S}}_{~j_{\rm S}}(p^2) = \left(\begin{array}{cccc} 0 & 0 & 0 & 0 \\ 0 & 0 & 0 & 0 \\ 0 & 0 & 0 & -1 \\ 0 & -\frac{2 p}{\bar{\mu}} & -1 & 0 \end{array} \right)
\end{equation}
Here and in the following the rows and columns of the matrices are labeled according to the sequence of fields given, i.\,e. the matrix \eqref{ScalarM}, for instance, acts on the column vector $(a,d,B,D)^{\rm T}$.

\noindent {\bf Vectors in the graviton sector ($b^m$, $c^m$, $A^m$, $D^m$, $d^m$, $B^m$, $C^m$):}
\begin{equation}
\big({\cal M}^{\rm V}\big)^{i_{\rm V}}_{~j_{\rm V}}(p^2) = \left(\begin{array}{ccccccc} 0 & 0 & 0 & 0 & 0 & 0 & 0\\ 0 & 0 & 0 & 0 & 0 & 0 & 0\\ 0 & 0 & 0 & 0 & - \frac{2 p}{\bar{\mu}} & 0 & -1 \\ 0 & 0 & 0 & 0 & 0 & -1 & 0 \\ 0 & 0 & 0 & 0 & 0 & 0 & 0 \\ 0 & 0 & 0 & -1 & 0 & 0 & 0 \\ 0 & - \frac{2 p}{\bar{\mu}} & -1 & 0 & 0 & 0 & 0\end{array} \right)
\end{equation}

\noindent {\bf Tensors in the graviton sector ($d^{mn}$, $B^{mn}$, $D^{mn}$):}
\begin{equation}
\big({\cal M}^{\rm T}\big)^{i_{\rm T}}_{~j_{\rm T}}(p^2) = \left(\begin{array}{ccc} 0 & 0 & 0 \\ 0 & 0 & -1 \\ - \frac{2 p}{\bar{\mu}} & -1 & 0 \end{array} \right)
\end{equation}
Note that these block matrices only depend on the ratio $\frac{p}{\bar{\mu}}$; in particular, ${\cal M}$ is independent of $\Lambda_k$, $\alpha_\Dr$, and $\alpha_\Lr^\prime$.

\section{Coefficient functions in \\sharp cutoff regularization}\label{Appendix:SharpCutoff}
If the sharp proper-time cutoff regularization is employed the coefficient functions appearing on the RHS of the flow equation (cf. eq. \ref{fluss-einheitlich}) read explicitly 
\begin{equation}
 \begin{aligned}
  I^{\rm grav}_{\rm F}(\lambda_k,\mu)&=\frac{1}{8\pi^2}\frac{n+3}{2}\!\int_0^\infty\!\!\dr y\:y^3\bigg[\!\!\sum_{\genfrac{}{}{0pt}{2}{\chi \in}{\{{\rm S}, {\rm V}, {\rm T}\}}}\!\!d_\chi \sum_{{\alpha_\chi} = 1}^{n_\chi} \eu^{-y^{2n}\big(\lambda^\chi_{\alpha_\chi} (y^2)\big)^2} - 40\,\eu^{-\mu^6 y^{2n}}\!\bigg]\\
  I^{\rm grav}_{\cal N}(\gamma_k)&=\frac{3}{8\pi^2} \:\Gamma\Big(\frac{2}{n}\Big) \bigg[\Big(1+\frac{1}{\gamma_k}\Big)^{-\frac{4}{n}} + \Big(1-\frac{1}{\gamma_k}\Big)^{-\frac{4}{n}} - 2\bigg]\\
  I^{\rm grav}_{\rm V(\pm)}(\lambda_k,\gamma_k,\mu)&=- \frac{n+3}{8\pi^2}\!\int_0^\infty\!\!\dr y\,y^{3+2n}\bigg[  \sum_{\genfrac{}{}{0pt}{2}{\chi \in}{  \{{\rm S}, {\rm V}, {\rm T}\}}}\sum_{{\alpha_\chi}=1}^{n_\chi}\sum_{i_\chi, j_\chi = 1}^{n_\chi}\eu^{- y^{2n} \big(\lambda^\chi_{\alpha_\chi}(y^2)\big)^2}\\
&\hspace{1.2cm} \big(\lambda^\chi_{\alpha_\chi}(y^2)\big)^2 \big(v^{\chi{\alpha_\chi}}\big)_{i_\chi}\big(\big(H^\chi\big)^{-1}\big)^{i_\chi}_{~j_\chi}(y^2)\big(\check{V}^{\chi(\pm)}_{\rm contr}\big)^{j_\chi}_{~k_\chi}(y^2)\big(v^\chi_{~{\alpha_\chi}}\big)^{k_\chi} \bigg]\\
  I^{\rm gh}_{\rm F}(\mu)&=\frac{1}{8\pi^2}\frac{n+1}{2} \!\int_0^\infty\!\!\dr y\:y^3 \bigg[\!\!\sum_{\genfrac{}{}{0pt}{2}{\chi \in}{\{{{\rm gh}\,{\rm S}}, {{\rm gh}\,{\rm V}}\}}}\!\! d_\chi \sum_{{\alpha_\chi} = 1}^{n_\chi} \eu^{-y^{2n}\big(\lambda^\chi_{\alpha_\chi} (y^2)\big)^2}- 10\,\eu^{-\mu^2 y^{2n}}\!\bigg]\\
  I^{\rm gh}_{\rm V}(\mu)&=-\frac{n+1}{8\pi^2}\!\int_0^\infty\!\!\dr y\,y^{3+2n}\bigg[\sum_{\genfrac{}{}{0pt}{2}{\chi \in}{\{{{\rm gh}\,{\rm S}}, {{\rm gh}\,{\rm V}}\}}}\sum_{{\alpha_\chi}=1}^{n_\chi}\sum_{i_\chi, j_\chi = 1}^{n_\chi}\eu^{- y^{2n} \big(\lambda^\chi_{\alpha_\chi}(y^2)\big)^2 }\\ &\hspace{4.8cm}\lambda^\chi_{\alpha_\chi} (y^2) \big(v^{\chi{\alpha_\chi}}\big)_{i_\chi}\big(\check{V}^\chi_{\rm contr}\big)^{i_\chi}_{~j_\chi}(y^2)\big(v^\chi_{~{\alpha_\chi}}\big)^{j_\chi} \bigg]
 \end{aligned}
\end{equation}
For a sharp proper-time cutoff, these coefficient functions replace those given in eq. \eqref{SmoothCutoffCoeffients} of the main text which apply to the $C^m_k$ and $f^m_k$ scheme, respectively.

\end{document}